\documentclass[12pt,a4paper]{extreport}
\usepackage{fullpage}
\usepackage{amssymb}
\usepackage[T1]{fontenc}
\usepackage[bitstream-charter]{mathdesign}
\usepackage{commath}
\usepackage{abstract}
\usepackage{graphicx}
\usepackage{color}
\usepackage{float}
\usepackage{subfig}
\usepackage{url}
\usepackage{xcolor}
\usepackage{rotating}
\usepackage{pdflscape}
\usepackage{pdfpages}
\usepackage[symbol]{footmisc}
\usepackage[font={small,it}]{caption}
\usepackage{tikz}
\usepackage{units}
\usepackage{amssymb}
\usepackage{multirow}
\usepackage{amsmath}
\usepackage{float}
\usepackage{mathtools, nccmath, textcomp}
\usepackage[numbers,sort&compress]{natbib}
\usetikzlibrary{calc}
\usepackage{filecontents,notoccite}
\usepackage{fancyhdr,layout}
\usepackage[titletoc]{appendix}
\usepackage[version=3]{mhchem}
\usepackage[refpage]{nomencl}
\usepackage[pageanchor]{hyperref}
\def\pagedeclaration#1{, \dotfill\hyperlink{page.#1}{#1}}
\makenomenclature
\usepackage{etoolbox}
\renewcommand\nomgroup[1]{%
  \item[\bfseries
  \ifstrequal{#1}{P}{Physics Constants}{%
  \ifstrequal{#1}{N}{Number Sets}{%
  \ifstrequal{#1}{O}{Other Symbols}{}}}%
]}

\renewcommand{\nompreamble}{The list describes several symbols that are used within the document.}

\usepackage{afterpage}

\newcommand\blankpage{
    \null
    \thispagestyle{empty}
    \addtocounter{page}{-1}
    \newpage
    }

\newcommand\blankpagewithoutnumberskip{
    \null
    \thispagestyle{empty}
    \addtocounter{page}{0}
    \newpage
    }

\begin{document}
\sloppy
\begin{titlepage}

\begin{tikzpicture}[overlay,remember picture]
\draw [line width=1.0pt,rounded corners=10pt,]
    ($ (current page.north west) + (1cm,-1cm) $)
    rectangle
    ($ (current page.south east) + (-1cm,1cm) $);       
\end{tikzpicture}
\begin{figure}
\includegraphics[scale=0.08]{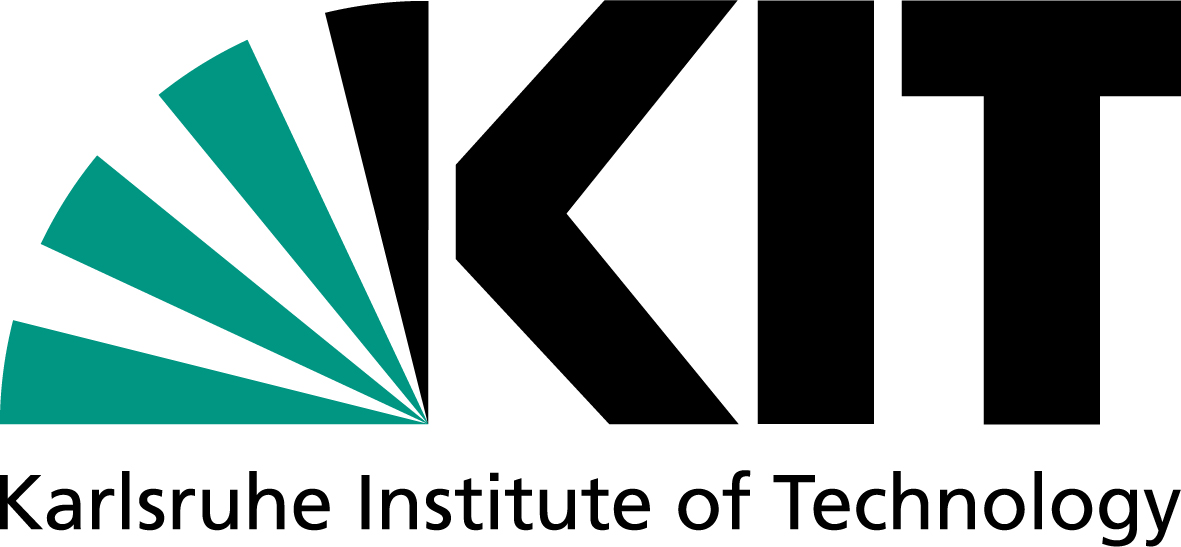}
\end{figure}\\
\Large\centering\textbf{Phase-field modeling on the diffusion-driven processes in metallic conductors and lithium-ion batteries}

\vspace*{2cm}
\large\centering{Zur Erlangung des akademischen Grades \\ \textbf{Doktor der Ingenieurwissenschaften} \\ von der KIT-Fakult{\"a}t f{\"u}r Maschinenbau des\\ Karlsruher Instituts f{\"u}r Technologie (KIT)}

\vspace*{1.5cm}
\large\centering{genehmigte \\ \textbf{Dissertation} \\ von}

\vspace*{1.5cm}
\large\centering\textbf{M. Tech. Jay Santoki}\\
\large\centering{geboren am 20.06.1992}\\
\large\centering{in Jam-Jodhpur (Indien)}\\

\vspace*{3cm}
\begin{flushleft}
 \large{Tag der m{\"u}ndlichen Pr{\"u}fung: } \hspace{1.25cm} 10.09.2020\\
 \large{Hauptreferent: \hspace{4.5cm} Prof. Dr. rer. nat. Britta Nestler}\\
 \large{Korreferent: \hspace{5.05cm} Prof. Dr.-Ing. habil. Marc Kamlah}
\end{flushleft}

\end{titlepage}

\afterpage{\blankpage}

\pagestyle{empty}

\setcounter{page}{1}
\setcounter{secnumdepth}{4} 
\setcounter{tocdepth}{4}

\pagenumbering{roman}
\cleardoublepage
\phantomsection


\pagestyle{fancy}
\fancyhf{}
\lhead[\thepage]{Abstract}      
\rhead[\thesection]{\thepage}
\cleardoublepage\phantomsection\addcontentsline{toc}{chapter}{Abstract}


\chapter*{Abstract}
Diffusion-driven processes are important phenomena of materials science in the field of energy conversion and transmission. During the conversion from chemical energy to electrical energy, the species diffusion is generally linked to the rate of exchange, and hence to the performance of the conversion device. Alternatively, the transmission of the electric field diffuses the species when it passes through any medium. The consequences of this effect can be regulated to attune surface nano-patterns. Otherwise, uncontrolled morphologies may lead to permanent degradation of the metallic conductors. Therefore, the understanding of the material behavior, in the presence of the driving forces of the diffusional species, is of scientific interest. The presented dissertation proposes to investigate one example of species diffusion in each case, during energy conversion and transmission. Specifically, the objective of the study is to explore the lithium insertion into the cathode electrode of lithium-ion batteries and the morphological evolution of inclusions, while propagating under the electromigration in the metallic conductors.

 During insertion, lithium manganese oxide spinel, a cathode electrode material of lithium-ion batteries, shows a coexistence of Li-rich and Li-poor phases. For an enhanced understanding of the mechanism of a two-phase coexistence, a mathematical model of phase separation  is derived, which is based on the Cahn-Hilliard equation. To begin with, the geometrical shape polydispersity of an isolated particle is considered to investigate the mesoscopic effect of the surface curvature. The simulation results show that the onset of the phase separation preferentially occurs in high-curvature regions of the particle. Furthermore, the elliptical particle with a higher aspect ratio is subjected to the onset of the phase separation, prior to the particles with a lower aspect ratio. Finally, the effect of the variation of the parameters on the charge dynamics is discussed. The study is further extended to multiple particle systems, so as to understand the influence of various microstructural descriptors, such as particle size, porosity, and tortuosity, on the transport mechanism. A linear dependence of the transportation rate is observed with tortuosity. The slope of this linear relation is independent of the particle size, but shows some interdependency with porosity. Furthermore, the presented results suggest that systems consisting of smaller particles are closely follow the surface reaction limited theory while larger particles tend towards the bulk-transport limited theory derived for planar electrodes. In order to identify the promising hierarchically structured electrodes, the presented simulation results could be utilized to optimize the experimental efforts.

The electromigration-induced morphological evolution of inclusions (voids, precipitates, and islands) has recently been scrutinized in terms of the efficient design of the interconnects and surface nanopatterns. To understand the morphological evolutions, a phase-field model is derived to account for the inclusions migrating under the external electric field. Initially, the insights gained from the numerical results of the isotropic inclusions corroborate the findings from the linear stability analysis. Additionally, the numerical results can elegantly elucidate the transition of a circular inclusion to a finger-like slit. The subsequent drift of the slit is characterized by shape invariance, along with a steady-state slit width and velocity,  which scale with the applied electric field as $E_{\infty}^{-1/2}$ and $E_{\infty}^{3/2}$, respectively. The results obtained from the phase-field simulations are critically  compared with the sharp-interface solution. The repercussions of the study, regarding the prediction of void migration in flip-chip Sn-Ag-Cu solder bumps and the fabrication of channels with desired micro/nano dimensions, are discussed. The study is further extended to anisotropic inclusions migrating in $\{110\}$, $\{100\}$, and $\{111\}$ crystallographic planes of face-centered-cubic crystals. Based on the numerical results, morphological maps are constructed in the plane of the misorientation angle and, the conductivity contrast between the inclusion and the matrix. The simulations predict a rich variety of morphologies, which includes steady-state and time-periodic morphologies, as well as zigzag oscillations and an inclusion breakup. Furthermore, the influence of the variation in conductivity contrast and the misorientation is observed to be influential on the morphological evolution of the time-periodic oscillations, steady-state shapes, and the way inclusions break apart. Finally, the numerical results of the steady-state dynamics, obtained for anisotropic inclusions, are critically compared with isotropic analytical and numerical results.

The presented dissertation demonstrates that the phase-field methods are able to elegantly capture the essential physics of the diffusion-driven phenomena discussed above.   

\clearpage

\pagestyle{fancy}
\fancyhf{}
\lhead[\thepage]{Kurzfassung}      
\rhead[\thesection]{\thepage}
\cleardoublepage\phantomsection\addcontentsline{toc}{chapter}{Kurzfassung}


\chapter*{Kurzfassung}
Diffusionsgetriebene Prozesse sind wichtige Ph{\"a}nomene der Materialwissenschaft im Bereich der Energieumwandlung und -{\"u}bertragung. W{\"a}hrend der Umwandlung von chemischer Energie in elektrische Energie ist die Speziesdiffusion im Allgemeinen mit der Austauschrate und folglich mit der Leistung der Umwandlungsvorrichtung verbunden. Alternativ diffundiert die {\"U}bertragung des elektrischen Feldes durch die Spezies, wenn sie durch irgendein Medium verl{\"a}uft. Die Konsequenzen dieses Effekts k{\"o}nnen reguliert werden, um Oberfl{\"a}chen-Nanomuster abzustimmen.
Andernfalls k{\"o}nnen die unkontrollierten Morphologien zu einer dauerhaften Verschlechterung der metallischen Leiter f{\"u}hren. Daher ist das Verst{\"a}ndnis des materiellen Verhaltens bei Vorhandensein der treibenden Kr{\"a}fte von Diffusionsspezies von wissenschaftlichem Interesse. Die vorgestellte Dissertation schl{\"a}gt eine Untersuchung von jeweils einem Beispiel der Speziesdiffusion w{\"a}hrend der Energieumwandlung und -{\"u}bertragung vor. Ziel der Studie ist es insbesondere, sowohl die Lithiumeinf{\"u}gung von Lithium-Ionen-Batterien in die Kathodenelektrode als auch die morphologische Entwicklung von Einschl{\"u}ssen zu untersuchen, w{\"a}hrend sie sich unter der Elektromigration in den metallischen Leitern ausbreiten.

Lithium-Manganoxid-Spinell, ein Kathodenelektrodenmaterial von Lithium-Ionen-Batterien, zeigt w{\"a}hrend des Einf{\"u}gens eine Koexistenz von Li-reichen und Li-armen Phasen. F{\"u}r ein besseres Verst{\"a}ndnis des Mechanismus einer zweiphasigen Koexistenz wird ein mathematisches Modell der Phasentrennung abgeleitet, das auf der Cahn-Hilliard-Gleichung basiert. Zun{\"a}chst wird die geometrische Formpolydispersit{\"a}t eines isolierten Partikels betrachtet, um den mesoskopischen Effekt der Oberfl{\"a}chenkr{\"u}mmung zu untersuchen.
Die Simulationsergebnisse zeigen, dass der Beginn der Phasentrennung bevorzugt in Bereichen auftritt, in denen das Partikel eine starke Kr{\"u}mmung aufweist.
Weiterhin wird das elliptische Teilchen mit einem h{\"o}heren Querschnittsverh{\"a}ltnis dem Einsetzen der Phasentrennung vor den Teilchen mit einem niedrigeren Querschnittsverh{\"a}ltnis ausgesetzt. Abschlie{\ss}end wird der Einfluss der Variation der Parameter auf die Ladungsdynamik diskutiert. Die Studie wird weiter auf mehrere Partikelsysteme ausgedehnt, um den Einfluss verschiedener mikrostruktureller Deskriptoren wie Partikelgr{\"o}{\ss}e, Porosit{\"a}t und Tortuosit{\"a}t auf den Transportmechanismus zu verstehen. Bei Tortuosit{\"a}t wird eine lineare Abh{\"a}ngigkeit der Transportrate beobachtet. Die Steigung dieser linearen Beziehung ist unabh{\"a}ngig von der Partikelgr{\"o}{\ss}e, zeigt jedoch eine gewisse Abh{\"a}ngigkeit von der Porosit{\"a}t.
Dar{\"u}ber hinaus legen die vorgestellten Ergebnisse nahe, dass Systeme, die aus kleineren Partikeln bestehen, der durch Oberfl{\"a}chenreaktionen begrenzten Theorie genau folgen, w{\"a}hrend gr{\"o}{\ss}ere Partikel zu der durch Massentransporte begrenzten Theorie tendieren, die f{\"u}r planare Elektroden abgeleitet wurde. Um die hierarchisch strukturierten Elektroden zu identifizieren, k{\"o}nnten die vorgestellten Simulationsergebnisse verwendet werden, um den experimentellen Aufwand zu optimieren.

Die durch Elektromigration induzierte morphologische Entwicklung von Einschl{\"u}ssen (Hohlr{\"a}ume, Ausf{\"a}llungen und Inseln) wurde k{\"u}rzlich im Hinblick auf die effiziente Auslegung der Verbindungen und Oberfl{\"a}chen-Nanomuster untersucht. Um die morphologischen Entwicklungen zu verstehen, wird ein Phasenfeldmodell abgeleitet, um Einschl{\"u}sse zu ber{\"u}cksichtigen, die unter dem externen elektrischen Feld wandern. Die Erkenntnisse aus den numerischen Ergebnissen zu isotropen Einschl{\"u}ssen best{\"a}tigen zun{\"a}chst die Ergebnisse der linearen Stabilit{\"a}tsanalyse. Zus{\"a}tzlich k{\"o}nnen die numerischen Ergebnisse den {\"U}bergang eines kreisf{\"o}rmigen Einschlusses zu einem fingerartigen Schlitz elegant erl{\"a}utern. Die nachfolgende Drift des Schlitzes ist durch eine Forminvarianz zusammen mit einer station{\"a}ren Schlitzbreite und -geschwindigkeit gekennzeichnet, die mit dem angelegten elektrischen Feld jeweils als $E_{\infty}^{-1/2}$ und $E_{\infty}^{3/2}$ skaliert werden. Die Ergebnisse aus Phasenfeldsimulationen werden kritisch mit der L{\"o}sung mit scharfen Grenzfl{\"a}chen verglichen. Die Auswirkungen der Studie auf die Vorhersage einer Hohlraumwanderung in Flip-Chip-Sn-Ag-Cu-L{\"o}tperlen und die Herstellung von Kan{\"a}len mit gew{\"u}nschten Mikro- / Nanodimensionen werden diskutiert. Die Studie wird weiter auf anisotrope Einschl{\"u}sse ausgedehnt, die in $\{110\}$, $\{100\}$ und $\{111\}$ kristallografischen Ebenen von fl{\"a}chenzentrierten kubischen Kristallen wandern. Basierend auf numerischen Ergebnissen werden morphologische Karten in der Ebene des Fehlorientierungswinkels und des Leitf{\"a}higkeitskontrasts zwischen dem Einschluss und der Matrix erstellt. Die Simulationen sagen eine Vielzahl von Morphologien voraus, darunter station{\"a}re und zeitperiodische Morphologien sowie Zick-Zack-Oszillationen und eine Einschlussaufl{\"o}sung. Dar{\"u}ber hinaus wird beobachtet, dass der Einfluss der Variation des Leitf{\"a}higkeitskontrasts und der Fehlorientierung Einfluss auf die morphologische Entwicklung der zeitperiodischen Schwingungen, der station{\"a}ren Formen und der Art und Weise hat, wie Einschl{\"u}sse auseinander brechen.
Schlie{\ss}lich werden die numerischen Ergebnisse der station{\"a}ren Dynamik, die f{\"u}r anisotrope Einschl{\"u}sse erzielt wurden, kritisch mit isotropen analytischen und numerischen Ergebnissen verglichen.

Die vorgestellte Dissertation zeigt, dass die Phasenfeldmethoden die wesentliche Physik der oben diskutierten diffusionsgetriebenen Ph{\"a}nomene elegant erfassen k{\"o}nnen.   


\pagestyle{fancy}
\fancyhf{}
\lhead[\thepage]{Preface}      
\rhead[\thesection]{\thepage}
\pagestyle{fancy}
\fancyhf{}
\lhead[\thepage]{No objection}      
\rhead[\thesection]{\thepage}

\pagestyle{fancy}
\fancyhf{}
\lhead[\thepage]{Declaration}      
\rhead[\thesection]{\thepage}

\pagestyle{fancy}
\fancyhf{}
\lhead[\thepage]{Acknowledgements}      
\rhead[\thesection]{\thepage}
\cleardoublepage\phantomsection\addcontentsline{toc}{chapter}{Acknowledgments}
\chapter*{Acknowledgements}

This thesis is possible due to the encouragement and support of countless people. Here, I take the opportunity to thank a few that have made this thesis possible with a plethora of memories to remember for life.

I begin my deepest gratitude to Prof. Dr. Britta Nestler for providing an opportunity to be a part of her research group. During technical discussions, her helpful insights are a source of inspiration, which keeps me learning and growing during this journey. The freedom provided by her, during the research work has garnered interest in me to investigate truly exceptional phenomena in materials science and ultimately in search of a better world. I am extremely grateful for her constructive comments, quality discussions, and support. I would like to extend my gratitude to Prof. Dr. Marc Kamlah to be the second referee of my thesis committee. His intriguing comments on my work have not only improved the quality of the research, but also helped me to develop scientific reasoning and consequently, to become an independent researcher.

I have had the good fortune to find some truly exceptional collaborators. Special thanks to Dr. Daniel Schneider for his continues feedback and numerous discussions. The interactions with Dr. Arnab Mukherjee has allowed me to observe the multi-facets of the phase-field modeling from close corners. I also acknowledge very thought-provoking discussions with Dr. Fei Wang, Dr. Michael Selzer, Dr. Oleg Tschukin, Prof. Leslie Mushongera, and Dr. Sebastian Schulz during the initial stages of PhD. I am extremely indebted to my colleagues and collaborators, Walter Werner, Andreas Reiter, Sumanth Nani, Dr. Prince Gideon, Paul Hoffrogge, Yinghan Zhao, Nishant Prajapati, Simon Daubner, Dr. Tao Zhang, Christoph Herrmann, Dr. Ephraim Schoof, Dr. Felix Schwab, Dr. Ramanathan Perumal, Dr. Amol Subhedar, Nikhil Kulkarni, Mehwish Huma Nasir, and Fridolin Haugg for many a technical discussions from which I have gained immensely. Furthermore, I am greatly obliged to all the group members for providing a thoughtful ambiance, which has inspired me to strive for continuous improvement.

I am grateful to the technical staff in the group, Christof Ratz and Halil Bayram for their support in both hardware and software related issues which helped me go about my work smoothly. A special thanks to Leon Geisen for his editorial assistance and the German translation of the abstract. I also express gratitude to the secretariat at IDM of the Hochschule Karlsruhe and IAM-CMS of the Karlsruhe Institute Technology, especially, Ms. Claudia Hertweck-Maurer, Ms. Stephanie Mueller and Ms. Inken Heise for all their help with the administrative work.

It would not have been possible to endure the rough patches alone without the support, love, and empathy of my family and friends. Thank you for being there for me every time I needed someone to count on. In addition, I request a sincere apology for not being able to spend as much time as you would have wanted in the last few years. I dedicate my thesis to my family and friends, who are the saviors of my life.

Finally, I acknowledge the financial supports during different stages of my PhD. The first phase of the research was funded by the cooperative graduate school "Gefuegestrukturanalyse und Prozessbewertung" of the Ministry of the State of Baden-Wuerttemberg and partially through the initiative "Mittelbau". The second phase of my research was funded by the German Research Foundation (DFG) under Project ID 390874152 (POLiS Cluster of Excellence).

\hypersetup{
    colorlinks=true, 
    linktoc=all,     
   linkcolor=blue,  
}

\pagestyle{fancy}
\fancyhf{}
\lhead[\thepage]{Contents}      
\rhead[\thesection]{\thepage}
\tableofcontents
\clearpage

\afterpage{\blankpagewithoutnumberskip}

\pagenumbering{arabic}
\setcounter{page}{1}
\pagestyle{fancy}
\fancyhf{}
\lhead[\thepage]{Chapter \thechapter .}      
\rhead[\thesection]{\thepage}

\newpage
\thispagestyle{empty}
\vspace*{8cm}
\phantomsection\addcontentsline{toc}{chapter}{I Introduction and Literature review}
\begin{center}
 \Huge \textbf{Part I} \\
 \Huge \textbf{Introduction and literature review}
\end{center}

\chapter{Introduction and literature review}

\begin{figure}
\centering
\includegraphics[scale=0.48]{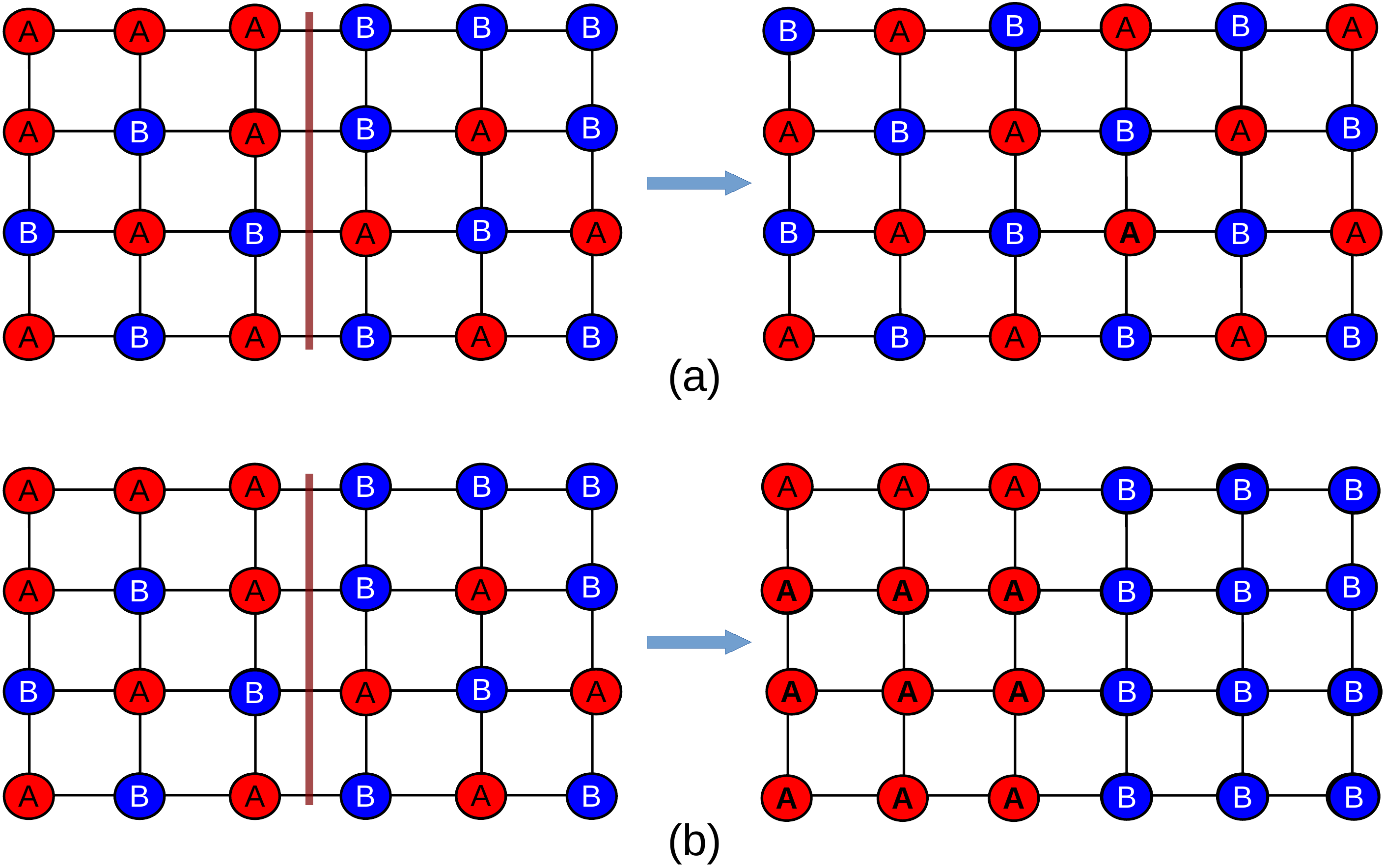}
\caption[Schematics of diffusion processes: In the direction of and against the concentration gradients.]{Schematics of diffusion processes: (a) in the direction of {the} concentration gradient and (b) against the concentration gradient.}\label{fig:IntroDiffusionSchematics}
\end{figure}

{During their operations, many} material systems encounter {a} movement of atoms, molecules or charge carriers. In response to the established gradient in {the} concentration of these species or the external driving force for diffusion, the movement in different materials is specific. Hence, the diffusion coefficient or {the} diffusivity is defined to reflect this material property. The diffusivity is generally measured for a given pair of species and pairwise for a multi-species system. For instance, carbon dioxide in the air has a diffusion coefficient of 16 mm$^2$/s, and in water, its diffusion coefficient is 0.0016 mm$^2$/s \cite{cadogan2014diffusion}. The higher the diffusivity of one species{,} with respect to another, the faster they diffuse into each other. These are the examples of diffusion{,} in order to minimize the presence of concentration gradients, as shown in Figure~\ref{fig:IntroDiffusionSchematics}(a). However, it is also possible that the diffusion in some cases can take place against the concentration gradients, where the region of high concentration accumulates further species from the lower one to grow even higher \cite{krishna2015uphill}, as shown in Figure~\ref{fig:IntroDiffusionSchematics}(b). As an example, consider a hot mixture of water and {phenol}. At high temperatures, the {phenol} and the water may mix to form a single thermodynamic phase. However, when the mixture is cooled, the clusters of the {phenol}-rich phase and the water-rich phase are expected. Hence, this necessitates rigorous investigations to understand diffusion in materials. 

Diffusion plays an important role in many processes of materials science{,} including its benefits as well as adverse nature. For instance, it can be applied as a process of diffusion bonding{,} in order to weld or join metals \cite{derby1982theoretical}{,} and sintering to shape materials in the powder metallurgy \cite{coble1961sintering}.
On the other hand, the diffusion can be critical in {processes} such as dimensional changes{,} due to {a} pore formation in the grain boundaries \cite{balluffi1955effect}. In addition, the diffusion {in adjoining layers} is prevented by {a} chemical isolation in integrated circuits{, so as} to avoid {a} transgranular movement of {the} species{,} while maintaining electrical connections with {the} help of barrier metal films \cite{holloway1990tantalum}. Therefore, it is important to understand {the} material behavior under the external driving forces for {the} species diffusion.

There are several external driving forces for the species diffusion. For instance, the appearance of a species flux{,} as a consequence of a thermal gradient, the so-called Soret effect{,} is important{,} where the large temperature gradients  are present such as {in} nuclear reactors \cite{rahman2014thermodiffusion}. In addition, the diffusion can be induced by centrifugal forces \cite{mashimo1988self}, gravity \cite{mashimo2008gravity}, hydrostatic pressure and mechanical stresses \cite{aziz1998pressure}. However, diffusion associated with the electric field is the most provocative field of research in many communities \cite{mehrer2007diffusion}. 

During {the} energy conversion from the chemical energy to the electrical energy, various phenomena can be observed{,} due to the species diffusion. {Gas diffusion layers, for instance,} are the key components to the various types of fuel cells, which essentially {act} as an electrode that facilitates {the} diffusion of reactants across the active area of {the} membrane \cite{cindrella2009gas}. In addition, diffusion-controlled electrodeposition \cite{lupo2019modeling} leads to {a} dendrite formation at the electrode surfaces. Furthermore, the diffusion is reported to be prominent in the battery systems \cite{panchmatia2014lithium,tian2015insight}.

In addition to energy conversion, the electric field diffuses the species when it passes through any medium. In several applications, this effect leads to significant consequences. For instance, the passage of {an} intense electric current in thin-film metallic conductors, known as electromigration, may cause open circuits \cite{sanchez1992slit}{,} due to the voids and short circuits \cite{lienig2003current, fantini1998electromigration}{, caused by the formation of hillocks}. On {the} contrary, the electric field can be helpful to determine the chloride diffusivity in concrete and the depth of {the} penetration in {cement pastes} \cite{luping1993rapid}. Furthermore, the microstructures in polymers can be monitored by the electric field to obtain a lamellar pattern \cite{amundson1994alignment}.

The presented dissertation proposes to investigate one example of each aspect of the species diffusion, {namely} during {the} energy conversion and during {the} transmission of the electric field in a medium. Specifically, the objective of the study is to explore the lithium insertion into the cathode particles of lithium-ion batteries and the morphological evolution of inclusions under the electromigration in metallic conductors. {In the following paragraphs, a} literature review on both topics is performed separately, which is necessary to establish a sound platform for the methods and results presented in the later chapters.

\section{Transport mechanism in lithium-ion batteries}
\label{section:twoPhaseCoexistence}

\begin{figure}[h]
\includegraphics[scale=0.4]{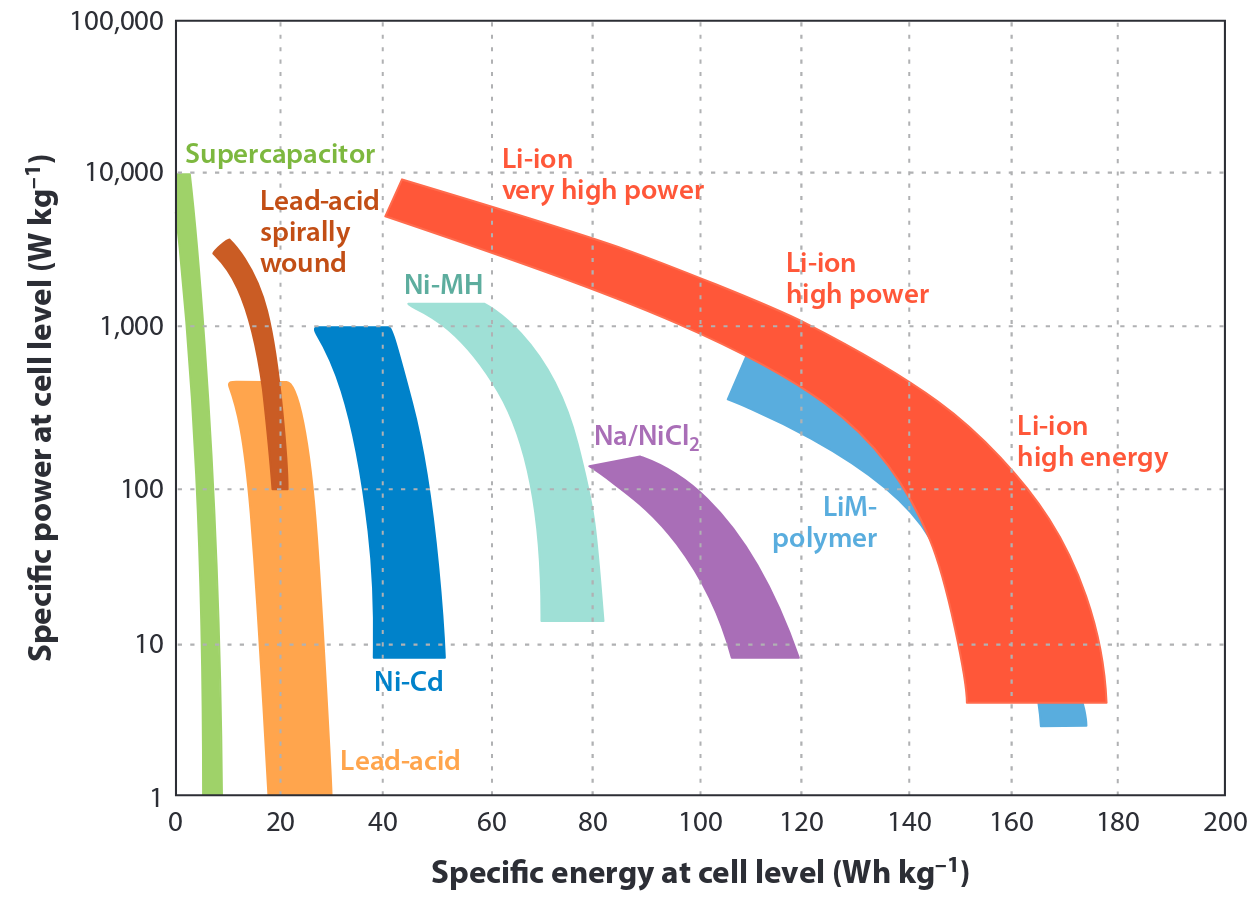}
\caption[Comparison of the specific energy and specific power of common rechargeable batteries]{Comparison of the specific energy and specific power of common rechargeable batteries{,} along with Li-ion batteries. Reprinted with permission from Ref. \cite{deforest2012advances}.} \label{fig:IntroBatteryComparison}
\end{figure}

{The m}iniaturization of portable electronic devices, {the} storage systems for renewable energy sources, and {achieving light weight} of the electrical vehicles require high specific energy, high specific power, and {a} low cost of the energy storage systems. One of the probable solutions for such future needs{,} in the form of electrochemical energy storage devices{,} is the battery {system} \cite{kordesch1981electrochemical}. Batteries are classified into two categories{,} based on its reversibility, the primary batteries and the secondary batteries. Primary batteries are non-rechargeable, while secondary batteries are rechargeable. A rechargeable battery can be used over many cycles throughout its lifespan. The increment in the number of possible charge/discharge cycles before performance degradation is a major concern{, with reference to the improvement of} battery life. Figure~\ref{fig:IntroBatteryComparison} shows a comparison of different rechargeable battery systems{,} with their typical operational range of specific power and specific energy.

{Due to their versatile applications with higher specific power and specific energy, lithium}-ion batteries are one of the promising candidates of rechargeable batteries \cite{deforest2012advances}. The lithium-ion battery consists of intercalated compounds, compared to the metallic lithium used in a non-rechargeable lithium battery \cite{dey1977lithium}. The three primary functional components of a lithium-ion battery are the positive and negative electrodes and electrolytes. Generally, the negative electrode (anode) of a conventional lithium-ion battery is made of graphite, silicon or carbon-based compounds \cite{hayner2012materials}. The positive electrode (cathode) is of a metal oxide, and the electrolyte is of a lithium salt in an organic solvent \cite{silberberg2014chemistry}. The lithium ions are stored as guest species in the crystal lattice of the electrode materials. The electrodes in lithium-ion batteries are a mixture of lithium-ion{-}conducting and electron-conducting particles which form a porous composition to sustain the transport of electrons {and} lithium ions to and from the active sites. 

\begin{figure}
\centering
\includegraphics[scale=0.2]{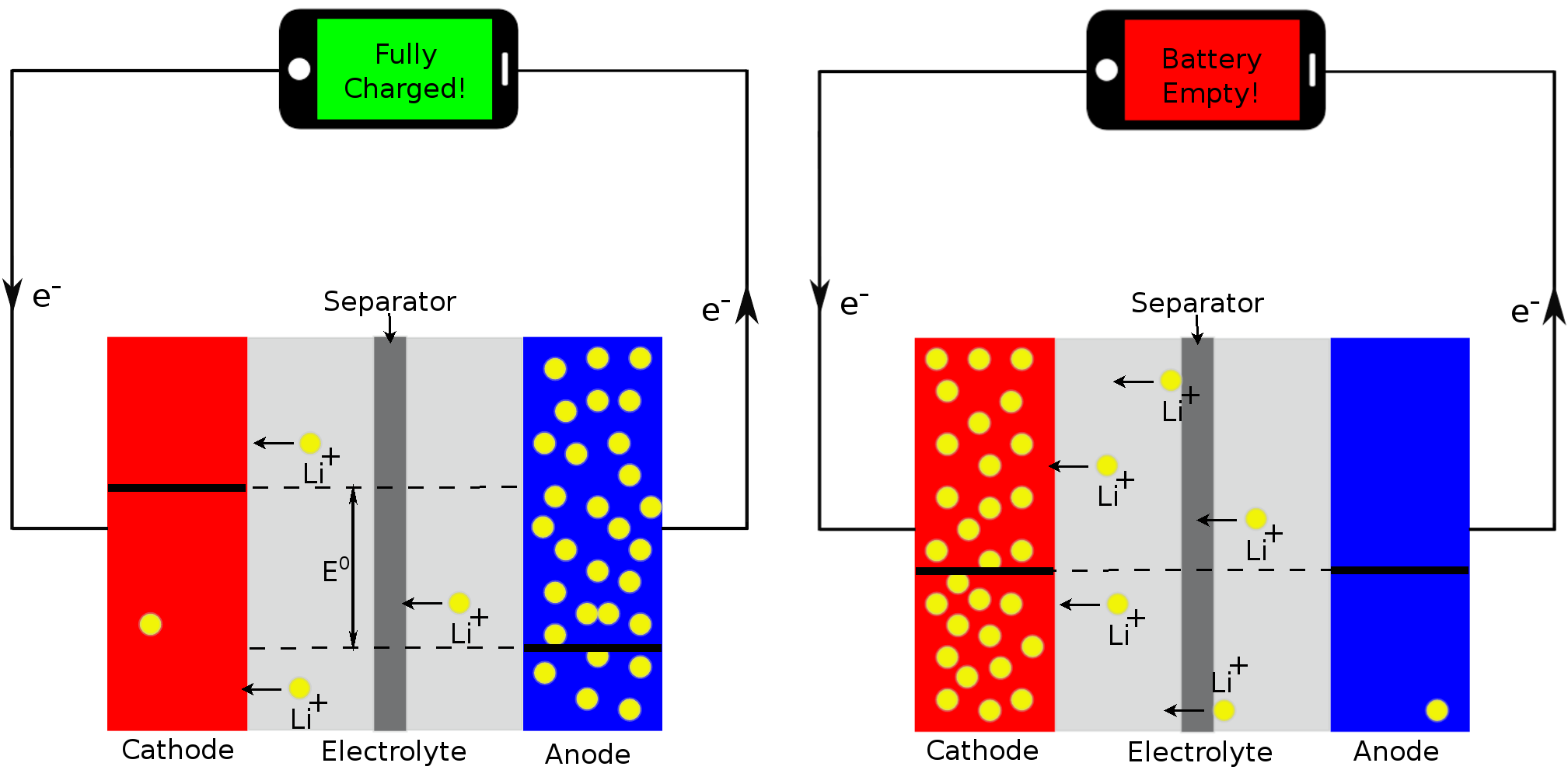}
\caption[Schematic diagram of a lithium-ion battery unit]{Schematic diagram of a lithium-ion battery unit{,} while discharging, attached to an external device, such as a mobile phone. In the fully charged state, the battery reaches its electrochemical voltage threshold ($E^0_{\textrm{max}}$) and the battery is ready for the application. During {the discharging process}, lithium-ions deport from the lithiated anode to the delithiated cathode{,} through the electrolyte. When the potential difference between the anode and cathode vanishes, the ionic transfer stagnates. Then, the battery is discharged and {requires an} external supply {to be charged again}.}\label{fig:IntroSchematicBattery}
\end{figure}

 During the discharging (or charging) process, the battery delivers (or consumes) a current spontaneously (or deliberately{)} from the external power supply{,} due to {a} electrochemical voltage difference between the anode and the cathode, 
\begin{equation}
E^0 = E^A -E^C,
\end{equation} 
where $E^0$ is the electrochemical voltage {of the open circuit and} $E^A$ and $E^C$ are the {respective} anode and cathode voltages. This delivered current is the transport of electrons from the anode to the cathode{,} through an external circuit, while lithium ions transport from the anode to the cathode inside the battery (Figure~\ref{fig:IntroSchematicBattery}). During discharge, lithium ions are transferred from the anode to the cathode{,} through the electrolyte. At the anode side:
\begin{center}
\ce{Li
<=>[\ce{Discharge}][\ce{Charge}]
${\ce{Li^+ +e^- }}${.}
}
\end{center}
The negatively charged electrons minimize their energy{,} when they {are transferred from the} anode to the cathode. At the cathode side:
\begin{center}
\ce{Li^+ +e^-
<=>[\ce{Discharge}][\ce{Charge}]
${\ce{Li}}${.}
}
\end{center}
The lithium-ion inflow continues until the potential difference between the anode and {the} cathode vanishes. Due to {the} reversibility, the battery can be charged up again through the external power supply.

However, the battery capacity decays upon electrochemical cycling{,} as a result of several non-reversible losses \cite{deshpande2012battery}. {In addition,} other aspects to improve the battery performance {with an improved battery life \cite{vetter2005ageing},} such as the energy density, {the stability of the operating temperature, the low self-discharge, the safety, the charging time, the output power, and the cost of the technology \cite{tarascon2001issues, kim2008spinel},} are of scientific interest to the increasing demand for high-performance lithium-ion batteries. The provision of high performance requires an understanding of the fundamental physical and chemical processes at various levels of battery operational and non-operational conditions. One such process is phase separation, which is the focus of the presented work.

\subsection{Phase separation mechanism}
Most of the popular anode (graphite) and cathode materials (one of three \cite{thackeray2000science}: a layered oxide such as lithium cobalt oxide, a polyanion such as lithium iron phosphate or a spinel such as lithium manganese oxide) show phase-separated states during lithium intercalation. The selection of {the} particular battery components, including the cathode material, the anode material, and the electrolyte, depends on various characteristics associated with a specific range, including the discharge rate, the battery cycle time, the battery weight, the battery life, the battery cost, and the operating temperature \cite{whittingham2004lithium, liu2010advanced}. {The t}ypical characteristics of few of the electrode materials are highlighted in the following paragraphs.

\begin{figure}
\centering
\includegraphics[scale=0.7]{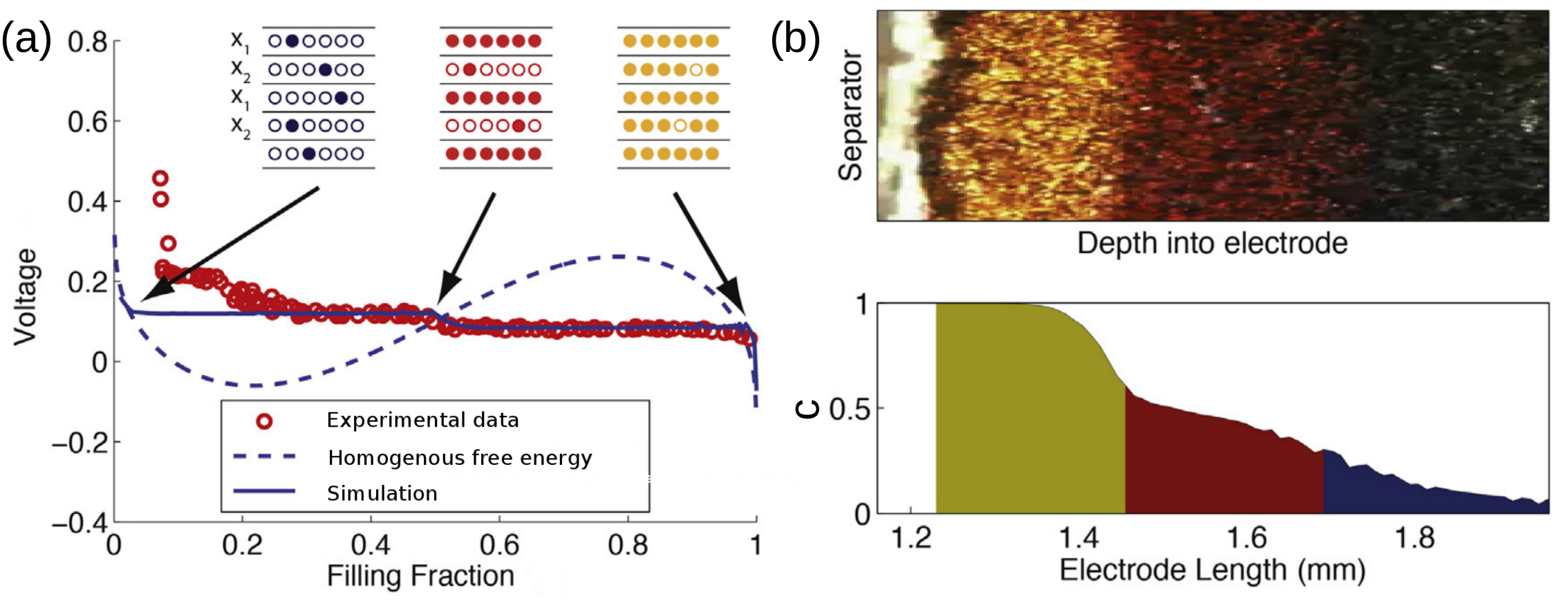}
\caption[{Multiple phase} coexistence in graphite materials.]{{Multiple phase} coexistence in graphite materials. (a) {O}pen{-}circuit voltage for {a} lithium insertion in{to} graphite{. The} experimental data from Ref. \cite{ohzuku1993formation} confirm multiple voltage plateaus, along with {the} results from {the} simulations (solid curve){,} and {the} voltage corresponds to {the} homogeneous free energy density {(dashed curve)}. (b) {Experimental image (above), along with simulation prediction (below)} \cite{ferguson2014phase}. {In the experiment and the simulation,} three distinct co-existent phases are observed at the same temporal position. Reprinted with permission from Ref. \cite{ferguson2014phase}.}\label{fig:IntroGraphite}
\end{figure}

Graphite exhibits multiple voltage plateaus during (de-)intercalation \cite{ohzuku1993formation}, as shown in Figure~\ref{fig:IntroGraphite}(a). It has complex thermodynamics \cite{ohzuku1990formation} and at least three clear stages with experimentally observed fronts of color-changing phase transformations, {which can be seen in the above inset of} Figure~\ref{fig:IntroGraphite}(b). In addition, graphite can form {multilayered structures} with different energies. In fact, each layer of the graphite may show {a} distinct phase separation behavior \cite{smith2017intercalation}. This complicates the mechanism of {the} phase separation even further{, when considered that a multilayered material and some complex morphologies, such as a checkerboard pattern \cite{smith2017intercalation},} can be observed. Therefore, this {multiple phase} coexistence is a unique characteristic of anode materials, in particular graphite.

\begin{figure}
\centering
\includegraphics[scale=0.60]{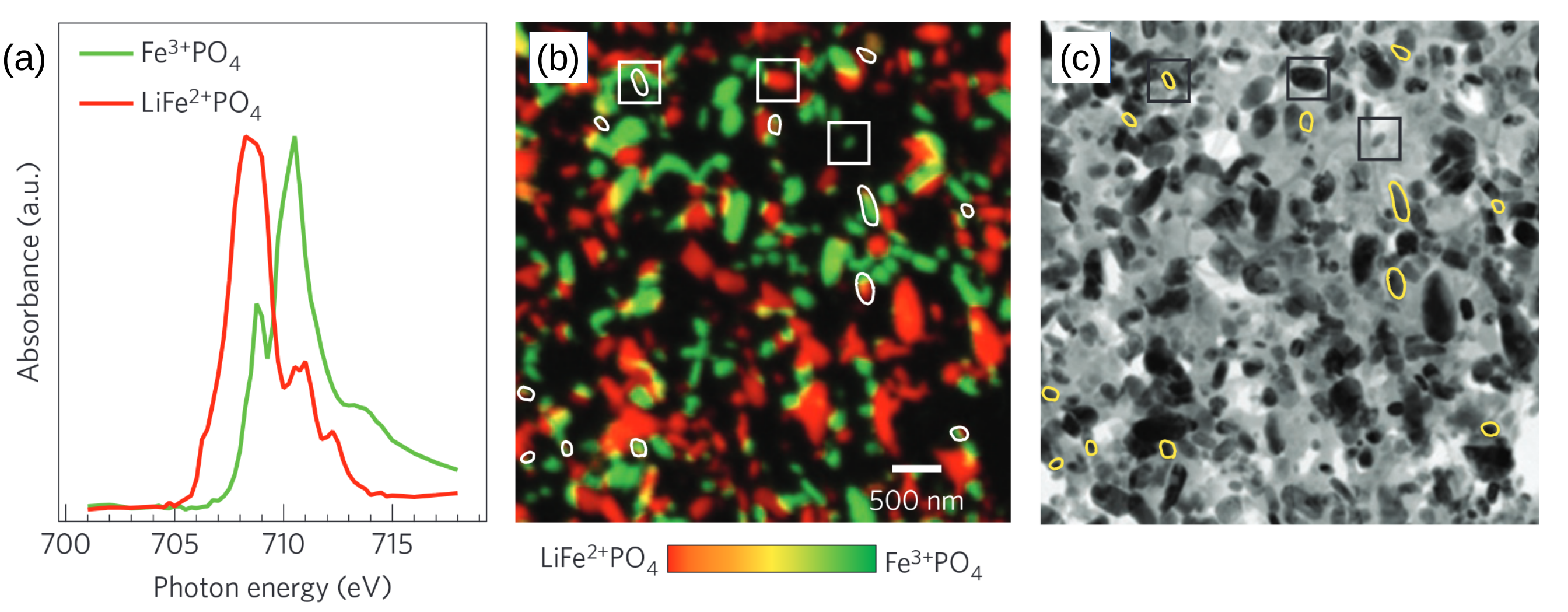}
\caption[Two-phase coexistence in lithium-iron-phosphate particles of {the} porous electrode]{Two-phase coexistence in lithium-iron-phosphate particles of the porous electrode. X-ray absorption spectra of {fully lithiated} and delithiated particles of LFP {are shown} in (a), which is considered for a reference{, so as} to identify these regions. The state-of-charge map {is shown} in (b), which is produced by fitting a linear combination of the spectrum to every {single pixel}. The red color represents the lithiated fraction, while the green color indicates the delithiated region. Transmission electron microscopy (TEM) image of the same electrode region, which can be utilized to identify the boundary of each particle. Reprinted with permission from Ref. \cite{li2014current}.}\label{fig:IntroLFP}
\end{figure}

Cathode material such as lithium-iron-phosphate (LFP) shows {a} two-phase coexistence \cite{tang2010electrochemically, padhi1997phospho}. {In a particle, photon energies of two phases, LiFePO$_4$ and FePO$_4$, have distinct behaviors under the X-ray absorption spectrum,} as shown in Figure~\ref{fig:IntroLFP}(a). This can be utilized for locating different regions of phase-separated states in (b), which {shows the experimental evidence of the phase separation behavior in the porous electrode consists} of several LFP particles. In the porous electrodes, the electrolytic solution permeates through the void spaces of porous matrices. Depending upon a particle situation, it can be regarded as completely lithiated, completely delithiated, or {as an} active particle under phase separation transition. {In the LFP electrode, t}he population of active particles varies widely, ranging from nearly particle-by-particle to concurrent (all particles simultaneously) intercalation. The active population depends on various factors influencing phase separation such as intercalation rates \cite{li2014current}. Therefore, understanding the influence of these factors is of technological relevance.

\begin{figure}[h]
\centering
\includegraphics[scale=0.70]{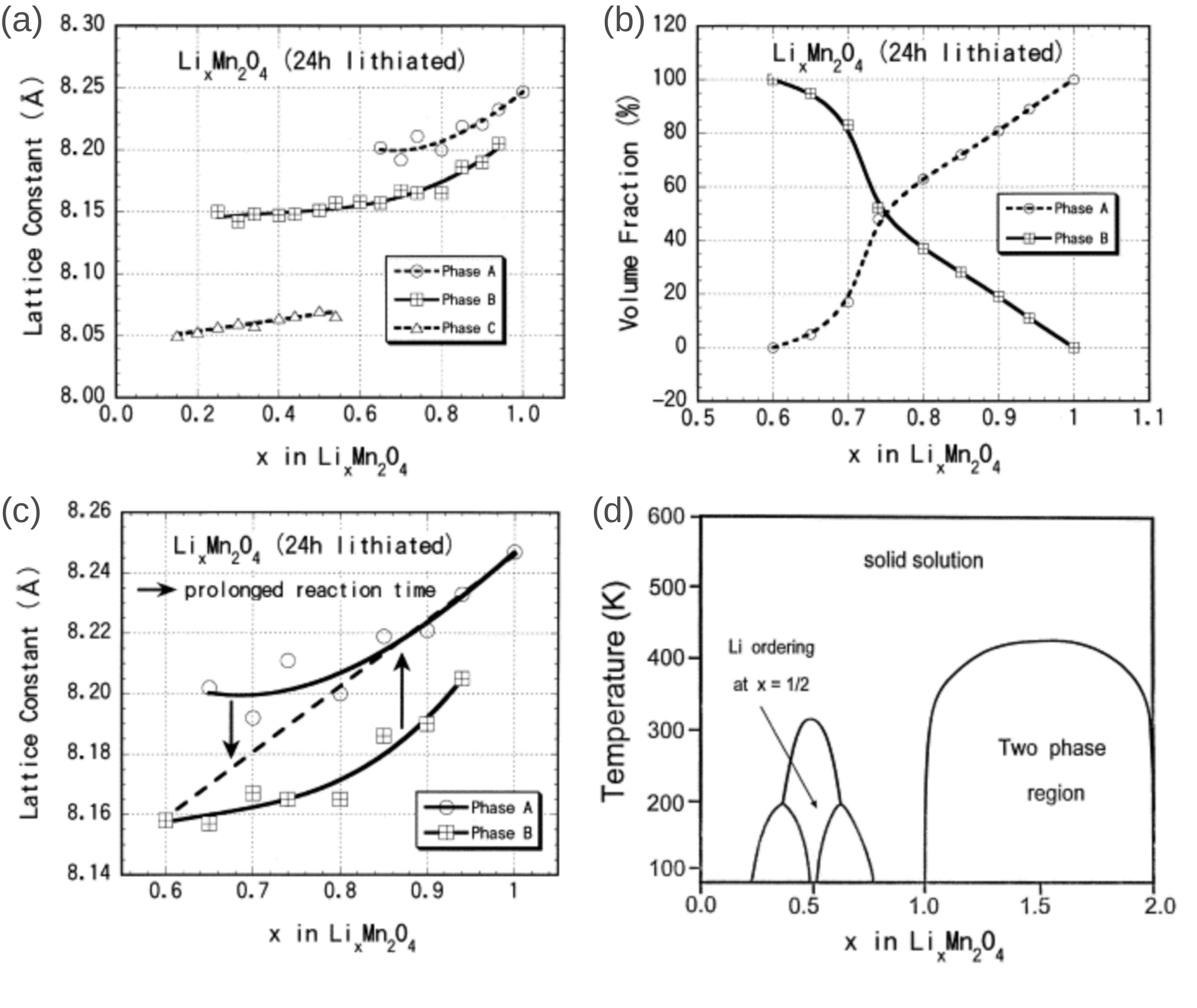}
\caption[Two-phase coexistence in lithium-manganese oxide spinel]{Two-phase coexistence in lithium-manganese oxide spinel. (a) Changes in the lattice parameters{,} with lithium concentration x {for }Li$_\textrm{x}$Mn$_2$O$_4$ $(0.15\leq \textrm{x} \leq 1.0)${.} (b) Changes in {the} volume fraction{,} with x for $(0.6\leq \textrm{x} \leq 1.0)${.} (c) Dependencies of {the} lattice parameters on x for $(0.6\leq \textrm{x} \leq 1.0)${.} (d) {The c}alculated phase diagram{,} as a function of temperature{,} and {the} lithium concentration x  for $(0.0\leq \textrm{x} \leq 2.0)$. Reprinted with permission from Refs. \cite{li2000phase, vanderven2000phase}.}\label{fig:IntroLMO}
\end{figure}

The mechanism of phase separation is also reported in other electrode materials, such as lithium cobalt oxide \cite{vanderven1998first}, silicon \cite{liu2012situ}, and lithium titanate \cite{vasileiadis2018toward}. However, the presented work explores lithium manganese oxide spinel (LMO) cathode material. For numerous reasons, LMO is a promising candidate for a cathodic material, including the fact that it is a cheaper, highly abundant material, which is environmentally favorable \cite{kim2008spinel}. However, a struggle with a capacity fade is a shortcoming \cite{vanderven2000phase}. An important source of the capacity fade is undergoing phase separation during battery operation, which leads to a premature battery failure~\cite{vanderven2000phase, ohzuku1990electrochemistry, liu1998mechanism, yang1999situ}. The finding of a two-phase coexistence is a breakthrough in scientific knowledge about LMO particles, obtained from ample experimental evidence \cite{li2000phase, yang1999situ}.

Figure~\ref{fig:IntroLMO} shows complex thermodynamics of phase separation in LMO materials. Substantial attempts are directed to understand the phase separation mechanisms of Li$_\textrm{x}$Mn$_2$O$_4$. There are several ranges of x, where two-phase coexistence is reported \cite{li2000phase, liu1998mechanism}. For instance, Phase B and Phase C are depicted from x$\approx$0.25 to 0.52, while Phase A and Phase B are visible from x$\approx$0.62 to 0.90 in (a). The findings of the latter case can be further corroborated by analyzing the volume fractions in (b), which shows an inflection point at x$\approx$0.75. There are strong dependencies of the phase separation mechanisms on various parameters such as {the} transportation rate{,} as shown in (c){,} and temperature{,} as shown in (d). For a very slow rate, the samples tend towards a single phase{,} instead of phase separation{,} and the lattice parameter varies linearly with x, see dashed line in (c). The phase diagram of LMO material as a function of x and temperature is shown in (d). Even though the phase separation exhibits several discrepancies in the literature \cite{li2000phase, vanderven2000phase, liu1998mechanism} and strong dependencies on various parameters (Figure~\ref{fig:IntroLMO}(b) and (c)), the two-phase coexistence is clearly eminent in LMO materials. Therefore, {the study of two-phase coexistence in LMO is considered} in the present work. 

\subsection{Numerical models of two-phase coexistence}
The coexistence of two phases in LMO{,} as a cathode electrode{,} is extensively explored in experiments \cite{ohzuku1990electrochemistry, vanderven2000phase, liu1998mechanism, yang1999situ}. Systematic numerical modeling of phase separation may complement to the scientific understanding of the mechanism and helps to optimize the process parameters efficiently and economically \cite{malik2013critical}.

 The modeling of two-phase coexistence in electrodes has a long-standing literature \cite{landstorfer2013mathematical, grazioli2016computational}. The intercalation in LMO follows shrinking-core kinetics, where the interface boundaries travel in the direction of lithium flux from the surface \cite{huttin2012phase}. Specific schemes, such as a flexible sigmoidal function \cite{huang2013stress}, a trapping concept \cite{drozdov2014a}, are employed for the interface tracking. However, the classical Cahn-Hilliard equation-based phase-field models are {mostly} employed to avoid cumbersome interface tracking \cite{cahnhilliard1958free} and embrace the existence of a high concentration gradient between the Li-rich and the Li-poor phases elegantly \cite{huttin2012phase}.  
 
 The phase-field models \cite{zhang2018nonlocal, stein2016effects, dileo2014a, welland2015miscibility, santoki2018phase, huttin2012phase, walk2014comparison} considered for the phase separation during intercalation are substantial for a single particle. These models focus on the isolated particles, which can be applicable to planar electrodes \cite{newman1975porous‐electrode}. In addition, the electrodes in many battery systems consist of a porous structure{,} composed of nanoparticles \cite{arico2011nanostructured}. There are some theoretical and numerical studies performed to account essential features of phase-separating porous electrodes \cite{bai2013statistical, li2014current, smith2017multiphase, chueh2013intercalation}. Even though extensive research is being carried out \cite{dreyer2011behavior, orvananos2014architecture} to understand the phase separation mechanism, the relative importance of the material properties, {the} operational conditions, and {the} microstructural properties on the phenomenon are {still} unclear. Therefore, this work is intended to postulate such technological insights {into the phase-separating LMO materials} through phase-field models.

\section{Electromigration in metallic conductors}

\begin{figure}[h]
\centering
\includegraphics[scale=0.55]{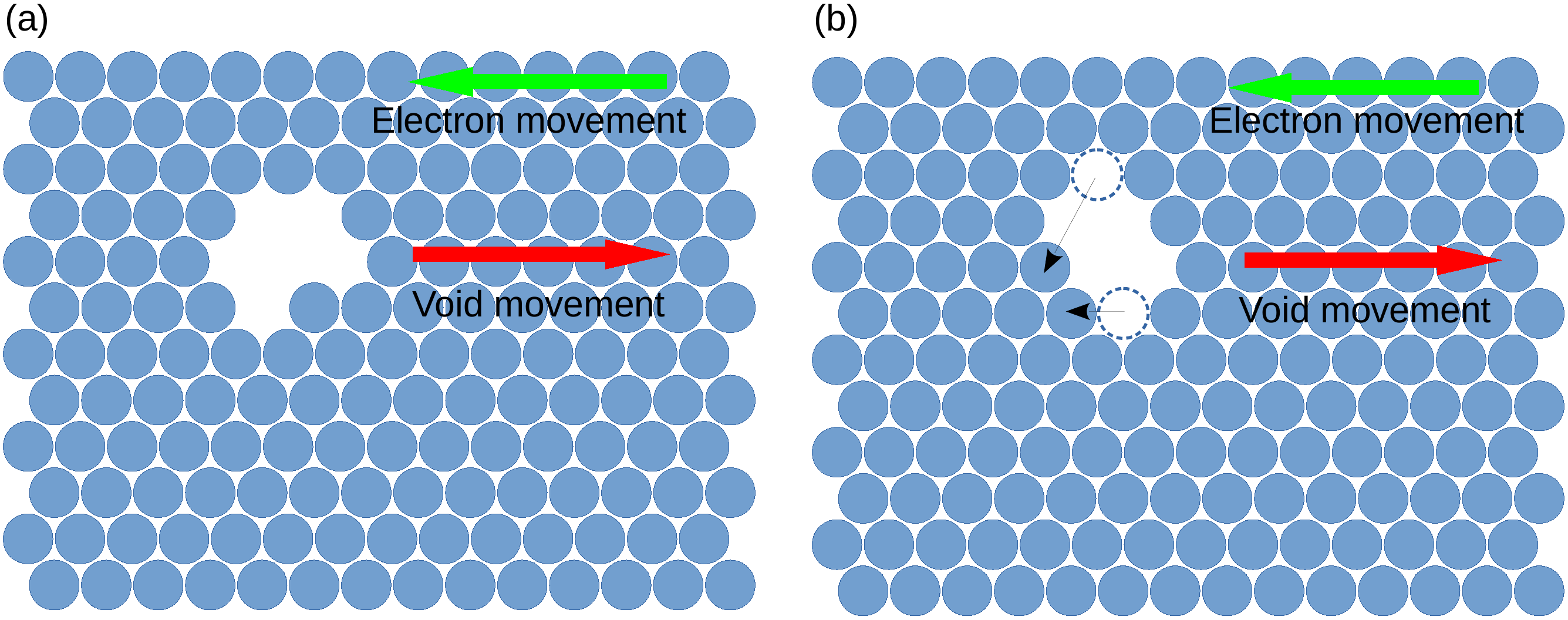}
\caption[Schematic of tiny void movement in {the} opposite direction of electric wind]{Schematic of tiny void movement in {the} opposite direction of {the} electric wind{,} due to {a} displacement of {the} species, (a) before {the} species move and (b) after {the} species move.}\label{fig:IntroAtomicDisplacement}
\end{figure}

When a metallic conductor {is} subjected to an external electric field, two distinct forces can be apparent in the medium. 
\begin{enumerate}
\item Direct Force ($\boldsymbol{F}_{\textrm{d}}$): Direct action of the external field on the charge of the migrating species.

\item Wind Force ($\boldsymbol{F}_{\textrm{w}}$): Electron{s} migrating in the opposite direction to electric field collide with the diffusing species and {transfer the} momentum to the species. This stimulates {the motion of the} species{,} in the direction of {the} electron propagation \cite{blech1976electromigration}.
\end{enumerate}
Both forces are proportional to the applied electric field {and} can be expressed as
\begin{equation}
\boldsymbol{F}_{\textrm{total}} =\boldsymbol{F}_{\textrm{d}}+ \boldsymbol{F}_{\textrm{w}} = eZ_s\boldsymbol{E}_\infty{,}
\end{equation}
where $e$ denotes the electron charge, $Z_s=Z_d+Z_w$ is the effective valence combining the valences of {the} electrostatic ($Z_d$) and {the} electron wind ($Z_w$), and $\boldsymbol{E}_\infty$ denotes the applied electric field. Therefore, the resultant {force} dictates the movement of the species. Generally, the later force is prominent in the field of ``electromigration (EM),''\nomenclature{EM}{electromigration} \cite{blech1976electromigration}, as shown in Figure~\ref{fig:IntroAtomicDisplacement}. {For simplicity, the term electron wind force therefore} often refers to the net effect of these two electrical forces.

\subsection{Presence of voids in the interconnect}
\label{sec:IntroVoidInterconnect}
In modern semi-conducting chips, a dense array of narrow, thin-film metallic conductors, called interconnects{,} are the crucial constituents of the integrated circuits (ICs){,} which connect active elements of microelectronics \cite{ho1989electromigration}. Due to the very large scale integrated (VLSI) circuits \cite{tu2003recent}, the interconnects are subject to intense electric current densities. This raises serious reliability concerns in the microelectronics industry{, noticeable in the form of an electrical failure of the interconnects,} due to electromigration. Furthermore, this is only expected to {worsen} with the current trend of continuing miniaturization \cite{deorio2010physically}. Therefore, understanding the {difficulties} associated with electromigration is important for the efficient design {of the batteries} and {the} prolongation of {the} lifetime of {the} interconnects.

\begin{figure}[h]
\centering
\includegraphics[scale=0.7]{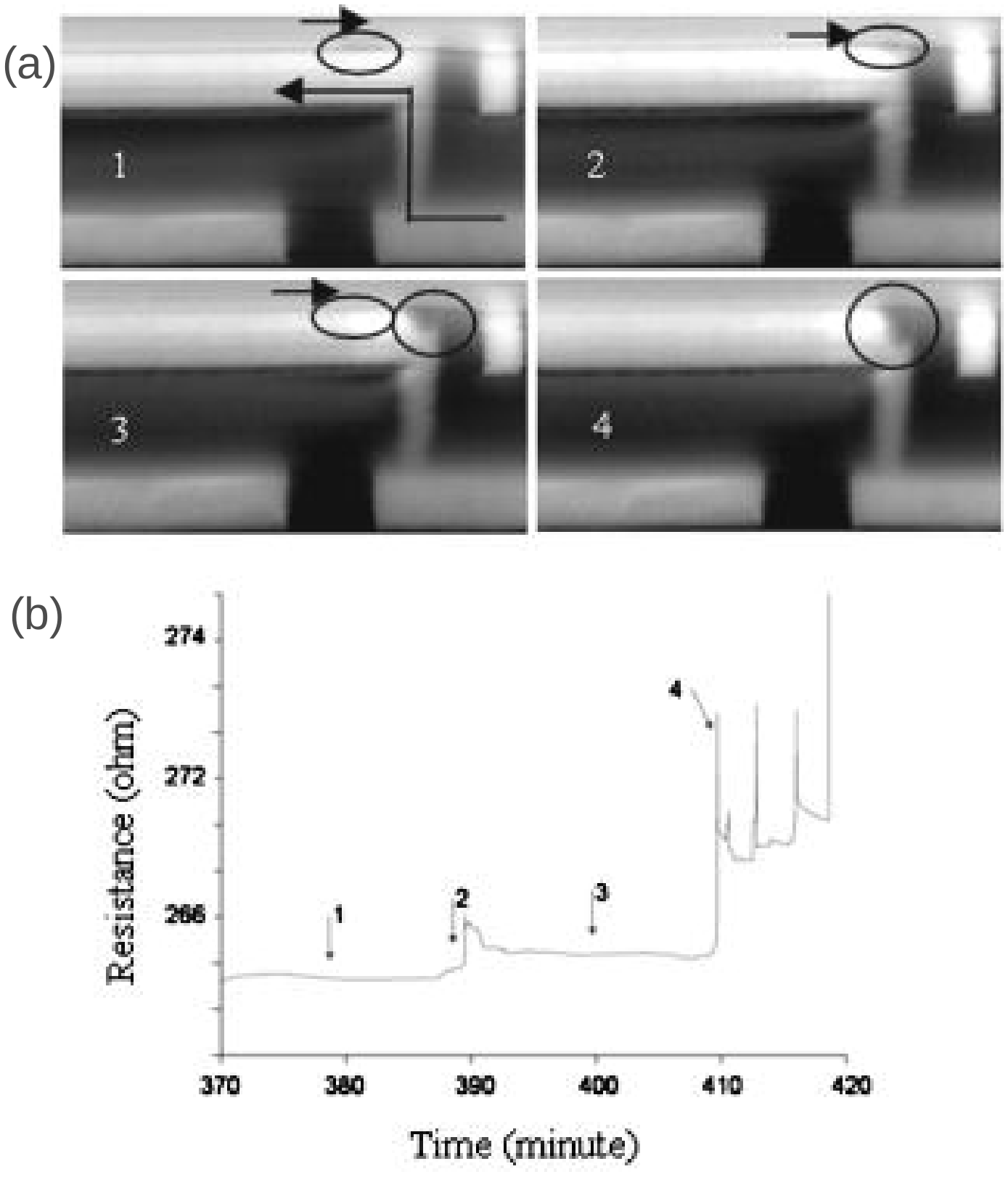}
\caption[SEM images of a cross section of {a} Cu interconnect with {a} void evolution and {a} change in resistance{,} at various time intervals]{(a) SEM images of a cross section of {a} Cu interconnect{,} at various time intervals{,} during {the characterization of} in situ electromigration. (b) Resistance of a structure shown in (a). The sequence of images in (a) indicates {an} electromigration-induced void movement with morphological shape changes, while (b) shows their relev{a}nce to the resistance. Reprinted with permission from Ref.\cite{vairagar2004situ}.}\label{fig:IntroVoidInterconnect}
\end{figure}

Postmortem analys{e}s of the damaged lines reveal the presence of multiple voids along the line, often of various shapes and sizes \cite{lin2017electromigration}, some of which have grown critical{,} leading to failure \cite{arzt1994electromigration}. Furthermore, \textit{in situ} studies endorse the failure to be a dynamic process \cite{riege1996influence, hunt1997healing}. Typically, voids {nucleate at current crowding zones} \cite{miyazaki2006electromigration} or at structural heterogeneous sites, such as grain boundaries (GBs)\nomenclature{GB}{grain boundaries}, or triple junctions \cite{riege1996influence, marieb1995observations, vairagar2004in}. Following nucleation, the voids may escape the grain boundary, migrate along or across the line in a self-similar manner or evolve into various time-dependent configurations \cite{arzt1994electromigration, riege1996influence, hunt1997healing}. Although {many} events precede the final failure, the lifetime of the interconnect is governed by the rate-limiting step \cite{lloyd2007black}. In addition, void shape changes, such as bifurcation \cite{gungor1999theoretical} and coalescence \cite{cho2007theoretical}, are a major lifetime{-}inhibiting factor. The physical dimensions of the voids are directly related to the electrical resistance{,} as shown in Figure~\ref{fig:IntroVoidInterconnect}{,} and in turn {to} the reliability of {the} interconnects. A large enough void may increase the resistance dramatically, and may even lead to the open circuits and {the eventual} failure of the interconnects \cite{cho2006current}. 
  
 The electromigration-induced degradation behavior can be observed by various \textit{in-situ} experimental methods, such as transmission electron microscopy \cite{liao2005situ}, scanning electron microscopy \cite{claret2006study}, and X-ray microscopy \cite{schneider2003situ}. However, accelerated lifetime tests \cite{nitta1993evaluating} are commonly employed to study reliability and performance predictions of the interconnects. Under these tests, the interconnect samples are stressed considerably {more} severe than their operating conditions{, in order} to speed up the process. After the Black's \cite{black1969electromigration} derivation, an empirical relation to predict the time to failure ($TTF$) can be expressed as
  \begin{equation}
  TTF = \frac{A_{it}}{j^{n}} e^{(E_a/k_BT)},
  \end{equation}
where $A_{it}$ represents a constant {which encompasses the material and geometrical properties of the} interconnect, $j$ is the current density, $k_B$ represents {the} Boltzmann constant, $T$ is the temperature, $E_a$ denotes {the} activation energy, and $n$ represents an exponent that describes the dependencies ($1\leq n \leq 2$) of {the} failure mechanism on the current density.

\subsection{Precipitates in the interconnects}

\begin{figure}[h]
\centering
\includegraphics[scale=0.60]{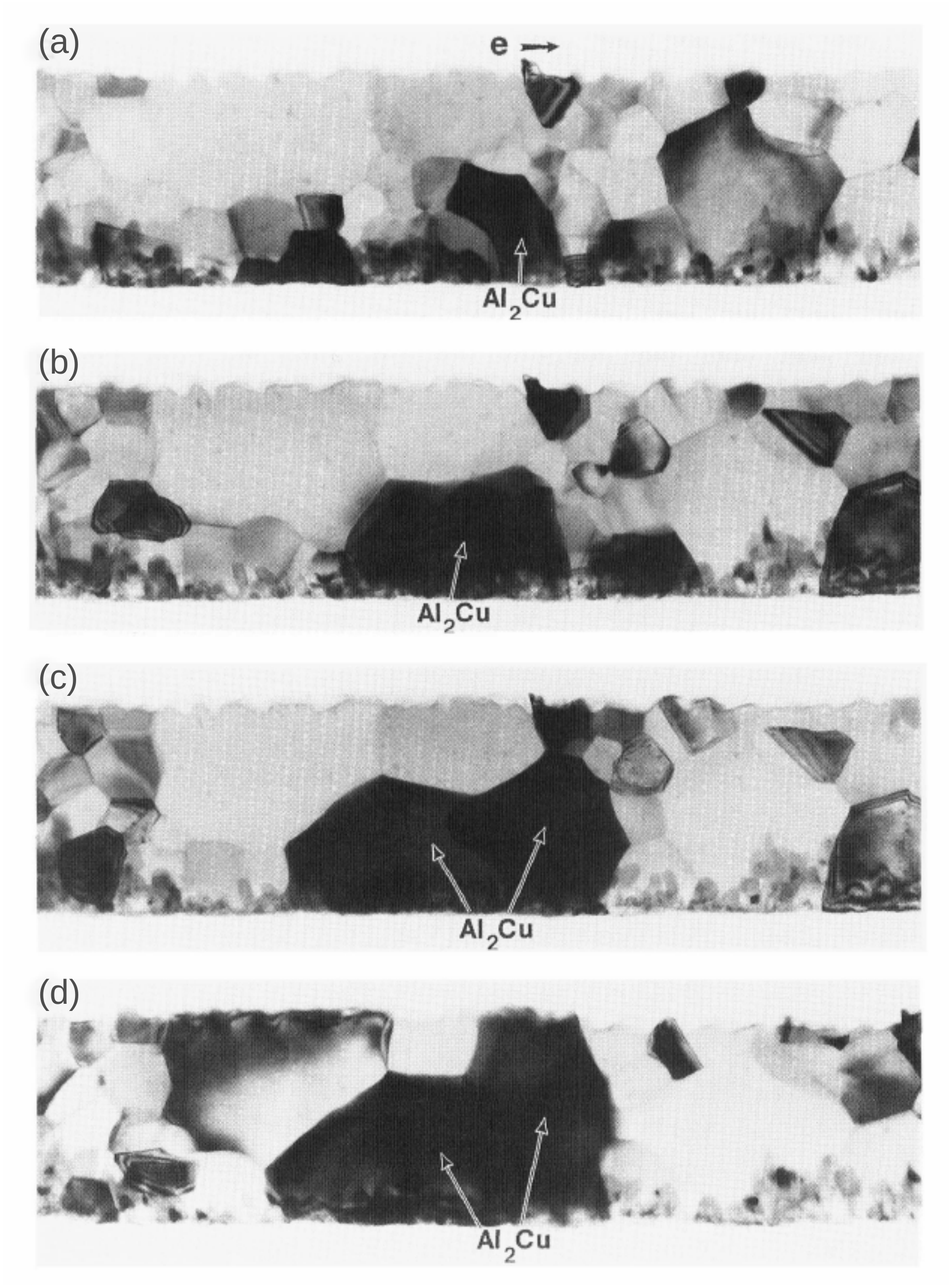}
\caption[TEM images of {the} growth of an Al$_2$Cu precipitate{,} at the cathode end]{TEM images of {the} growth of an Al$_2$Cu precipitate{,} at the cathode end{,} during electromigration testing. The sequence (a) to (d) indicates {the} progress of time. Reprinted with permission from Ref. \cite{shaw1996copper}}\label{fig:IntroPrecipitate}
\end{figure}

Contrary to the {disturbance} of voids in interconnects, the second application of electromigration, an absorption of metallic substances{,} in the form of precipitates{,} provides a constructive advantage to the performance of the interconnects. A strategical introduction of the desired metal precipitates in{to} the interconnect{s significantly} enhances the material properties related to electromigration \cite{ma1993precipitate}. For instance, a small number of precipitates of Al$_2$Cu are known to improve the reliability of Al-Cu interconnects \cite{witt2003electromigration}. These precipitates serve as reservoirs of Cu{, for a} depleted interconnect solution \cite{theiss1996situ}{. Thus,} it is expected that {the} precipitates increase the reliability of {the} interconnects by redistribution \cite{rosenberg1972inhibition}.

 Figure~\ref{fig:IntroPrecipitate} shows the growth of a precipitate at {a} cathode end. Furthermore, it is possible that all precipitates are accumulated at one end of the electrode{,} during operation \cite{witt2000electromigration}. Then, a current reversal can be considered for the redistribution of {the} precipitates. This {current reversal process} further enhances the effectiveness of the precipitates \cite{witt2000electromigration}. The presence of {the} precipitates{, in the interconnect lines,} modifies {the} actual time to failure \cite{spolenak1998effects}, which can be expressed as
\begin{equation}
TTF_{wp} = TTF + \frac{d_p}{A_p},
\end{equation}
where $TTF$ is {the} time to failure{,} without precipitates{,} including {the} incubation period of precipitate drifts, $d_p$ denotes the factor associated with the geometrical properties of the precipitates, and $A_p$ reflects {the} operating conditions such as temperature and current density.

\subsection{Surface nano-patterning by single-layer atomic clusters}

\begin{figure}[hbt!]
\centering
\includegraphics[scale=0.8]{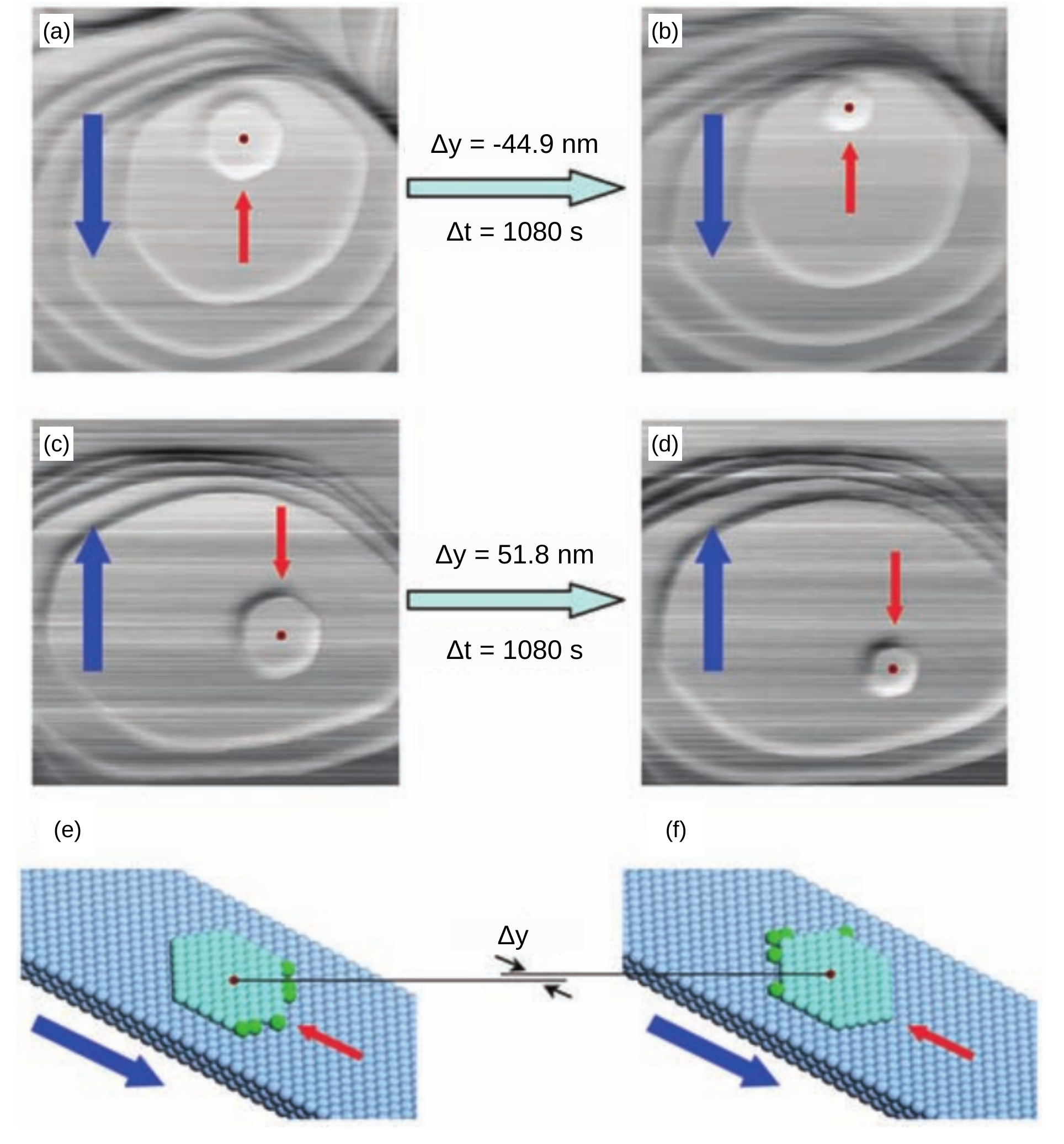}
\caption[Experimental evidence of the movement of monolayer islands{,} due to electromigration]{Experimental evidence of the movement of monolayer islands{,} due to electromigration. (a) and (b) show {the upward displacement of the island and the downward movement of the current}, while (c) and (d) show {the downward displacement of the island and the upward movement of the current}. {the s}chematics of the island in (e) and (f) illustrate {the} displacement{,} due to electromigration. {The b}lue arrows indicate a direct current direction, {while the red arrows indicate the direction of the} electron wind. Reprinted with permission from Ref. \cite{tao2010visualizing}}\label{fig:IntroIsland}
\end{figure}

The pattern formation{, directed by an electric field,} has experienced great success in applications like block copolymers \cite{mukherjee2016influence, mukherjee2016electric}, EM-induced liquid metal flow \cite{dutta2009electric}, growth of intermetallic compound \cite{park2013phase, zhou2011a, attari2018on}, and grain-boundary grooving \cite{mukherjee2016phase, chakraborty2018phase}. In addition, the surface nanopatterning is the third intriguing application of electromigration{,} in the rapidly advancing field of surface engineering \cite{yamanaka1989surface, yongsunthon2011surface}, which has similar characteristics to void-interconnect mechanisms. {This means that} the atomic clusters (islands) are adsorbed on a crystalline substrate{,} under the action of an external electric field. The islands evolve due to {the} momentum transfer by the collision of the electrons \cite{metois1997an}, as shown in Figure~\ref{fig:IntroIsland}. Adatom migration along the surface of the island is reported to be a dominant mass transport mechanism \cite{yamanaka1989surface, hedouin2000relationship}. 

Understanding the influence of the materialistic properties and the characteristics of the external fields {is} of scientific importance for the desired dynamics of the islands, to tailor the surface resistivity. Surprisingly, the islands are reported to glide  in or against the current direction, {at a constant velocity,} depending on the {reconstruction of the island} \cite{metois1999steady}{,} and whether the electron flow interacts with the substrate or the island. To be consistent with the direction of {the} void migration{,} defined in section~\ref{sec:IntroVoidInterconnect}, {the islands in the following work} are presumed to migrate in the opposite direction of {the} electron flow. {Hereafter, `inclusion' is used as a general terminology to refer to the voids (insulator), precipitates (conductivity different from the matrix), and islands (conductivity identical to the matrix),} otherwise stated explicitly.

\subsection{Numerical models of inclusion migration}

The analytical \cite{wang1996simulation, schimschak1998electromigration, hao1998linear} and numerical models \cite{mehl2000electromigration, pierre2000electromigration, kuhn2005complex, dasgupta2013surface, kumar2016surface, dasgupta2018analysis, santoki2019phase, hauser2010the, banas2009phase, li2012the, bhate2000diffuse, mahadevan1999simulations} have been employed {extensively} for the electromigration of inclusions. Sharp interface analyses are presented to analyze the characteristics of {a} presumed shape in Refs. \cite{suo1994electromigration,  yao2015an}. Linear stability analyses 
are exerted for the prediction of {the} inclusion morphology \cite{wang1996simulation, schimschak1998electromigration, hao1998linear}. Pertaining to the complexity, theoretical analyses have focused on the stability of {a} presumed shape \cite{hao1998linear}, which lacks to provide any information regarding the shape after {an} instability, or the prediction accuracy decreases as a geometry deviates from a presumed shape. Alternately, the sharp-interface numerical models, including Monte-Carlo \cite{mehl2000electromigration}, continuum \cite{pierre2000electromigration, kuhn2005complex, dasgupta2013surface}, and direct dynamical models \cite{kumar2016surface, dasgupta2018analysis}, provide important technological insights {into} the phenomena such as many-inclusion pattern{s} \cite{dasgupta2013surface}, a phase diagram of anisotropy strength{, the} inclusion size \cite{kuhn2005complex}, nanowire pattern{s} \cite{kumar2016surface}, and {the} analysis of {the} oscillatory dynamics of inclusions \cite{dasgupta2018analysis}.

Due to its capability to {implicitly} track the interface, the phase-field methods are extensively employed to study {the motion of} electromigration-induced isotropic inclusion in {single crystals} \cite{bhate2000diffuse, mahadevan1999simulations, barrett2007a, santoki2019phase, hauser2010the, banas2009phase}. In addition, relatively few studies are performed to understand the {motion of anisotropic inclusions} \cite{li2012the, santoki2019role}. However, the literature is devoid of exquisite comparison{s} of analytical findings with the phase-field results. In addition, the {shape-changing} mechanism{, observed during the migration of inclusions,} is complex and obscure. Therefore, the presented work is intended to understand {this mechanism} through the electromigration-based phase-field model.


\section{Conclusion}

In recent decades, phase-field models have become an ideal simulation tool for numerous applications. In the phase-field approach, a continuous variable is defined to represent the entire microstructure, which is also known as the order parameter or phase-field parameter. The order parameter itself holds a nearly constant value in the bulk domains and varies continuously across a finite but narrow interface. Due to the presence of a most elegant feature, the diffuse interface, these models provide an efficient framework to address the time-dependent free boundary problems. In addition, the boundary conditions are incorporated implicitly to avoid cumbersome interface tracking. Therefore, the movement of complex boundaries and stringent boundary conditions can be managed readily. The transformation of the order parameter works on optimization principles, such as the minimization of the free energy or the maximization of the system entropy, which leads to morphological changes in the microstructure. The Cahn-Hilliard model \cite{cahnhilliard1958free} is important in the field of conserved order parameters, while the Allen-Cahn model \cite{allen1979a} is relevant in the field of non-conserved order parameters. 


The thesis is organized as follows. Chapter~\ref{chapter:PhaseField} presents a review of the basic framework of the phase-field models, by considering theoretical advancements from {the} microscopic properties to the mesoscopic description. Thereafter, Chapters~\ref{chapter:phaseFieldModelLIB} and \ref{chapter:phaseFieldModelElectromigration} {respectively} derive the models of two {previously} described phenomena, namely, two-phase coexistence in LMO materials and {the motion of inclusions,} due to electromigration. The numerical results of phase separating LMO electrodes are presented in Chapters~\ref{chapter:NeumannConstantFlux} and \ref{chapter:Potentiostatic}. Subsequently, the study of various aspects of {the electromigration-induced motion of inclusions is presented} in Chapters~\ref{chapter:EM1}, \ref{chapter:EM2}, and \ref{chapter:4Fold6FoldEM}. {In Chapter~\ref{chapter:conclusion},} the thesis concludes with a brief summary and a few possible future directions.

\newpage
\thispagestyle{empty}
\vspace*{8cm}
\phantomsection\addcontentsline{toc}{chapter}{II Methods : Phase-field formulation}
\begin{center}
 \Huge \textbf{Part II} \\
 \Huge \textbf{Methods : Phase-field formulation}
\end{center}

\afterpage{\blankpagewithoutnumberskip}
\clearpage

\chapter{Phase-field model}
\label{chapter:PhaseField}

In {the} past decades, the understanding of many natural phenomena {has been} obtained through experimental investigations. Afterward{s}, these experimental findings have been able to motivate the development of theories one way or the other. Thereafter, the exponential increase in computational resources {has recently encouraged} the advancement of numerical techniques based on a unique theoretical framework, which may untangle the intricacies associated with the physical process under investigation. The resulting enhanced understanding may assure {an optimal design, a better performance prediction, and an} efficient employment of the resources.

\begin{figure}
\begin{center}
\includegraphics[scale=1.0]{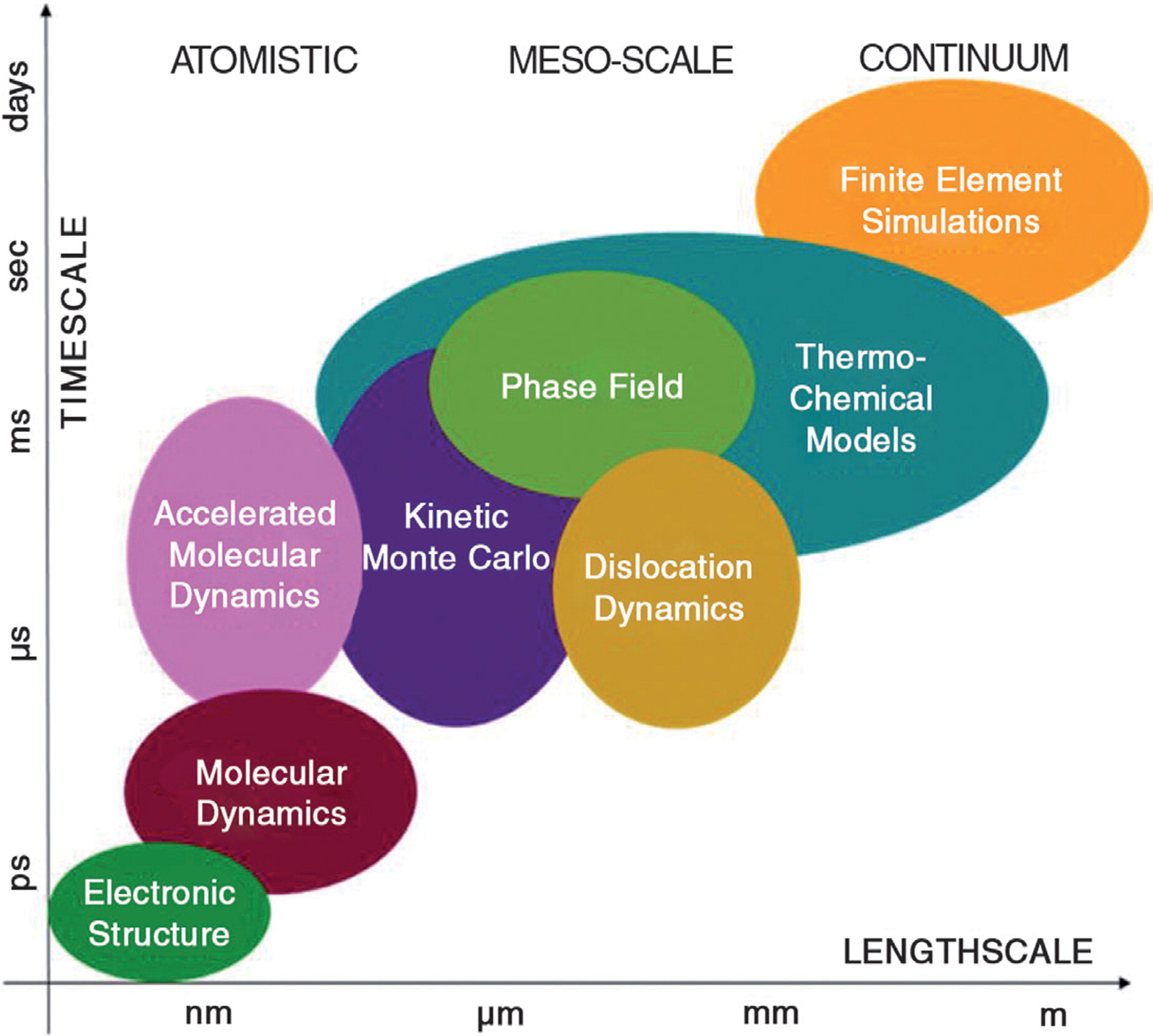}
\caption[Estimated physical length scale and time scale of the various computational methods{,} utilized for the numerical modeling]{Estimated physical length scale and time scale of the various computational methods{,} utilized for numerical modeling. Reprinted from the work of Stan \cite{stan2009discovery} with permission from Elsevier.} \label{fig:numericalModels}
\end{center}
\end{figure}

The applicability of the numerical modeling ranges from several pico-seconds to thousands of years in time scale and {on the} atomistic level (nanometers) to the level of the universe (kilometers) in length scale. Figure~\ref{fig:numericalModels} illustrates some of the computational methods with their typical range of length and time scale used in the literature. In addition, communication across the models is often considered to improve the accuracy of the employed method. For instance, simulations of molecular dynamics are considered in the fiber-reinforced thermosets to estimate material properties{. Then,} the curing process determines the eigenstresses. On the basis of these local eigenstresses, the phase-field fracture simulations are considered to predict the occurrence of {microcracks} in Ref. \cite{schwab2019fibre}. The selection of a specific computational method for a concerned phenomenon is a matter of choice{,} based on its feasibility and effectiveness.

{Phase-field} modeling is one of such computational technique recognized for the past couple of decades. This mesoscopic approach is employed in diverse avenues of research{,} including multicellular systems in biology \cite{nonomura2012study}, multicomponent {fluid flow} in fluid dynamics \cite{kim2012phase}, alloy solidification in materials science \cite{nestler2005multicomponent}, among others. As depicted from Figure~\ref{fig:numericalModels}, the typical dimensions of the utilized phase-field methods are {on the} mesoscale level. Therefore, before introducing the phase-field model, a preface is presented to relate the mesoscale description with the microscopic properties. 

\section{Mean-field theory}
\label{sec:phaseFieldMeanFieldTheory}

The origins of the phase-field methodology have been considerably influenced by the mean-field theory of phase transformation. Thus, some of the foundations can be interpreted by the mean-field theory{,} which is based on the framework of statistical thermodynamics. This theory provides {the} necessary groundwork for the concept of {an} order parameter and its relation to the system free energy, commonly adopted in the phase-field theories. The present section is motivated by Refs. \cite{huttin2014phase, porter2009phase, provatas2011phase} and some parts of their works are represented here, according to the present context.

{For simplicity}, consider a binary mixture of two species{,} A and B. This mixture is scattered on a region {which} consists of $N_m$ \nomenclature{$N_m$}{total number of lattice sites} {small volume elements,} known as lattice sites, which {are} identified by an index \nomenclature{$\alpha$}{index of a lattice site} $\alpha$. These lattice sites carry either {the A or B species}. {The state variable} $c_\alpha$ \nomenclature{$c_\alpha$}{state of the lattice site $\alpha$} is defined to identify the state of the cell, i.e. $c_\alpha = 0$ for a cell occupied by the A species and 1 for the B species. Therefore, {the} total number of B species, present in the domain, can be given by \nomenclature{$N_B$}{the total number of specified species, occupied in the lattice sites} $N_B=\sum_{\alpha=1}^{N_m} c_\alpha$. In a simple lattice model, the total energy (Hamiltonian) \nomenclature{$H$}{Hamiltonian of the lattice model} $H$ of such a system with A and B states is expressed as
\begin{equation}
H\{c_\alpha\}= B_{0m} \sum_{\alpha=1}^{N_m} c_\alpha + B_{m}\sum_{\alpha=1}^{N_m} \sum_{\alpha' \neq \alpha}^{M_m} c_\alpha c_{\alpha'}, \label{eq:hamiltonian}
\end{equation}
where $\alpha'$ \nomenclature{$\alpha'$}{the neighboring interstitial lattice site} denotes {the} neighboring interstitial site, $c_{\alpha'}$ \nomenclature{$c_{\alpha'}$}{state of {the} neighboring interstitial lattice state $\alpha'$} indicates the state of the neighboring interstitial lattice site $\alpha'$, and $M_m$ \nomenclature{$M_m$}{co-ordination number} represents {the} coordination number, which accounts {for} the number of neighboring lattice sites. In addition, $B_{0m}$ \nomenclature{$B_{0m}$}{the internal system energy parameter} represents the internal system energy coefficient, which is independent from neighboring interactions amongst interstitial lattice sites. The interaction between the neighboring lattices $\alpha$ and $\alpha'$ is quantified by the two-body interaction field $B_{m}$\nomenclature{$B_{m}$}{two-body interaction field from neighboring lattices}. The first term in Eq. \eqref{eq:hamiltonian} comprises {the} contribution of {the} individual lattices and {the} second term represents {the} interaction between {the} neighboring lattices. The estimation of the system free energy, by means of the Hamiltonian Eq. \eqref{eq:hamiltonian}, requires the calculation of the \nomenclature{$Z_{gc}$}{grand canonical partition function}grand canonical partition function
\begin{equation}
Z_{gc} (T,\mu) = \sum_{c_\alpha \in \{ 0,1\}} \textrm{exp}\left( -\frac{H\{ c_\alpha \}-\mu \sum_{\alpha} c_\alpha}{k_B T} \right){,} \label{eq:grandCanonicalPartitionFunction}
\end{equation} 
where $\mu$\nomenclature{$\mu$}{chemical potential} is the chemical potential, $T$\nomenclature{$T$}{absolute temperature} denotes the absolute temperature, and $k_B$\nomenclature{$k_B$}{Boltzmann constant} is the Boltzmann constant. In addition, $Z_{gc}$ is related to {the} grand canonical potential function $g(T,\mu)$\nomenclature{$g$}{grand canonical potential function} and the free energy can be calculated from {the} following expressions{:}
\begin{eqnarray}
g(T,\mu)=&-k_B T \ln{Z_{gc}(T,\mu)}, \label{eq:meanFieldFlux}\\
N_B=&-\frac{\partial g(T,\mu)}{\partial \mu},\label{eq:meanFieldConcentration}\\
F(T,N_B)=&g(T,\mu)+N_B\mu(T,N_B) \label{eq:meanFieldEnergyFunction}.
\end{eqnarray}
Note that the primary objective of the presented analysis is to estimate the free energy $F(T,N_B)$\nomenclature{$F$}{system free energy}, which can also be accomplished from the canonical partition function. However, the estimation of {the} canonical partition function imposes mathematical intricacies. To overcome these difficulties, instead of using {the} canonical partition function{,} which is a function of {the} concentration, the free energy is calculated from {the} grand canonical partition function, a function of {the} chemical potential. 

 Landau \cite{landau1965collected} postulated a phenomenological approach for phase transformation to analyze the transition from one state to the other state, which can be viewed as a growth of a state variable. After that, an important scalar state variable, referred {to} as \textit{order parameter}, $c$\nomenclature{$c$}{concentration/order parameter} is introduced to describe the transition and state of collective sites. This parameter is treated as an average thermodynamic property of the state, such as symmetry, concentration, and density, which differentiates different states,
\begin{equation}
c  = \frac{N_B}{N_m}. \label{eq:meanFieldAverageConcentration}
\end{equation}
Due to {the} two-body interaction term in {the} Hamiltonian Eq. \eqref{eq:hamiltonian}, the calculation of the grand canonical partition function is tedious. To overcome the difficulties, the mean-field approximation allows to replace {the} two-body term in the Hamiltonian \eqref{eq:hamiltonian} by a one-body term{,} considering a small fluctuation in $c_\alpha$ around the average concentration $c${,} by utilizing the relation $c_\alpha = c + \delta c_\alpha$, where $\delta c_\alpha / c\ll 1$ and $\delta c_\alpha$\nomenclature{$\delta c_\alpha$ (or $\delta c_{\alpha'}$)}{fluctuation of $c_\alpha$ (or $c_{\alpha'}$) compared to the average} {are} the fluctuation{s} of $c_\alpha${,} compared to the average. After substituting this relation into {the} two-body term of Eq.~\eqref{eq:hamiltonian}, the approximation {consisting of the} one-body term can be expressed as
\begin{equation}
\sum_{\alpha=1}^{N_m} \sum_{\alpha' \neq \alpha}^{M_m} (c + \delta c_\alpha) (c + \delta c_{\alpha'}) = - \frac{N_m M_m c^2}{2} + M_m c \sum_{\alpha=1}^{N_m} c_\alpha. \label{eq:hamiltonianFluctuations}
\end{equation}
As $\delta c_\alpha / c\ll 1$, the second{-}order fluctuations $\delta c_{\alpha}\delta c_{\alpha'}$ are neglected. In addition, the relations $\Sigma_{\alpha=1}^{N_m}\Sigma_{\alpha' \neq \alpha}^{M_m} 1 = N_m M_m/2$ and $2\Sigma_{\alpha=1}^{N_m}\Sigma_{\alpha' \neq \alpha}^{M_m} \delta c_\alpha = - N_m M_m c + M_m \Sigma_{\alpha=1}^{N_m} c_\alpha$ are utilized. Substituting Eq. \eqref{eq:hamiltonianFluctuations} into the Hamiltonian Eq. \eqref{eq:hamiltonian}, the equivalent Hamiltonian{, consisting of the one-body term,} can be written as
\begin{equation}
H\{c_\alpha\} = (B_{0m} + M_m c B_m) \sum_{\alpha=1}^{N_m} c_\alpha - \frac{N_m M_m c^2}{2} B_m.\label{eq:hamiltonianEffective}
\end{equation}
The two-body term is eliminated from the Hamiltonian \eqref{eq:hamiltonian}, by introducing a field variable~$c$. Therefore, the effect{s} of {the} neighboring sites are incorporated through the averaged field variable. By substituting Eqs. \eqref{eq:hamiltonianEffective} and \eqref{eq:meanFieldAverageConcentration} into Eq. \eqref{eq:grandCanonicalPartitionFunction}, the resultant grand canonical partition function can be obtained as
\begin{align}
Z_{gc}(T,\mu)=& \textrm{exp}\left(  \frac{N_m M_m c^2}{2k_BT}B_m \right) \prod_{\alpha=1}^{N_m} \sum_{c_\alpha \in \{0,1\}}  \textrm{exp} \left( - \frac{(B_{0m} + B_m M_m c - \mu)}{k_B T} c_{\alpha}\right) \nonumber \\
=& \textrm{exp}\left(  \frac{N_m M_m c^2}{2k_BT}B_m \right) \left( 1 + \textrm{exp} \left( - \frac{B_{0m} + B_m M_m c - \mu}{k_B T} \right) \right)^{N_m} \label{eq:grandCanonicalLast}
\end{align}
{Inserting} this relation into Eq. \eqref{eq:meanFieldFlux}, the grand canonical potential can be expressed as
\begin{equation}
g(T,\mu)= -\frac{N_m B_m M_m c^2}{2} - N_m k_B T \ln\left( 1 + \textrm{exp} \left(- \frac{B_{0m} + B_m M_m c - \mu}{k_B T} \right) \right). \label{eq:meanFieldFinalG}
\end{equation}
Substituting this relation and Eq. \eqref{eq:meanFieldConcentration} into Eq. \eqref{eq:meanFieldAverageConcentration}, the expression for {the} chemical potential can be obtained as
\begin{equation}
\mu = B_{0m} + B_m M_m c +k_B T \ln\left(\frac{c}{1-c}\right). \label{eq:meanFieldFinalMu}
\end{equation}
Finally, from Eqs. \eqref{eq:meanFieldEnergyFunction}, \eqref{eq:meanFieldFinalG}, and \eqref{eq:meanFieldFinalMu}, the free energy of the system can be expressed as
\begin{align}
F(T,N_B)/N_m = \left(B_{0m} + \frac{ B_m M_m}{2}\right) c &- \left( \frac{ B_m M_m}{2} \right) c (1-c) \nonumber \\
&+ k_B T  \left(c \ln(c) + (1-c) \ln(1-c) \right).
\end{align}
To simplify further discussions on the free energy, considering the relations $B_{0m} + B_m M_m/2 = X_1 k_B T_{\textrm{ref}}$, $ B_m M_m/2=X_2 k_BT_{\textrm{ref}}$, and $F(T,N_B)/(N_mk_BT_{\textrm{ref}})=f(T,c)$, \nomenclature{$X_1$}{first free energy density parameter}\nomenclature{$X_2$}{second free energy density parameter}\nomenclature{$f$}{system free energy density}\nomenclature{$T_{\textrm{ref}}$}{reference temperature},the system free energy density is expressed as
\begin{equation} \label{eq:freeEnergyLandauFinal}
f(T,c) = X_1 c + X_2 c(1-c) + \frac{T}{T_{\textrm{ref}}}\left(c \ln(c) + (1-c) \ln(1-c) \right),
\end{equation}
which is in normalized form. The term energy density, is {used to refer to its} normalized form throughout the document.

\begin{figure}
\begin{center}
\includegraphics[scale=0.70]{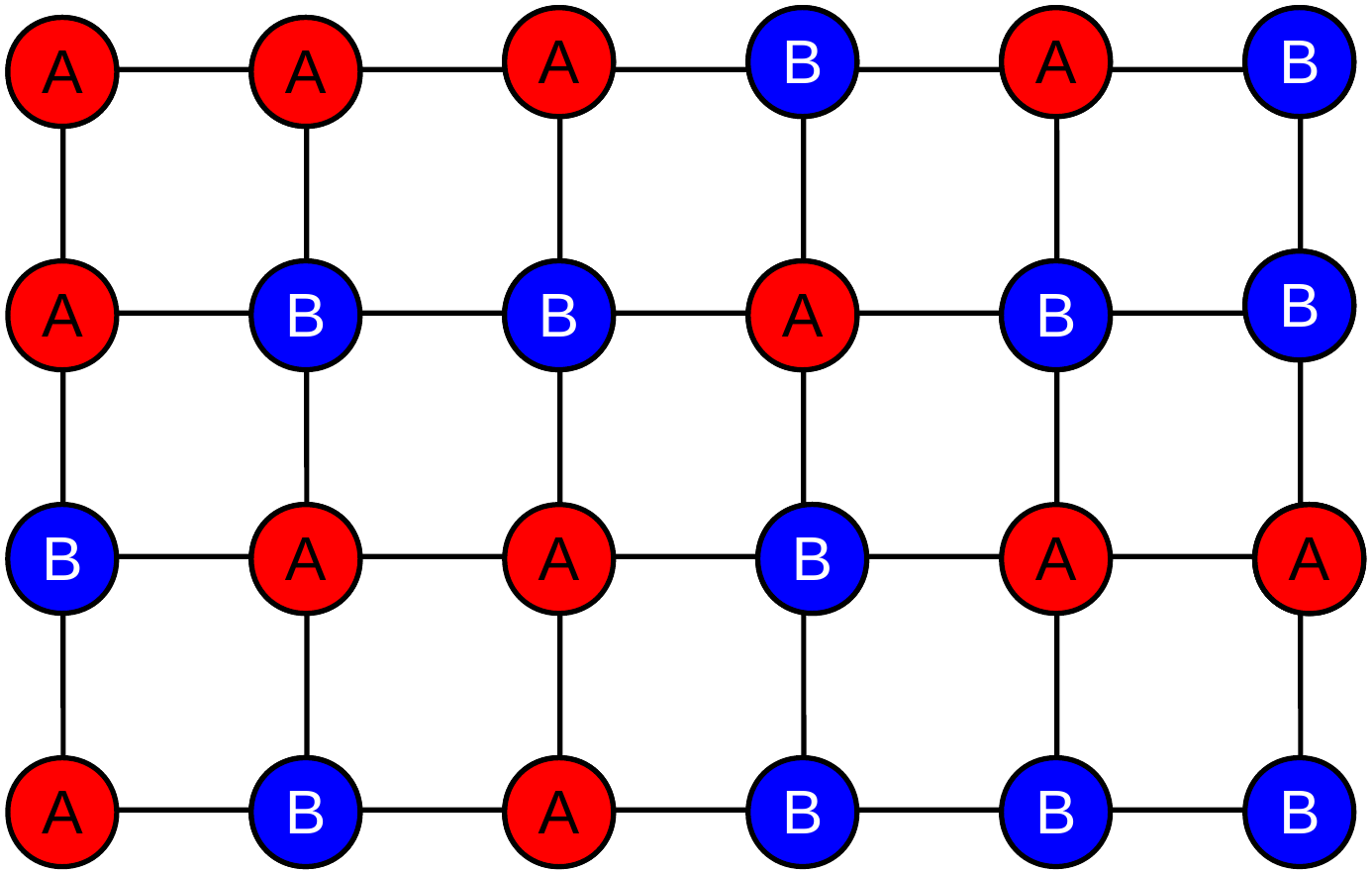}
\caption[Schematic of a lattice consisting of a mixture of A and B species]{Schematic of a lattice consisting of a mixture of A and B species, which forms bonds between A--A, B--B, and A--B.}\label{fig:interAtomicBonds}
\end{center}
\end{figure}

{T}o understand the different characteristics of the free energy equation~\eqref{eq:freeEnergyLandauFinal}, {we} consider $X_1=0$ and $T=T_{\textrm{ref}}$ {for simplicity, and additionally designate the two terms} $\Delta H = X_2c(1-c)$ and $-\Delta S =c \ln(c) + (1-c) \ln(1-c)$ in Eq. \eqref{eq:freeEnergyLandauFinal}, which {respectively} are the change in the\nomenclature{$\Delta H$}{change in internal energy of the system} internal energy and the\nomenclature{$\Delta S$}{change in entropy of the system} entropy of the system. In the forthcoming paragraphs, these terms are extensively utilized to distinguish typical features of the free energy. In order to describe these characteristics, a simple lattice system of {a} binary solution is considered{,} as shown in Figure~\ref{fig:interAtomicBonds}.

 The structure of a binary mixture contains three types of bonds, A--A, B--B, and A--B{, with the} bond energies $\varepsilon_{AA}$\nomenclature{$\varepsilon_{AA}$}{bond energy between A--A species}, $\varepsilon_{BB}$\nomenclature{$\varepsilon_{BB}$}{bond energy between B--B species}, and $\varepsilon_{AB}$\nomenclature{$\varepsilon_{AB}$}{bond energy between A--B species} respectively, which are illustrated schematically. {After mixing, the} change in {the} internal energy of the system can be expressed as 
 \begin{equation}
 \Delta H =  \varepsilon_t P_{AB},
 \end{equation}
where $P_{AB}$\nomenclature{$P_{AB}$}{total number of A--B bonds} is the total number of A--B bonds and $\varepsilon_t$\nomenclature{$\varepsilon_t$}{difference between the A--B bond energy and the average of A-A and B-B bond energies} is the difference between the A--B bond energy and the average of the A-A and B-B bond energies{:} 
\begin{equation}
 \varepsilon_t = \varepsilon_{AB}- \frac{1}{2}(\varepsilon_{AA}+\varepsilon_{BB}).
\end{equation}  
As the spontaneous change in {the} entropy is always positive $\Delta S>0$, {the} following cases can be obtained {by} considering the value of $\Delta H$: 
 
\begin{figure}[t]
\begin{center}
\includegraphics[scale=0.70]{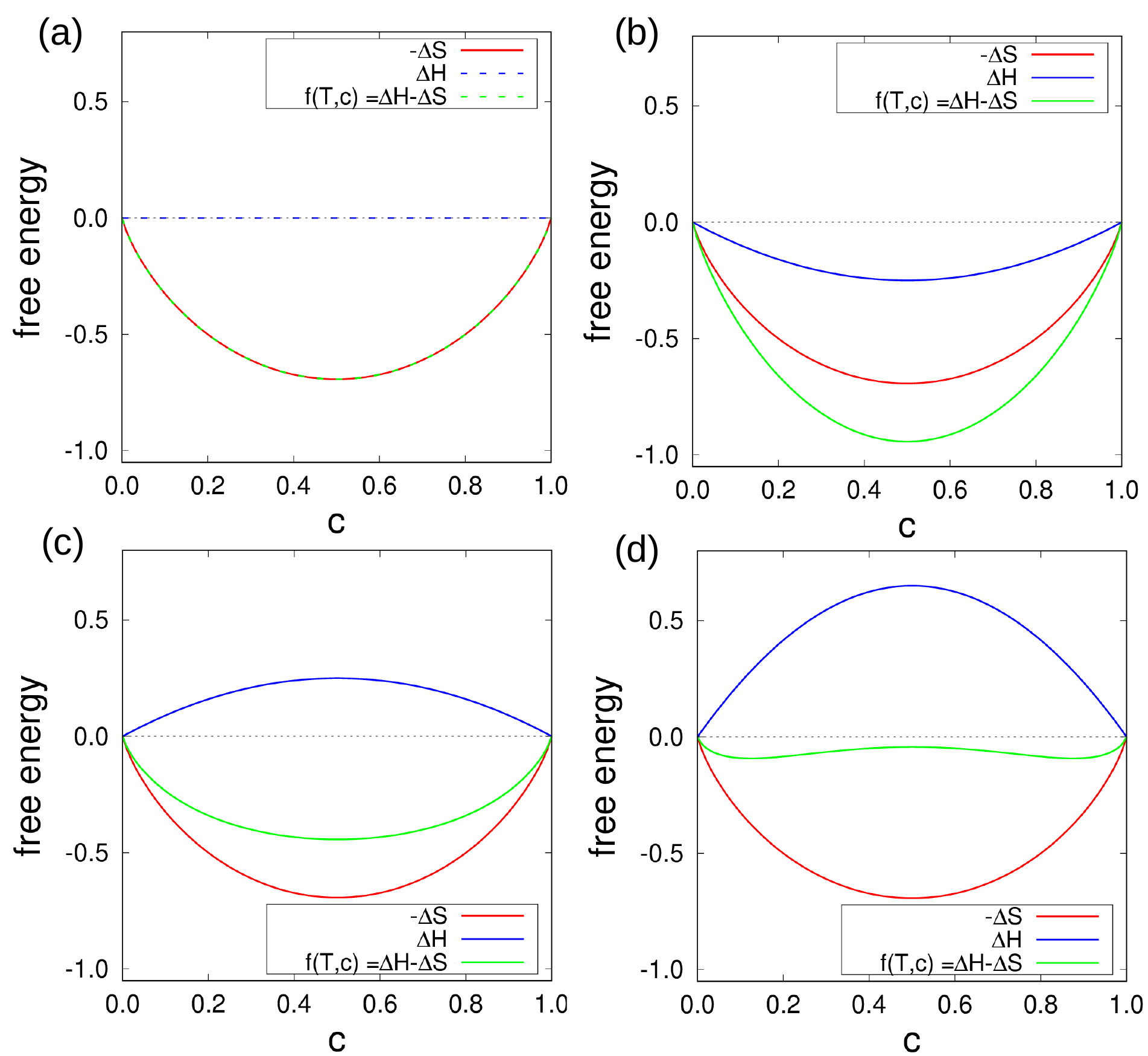} 
\caption[Free energy curves for different values of $\varepsilon_t$.]{Free energy curves for different values of $\varepsilon_t$. The plots correspond to (a) $X_2=0$, (b) $-1$, (c) $1$, and (d) $2.6$ in Eq. \eqref{eq:freeEnergyLandauFinal}, while considering $X_1=0$ and $T=T_{\textrm{ref}}$.}\label{fig:freeEnergyCurvesPhaseField}
\end{center}
\end{figure}

 \begin{enumerate}
 \item If $\varepsilon_t=0${, it implies that} $\Delta H=0$, {which means that} the solution is ideal. In this case, A--B bonds are equally favorable to A--A and B--B bonds{,} based on {the} internal energy argument, which indicates {that the} species are {arranged randomly}. As the change in the internal energy is treated to be negligible for an ideal solution $\Delta H = 0$, the system free energy only contains the entropy term, which is shown schematically in Figure \ref{fig:freeEnergyCurvesPhaseField}(a). However, this type of behavior is seldom observed in practice and usually, the internal energy changes with the mixing, i.e. $\Delta H \neq 0$.
 \item If $\varepsilon_t<0$ implies $\Delta H <0$ and $f(T,c)$, the species prefer to be surrounded by species of another type, which results in mixing{, which increases} the number of A--B bonds $P_{AB}${,} compared to the ideal solution. The internal energy and the enthalpy are in consensus to keep {the} free energy of the mixture at the lowest{,} as shown in Figure~\ref{fig:freeEnergyCurvesPhaseField}(b){. This} signifies that maintaining {the} isolated regions of {the A or B species} is energetically unfavorable, {and that the system rather} prefers a state of a homogeneous mixture. 
 \item More complicated situations arise for $\varepsilon_t>0$ (implies $\Delta H>0$). On the one hand, even though $P_{AB}$ tends to decrease, compared to an ideal solution, for $-\Delta S \gg \Delta H$, the system still prefers mixing, as shown in Figure~\ref{fig:freeEnergyCurvesPhaseField}(c). This is due to the fact that the system free energy curve has a positive curvature at all points. However, on the other hand, when $-\Delta S$ is comparable or lower than $\Delta H$, the competition between these two establishes {a \textit{double well}} in the resultant free energy curve, as shown in Figure~\ref{fig:freeEnergyCurvesPhaseField}(d). The existence of the concavity in the curve indicates {that} a homogeneous mixture is energetically unfavorable in a certain region. Instead, phase-separated states of two distinguished colonies are expected in the system. This behavior of concavity is a {focus of} the presented work, which is extensively explored in the forthcoming sections.
 \end{enumerate}

Based on {the} mean-field approximation, the order parameter $c$ is assumed to {be} homogeneous over the entire system. However, {the} concavity in Figure~\ref{fig:freeEnergyCurvesPhaseField}(d) indicates that the homogeneous distribution does not ensure {the} system free energy to be minimal in a certain region. Therefore, the mean-field theory alone is inadequate to confiscate {the} non-homogeneous behavior. As an extension, {the} Ginzburg-Landau free-energy functional can be considered to address the concavity in the system free energy.

\section{Total free energy functional}
\label{sec:phaseFieldTotalFreeEnergyFunctional}
The mean-field-free energies discussed in the previous section \ref{sec:phaseFieldMeanFieldTheory} only consider an average of the bulk properties in a material. {This} type of free energy can be employed for a system that is infinite in extent and has uniform thermodynamic properties throughout. {This} restricts its applicability to {the extent of considering the} finiteness of the phase\footnote[2]{{A p}hase can be defined as a distinct and homogeneous form of matter{,} separated from other forms by its surface.}. More importantly, the relevance of {this} approach is devoid of the multiple phases and interfaces separating these phases, which are inexplicable with the mean-field approximation. However, {the} interfaces are the most important features, governing the formation of microstructures in most materials. Their migration and interaction are {central} to the manifold applications \cite{moelans2008an}. Therefore, this section aims to incorporate interfacial properties into the mean-field free energy, resulting in {the} Ginzburg-Landau free energy functional.

To facilitate {the incorporation of} interfacial properties into the system free energy and enable spatial{ly} dependent phase changes, a spatial dependency is incorporated into the order parameter field, $c(\boldsymbol{x}_\alpha)$\nomenclature{$\boldsymbol{x}_\alpha$}{spatial position of the $\alpha$ site}{,} defined in section \ref{sec:phaseFieldMeanFieldTheory}, which {was only} associated with bulk properties. To formulate a free energy which encompasses the interface, the Hamiltonian introduced in Eq. \eqref{eq:hamiltonian} must be redefined extensively. Correspondingly, the spatial-dependent Hamiltonian reads {as}
\begin{equation}
H\{c (\boldsymbol{x}_\alpha) \}= B_{0m} \sum_{\alpha=1}^{N_m} c(\boldsymbol{x}_\alpha) + B_{m}\sum_{\alpha=1}^{N_m} \sum_{\alpha' \neq \alpha}^{M_m} c(\boldsymbol{x}_{\alpha}) c(\boldsymbol{x}_{\alpha'}). \label{eq:hamiltonianGinzburgLandau}
\end{equation}
By considering the two-body term, the interaction between the phases in Eq. \eqref{eq:hamiltonianGinzburgLandau} encompasses both the bulk phases and {the} interface. Note that the contribution of the interface is not separate and implicit in the expression. The two-body term{,} represented as an interaction between two different interstitial sites $c(\boldsymbol{x}_{\alpha})$ and $c(\boldsymbol{x}_{\alpha'})$\nomenclature{$\boldsymbol{x}_{\alpha'}$}{spatial position of the $\alpha'$ site}{,} can be simplified {to}
\begin{align}
H\{c (\boldsymbol{x}_\alpha) \}&= B_{0m} \sum_{\alpha=1}^{N_m} c(\boldsymbol{x}_\alpha) + \frac{B_{m}}{2} \sum_{\alpha=1}^{N_m} \sum_{\alpha' \neq \alpha}^{M_m} \left(c^2(\boldsymbol{x}_\alpha)+ c^2(\boldsymbol{x}_{\alpha'}) - \left(c (\boldsymbol{x}_\alpha) -c (\boldsymbol{x}_{\alpha '}  )\right)^2 \right) \nonumber \\
&= B_{0m} \sum_{\alpha=1}^{N_m} c(\boldsymbol{x}_\alpha) + \frac{B_{m} M_m}{2} \sum_{\alpha=1}^{N_m} c^2(\boldsymbol{x}_\alpha) - \frac{B_{m}}{2} \sum_{\alpha=1}^{N_m} \sum_{\alpha' \neq \alpha}^{M_m} \left(c (\boldsymbol{x}_\alpha) -c (\boldsymbol{x}_{\alpha '}  )\right)^2. \label{eq:hamiltonianSpatialModified}
\end{align}
The first two terms on the right{-hand} side of Eq.~\eqref{eq:hamiltonianSpatialModified} represent the contribution of the individual phases, while the third term corresponds to the interaction across the phases, which can be attributed {to} the neighbor sites. In a cubic lattice setup, wherein the coordination number $M_m =6$, the interaction of any site $c_\alpha$ is restricted to its neighbor sites, front $c(\boldsymbol{x}_{\alpha F})$\nomenclature{$\boldsymbol{x}_{\alpha F}$}{neighbor on the front side of the $\alpha$ site}, back $c(\boldsymbol{x}_{\alpha B})$\nomenclature{$\boldsymbol{x}_{\alpha B}$}{neighbor on the back side of the $\alpha$ site}, top $c(\boldsymbol{x}_{\alpha T})$\nomenclature{$\boldsymbol{x}_{\alpha T}$}{neighbor on the top side of the $\alpha$ site}, bottom $c(\boldsymbol{x}_{\alpha O})$\nomenclature{$\boldsymbol{x}_{\alpha O}$}{neighbor on the bottom side of the $\alpha$ site}, right $c(\boldsymbol{x}_{\alpha R})$\nomenclature{$\boldsymbol{x}_{\alpha R}$}{neighbor on the right side of the $\alpha$ site}, and left $c(\boldsymbol{x}_{\alpha L})$\nomenclature{$\boldsymbol{x}_{\alpha L}$}{neighbor on the left side of the $\alpha$ site} sites. Therefore, the expression in the third term of Eq. \eqref{eq:hamiltonianSpatialModified} can be expanded as
\begin{align}
\sum_{\alpha' \neq \alpha}^{M_m} \left(c (\boldsymbol{x}_\alpha) -c (\boldsymbol{x}_{\alpha '}  )\right)^2 = \mathring{a}^2 \Bigg{\{} &\Bigg{(}\frac{\left( c (\boldsymbol{x}_\alpha) - c (\boldsymbol{x}_{\alpha F})  \right)^2}{\mathring{a}^2} + \frac{\left( c (\boldsymbol{x}_\alpha) - c (\boldsymbol{x}_{\alpha R})  \right)^2}{\mathring{a}^2} \nonumber\\
 + &\frac{\left( c (\boldsymbol{x}_\alpha) - c (\boldsymbol{x}_{\alpha T})  \right)^2}{\mathring{a}^2} \Bigg{)} + \Bigg{(} \frac{\left( c (\boldsymbol{x}_\alpha) - c (\boldsymbol{x}_{\alpha B})  \right)^2 }{\mathring{a}^2}  \nonumber \\
 + &\frac{\left( c (\boldsymbol{x}_\alpha) - c (\boldsymbol{x}_{\alpha L})  \right)^2}{\mathring{a}^2} + \frac{\left( c (\boldsymbol{x}_\alpha) - c (\boldsymbol{x}_{\alpha O})  \right)^2 }{\mathring{a}^2} \Bigg{)} \Bigg{\}}, \label{eq:twoTermsGinzburgLandau}\\
 \approx \mathring{a}^2 \big{\{} &2  \vert \boldsymbol{\nabla} c(\boldsymbol{x}_\alpha)  \vert^2 \big{\}}. \label{eq:twoTermsGinzburgLandauFinal}
\end{align}
Here $\boldsymbol{\nabla}(\bullet)$ indicates {the} gradient of the respective field and $\mathring{a}$\nomenclature{$\mathring{a}$}{lattice parameter} is the lattice parameter, which is the distance of {the} neighboring lattice sites. In the present approximation, it is assumed to be a constant. Such an approximation might lead to the omission of several spatial microscopic details and misestimating properties associated with it. However, for the materials which consist of a direction{ally} independent lattice parameter, which is of interest to the present work, the employed assumption is valid. Each term in the round brackets of Eq.  \eqref{eq:twoTermsGinzburgLandau} represents the magnitudes of the one-sided gradients of the order parameter in discrete form, at the point $\alpha$, which is replaced by a continuous description in Eq.~\eqref{eq:twoTermsGinzburgLandauFinal}.

Based on the approach provided by Ginzburg and Landau \cite{ginzburg1950k}, the order parameter, which is defined non-continuously, considering the microscopic properties in the Hamiltonian Eq.~\eqref{eq:hamiltonianGinzburgLandau}, can be transformed to spatially continuous, in the mesoscopic length scale. {In the mesoscopic description}, the discrete microscopic summation over "$\alpha$" can therefore be replaced by the continuous operators of the form
\begin{equation}
\sum_{\alpha=1}^{N_m} \Rightarrow \int_{V_\Omega} \frac{\textrm{d}{V_\Omega}}{\mathring{a}^3}{,} \label{eq:mesoscopicDescription}
\end{equation}
where $V_\Omega$\nomenclature{$V_\Omega$}{volume of the system or simulation domain} is the volume of the system and $\int_{V_\Omega} (\bullet) \textrm{d}V_\Omega$ is the integral over {the} volume of the respective function. The devision by $\mathring{a}^3$ is intended to encapsulate the microscopic properties in the mesoscopic limit, by considering the volume of an element. This description {enables} the transition of {the} piecewise description of the properties to the continuum limit. By substituting Eqs.  \eqref{eq:twoTermsGinzburgLandauFinal} and  \eqref{eq:mesoscopicDescription} into the Hamiltonian Eq.~\eqref{eq:hamiltonianGinzburgLandau}, the resultant continuously defined Hamiltonian, which is identified as an internal energy, can be expressed as
\begin{equation}
E(c (\boldsymbol{x})) = \int_{V_\Omega} \Big{\{}B_{0m} c(\boldsymbol{x}) + \frac{B_{m} M_m}{2}  c^2(\boldsymbol{x}) - B_{m} \mathring{a}^2 \vert \boldsymbol{\nabla} c(\boldsymbol{x})  \vert^2 \Big{\}} \frac{\textrm{d}{V_\Omega}}{\mathring{a}^3}.\label{eq:hamiltonianFinalContinuumGinzburgLandau}
\end{equation}
{The first two terms in the equation are the bulk terms, which only consist of the current position dependency}, while the third term is the gradient term, which encloses the interfacial properties across the neighbor phases. Note that the two-body interaction term in the Hamiltonian Eq.~\eqref{eq:hamiltonianGinzburgLandau} is replaced by the gradient term in Eq.~\eqref{eq:hamiltonianFinalContinuumGinzburgLandau}. In addition, this term is nearly vanishing in the bulk phases and only varies significantly at the interfaces, where there is appreciable change of the order parameters. Therefore, this term can be linked with the surface energy of the phases.  In addition to this contribution, the role of the entropy $\Delta S (c(\boldsymbol{x}), T(\boldsymbol{x}))$ should be included to translate the internal energy into system free energy. Conventionally, the total number of microscopic configurations correlates to the entropy term, which can be formulated by employing Sterling's approximation. If the free energy contribution of the bulk phase is represented by $f (c(\boldsymbol{x}), T(\boldsymbol{x}))$, after the inclusion of the entropy, the resulting free energy functional is expressed as
\begin{eqnarray}
F\left(c(\boldsymbol{x}),T(\boldsymbol{x}),\boldsymbol{\nabla}c(\boldsymbol{x})\right)=& \int_{V_\Omega} \Big(  X_1 c(\boldsymbol{x}) + X_2 c(\boldsymbol{x}) \left(1-c(\boldsymbol{x})\right) + \frac{1}{2} \kappa\vert \boldsymbol{\nabla}c(\boldsymbol{x}) \vert^2 \nonumber \\ 
& + \frac{T(\boldsymbol{x})}{T_{\textrm{ref}}}\left(c \ln(c) + (1-c) \ln(1-c) \right) \Big) \textrm{d}{V_\Omega}.\\
=& \int_{V_\Omega} \left(f(c(\boldsymbol{x}),T(\boldsymbol{x}))+ \frac{1}{2} \kappa \vert \boldsymbol{\nabla}c(\boldsymbol{x}) \vert^2 \right) \textrm{d}{V_\Omega}{,} \label{eq:GinzburgLandauFreeEnergyFunctional}
\end{eqnarray}
where $\kappa k_B T_{\textrm{ref}} = -2B_m \mathring{a}^2 $\nomenclature{$\kappa$}{gradient energy coefficient} is the gradient energy coefficient. This formulation, which is based on a mesoscopic description, consistently allows to represent the properties of the bulk phases ($f(c(\boldsymbol{x}),T(\boldsymbol{x}))$) and the interfaces separating them ($ \kappa/2 \vert \boldsymbol{\nabla}c(\boldsymbol{x}) \vert^2$). The derived Eq.~\eqref{eq:GinzburgLandauFreeEnergyFunctional} is commonly referred to as total free energy functional, where the order parameter spatially varies continuously. In this sense, the discrete description of microscopic properties, in the form of the Hamiltonian \eqref{eq:hamiltonianGinzburgLandau}, are transformed to the mesoscopically defined, spatially continous free energy functional \eqref{eq:GinzburgLandauFreeEnergyFunctional}, in the present section. However, this form of Hamiltonian is not followed further. Instead, the free energy functional serves as a starting point for many phenomena which are modeled using the phase-field methodology \cite{chen2002phase, nestler2011phase}.

\section{Chemical potential and equilibrium interfaces}
In thermodynamics, it is of interest to know the change in free energy, $\delta F(c,\boldsymbol{\nabla}c)$, of a given system with some \textit{activity}, $\delta c$. The activity can be regarded as either an addition (or removal) of species or a growth (or shrinkage) in at least one of the phases. A chemical potential is introduced as a proportionality constant{, so as} to relate the activity with its effect on the free energy \cite{porter2009phase}. Based on a similar rationale, the chemical potential can be expressed as
\begin{equation}
\mu = \frac{\delta F(c,\boldsymbol{\nabla}c)}{\delta c} = \frac{\partial \left(f(c) + \frac{1}{2} \kappa \vert \boldsymbol{\nabla} c \vert^2\right)}{\partial c} - \boldsymbol{\nabla} \cdot \frac{\partial \left(f(c) + \frac{1}{2} \kappa \vert \boldsymbol{\nabla} c \vert^2\right)}{\partial \boldsymbol{\nabla} c}. \label{eq:phaseFieldChemicalPotentialGeneral}
\end{equation}
Here $\delta (\bullet)/\delta c$ is the variational derivative of the free energy functional
\begin{equation}
F(c,\boldsymbol{\nabla}c) =\int_{V_\Omega} \left(f(c) + \frac{1}{2} \kappa \vert \boldsymbol{\nabla} c \vert^2\right) \textrm{d}{V_\Omega},
\end{equation} 
 with the order parameter $c$. Even though the functional still varies spatially and may contain temperature dependencies similar to Eq.~\eqref{eq:GinzburgLandauFreeEnergyFunctional}, these variables are implicitly incorporated and will be omitted from the expression for a convenience, without loss of generality.
 
 Since the activity is related to the chemical potential from Eq.~\eqref{eq:phaseFieldChemicalPotentialGeneral}, the activity is another means of describing the state of a system. If the activity or chemical potential is very low, the matter is reluctant to change its form, place or orientation, and the system is in equilibrium. In equilibrium, {in which all competing influences are balanced,} the system is stable. In the present sense, this is due to {the} negligible chemical potential{. T}he variational derivative of the free energy functional\cite{boettinger2002phase} is expressed as
 \begin{equation}
0 = \frac{\delta F(c,\boldsymbol{\nabla}c)}{\delta c} = \frac{\partial f(c)}{\partial c} - \kappa \boldsymbol{\nabla} \cdot \boldsymbol{\nabla}c.
 \end{equation}
 This equilibrium is sustained for a particular system configuration. However, any finite activity might interrupt the developed equilibrium and the chemical potential will be nonzero. {This} demands an investigation to minimize the free energy functional. Eventually, the system might attain a new equilibrium with an improvised configuration. To dissect these characteristics, it is crucial to understand some specific characteristics of the order parameter, which is described in the next section.

\section{Phase field order parameters}
It is important to note that the scalar state variable{, which is identified as an order parameter $c$, in the Ginzburg-Landau theory \cite{ginzburg1950k},} possesses a characterizing property which aids in distinguishing different phases. This parameter represents the fraction of the system, exhibiting a particular property or a state, which changes spatially along with the system. For instance, in the solid to liquid transformation, an order parameter can be assigned to differentiate the solid and the liquid phase. Eventually, considering the transformation, wherein the solid changes entirely to liquid, the order parameter, pertaining to the original system, completely vanishes. In other words, the order parameter {is considered as a non-conserved parameter}. Consequently, it becomes unphysical to directly relate this order parameter with the concentration, where the conservation of mass should be achieved. 

In order to formulate {the} free energy functional for a conserved parameter, Cahn and Hilliard \cite{cahnhilliard1958free} derived an expression, identical to Eq.~\eqref{eq:GinzburgLandauFreeEnergyFunctional}, for concentration, , in their pioneering work, similar to the Ginzburg-Landau \cite{ginzburg1950k} free energy functional, by replacing the non-conserved order parameter. Despite the differences in the nature of the thermodynamical properties, the free energy functional alone does not discriminate against the conservancy of the order parameter. Therefore, the expressions derived in sections~\ref{sec:phaseFieldMeanFieldTheory} and  \ref{sec:phaseFieldTotalFreeEnergyFunctional} are equally applicable to both conserved and non-conserved order parameters. In addition, it is important to note that some salient features remain identical to these order parameters, such as the fact that it assumes a definite value in the bulk phases, while it spatially varies in the narrow, but finitely defined interface regions, which separate the phases.

As the free energy functionals could be considered identical to conserved and non-conserved order parameters, for non-equilibrium dynamics, there should be another means to separate these properties. This can be achieved through the dynamic equations, which are the time-dependent evolution of the order parameter in a non-equilibrium. The kinetics of these quantities are typically formulated as a Langevin-type equation \cite{lemons1997paper}, which evolves to minimize the free energy functional. In the next sections, these kinetic equations are consequently described for conserved and non-conserved order parameters.

\section{Cahn-Hilliard equation}
\label{sec:phaseFieldCahnHilliard}
\begin{figure}
\begin{center}
\includegraphics[scale=0.60]{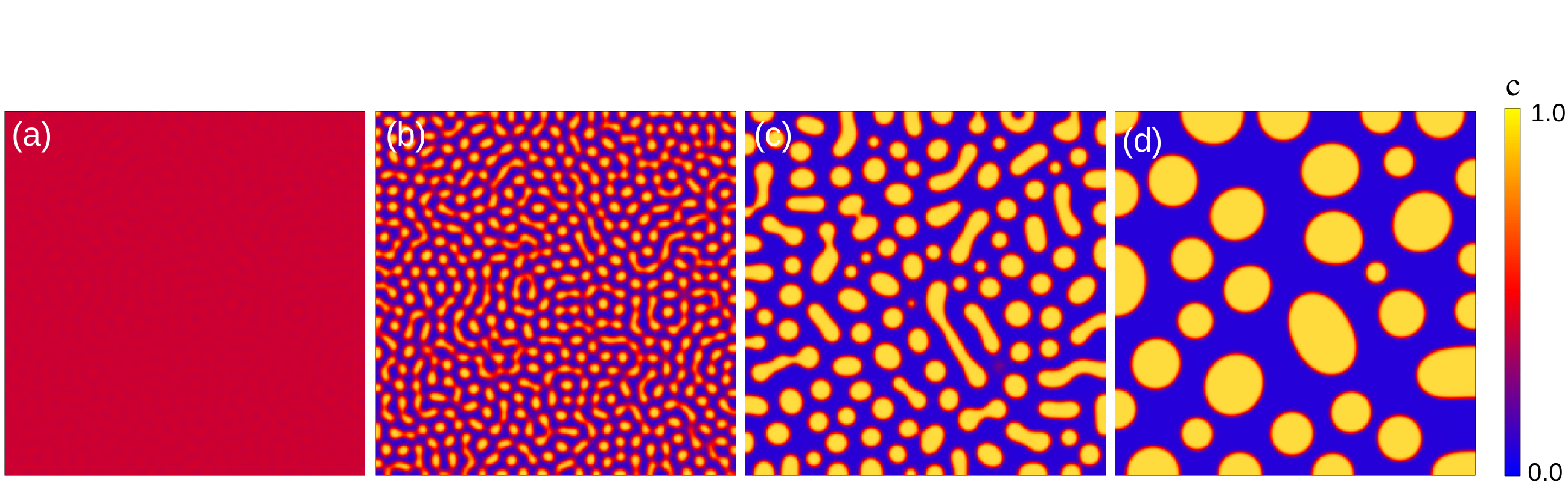}
\end{center}
\caption[Schematic of an evolution, containing a conserved order parameter, concentration $c$.]{Schematic of an evolution, containing a conserved order parameter, concentration $c$, during spinodal decomposition. {The figures from left to right} show {the evolution of time,} with the coarsening of smaller grains to larger ones.} \label{fig:phaseFieldCahnHilliard}
\end{figure}

Dynamic equations for the order parameter fields are called conserved if they take to be of a flux-conserving form. For instance, when the order parameter is required to be conserved, in terms of variables, like the concentration of the species during spinodal decomposition, as shown in Figure~\ref{fig:phaseFieldCahnHilliard}. This implies that a spatial integral of the field on the entire volume should a constant for a closed system. Therefore, the \textit{concentration} will be assigned in the forthcoming descriptions to identify the conservative property of the order parameter. Under a non-equilibrium condition, in addition to spatial dependency, the concentration exhibits a temporal evolution. The concentration evolution is modeled through the diffusion equation. In his pioneering work, Fick \cite{fick1855on} provided a time-dependent concentration evolution in which the fluxes $\boldsymbol{J}$\nomenclature{$\boldsymbol{J}$}{effective diffusional mass fluxes} should obey the conservation of mass in the form
\begin{equation}
\frac{\partial c}{\partial t} = -\boldsymbol{\nabla}\cdot \boldsymbol{J}{,} \label{eq:phaseFieldFicksTimeEvolution}
\end{equation}
where $t$\nomenclature{$t$}{time} is the time and $\boldsymbol{\nabla} \cdot (\bullet)$ denotes the divergence of the vector. The diffusional fluxes are the driving force that stimulates the matter movement. In Fick's analogy, these forces are directly associated with a gradient of the chemical potential, which is linked to a gradient of the concentration. Hence, the fluxes can be expressed as
\begin{equation}
\boldsymbol{J} = -D_0 \boldsymbol{\nabla}c{,} \label{eq:phaseFieldFicksLaw}
\end{equation}
where $D_0$ denotes the diffusion coefficient, which is a material property. This equation is crucial in the dilute solution limit. Note that similar to these expressions, Fourier's law of heat conduction in pure materials can be derived by considering the temperature and the thermal conductivity. However, this lacks to separate the bulk phase and is inadequate to provide any information about the interfaces, which is central to the phase-field community \cite{nestler2011phase}. For such a system, the chemical potential should be based on {the} free energy functional in Eq.~\eqref{eq:phaseFieldChemicalPotentialGeneral}.

To account for such effects, Cahn and Hilliard \cite{cahnhilliard1958free} proposed a model of phase separating binary systems. This equation can be derived from the basic laws of classical thermodynamics, with the consideration of the diffusion between two types of A and B species, which consist of the concentrations $c_A$\nomenclature{$c_A$}{concentration of A species} and $c_B$\nomenclature{$c_B$}{concentration of B species}, respectively. According to the linear phenomenological relations of the thermodynamics of fluxes\nomenclature{$\boldsymbol{J}_A$}{mass flux of A species}\nomenclature{$\boldsymbol{J}_B$}{mass flux of B species},
\begin{eqnarray}
\boldsymbol{J}_A = -M_{AA}\boldsymbol{\nabla} \mu_A - M_{AB} \boldsymbol{\nabla} \mu_B,\label{eq:CHFluxComponentA}\\
\boldsymbol{J}_B = -M_{BA}\boldsymbol{\nabla} \mu_A - M_{BB} \boldsymbol{\nabla} \mu_B,\label{eq:CHFluxComponentB}\
\end{eqnarray}
where $M_{AA}, M_{AB}, M_{BA}, $\nomenclature{$M_{AA}$}{mobility of the A species, due to the interaction with the A species}\nomenclature{$M_{AB}$}{mobility of the A species, due to the interaction with the B species}\nomenclature{$M_{BA}$}{mobility of the B species, due to the interaction with the B species}\nomenclature{$M_{BB}$}{mobility of the B species, due to the interaction with the B species} and $M_{BB}$ are the mobilities, due to interaction among the A and B species, $\mu_A$\nomenclature{$\mu_A$}{chemical potentials of the A species} and $\mu_B$\nomenclature{$\mu_B$}{chemical potentials of the B species} are the chemical potentials of A and B, respectively. The net flux of the component B, opposed by component A, can be expressed as
\begin{equation}
\boldsymbol{J} = \boldsymbol{J}_B-\boldsymbol{J}_A. \label{eq:CHNetFlux}
\end{equation}
A convenient notation for the binary system reads as $c=c_B=1-c_A$, where $c$ refers to the concentration. Then the Gibbs-Duhem relationship reads as
\begin{equation}
(1-c)\boldsymbol{\nabla} \mu_A + c \boldsymbol{\nabla} \mu_B = 0. \label{eq:CHGibbsDuhemRelation}
\end{equation}
Substituting Eqs. \eqref{eq:CHFluxComponentA}, \eqref{eq:CHFluxComponentB}, and \eqref{eq:CHGibbsDuhemRelation} into Eq. \eqref{eq:CHNetFlux}, the net flux can be expressed as
\begin{equation}
\boldsymbol{J} = -M \boldsymbol{\nabla} \mu, \label{eq:phaseFieldCahnHilliardFlux}
\end{equation}
where the mobility $M$\nomenclature{$M$}{effective diffusional mobility} is given by $M=(1-c)(M_{BB}-M_{AB})+c(M_{AA}-M_{BA})$ and the chemical potential, $\mu=\mu_B -\mu_A$ can be expressed as the difference from B to A. The chemical potential can be obtained from the variational derivative of the system free energy functional, expressed in Eq.~\eqref{eq:phaseFieldChemicalPotentialGeneral}. Substituting the flux equation \eqref{eq:phaseFieldCahnHilliardFlux} into the conservation equation \eqref{eq:phaseFieldFicksTimeEvolution}, the resultant expression can be written in the form of
\begin{equation}
\frac{\partial c}{\partial t} = \boldsymbol{\nabla}\cdot \left(M \boldsymbol{\nabla} \mu \right), \label{eq:phaseFieldCahnHilliard}
\end{equation}
which is designated as Cahn-Hilliard equation. To understand the analogy of this equation, two components are considered in the present section, for simplicity. However, this equation can be extended to consider the multiple components along the lines of Ref. \cite{nauman1994morphology}, employed for a ternary system. Thereafter, the Cahn-Hilliard equation became prominent in numerous applications, such as spinodal decomposition \cite{cahn1961spinodal}, intercalation in lithium-ion batteries \citep{anand2012a}, and electromigration in metallic conductors \cite{bhate2000diffuse}.

\section{Allen-Cahn equation}
\label{sec:phaseFieldAllenCahn}
\begin{figure}
\begin{center}
\includegraphics[scale=0.56]{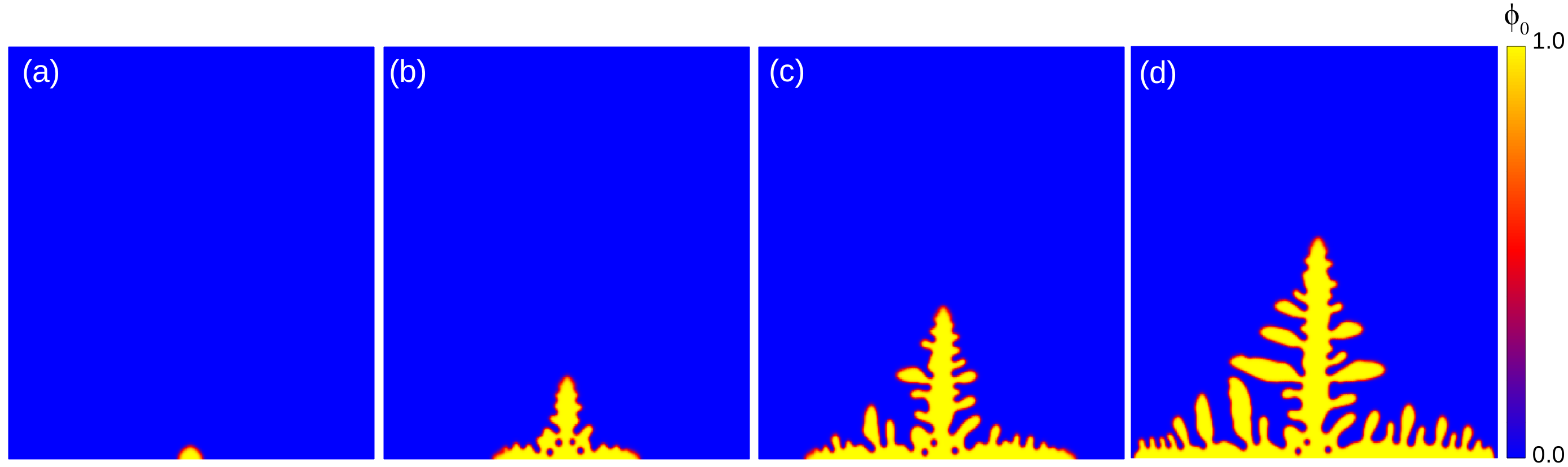}
\end{center}
\caption[Schematic of an evolution containing a non-conserved domain parameter $\phi_0$]{Schematic of an evolution containing a non-conserved domain parameter $\phi_0$ during crystal growth, as the solidification of a pure melt. The figures from left to right show the time evolution, with the formation and growth of a dendritic pattern.}\label{fig:phaseFieldAllenCahn}
\end{figure}

The conservation of an order parameter is not always desired. In fact, some of the phase transformations are driving to the energy minimum, through the loss (or gain) of phase fractions. For instance, crystal growth, due to the solidification of a pure melt, is one of the examples of a non-conserved order parameter, as shown in Figure~\ref{fig:phaseFieldAllenCahn}. Therefore, the fraction of a crystal over the entire domain might not be constant during solidification. To associate the non-conservative properties, {this parameter} will be referred to as the domain (indicator) parameter $\phi_0$. In accordance with the non-equilibrium thermodynamics \cite{prigogine1961introduction}, the temporal evolution of the domain parameter is directly related to the variation derivative of the functional expressed in Eq.~\eqref{eq:GinzburgLandauFreeEnergyFunctional}. A relaxation coefficient $\tau_p$\nomenclature{$\tau_p$}{relaxation coefficient} can be introduced as a propotionality constant. Therefore, the resultant time-dependent Ginzburg-Landau equation  \cite{schmid1966time, schmidt1968onset} can be expressed as
 \begin{equation}
 \frac{\partial \phi_0}{\partial t} = -\frac{1}{\tau_p} \frac{\delta F(\phi_0, \boldsymbol{\nabla}\phi_0)}{\delta \phi_0}. \label{eq:phaseFieldAllenCahn}
\end{equation}  
Note that the free energy functional $F(\phi_0, \boldsymbol{\nabla}\phi_0)$ can be obtained similar to Eq.~\eqref{eq:GinzburgLandauFreeEnergyFunctional}. Independantly, Allen and Cahn \cite{allen1979a} derived a similar expression to the time-dependent Ginzburg-Landau equation \eqref{eq:phaseFieldAllenCahn}. In the phase-field community, this equation is thus referred to as the Allen-Cahn equation \cite{moelans2008an}.

{As discussed} in section~\ref{sec:phaseFieldCahnHilliard}, the Cahn-Hilliard equation \eqref{eq:phaseFieldCahnHilliard} seems to be the obvious choice for the order parameters that observe the mass conservation. However, the Allen-Cahn equation \eqref{eq:phaseFieldAllenCahn} is employed for a few of such cases, by incorporating specific features like the redistribution-energy technique \cite{amos2019limitations}. This provision contributes volume-preserving properties to the order parameter, by compensating a bulk-phase fraction by means of addition or subtraction \cite{nestler2008phase}. One of the primary hypotheses for such an approach is that the former contains a fourth-order spatial derivative, while the latter has second-order spatial derivative. In many numerical techniques like finite-difference schemes, the computational time multiplies by manifolds, with an increase in the order of the spatial derivatives. Therefore, the Allen-Cahn equation might reduce the computational efforts and improves the efficiency of the algorithms. However, to preserve the volume anomalously, an additional term in the free energy allows counteraction to the source or the sink of the phases. It is identified that this approach might constitute a physically improbable and thermodynamically incoherent transformation mechanism \cite{amos2019limitations}. However, this establishes an appropriate and theoretically consistent final microstructure in equilibrium. Therefore, the volume-preserving Allen-Cahn equation can only be employed where the final microstructure is of a prime objective, compared to the transformation mechanism.

The elegance of the Cahn-Hilliard and the Allen-Cahn equations lies in their applicability. As they are directly related to the free energy, which can be readily extended to the multiphysical phenomenon, by considering a respective contribution in a free energy Liaponuv functional \cite{novick1984nonlinear}. The challenging research goal is then to derive such functionals, which can sufficiently capture the physical phenomenon. In addition, it is also plausible to compose an algorithm that embraces both conserved and non-conserved order parameters simultaneously \cite{halperin1974renormalization} for a phase transformation such as solidification of a binary Fe-C melt, which requires a non-conserved state variable and conserved concentrations of Fe and C components or electromigration-induced grain boundary grooving \cite{mukherjee2019electric}. Therefore, these two equations, commonly known as phase-field equations, are employed extensively for numerous applications \cite{chen2002phase, nestler2011phase, moelans2008quantitative}.

\section{Conclusion}

The specific feature of phase-field modeling is the introduction of the diffuse interface, replacing the sharp one. One common practice in all phase-field models is that they consider one or several phase-fields, which describe the physical states such as the density, the form, the orientation or the order of any geometry under investigation. The physical state variable, the phase field, assumes a fixed and predetermined value in the bulk phases, while varying smoothly from one bulk value to the other, along with the interface of finitely defined width. This uniquely defined diffuse interface provides a remarkable advantage of implicit interface tracking, by averting the cumbersome techniques, as in the case of the sharp-interface models. In the next chapters, the phase-field models are derived on the basis of the principles described in the present chapter for the concerned phenomena, i.e., two-phase coexistence in cathode materials of lithium-ion batteries in Chapter~\ref{chapter:phaseFieldModelLIB} and inclusion motion under electromigration in Chapter~\ref{chapter:phaseFieldModelElectromigration}. 
\afterpage{\blankpagewithoutnumberskip}
\clearpage
\chapter{Phase-field model for cathode particles of Lithium-ion battery}
\label{chapter:phaseFieldModelLIB}

During the charging and discharging of a battery, electrons transfer from the outer circuit {and} the positively charged lithium ions shuttle from one electrode to the other electrode through the electrolyte internally. The three-dimensional structural network of the electrode hosts the Li-ions, and the reversible insertion process of the Li-ions into the host compounds takes place at both ends of the electrodes \cite{singh2008intercalation}. As discussed in section~\ref{section:twoPhaseCoexistence}, most of the electrodes of lithium-ion batteries, such as lithium manganese oxide (LMO) \cite{ohzuku1990electrochemistry} observe the mechanism of phase separation during the insertion, which is of scientific interest due to its direct implications as an important source of capacity fade. {Also}, the coexistence of two phases in LMO is extensively explored in experiments \cite{ohzuku1990electrochemistry, vanderven2000phase, liu1998mechanism, li2000phase, yang1999situ}. Systematic numerical modeling of phase separation may complement to the scientific understanding of the mechanism and helps to optimize the process parameters efficiently and economically \cite{malik2013critical}.  

To represent the two-phase coexistence in LMO cathode particles, a phase-field model is derived based on Cahn-Hilliard equation in Section~\ref{section:phaseFieldModelLithiumIonBattery}. Thereafter, the boundary conditions are explained in Sections~\ref{sec:smoothedBoundaryMethod} and~\ref{section:boundaryConditionsSeparator}. Followed by the implementation strategies, which are presented in Section~\ref{section:LIBimplementationStrategies}. Finally, the chapter concludes with the outline and model applicability in Section~\ref{section:lithiumIonBatteryPFMConclusion}.

\section{Model description for two-phase coexistence}
\label{section:phaseFieldModelLithiumIonBattery}
The lithium species migrate inside the particles after being deposited at the interface of the electrodes. In a closed system like lithium-ion batteries, these species obey mass conservation laws. Therefore, a concentration parameter ${c}(\boldsymbol{x},t) \in (0,1)$ can be introduced to indicate the local state of the diffusion with time $t$ and spatial coordinates $\boldsymbol{x}$. Note that the concentration $c$ in the present case is the mole fraction of occupied sites by the lithium species to the total available sites in the electrode particles \footnote[2]{Alternatively, the concentration can be defined as a molar concentration of the total lithium species expressed in moles divided by the volume under consideration \cite{santoki2018phase}. In this case, the concentration $c$ varies from $0 <c < c_{\textrm{max}}$, where $c_{\textrm{max}}$ denotes the maximum concentration in the cathode materials.}. The two-phase {coexistence} in the particles is due to the nature of phase-separating systems. Within the miscibility gap of a phase-separating system, spinodal decomposition can take place at intermediate values of $c$, in which the system is unconditionally unstable {with} spontaneous phase decomposition, which initiates Li-rich and Li-poor phases. The free energy densities of such a system are described in the following paragraphs.

\subsection{Free energy densities}
{The occurrence of such phase decomposition can be understood by the diffusion of Li in a manner that allows it to instigate a Li-poor and a Li-rich region. These two regions are separated by a steep concentration gradient called an interface.} Furthermore, the motion of the interface is governed by the Li flux at the surface of the particle. Therefore, in the electrochemical systems, the change in the concentration distribution leads to the alteration of the concentration gradient and, as a consequence, to the reshaping of the concentration into a changed minimum energy state. In turn, the concentration and the concentration gradient are implicitly coupled. The combined change in the concentration gradient $\boldsymbol{\nabla} c$ and the concentration $c$ are related by the system free energy functional by integrating over the volume ${V_\Omega}$ and the surface ${S_\Omega}$\nomenclature{$S_\Omega$}{surface of the domain} of the {form,}
\begin{equation}
F(c,\boldsymbol{\nabla}c) =\int_{V_\Omega} \left\{f(c) + \frac{1}{2} \kappa \vert \boldsymbol{\nabla} c \vert^2\right\} \textrm{d}{V_\Omega} + \oint_{S_\Omega} f^{\textrm{S}}(c) \mathrm{ d }{S_\Omega}, \label{eq:libFreeEnergyFunctional}
\end{equation} 
to model the process of phase separation. Here ${f}({c})$ is the chemical free energy density, $1/2\kappa \vert \boldsymbol{\nabla} {c}\vert^2$ denotes the gradient energy density, and $f^{\textrm{S}}(c)$\nomenclature{$f^{\textrm{S}}$}{surface energy density} represents the surface energy. In addition, $\kappa$ is the gradient energy coefficient, which controls the interface thickness between the adjacent Li-rich and Li-poor phases.

The model focuses on a representative elementary volume (REV)\nomenclature{REV}{representative elementary volume} of an electrode. The set of indicator parameters \nomenclature{$\boldsymbol{\psi}$}{indicator parameter set}$\boldsymbol{\psi} = [\psi_1,...,\psi_N] $, $\psi_{a} \in \{0,1\}$, $ \forall a \in \{1,\ldots,N\}$\nomenclature{$a$}{index of the phase}\nomenclature{$N$}{total number of phases} is defined to recognize different phases. For instance, $\psi_1=1$\nomenclature{$\psi_a$}{stationary indicator parameter for phase $a$} (and $\psi_a=0$, where $a \ne 1$) represents electrode particles, while $\psi_2=1$ (and $\psi_a=0$, where $a \ne 2$) indicates the electrolyte. Thus, the indicator parameter identifies multiple regions of the same physical properties as a single phase. In general, the chemical free energy density can be expressed as 
\begin{equation}
{f}({c}) = \sum_{a=1}^N h(\psi_{a}) {f}_{a}^{\mathrm{chem}}({c})
\end{equation}
where $h(\psi_{a})$\nomenclature{$h$}{interpolation function} is the interpolation function and ${f}_{a}^{\mathrm{chem}}(c)$\nomenclature{${f}_{a}^{\mathrm{chem}}$}{regular solution free energy density of phase $a$} denotes the chemical-free energy density of the phase $a$. Here, as the phases are defined by a step function of indicator parameters, meaning, the interface between the different phases is sharply defined, a first-order interpolation $h(\psi_a) =\psi_a$ can be utilized \cite{santoki2019role}.

\begin{figure}[t]
\begin{center}
\includegraphics[scale=0.70]{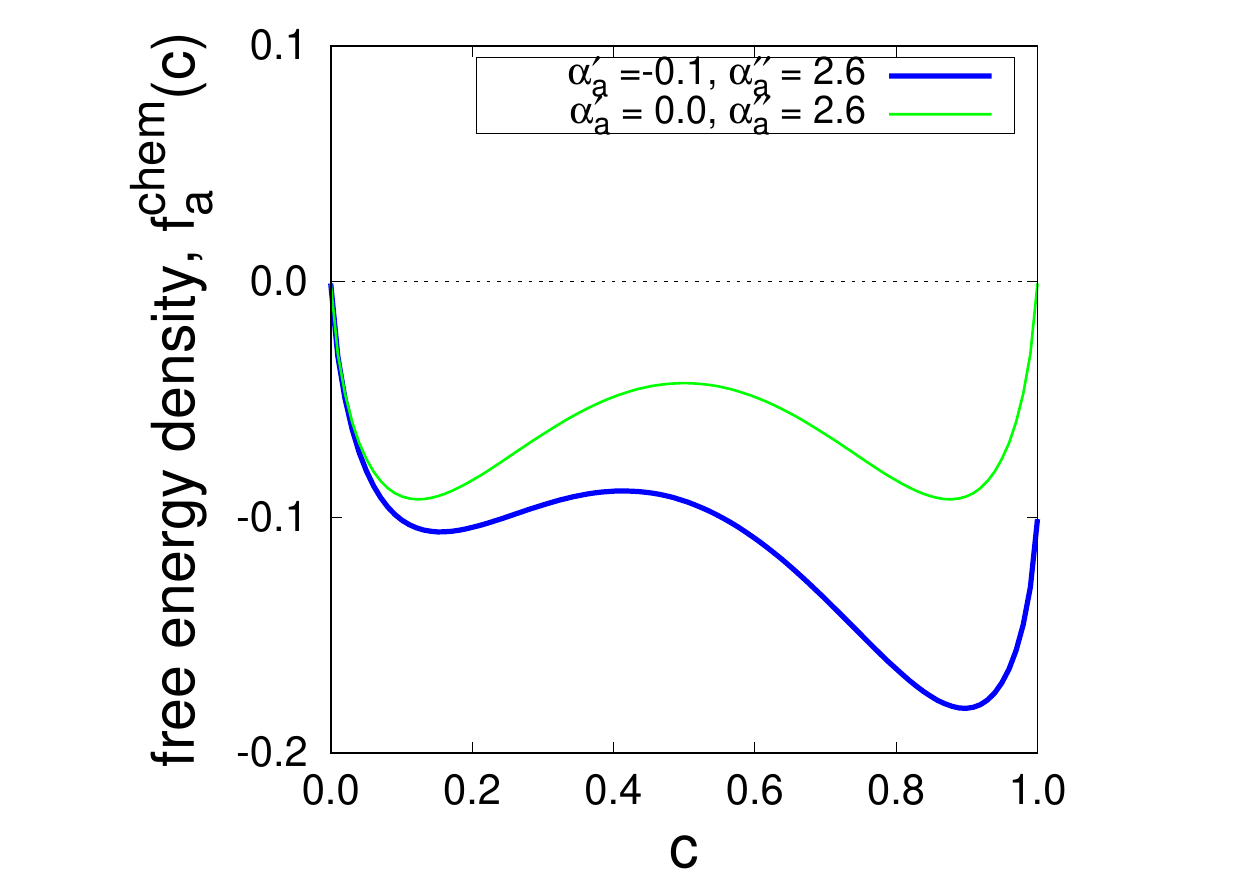} 
\caption[Effect of the parameter $\alpha'_{a}$ on the free energy curves.]{(a) Effect of the parameter $\alpha'_{a}$ on the free energy curves. The green curve represents the diminishing $\alpha'_{a}$ where two energy wells are at same energy height and the curve is symmetric with $c=0.5${. The blue curve indicates} a nonzero free energy coefficient $\alpha'_{a}$ considered for the two-phase coexistence in intercalating materials, otherwise stated.}\label{fig:freeEnergyCurvesLIB}
\end{center}
\end{figure}

The {chemical-free} energy density ${f}_{a}^{\mathrm{chem}}(c)$ is a function of local field parameters, such as the local lithium-ion concentration $c$, externally fixed parameters like the absolute {temperature $T$,} and the material properties like internal energy coefficients. %
The coexistence of lithium-rich and poor phases in several Li intercalation compounds is of scientific importance to {investigate} the phase separating behavior. To capture this effect, in the present study, the regular solution model is considered for the expression of the chemical free energy density equivalent to Eq.~\eqref{eq:freeEnergyLandauFinal} of the form,
\begin{equation} \label{eq:libFreeEnergyDensity}
{f}_{a}^{\mathrm{chem}}({c}) = \alpha'_{a} c + \alpha''_{a} c(1-c) + \frac{T}{T_{\textrm{ref}}}\left\{c \ln(c) + (1-c) \ln(1-c) \right\}.
\end{equation}
where $\alpha'_a$\nomenclature{$\alpha'_a$ and $\alpha''_a$}{regular solution parameters of phase $a$} and $\alpha''_a$ are the regular solution parameters associated with the internal energy, $T_\mathrm{ref}$ denotes the reference temperature and $\mathrm{ln}(\bullet)$ is the natural logarithm of the respective system variable. The effect of variation in $\alpha''_{a}$ on the free energy density curves is discussed in section~\ref{sec:phaseFieldMeanFieldTheory}. However, to {perceive} the significance of the parameter $\alpha'_a$, consider a Figure~\ref{fig:freeEnergyCurvesLIB}. When $\alpha'_a=0$, the free energy curve is symmetric to the center $c=0.5$. In addition, the two wells are situated at the same height. This can be justified when the free energy density of the all vacant sites $(c=0)$ is equals to the totally occupied sites $(c=1)$ by the lithium in the intercalating material. Otherwise, a nonzero $\alpha'_{a}$ should be considered, which regulates the relative heights of the two wells of the free energy, as shown in Figure~\ref{fig:freeEnergyCurvesLIB}. This free energy density \eqref{eq:libFreeEnergyDensity} is considered \cite{huttin2014phase} with nonzero $\alpha'_{a}$ and $\alpha''_{a}$ for the study of two-phase coexistence in the active particles of cathode material in lithium-ion batteries. Furthermore, the gradient energy coefficient can be expressed as an interpolation between the phases of the form,
\begin{equation}
\kappa = \sum_{a=1}^N  h(\psi_a) \kappa_{a} \label{eq:LIBGradientEnergyCoefficeintInterpolation}
\end{equation}  
where $\kappa_a$\nomenclature{$\kappa_a$}{gradient energy coefficient of phase $\alpha$} is the gradient energy coefficient of the phase $a$.
 
 \subsection{Surface Wetting and Chemical Potential}
In the phase-field formulation, the optimization of an objective functional \eqref{eq:libFreeEnergyFunctional}, formulated in terms of the energy densities is of primary concern. Towards such intent, performing the variational derivative of the free energy functional and rearranging the terms,
\begin{eqnarray}
\delta F(c,\boldsymbol{\nabla}c) &=& \int_{V_\Omega} \Bigg[ \frac{\delta  f(c)}{\delta c} \delta c + \frac{\delta  \left\{1/2 \kappa \vert  \boldsymbol{\nabla}c \vert^2\right\}}{\delta c} \delta c  \Bigg] \mathrm{d}{V_\Omega} + \oint_{S_\Omega} \frac{\delta  f^{\textrm{S}}(c)}{\delta c} \delta c \mathrm{ d }{S_\Omega}  \nonumber\\
&=& \int_{V_\Omega} \Bigg[ \frac{\partial  f(c)}{\partial c} \delta c +  \frac{\kappa}{2} \frac{ \partial (\boldsymbol{\nabla} c)^2}{\partial \boldsymbol{\nabla} c} \cdot \boldsymbol{\nabla}  \delta c \Bigg] \mathrm{d}{V_\Omega} + \oint_{S_\Omega} \frac{\partial  f^{\textrm{S}}(c)}{\partial c} \delta c \mathrm{ d }{S_\Omega}   \nonumber\\
&=&  \int_{V_\Omega} \Bigg[ \frac{\partial  f(c)}{\partial c} \delta c -  \frac{\kappa}{2} \boldsymbol{\nabla} \left(\frac{ \partial (\boldsymbol{\nabla} c)^2}{\partial \boldsymbol{\nabla} c}\right) \cdot \delta c + \frac{\kappa}{2} \boldsymbol{\nabla} \cdot \left(\frac{ \partial (\boldsymbol{\nabla} c)^2}{\partial \boldsymbol{\nabla} c} \delta c \right) \Bigg] \mathrm{d}{V_\Omega}  \nonumber \\
&&\hspace{8cm}+ \oint_{S_\Omega} \frac{\partial  f^{\textrm{S}}(c)}{\partial c} \delta c \mathrm{ d }{S_\Omega}   \label{eq:LIBWettingConditionInterMideate} \\
&=&  \int_{V_\Omega} \Bigg[ \frac{\partial  f(c)}{\partial c}  - \kappa \boldsymbol{\nabla} \cdot \boldsymbol{\nabla}c \Bigg] \delta c\mathrm{d}{V_\Omega} + \oint_{S_\Omega} \left[\frac{\partial  f^{\textrm{S}}(c)}{\partial c} - \kappa \left({ \boldsymbol{n} \cdot \boldsymbol{\nabla}c}  \right)\right]  \delta c \mathrm{ d }{S_\Omega} \label{eq:LIBWettingConditionEulerLagrange}
\end{eqnarray}
where $\boldsymbol{n}$\nomenclature{$\boldsymbol{n}$}{inward pointing unit normal to the surface} denotes the inward pointing unit normal to the particle surface and divergence theorem is operated to the third term of Eq.~\eqref{eq:LIBWettingConditionInterMideate}, which converts the volume integral to a surface integral in Eq.~\eqref{eq:LIBWettingConditionEulerLagrange}. It is important to differentiate the gradient energy and the surface energy. The second term in the volume integral corresponds to interface width between Li-rich and Li-poor phases, while the terms in the surface integral associated with the surface energy. 

The understanding of the surface energy is significant to consider heterogeneous nucleation \cite{welland2015miscibility}. As the lithiation takes place at the surface of the particles, the phase separation initiates from the surface. However, it might be probable to observe heterogeneous nucleation of phase separation at the interface. This behavior can be incorporated from the surface energy, $f^{\textrm{S}}(c)$. The minimization of the system free energy is ensured by equating the functional derivative to zero, $\delta F(c,\boldsymbol{\nabla}c) =0$. As the variations in the domain and on the boundary are represented by the spatially smooth functions, both terms multiplying $\delta c$ should vanishes independently. Therefore, Eq. \eqref{eq:LIBWettingConditionEulerLagrange} yields the boundary condition,
\begin{equation}
\left[\frac{\partial  f^{\textrm{S}}(c)}{\partial c} - \kappa \left({ \boldsymbol{n} \cdot \boldsymbol{\nabla}c}  \right)\right]  \delta c = 0 \textrm{ on } {S_\Omega}, \label{eq:LIBSurfaceWettingBoundaryCondition}
\end{equation}
 on the particle surface. Furthermore, the quantity in the volume integral of Eq.~\eqref{eq:LIBWettingConditionEulerLagrange} is referred as the chemical potential,
\begin{equation}
 \mu = \frac{\partial f(c)}{\partial c} - \kappa \boldsymbol{\nabla} \cdot \boldsymbol{\nabla} c,
   \label{eq:LIBVariationalDerivative}
 \end{equation}
which is an objective functional to optimize~\cite{gurtin1996generalized}. The first term refers to the chemical potential corresponding to the bulk free energy density $\mu^{\mathrm{chem}}$\nomenclature{$\mu^{\mathrm{chem}}$}{chemical potential corresponding to the bulk free energy density}. For $a$\nomenclature{$\mu_{a}^{\mathrm{chem}}$}{chemical potential corresponding to the bulk free energy density of $a$ phase} phase, the bulk chemical potential is expressed as,
\begin{equation}
\hspace{-1.5cm} \mu_{a}^{\mathrm{chem}}(c)=\frac{\partial f_{a}^{\mathrm{chem}}(c)}{\partial c}=  \alpha'_a + \alpha''_a (1-2c) + \frac{T}{T_{\textrm{ref}}}\left\{\ln(c) - \ln(1-c) \right\}. \label{eq:LIBmuChemCh}
 \end{equation}
 Furthermore, the second term denotes the chemical potential corresponding to the gradient free energy density, $\mu^{\mathrm{grad}}$\nomenclature{$\mu^{\mathrm{grad}}$}{chemical potential corresponding to the gradient free energy density}. For $a$\nomenclature{$\mu_{a}^{\mathrm{grad}}$}{chemical potential corresponding to the gradient free energy density of $a$ phase} phase, the gradient chemical potential is expressed as,
 \begin{equation}
 \mu_{a}^{\mathrm{grad}}(\boldsymbol{\nabla}c)= - \kappa_{a} \boldsymbol{\nabla} \cdot \boldsymbol{\nabla} c. \label{eq:LIBmuGradCH}
 \end{equation}

\subsection{Kinetic evolution}

The driving force for the diffusion of each species in the electrochemical systems is the gradient of the chemical potential of the species. The Onsager relation defines the flux distribution inside the particle of the form
\begin{equation}
\boldsymbol{J} = - M \boldsymbol{\nabla}\mu, \label{eq:LIBcahnHilliardFlux}
\end{equation} 
where $\mu$ {denotes} the chemical potential of the system expressed in Eq.~\eqref{eq:LIBVariationalDerivative} and $M$ is the atomic mobility in the form of a diagonal matrix. If the mobility is direction-dependent, as for instance in lithium-iron phosphate (LFP)\nomenclature{LFP}{lithium iron phosphate, LiFePO$_4$} \cite{hong2016anisotropic}, $M$ is represented by unequal components of the diagonal matrix. Since the mobility in LMO is isotropic, the matrix reduces to a scalar prefactor of the form \cite{santoki2018phase},
\begin{equation}
M=c(1-c) \left( \frac{T_{\textrm{ref}}}{T}\sum_{a=1}^N h(\psi_a)D_a \right), \label{eq:LIBmobilityEquationInterpolation}
\end{equation}
where $D_a$\nomenclature{$D_a$}{diffusion coefficient of phase $a$} is the diffusion coefficient of phase $a$. 

The time evolution of the lithium-ion diffusion can be obtained from the Onsager's relation for non-equilibrium thermodynamics. As the diffusion follows mass conservation of species, the continuously-defined concentration $c(\boldsymbol{x},t)$ takes spatially and temporally dependent form,
\begin{equation}
\frac{\partial {c}}{\partial t} = - \boldsymbol{\nabla} \cdot \boldsymbol{J} , \label{eq:LIBconservationEquation}
\end{equation}
Note that by substituting the flux Eq.~\eqref{eq:LIBcahnHilliardFlux} in the mass conservation Eq. \eqref{eq:LIBconservationEquation} yields the classical Cahn-Hilliard equation of the form
\begin{equation}
\frac{\partial {c}}{\partial t} = \boldsymbol{\nabla} \cdot (M \boldsymbol{\nabla} \mu) \textrm{\hspace{3cm} on $V_\Omega$}. \label{eq:LIBCahnHilliard}
\end{equation}
This partial differential equation determines the evolution of concentration $c(\boldsymbol{x},t)$ for given initial and boundary conditions. 

\section{Smoothed-boundary method}
\label{sec:smoothedBoundaryMethod}

The solution of the partial differential equation~\eqref{eq:LIBCahnHilliard} under the boundary conditions is of a {primary} objective to represent the lithium diffusion. It can be feasible to utilize analytical methods to solve non-complex partial differential equations under trivial boundary conditions. However, with the increase in the complexity of the equations, it became arduous to employ analytical theories. In such cases, {these equations need to solve with numerical algorithms.}
Even for numerical schemes like Fourier-spectral and finite-difference methods, it is challenging to employ a boundary condition on the surfaces which are inside the simulation box. Furthermore, the complications escalate with considering particle of a curvaceous geometry. For such cases, the general boundary condition of arbitrary geometrical {shapes need to implement through some special techniques such as the smoothed boundary method} \cite{hong2016anisotropic,yu2012extended}.
\subsection{Neumann flux condition at the particle surface}

\begin{figure}[h]
\begin{center}
\includegraphics[scale=1.0]{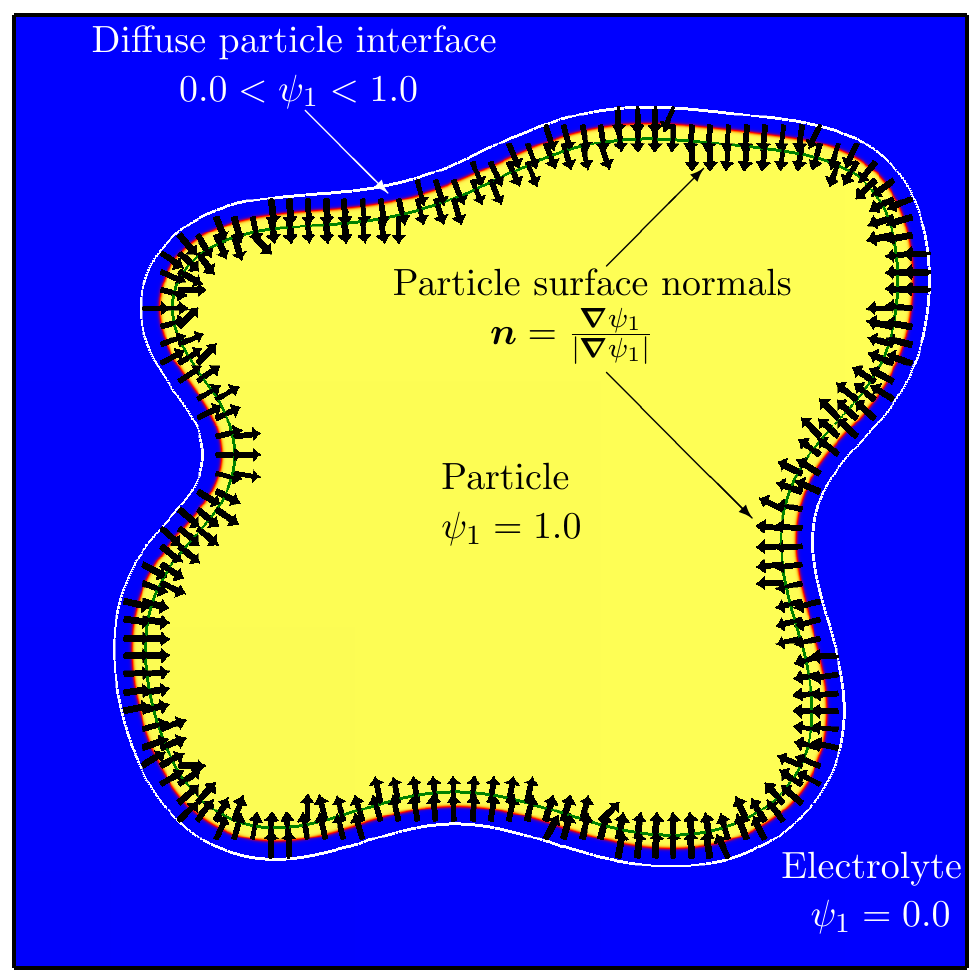}
\end{center}
\caption[Schematic diagram of a cathode particle, surrounded by the electrolyte in a lithium-ion battery.]{Schematic diagram of a cathode particle, surrounded by the electrolyte in a lithium-ion battery. The domain parameter $\psi_1$ indicates the particle by $\psi_1=1.0$, the electrolyte by $\psi_1=0.0$ and the diffuse particle surface region by $0.0<\psi_1<1.0$. The surface normals in the diffuse region are used to implicitly incorporate the boundary condition into the evolution equation.}
\label{fig:LIBGalvanostaticSchematic}
\end{figure}
 In smoothed-boundary method, a domain parameter $\psi_1$, is registered to differentiate the cathode particle and the electrolyte, and interpolate between different phases. The domain parameter $\psi_1$ indicates the particle by $\psi_1=1.0$, the electrolyte by $\psi_1=0.0$ and the diffuse particle surface region by $0.0<\psi_1<1.0$. The smoothed boundary method is applied to the classical Cahn-Hilliard equation~\eqref{eq:LIBCahnHilliard}, which modifies the main evolution equation. Multiplying both sides by the domain parameter $\psi_1$,
\begin{equation}
\psi_1 \frac{\partial c}{\partial t} = \psi_1 \boldsymbol{\nabla} \cdot (M\boldsymbol{\nabla} \mu),
\end{equation}
to incorporate the smoothed boundary method. Using the relation $-M\boldsymbol{\nabla} \mu = \boldsymbol{J}$, the evolution equation can be rearranged as
\begin{eqnarray}\label{eq:LIBnormaltoFlux}
 \psi_1 &\frac{\partial c}{\partial t}  &= \boldsymbol{\nabla} \cdot (\psi_1 M \boldsymbol{\nabla} \mu ) + \boldsymbol{\nabla} \psi_1 \cdot \boldsymbol{J} \nonumber \\
 &\frac{\partial c}{\partial t}  &= \frac{1}{\psi_1} \boldsymbol{\nabla} \cdot (\psi_1 M\boldsymbol{\nabla} \mu) + \frac{\vert \boldsymbol{\nabla} \psi_1 \vert}{\psi_1} \boldsymbol{n} \cdot\boldsymbol{J}\\ 
&  &= \frac{1}{\psi_1} \boldsymbol{\nabla} \cdot (\psi_1 M \boldsymbol{\nabla} \mu) + \frac{\vert \boldsymbol{\nabla} \psi_1 \vert}{\psi_1} J_n \nonumber,
\end{eqnarray}
where $\boldsymbol{n} = \boldsymbol{\nabla} \psi_1 /\vert \boldsymbol{\nabla} \psi_1 \vert$ is the inward normal to the particle surface, and $J_n$\nomenclature{$J_n$}{Neumann flux boundary condition at the particle surface} is the boundary flux in normal direction to the particle.

For the smoothed boundary method, the chemical potential corresponding to the chemical free energy density remains unchanged from Eq.~\eqref{eq:LIBmuChemCh}, while the chemical potential corresponding to the gradient energy density changes from Eq.~\eqref{eq:LIBmuGradCH} to
\begin{equation}
\mu^{\mathrm{grad}} (\boldsymbol{\nabla}c)= - \kappa \boldsymbol{\nabla} \cdot \boldsymbol{\nabla} c = - \left(\frac{\kappa}{\psi_1}\boldsymbol{\nabla} \cdot \psi_1 \boldsymbol{\nabla} c -\frac{\kappa}{\psi_1} \boldsymbol{\nabla} \psi_1 \cdot \boldsymbol{\nabla} c \right). \label{eq:LIBLaplaceModified}
\end{equation}
 The second term on the right side of  Eq.~\eqref{eq:LIBLaplaceModified} can be further simplified as $\boldsymbol{\nabla} \psi_1 \cdot \boldsymbol{\nabla} c = \vert \boldsymbol{\nabla} \psi_1 \vert \boldsymbol{n} \cdot \boldsymbol{\nabla} c $. According to Eq. \eqref{eq:LIBSurfaceWettingBoundaryCondition}, $\boldsymbol{n} \cdot \boldsymbol{\nabla} c $ is linked to preferential surface wetting energy, $f^S(c)$ responsible for heterogeneous nucleation. However, in the case of isotropic surface energy and homogeneous nucleation, this term vanishes identically, i.e.,
\begin{equation}
\boldsymbol{\nabla} \psi_1 \cdot \boldsymbol{\nabla} c =0. \label{eq:LIBisotropicSurfaceEnergy}
\end{equation} 
  Finally, considering Eqs.~\eqref{eq:LIBVariationalDerivative}, \eqref{eq:LIBnormaltoFlux}, and \eqref{eq:LIBLaplaceModified}, the resultant evolution equation is expressed of the form,
 \begin{equation}
\frac{\partial {c}}{\partial {t}}  = \frac{1}{\psi_1} \boldsymbol{\nabla} \cdot \left[\psi_1 {M} \boldsymbol{\nabla} \left\{ \frac{\partial {f(c)}}{\partial {c}}  - \left(\frac{\kappa}{\psi_1} \boldsymbol{\nabla} \cdot \psi_1 \boldsymbol{\nabla} {c} -\frac{\kappa}{\psi_1} \boldsymbol{\nabla} \psi_1 \cdot \boldsymbol{\nabla} {c} \right) \right\} \right] + \frac{\vert \boldsymbol{\nabla} \psi_1 \vert}{\psi_1} {J_n},\label{eq:LIBNeumannFluxEvolutionEquation}
\end{equation}
here the Neumann flux condition can be operated in the last  term of the equation by considering appropriate choice of function, $J_n$. When a natural (no-flux) boundary condition is feasible, i.e. $J_n=0$, the last term vanishes.

\subsection{Dirichlet concentration condition at the particle surface}
Similar to Neumann condition, the Dirichlet condition can be implemented on the particle surface through smoothed-boundary method by manipulating the classical Cahn-Hilliard equation~\eqref{eq:LIBCahnHilliard}. To incorporate the Dirichlet condition, the chemical potential corresponds to the gradient energy Eq.~\eqref{eq:LIBLaplaceModified} can be rewritten as,
\begin{align}
\mu^{\mathrm{grad}} (\boldsymbol{\nabla}c) &= - \left[\frac{\kappa}{\psi_1}\boldsymbol{\nabla} \cdot \psi_1 \boldsymbol{\nabla} c -\frac{\kappa}{\psi_1} \boldsymbol{\nabla} \psi_1 \cdot \boldsymbol{\nabla} c \right], \nonumber \\
&= - \left[\frac{\kappa}{\psi_1}\boldsymbol{\nabla} \cdot \psi_1 \boldsymbol{\nabla} c -\frac{\kappa}{\psi_1^2} \Big\{ \boldsymbol{\nabla} \psi_1 \cdot \boldsymbol{\nabla} (c \psi_1) - c \boldsymbol{\nabla} \psi_1 \cdot \boldsymbol{\nabla} \psi_1 \Big\} \right], \nonumber \\
&= - \left[\frac{\kappa}{\psi_1}\boldsymbol{\nabla} \cdot \psi_1 \boldsymbol{\nabla} c -\frac{\kappa}{\psi_1^2} \Big\{ \boldsymbol{\nabla} \psi_1 \cdot \boldsymbol{\nabla} (c \psi_1) - c_n \vert \boldsymbol{\nabla} \psi_1 \vert^2 \Big\} \right].  \label{eq:LIBSmoothedBoundaryMethodDirichlet}
\end{align}
where $c_n$\nomenclature{$c_n$}{Dirichlet concentration boundary condition at the particle surface} is the Dirichlet boundary condition imposed at the location where $\vert \boldsymbol{\nabla} \psi_1 \vert$ {contains} finite value, i.e. particle surface. By substituting Eq.~\eqref{eq:LIBSmoothedBoundaryMethodDirichlet} into Eq.~\eqref{eq:LIBVariationalDerivative}, the final evolution equation can be written as,
\begin{equation}
\frac{\partial {c}}{\partial {t}}  = \frac{1}{\psi_1} \boldsymbol{\nabla} \cdot \left[\psi_1 {M} \boldsymbol{\nabla} \left\{ \frac{\partial {f(c)}}{\partial {c}}  - \left(\frac{\kappa}{\psi_1}\boldsymbol{\nabla} \cdot \psi_1 \boldsymbol{\nabla} c -\frac{\kappa}{\psi_1^2} \left( \boldsymbol{\nabla} \psi_1 \cdot \boldsymbol{\nabla} (c \psi_1) - c_n \vert \boldsymbol{\nabla} \psi_1 \vert^2 \right) \right) \right\} \right].\label{eq:LIBDirichletConcentrationEvolutionEquation}
\end{equation}
Note that the Dirichlet concentration boundary condition can be employed at the last term of the form, $c_n$.

It is worth noticing that these equations \eqref{eq:LIBNeumannFluxEvolutionEquation} and \eqref{eq:LIBDirichletConcentrationEvolutionEquation} can be combined to specify the Neumann flux and Dirichlet concentration boundary conditions, $J_n$ and $c_n$ simultaneously, which can be referred as Robin boundary condition. The final form of the evolution equation is expressed as,
\begin{align}
\frac{\partial {c}}{\partial {t}}  = \frac{1}{\psi_1} \boldsymbol{\nabla} \cdot \Bigg[\psi_1 {M} \boldsymbol{\nabla} \Bigg\{ \frac{\partial {f(c)}}{\partial {c}}  - \bigg(\frac{\kappa}{\psi_1}\boldsymbol{\nabla} \cdot \psi_1 \boldsymbol{\nabla} c -\frac{\kappa W_c}{\psi_1^2} \big( \boldsymbol{\nabla} \psi_1 \cdot & \boldsymbol{\nabla} (c \psi_1) - c_n \vert \boldsymbol{\nabla} \psi_1 \vert^2 \big) \bigg) \Bigg\} \Bigg]\nonumber \\ 
&+ W_J \frac{\vert \boldsymbol{\nabla} \psi_1 \vert}{\psi_1} {J_n},\label{eq:LIBRobinEvolutionEquation}
\end{align}
where $W_J$\nomenclature{$W_J$}{spatially dependent weight for Neumann flux boundary condition} and $W_c$\nomenclature{$W_c$}{spatially dependent weight for Dirichlet concentration boundary condition} are the spatially dependent weights for Neumann flux and Dirichlet concentration boundary conditions respectively. In addition, $W_c +W_J=1$, these factors discriminate Neumann flux and Dirichlet concentration boundary conditions imposed on different regions of particle surface.

The evolution equations \eqref{eq:LIBNeumannFluxEvolutionEquation}, \eqref{eq:LIBDirichletConcentrationEvolutionEquation}, and \eqref{eq:LIBRobinEvolutionEquation} are solved everywhere in the domain, i.e. in the region of the particle ($\psi_1=1$) as well as in the electrolyte ($\psi_1=0$). Hence, a small positive value, $\nu=1 \times 10^{-4}$\nomenclature{$\nu$}{denominator coefficient}, is added to the denominators to {prevent} division by zero in the region $\psi_1 = 0$. Yu et al.  \cite{yu2012extended} described the effect of different values of $\nu$ on the simulation results in detail.

\section{Boundary conditions at the separator}
\label{section:boundaryConditionsSeparator}

\begin{figure}
\begin{center}
\includegraphics[scale=0.25]{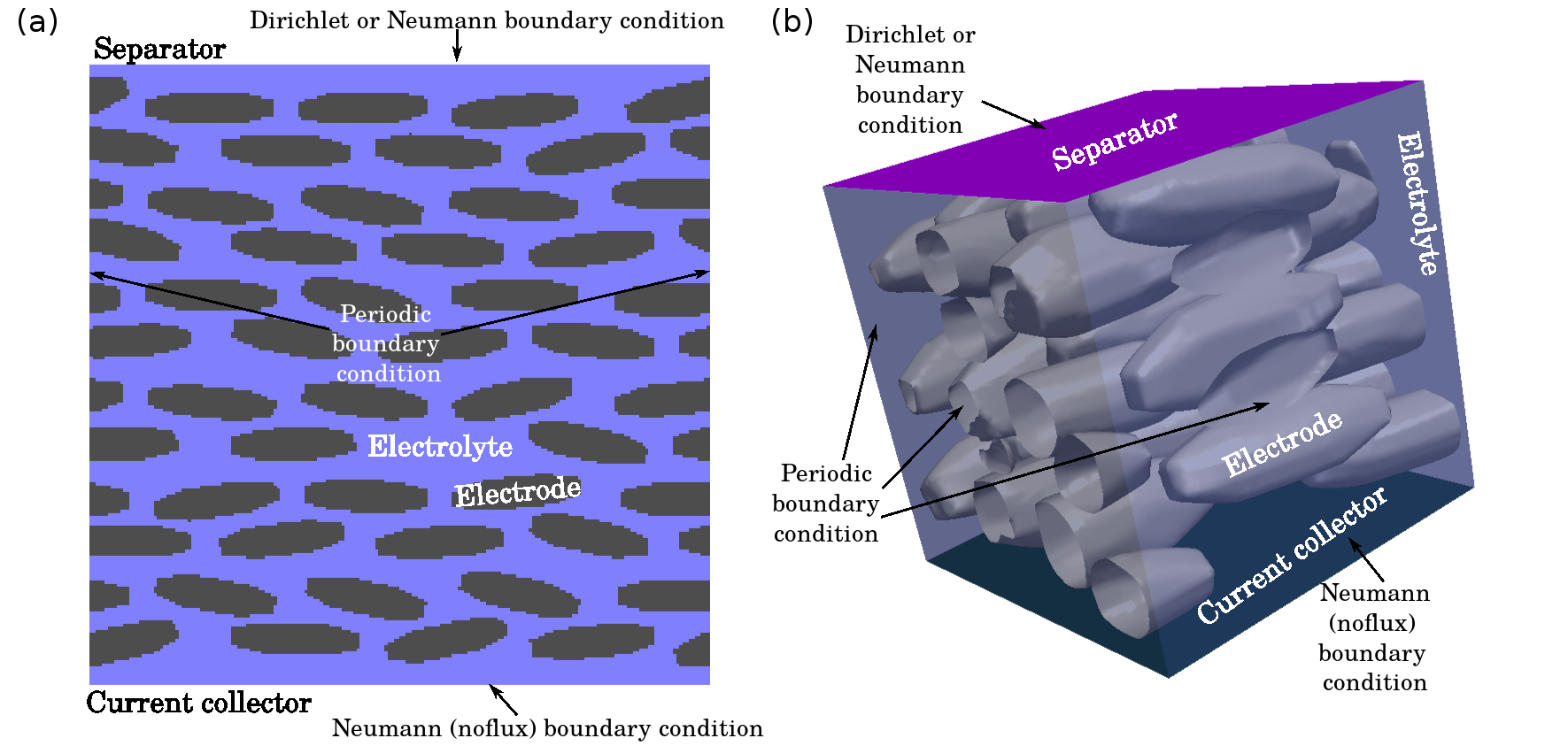}
\end{center}
\caption[Simulation setup of Dirichlet concentration or Neumann flux boundary conditions at the separator]{Simulation setup of Dirichlet concentration or Neumann flux boundary conditions at the separator on 2D microstructure in (a) and 3D microstructure in (b) for the diffusion of lithium species in the electrode.}\label{fig:boundaryConditionsSeparator}
\end{figure}

The boundary conditions at the particle surface can be encompassed into the main evolution equation through the smoothed-boundary method, as described in section~\ref{sec:smoothedBoundaryMethod}. However, the boundary conditions on the separator end can be realized by considering a REV element, as shown in Figure~\ref{fig:boundaryConditionsSeparator}. As the conditions are provided at the boundary of the simulation domain, the main evolution equation~\eqref{eq:LIBCahnHilliard} need not to be modified. A concentration equation in the form of Dirichlet condition is prescribed on one side of the boundary, while on the opposite end a Neumann condition is considered. The remaining two boundaries for 2D (four boundaries for 3D) simulations are considered periodically repetitive. The complete expression can be written in the form,

\begin{eqnarray}
{c}(t,i=N_x,j, k) &=& {c}_{ps},\nonumber\\
\partial {c}/\partial x (t,i=0, j,k) &=& 0, \textrm{\hspace{3cm}on $\partial V_\Omega$}\label{eq:boundaryConditionSeparator}\\
{c}(t,i,j+N_y,k) &=& {c}(t,i,j,k)\nonumber\\
{c}(t,i,j,k+N_z) &=&{c}(t,i,j,k).\nonumber
\end{eqnarray}
Where \nomenclature{$\partial V_\Omega$}{simulation domain boundary surfaces}${c}_{ps}$\nomenclature{${c}_{ps}$}{Dirichlet concentration boundary condition at the separator} is the prescribed concentration equation at the boundary and $N_x$\nomenclature{$N_x$, $N_y$, and $N_z$}{number of grid points in x-, y- and z-directions}, $N_y$, and $N_z$ are the number of grid points in x-, y- and z-directions ($i,j,k \in \boldsymbol{x}$)\nomenclature{$i,j,k \in \boldsymbol{x}$}{spatial location of a simulation grid point in the x-, y-, and z-directions respectively} respectively. Similarly, a flux equation in the form of Neumann condition can be prescribed on one of the boundary instead of the first provision in the equation set~\eqref{eq:boundaryConditionSeparator} in the form,
\begin{equation}
M \frac{\partial \mu }{\partial x}(t,i=N_x, j, k) = J_{ps}, \label{eq:boundaryConditionSeparatorFlux}
\end{equation}
where $J_{ps}$\nomenclature{$J_{ps}$}{Neumann flux boundary condition at the separator} {denotes} the prescribed flux at the boundary. Similar to Eq.~\eqref{eq:LIBRobinEvolutionEquation}, it is also possible to combine the Dirichlet concentration (first equation of the set~\eqref{eq:boundaryConditionSeparator}) and Neumann flux (Eq.~\eqref{eq:boundaryConditionSeparatorFlux}) boundary conditions to obtain the Robin condition at the separator.

It is worth noticing that the boundary conditions at the particle surface $c_n$ and $J_n$ and at the separator $c_{ps}$ and $J_{ps}$ are not confined to be constant values. {Alternatively, these might be variables, which encompass} the function of local concentration, time, space, {and} other parameters.

\section{Implementation strategies}
\label{section:LIBimplementationStrategies}

For computational convenience, the variables in the evolution equations and the boundary conditions are considered non-dimensional as,
\begin{equation}\label{eq:LIBNondimensionalization}
\hat{\boldsymbol{x}} = \frac{\boldsymbol{x}}{L},   \hat{t} = \frac{D_1}{{L}^2}t, \hat{J}_n = \frac{L}{D_1} J_n.
\end{equation}
where $L$\nomenclature{$L$}{reference length-scale for normalization} is the reference length for normalization, $D_1$ is the diffusion coefficient of lithium species in the electrode material, and the notation $\hat{\bullet}$\nomenclature{$\hat{\bullet}$}{normalized quantity of the respective entities} defines the normalized quantity of the respective entities.

 \begin{figure}
 \begin{center}
 \includegraphics[scale=1.0]{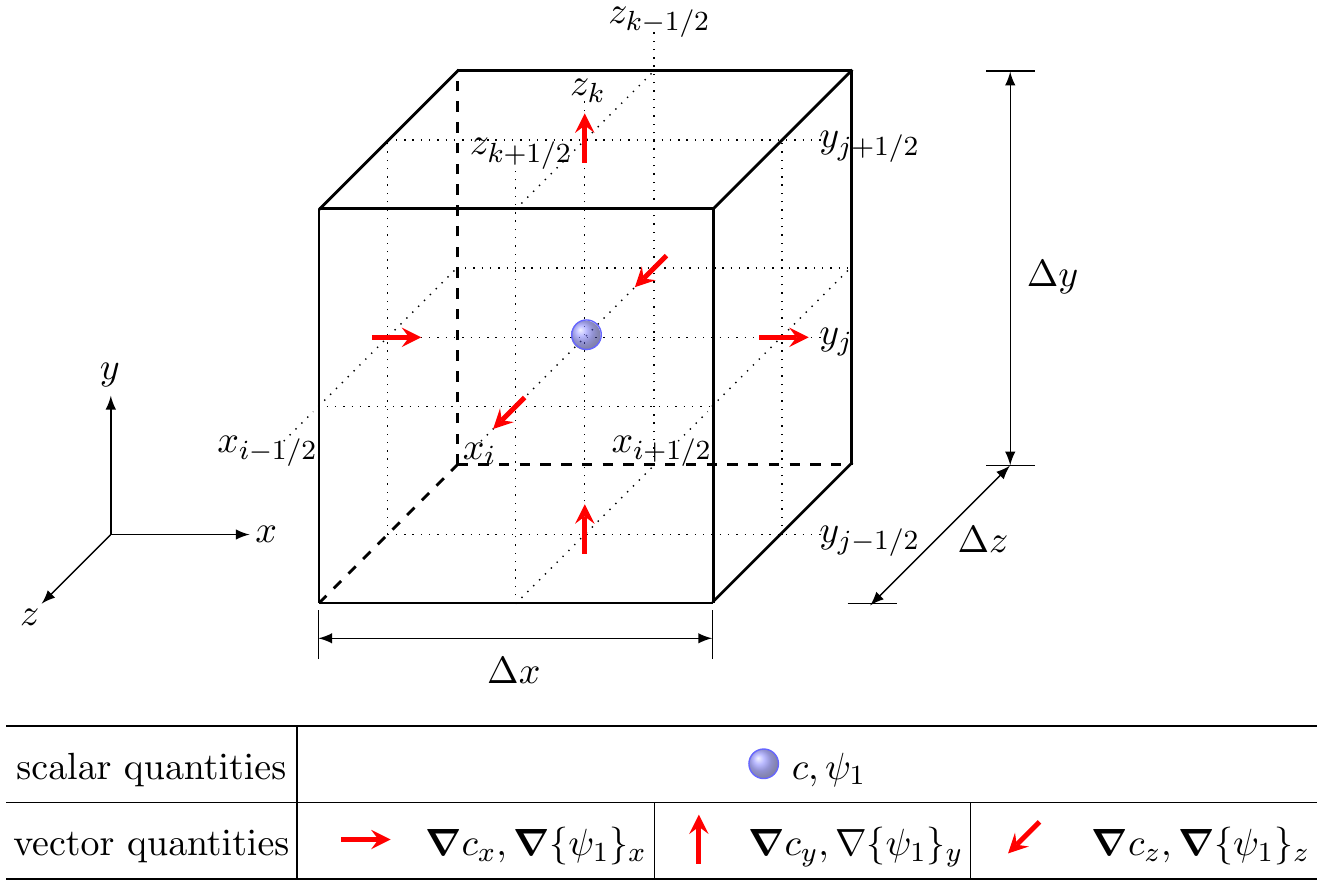}
\caption[Arrangement of the scalar and vector quantities on a staggered grid.]{Arrangement of the scalar and vector quantities on a staggered grid. A single simulation cell is shown with employed notation.}
\label{fig:LIBStaggeredGrid}
\end{center}
\end{figure} 
 Standard finite-difference numerical schemes are employed to solve the partial differential equations on a rectangular grid. A single simulation cell is displayed in Figure~\ref{fig:LIBStaggeredGrid} with utilized nomenclature in the present study at position $(i,j,k)$. The spatial derivatives are discretized by the finite-difference scheme. The gradient of concentration in all directions $x,$ $y,$ and $z$ at a discrete time $n$\nomenclature{$n$}{discrete time}, for example, is computed as
\begin{eqnarray}
&\left\{\boldsymbol{\nabla} {c}_x \right\}^{n}_{i+1/2,j,k} =  \frac{{c}^n_{i+1,j,k} - {c}^n_{i,j,k}}{\Delta x}  + \mathcal{O}(\Delta x ),\\
&\left\{\boldsymbol{\nabla} {c}_y \right\}^{n}_{i,j+1/2,k} =  \frac{{c}^n_{i,j+1,k} - {c}^n_{i,j,k}}{\Delta y} + \mathcal{O}(\Delta y ),\\ 
&\left\{\boldsymbol{\nabla} {c}_z \right\}^{n}_{i,j,k+1/2} =  \frac{{c}^n_{i,j,k+1} - {c}^n_{i,j,k}}{\Delta z}  + \mathcal{O}(\Delta z ),
\end{eqnarray}  
where, $ \Delta x $, $\Delta y $, and $\Delta z$\nomenclature{$ \Delta x $, $\Delta y $, and $\Delta z$}{simulation cell widths in $x$-, $y$- and $z$-directions respectively} are the simulation cell widths in $x$-, $y$- and $z$-directions, respectively and $\mathcal{O}(\bullet)$\nomenclature{$\mathcal{O}$}{order of approximation} represents order of approximation. The chosen finite difference formulation approximates the first derivatives to first order. Here, an increment (or decrement) in the subscript indicates the next (or previous) cell in the respective direction. The superscript denotes the time step $n$. This scheme is implemented on a staggered grid, in which the scalar variables are stored in the cell centers, whereas the vectors are located at the cell faces (see Figure \ref{fig:LIBStaggeredGrid}). As a consequence, a scalar, at the cell centers to be multiplied with a vector, is translated to the cell faces. For example, a contribution of $\boldsymbol{\nabla} \cdot (\psi_1 \boldsymbol{\nabla} {c})$ in an $x$-direction at the position $(i,j,k)$, is calculated as
 \begin{eqnarray}
\hspace{-1.8cm} \left[\boldsymbol{\nabla} \cdot \big{(}{\psi_1} \boldsymbol{\nabla} {c}\big{)}\right]^n_x = \Bigg\{ &\frac{ \big{(} \{\psi_1\}_{i+1,j,k} +\{\psi_1\}_{i,j,k}\big{)}{c}^n_{i+1,j,k} }{2\Delta x^2}\nonumber\\
 &-\frac{\big{(} \{\psi_1\}_{i+1,j,k} +2\{\psi_1\}_{i,j,k}+\{\psi_1\}_{i-1,j,k}\big{)} {c}^n_{i,j,k} }{2\Delta x^2} \nonumber\\
&+ \frac{ \big{(} \{\psi_1\}_{i,j,k} +\{\psi_1\}_{i-1,j,k}\big{)}  {c}^n_{i-1,j,k} }{2\Delta x^2}\Bigg\}.
 \end{eqnarray}
 The contributions in $y$- and $z$-direction are calculated analogously and the sum of all resultants provides the value of $\boldsymbol{\nabla} \cdot {(}\psi_1 \boldsymbol{\nabla} {c}{)}$ of the form,
 \begin{eqnarray}
\hspace{-1.0cm} \left[\boldsymbol{\nabla} \cdot \big{(}{\psi_1} \boldsymbol{\nabla} {c}\big{)}\right]^n = \Bigg[ \Bigg\{ &\frac{ \big{(} \{\psi_1\}_{i+1,j,k} +\{\psi_1\}_{i,j,k}\big{)}{c}^n_{i+1,j,k} }{2\Delta x^2} + \frac{ \big{(} \{\psi_1\}_{i,j,k} +\{\psi_1\}_{i-1,j,k}\big{)}  {c}^n_{i-1,j,k} }{2\Delta x^2} \nonumber\\
 &-\frac{\big{(} \{\psi_1\}_{i+1,j,k} +2\{\psi_1\}_{i,j,k}+\{\psi_1\}_{i-1,j,k}\big{)} {c}^n_{i,j,k} }{2\Delta x^2} \Bigg\}\nonumber\\
+ \Bigg\{ &\frac{ \big{(} \{\psi_1\}_{i,j+1,k} +\{\psi_1\}_{i,j,k}\big{)}{c}^n_{i,j+1,k} }{2\Delta y^2} + \frac{ \big{(} \{\psi_1\}_{i,j,k} +\{\psi_1\}_{i,j-1,k}\big{)}  {c}^n_{i,j-1,k} }{2\Delta y^2} \nonumber\\
 &-\frac{\big{(} \{\psi_1\}_{i,j+1,k} +2\{\psi_1\}_{i,j,k}+\{\psi_1\}_{i,j-1,k}\big{)} {c}^n_{i,j,k} }{2\Delta y^2} \Bigg\}\nonumber\\
+ \Bigg\{ &\frac{ \big{(} \{\psi_1\}_{i,j,k+1} +\{\psi_1\}_{i,j,k}\big{)}{c}^n_{i,j,k+1} }{2\Delta z^2} + \frac{ \big{(} \{\psi_1\}_{i,j,k} +\{\psi_1\}_{i,j,k-1}\big{)}  {c}^n_{i,j,k-1} }{2\Delta z^2}\nonumber\\
 &-\frac{\big{(} \{\psi_1\}_{i,j,k-1} +2\{\psi_1\}_{i,j,k}+\{\psi_1\}_{i,j,k-1}\big{)} {c}^n_{i,j,k} }{2\Delta z^2} \Bigg\} \Bigg].
\end{eqnarray}  
Similarly, the terms containing spatial derivative in the evolution equation \eqref{eq:LIBRobinEvolutionEquation} can be calculated.

 The temporal derivative in the evolution equation is discretized by an explicit Euler scheme, 
\begin{equation}
\frac{\partial {c}}{\partial {\hat{t}}} = \frac{{c}^{n+1}_{i,j,k} - {c}^n_{i,j,k}}{\Delta {t}} + \mathcal{O}(\Delta t),
\end{equation} 
 which approximates the time derivative by a forward difference of first order. Here,  superscripts $n+1$ and $n$ {represent} the values at the next and the current time steps, the subscripts $i,j,k$ indicate the spatial position and $\Delta {t}$\nomenclature{$\Delta {t}$}{time step increment} is the difference between the current and the next time step. The numerical stability is ensured by a limiting time step \cite{provatas2011phase},
\begin{equation}
\Delta {t} <\frac{\Delta \boldsymbol{x}^4}{2^{2d+1}\hat{M}_{\mathrm{max}} \hat{\kappa}},
\end{equation} 
 where, $\hat{\kappa}$ is the dimensionless gradient energy coefficient, $d$\nomenclature{$d$}{dimensions of the simulation study} {denotes} the dimensions of the simulation study, $\hat{M}_{\mathrm{max}}$\nomenclature{$\hat{M}_{\mathrm{max}}$}{maximum of dimensionless mobility $\hat{M}$} {represents} the maximum of dimensionless mobility $\hat{M}$ defined in Eq.~\eqref{eq:LIBmobilityEquationInterpolation}. $\Delta\boldsymbol{x}$ is the width of the unit cubic simulation cell.

\section{Conclusion}
\label{section:lithiumIonBatteryPFMConclusion}

 In this chapter, the phase-field method for the phase separation in LMO cathode particles is described. The regular solution free energy density is considered to represent the phase separation. Above a critical value of the regular solution parameter, the phase separation is more favorable than the homogeneous mixture, which explains the two-phase coexistence in LMO cathode particles. The driving force for the movement of the interface separating these two phases is considered through boundary conditions. These conditions on the particle surfaces are employed through the smoothed-boundary method, which is expected to capture the geometry of complex-shaped particles elegantly. The validation of the approach and the numerical results obtained for an isolated single particle is explained in Chapter~\ref{chapter:NeumannConstantFlux}. The model is further extended to consider boundary conditions at the separator, which is intended to capture multiple particle configurations irrespective of particle size, shape, and orientations in an electrode. The numerical results of this model are presented in Chapter~\ref{chapter:Potentiostatic}. 
\afterpage{\blankpagewithoutnumberskip}
\clearpage
\chapter{Phase-field model for inclusion morphology under electromigration}
\label{chapter:phaseFieldModelElectromigration}

Most of the {present} analytical and numerical theories which are based on sharp interface description suffer from three {significant} limitations.
 Firstly, analytical theories only {allow} the criteria of the onset of bifurcation or can deduce the characteristics of the assumed steady-state shape \cite{yang1994cavity, suo1994electromigration}.
 {Moreover}, even if a numerical technique is employed, it requires an explicit tracking of the interface boundaries \cite{gungor1999theoretical, xia1997a}. 
 This is a cumbersome theoretical challenge, especially since the interface evolves continuously and with complex geometrical shapes \cite{gungor1998electromigration}. 
 Finally, the responses in a local electric field due to the dynamic change in the inclusion shape are neglected \cite{cho2007theoretical, cho2006current}.  
In the present work, a phase-field model is derived to {investigate} the temporal evolution of the inclusions propagating in the conductor. 
The elegance of the phase-field method lies in its ability to simulate moving boundary problems without having to track the interfaces explicitly. In the {following} section, a free energy functional is described to track a morphologically evolving inclusion in a conductor.

\section{Free energy functional}
\begin{figure}[h]
\centering
  \includegraphics[scale=0.205]{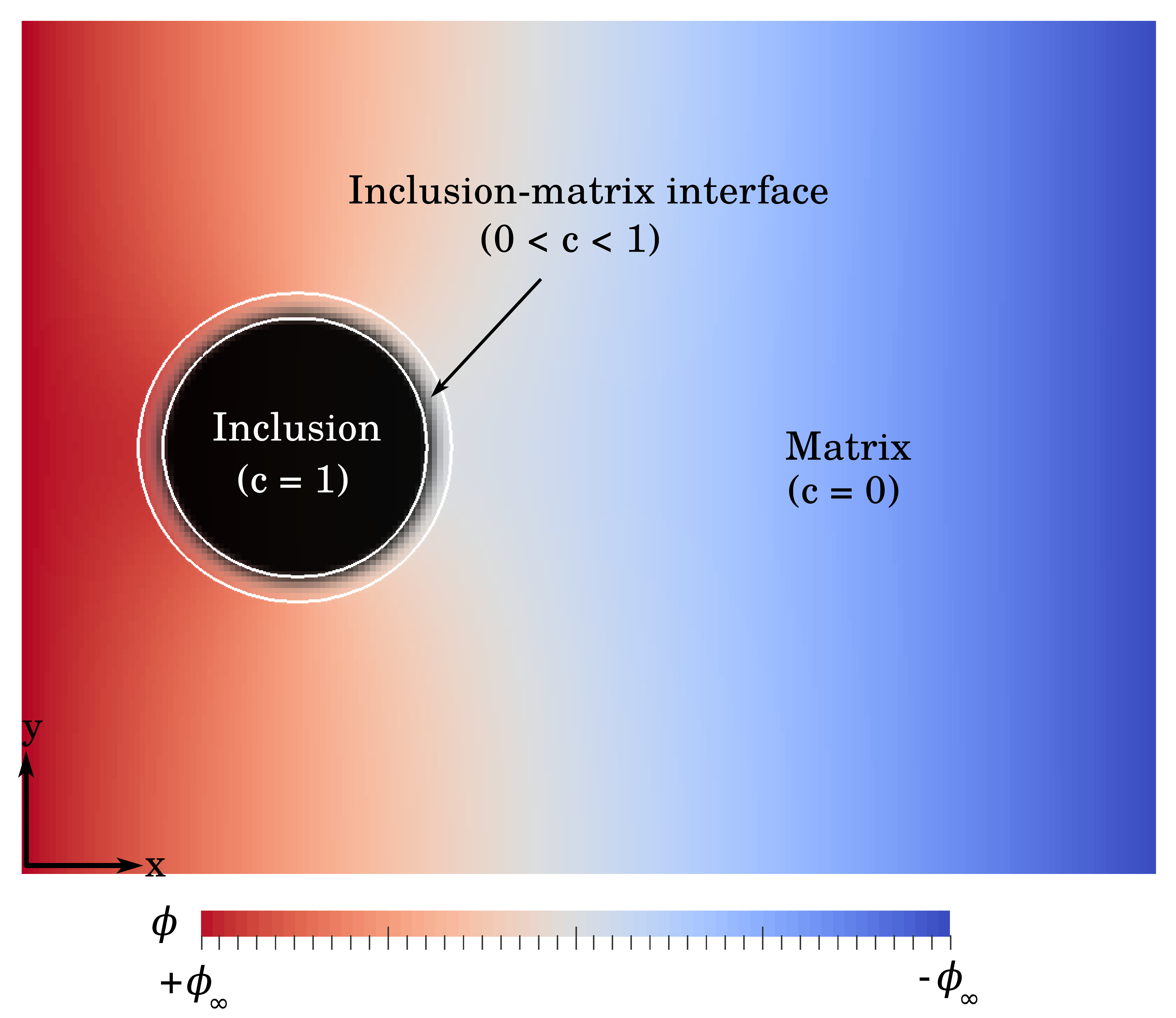}
\caption[Schematic diagram of the simulation setup {describes} the inclusion in the matrix subjected to external electric potential]{A schematic diagram of the simulation setup {describes} the inclusion with the order parameter $c=1$, in the interconnect material $c=0$, which is subjected to the electric potential $+\phi_\infty$\nomenclature{$\phi_\infty$}{electrical potential boundary condition} at the anode, and $-\phi_\infty$ at the cathode.}
\label{fig:EMSchameticBasic}     
\end{figure}

A phase-field model is employed to {examine} the electromigration-driven dynamics of transgranular inclusion. The schematic of an inclusion subjected to an external electric field is presented in Figure~ \ref{fig:EMSchameticBasic}. A conserved order parameter $c$ is introduced to demarcate the matrix denoted by $c = 0$ and the inclusion where it assumes a value of $c = 1$. The traditional sharp interface between the matrix and the inclusion is replaced by a narrow region of finite {thickness. In this region, the order parameter} varies smoothly between 0 and 1. {In this way}, the variable $c$ {fulfills} a dual role in tracking the species concentration and the interface between the matrix and the inclusion. The smooth interface provides a remarkable advantage in phase-field modeling to avoid the tedious task of explicit tracking as opposed to the sharp interface counterpart.

The free energy of the system $F(c,\boldsymbol{\nabla} c)$ is expressed as a function of the order parameter and its gradient as
\begin{equation}
F(c,\boldsymbol{\nabla} c) = \int_{V_\Omega} \left[ f(c) + \frac{\kappa}{2} \vert \boldsymbol{\nabla} c \vert^2 \right] \textrm{d}{V_\Omega}, \label{eq:EMfreeEnergyFunctional}
\end{equation}
where $f(c)$ {represents} the bulk free energy, $\kappa$ is the gradient energy coefficient, and $ \vert \bullet \vert$ {denotes} the norm of the vector.

It is worth noticing that the choice of the free energy function, $f(c)$, can {produce} a significant effect on the physical behavior of the {interface. Hence, it} should be selected appropriately. The regular solution model derived in section~\ref{sec:phaseFieldMeanFieldTheory} is a well-known free energy function for a {broad} range of applications. In that, the position of the two wells is decisive to obtain equilibrium values based on optimal energy state. %
For symmetric energy curve ($X_1=0$) and $T=T_{\textrm{ref}}$ in Eq.~\eqref{eq:freeEnergyLandauFinal}, the minima of the free energy are the equilibrium values \footnote[2]{For asymmetric curves, a Maxwell construction can be employed to determine the equilibrium values, which is further discussed in details in section~\ref{sec:GalvanostaticBasicFeatures}. Furthermore, these values are pertinent to flat interfaces. Alternatively, the curvature correction should be added according to Gibbs-Thomson law \cite{plapp2015phase}.}. {For instance, at} $X_2=2.6$, two equilibrium values are $c^L=0.124$ and $c^H=0.876$. In addition, $X_2=2$ {represents} the critical limit below which there is no concavity and the free energy plot resembles to Figure~\ref{fig:freeEnergyCurvesPhaseField}(c). As the $X_2$ increases above a critical {limit}, the minima starting to drift apart progressively. For a very high $X_2$, the two equilibrium values coincide $c^L=0$ and $c^H=1$. These values are convenient to indicate the bulk phases \cite{moelans2008an, steinbach2013phase, emmerich2012phase, nestler2011phase}.
However, the numerical calculations are unstable at these points due to logarithmic terms in Eq.~\eqref{eq:freeEnergyLandauFinal}. An alternative form of the free energy can be expressed as Landau polynomials \cite{moelans2008an, bhate2000diffuse} of the form,
\begin{equation}
f^L(c) = \sum_{l=0}^{N_l} L_l c^l,
\end{equation}
where $l$ indicates the index, $N_l$ represents the last term, and $L_l$ is the coefficient of the $l$-th term of the Landau polynomial. The simplest form of the polynomial is the double-well function, $f^{dw}(c)=c^2(1-c)^2$.

 The double-well function represents an approximation of the Van der Waals \cite{van1894thermodynamische, van1979thermodynamic} near the critical point, and has been employed extensively in the phase-field models. {However, when the model is developed solely for interface tracking purposes, this has led to the frequently observed spontaneous volume shrinkage phenomenon. Whereby,} the high volume phase allows significant infiltration into the low volume phases and can {ultimately cause} the complete disappearance of the lower volume phases \cite{yue2007spontaneous}. In addition, the movement of the lower volume phase inside the higher one may further enhance this effect. Therefore, to minimize these losses over the duration of a simulation, it requires a reconsideration of the free energy function. An alternative energy function for interface tracking applications in the form of the double-obstacle can be considered. The diffuse interface of the order parameter follows cosine function in the double-obstacle, while hyperbolic-tangent is expected in the former case, see Appendix \ref{app:interfaceProfiles}. This provides an extremely controlled diffuse interface width and less dispersion of inclusion volume. Consequently, lower change in inclusion volume leads to a stoppage of the spontaneous volume shrinkage phenomena. In addition, this function is adequate to track larger movements of the lower volume phase. Therefore, the obstacle-type function may prove useful for interface tracking applications of the phase-field model where the nature of the simulated phenomena introduces phase continuity concerns like the inclusion migration in the metal conductors.

\begin{figure}[t]
\begin{center}
\includegraphics[scale=0.90]{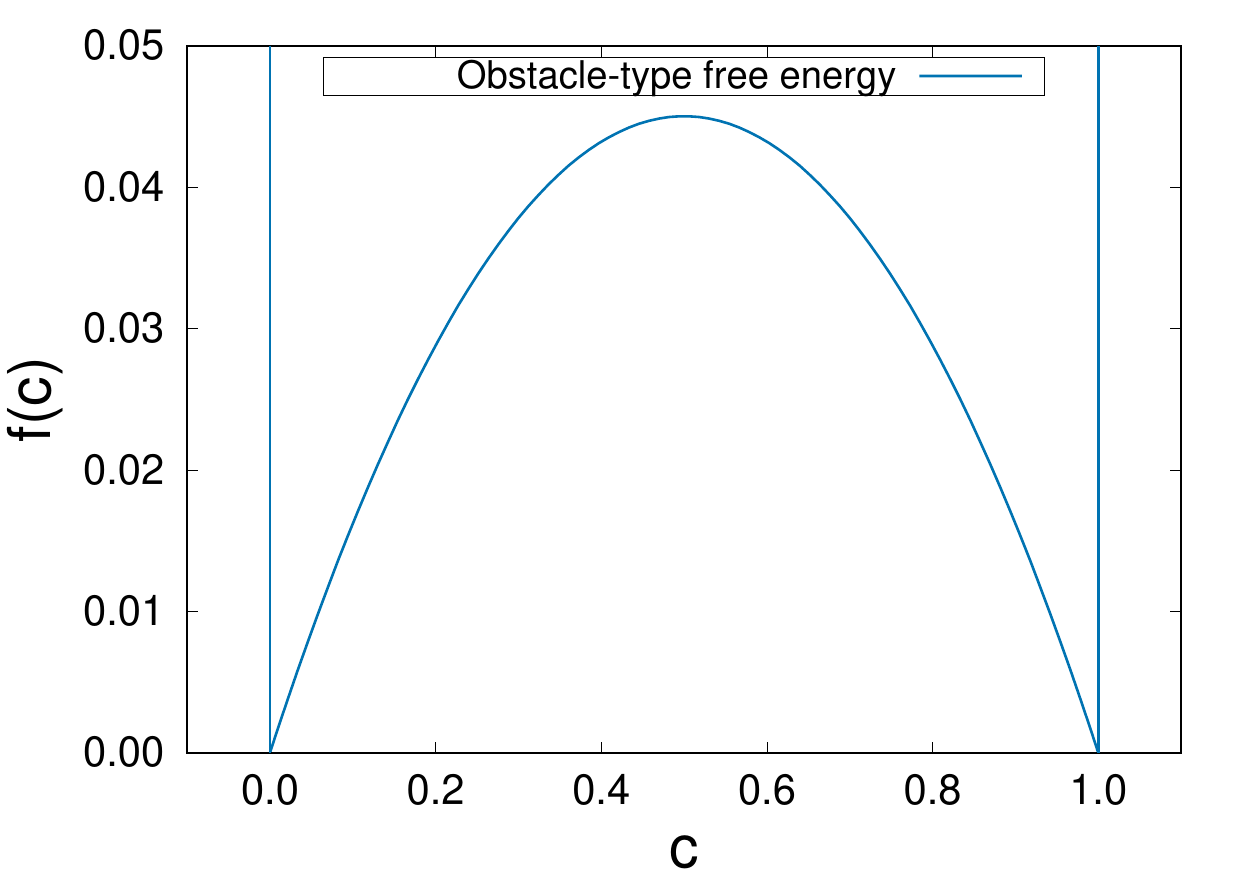}
\end{center}
\caption[Double-obstacle type free energy density $f^{\textrm{ob}}(c)$ as a function of $c$]{{Double-obstacle type} free energy $f^{\textrm{ob}}(c)$ as a function of $c$. The function is symmetric about center $c=0.5$ and has two free energy minima at 0 and 1.}\label{fig:EMObstacleTypeFreeEnergy}
\end{figure}

In addition, the obstacle-type free energy gains a computational advantage, by allowing the solution needs to be determined {merely} in the interface region, over the {double-well type} formulation \cite{nestler2005multicomponent}. The obstacle-type free energy density $f^{\textrm{ob}}(c)$\nomenclature{$f^{\textrm{ob}}$}{obstacle-type free energy density}, as shown in Figure~\ref{fig:EMObstacleTypeFreeEnergy}, is considered with equal minima at $c=0$ and 1:
 \begin{equation}
 f(c)= f^{\textrm{ob}}(c) = X_A c (1-c) + I(c),
 \end{equation}
where $X_A$\nomenclature{$X_A$}{barrier height of the obstacle-type free energy density} sets the barrier height of the free energy density, and $I(c)$\nomenclature{$I$}{indicator function for the obstacle-type free energy density} is the indicator function expressed as
\begin{eqnarray}
 I_{[0,1]} =     \left\{ \begin{array}{ccl}
0, & \mbox{  for}& 0 \le c \le 1 , \\ 
\infty,  & \mbox{  for} & c<0  \mbox{   \&   } c>1. 
\end{array}\right.%
\end{eqnarray}%

The morphological evolution of inclusion can be recognized by the competition between the electromigration force and the capillary force. The capillarity alone prefers circular shape inclusion. While the external electric field instigates species transport, which leads to shape alterations. In addition, the capillary force seeks uniform curvature and consequently diffuses the species to reduce any curvature gradient along the inclusion surface. Therefore, the shape of the inclusion is governed by the relative strength of these two forces, which can be incorporated into the diffusional fluxes.

\section{Diffusional fluxes and time evolution}

The evolution of the diffusing species follows a continuity equation
\begin{equation}\label{eq:EMFluxEvolution}
\frac{\partial c}{\partial t} = - \boldsymbol{\nabla} \cdot \boldsymbol{J}_i,
\end{equation}
where $t$ denotes time, $\boldsymbol{\nabla} \cdot (\bullet)$ {represents} the divergence of the vector and $\boldsymbol{J}_i$\nomenclature{$\boldsymbol{J}_i$}{flux of the diffusing species $i$} is the net flux of the diffusing species $i$. The flux $\boldsymbol{J}_i$ is expressed as a linear combination of the flux and driving force within the framework of irreversible thermodynamics as \cite{ho1989electromigration, tu1992electromigration} 
\begin{equation}\label{eq:EMFluxVsChemicalPotential}
\boldsymbol{J}_i = -M_{ii}\boldsymbol{\nabla}(\mu + eZ_i\phi) -M_{ie}eZ_w\boldsymbol{\nabla}\phi
\end{equation}
where $M_{ii}$\nomenclature{$M_{ii}$}{mobility of species $i$ due to the interaction with the species $i$} and $M_{ie}$\nomenclature{$M_{ie}$}{mobility of species $i$ due to the interaction with the electrons} are the mobilities related to the diffusivity of the species $i$ and interaction between species $i$ and electron respectively. $\boldsymbol{\nabla} (\bullet)$ represents the gradient of a scalar, $\mu$ is the chemical potential, $Z_i$\nomenclature{$Z_i$}{valence of the diffusing metal species $i$} denotes the valence of the diffusing species, $Z_w$\nomenclature{$Z_w$}{momentum exchange effect between the electrons and the diffusing species} denotes the momentum exchange effect between the electrons and the species, $e$ denotes the electric charge and $\phi$\nomenclature{$\phi$}{local electrical potential} is the electrical potential. The term $(\mu + eZ_i\phi)$ combined is referred as the electrochemical potential \cite{guyer2004phase}. To {perceive} the driving forces arising due to the imposed electric field, Eq.~\eqref{eq:EMFluxVsChemicalPotential} can be rearranged to give
\begin{equation}\label{eq:EMFluxModified}
\boldsymbol{J}_i = -M_{ii}\boldsymbol{\nabla}\mu - M_{ii}eZ_i\boldsymbol{\nabla}\phi - M_{ie}eZ_w\boldsymbol{\nabla}\phi
\end{equation}
The second term in the {previous} equation is the result of the direct electrostatic force, while the third term is the electron wind force, which reflects the cross-effect arising from the interaction between the diffusing species and the conducting electrons. In conductors, metal species are shielded by the negative electrons so that the direct electrostatic force is much less than the wind force \cite{ho1989electromigration, ceric2011electromigration}. The dominance of the wind force is further corroborated by experimental observation which suggests the movement of the diffusing species in the direction of the electron flow \cite{lau1975iron, blech1976electromigration}. Therefore, the second and third terms are combined through effective valence, $Z_s$. Hence considering the effective valence in Eq.~\eqref{eq:EMFluxModified} and substituting in Eq.~\eqref{eq:EMFluxEvolution}, the modified Cahn-Hilliard equation can be expressed as,
\begin{equation}\label{eq:EMCahnHilliardModified}
\frac{\partial c}{\partial t} = \boldsymbol{\nabla} \cdot \left(M_{ii}\boldsymbol{\nabla}\left(\mu +eZ_s \phi \right)\right) 
\end{equation}
The chemical potential $\mu$ can be obtained from the variational derivative of the system free energy,
\begin{equation}
\mu = \frac{\delta F}{\delta c}.
\end{equation}

\begin{figure}[hbt!]
\begin{center}
\includegraphics[scale=0.20]{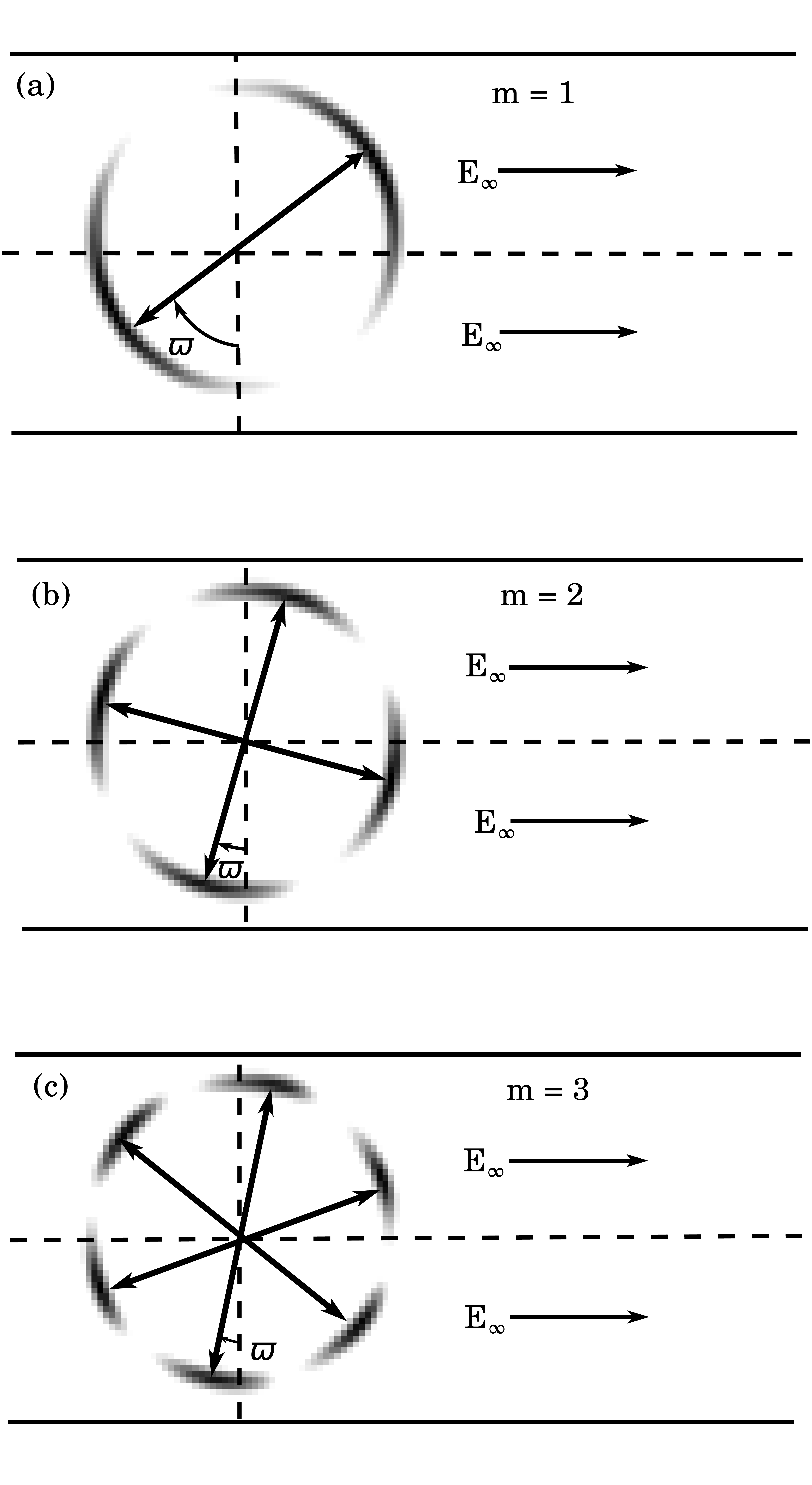}
\caption[Schematic of anisotropy in surface diffusivity of inclusions]{Schematic of anisotropy in surface diffusivity with (a) twofold, m=1, (b) fourfold, m=2, and (c) sixfold, m=3 symmetrical inclusions. The double-headed arrows define positions of maximum surface diffusivity. A misorientation angle, $\varpi$ for each case is defined with respect to the direction of the perpendicular to the external electric field.}\label{fig:EMsymmetryFolds}
\end{center}
\end{figure}

The surface diffusion is dominant compared to the other forms of mass transport mechanisms at the operating conditions. Therefore, the surface effective charge $Z_s$ and the atomic mobility $M_{ii}$ are restricted at the surface by selecting a bi-quadratic form in $c$, expressed as,
\begin{equation}
M_{ii}(c)= D_s f^{\theta}(\theta) c^2(1-c)^2,\label{eq:EMmobilityFunction}
\end{equation}
where $D_s$ denotes the surface diffusion coefficient\nomenclature{$D_s$}{surface diffusion coefficient}. Here $f^{\theta}(\theta)$\nomenclature{$f^{\theta}$}{anisotropy function} is the anisotropy function, given by \cite{dasgupta2013surface, kuhn2005complex},


\begin{equation}
f^{\theta}\left(\theta\right) =\frac{\left(1+A \, \textrm{cos}^2\left(m\left(\theta+ \varpi\right)\right)\right)}{\left(1 +A\right)}. \label{eq:anisotropy}
\end{equation}
In the above equation, $A$\nomenclature{$A$}{strength of anisotropy} {represents} the strength of anisotropy, which describes the superiority of the maximum value of diffusion compared to the minimum along the surface. In addition, $m$\nomenclature{$m$}{grain symmetry parameter} is the parameter related to the grain symmetry. Specifically, $2m$ denotes the number of crystallographic directions of fast diffusion sites in the plane of inclusion migration. Therefore, based on a total number of locations for fast diffusivity, $m=1,\,2,$ and 3 are characterized by twofold, fourfold, and sixfold symmetry respectively, as shown in Figure~\ref{fig:EMsymmetryFolds}, while $m=0$ indicates isotropic inclusions. Furthermore, $\theta = \tan^{-1} \big( \frac{\partial c}{\partial y}/\frac{\partial c}{\partial x}\big)$\nomenclature{$\theta$}{angle formed by the local tangent at the inclusion surface} is the angle formed by the local tangent at the inclusion surface and $\varpi$\nomenclature{$\varpi$}{misorientation angle} is the misorientation angle formed in the clockwise direction by the position of the maximum diffusion site with the perpendicular to the external electric field.

\section{Electrical potential}
The flux of the charge carrier, i.e. electron, $\mathbf{J}_{e}$\nomenclature{$\boldsymbol{J}_{e}$}{flux of the charge carrier} is given by
\begin{equation}
\mathbf{J}_e = -M_{ei} \boldsymbol{\nabla}(\mu + eZ_i\phi) - M_{ee}eZ_w\boldsymbol{\nabla}\phi
\end{equation}
where $M_{ei}$\nomenclature{$M_{ei}$}{mobility of electrons due to interaction with species $i$} and $M_{ee}$\nomenclature{$M_{ee}$}{mobility of electrons due to interaction with the electrons} are the phenomenological coefficients and $e$\nomenclature{$e$}{electron charge} is the electron charge. The current inside the conductor is entirely due to the imposed electron wind and the cross effect due to mass flux i.e. the first term is negligible. In addition, the timescale of relaxation of charge is much faster compared to the diffusion process. Hence, the current continuity equation translates into Laplace equation as
\begin{equation}
\boldsymbol{\nabla} \cdot \mathbf{J}_e = \boldsymbol{\nabla} \cdot [M_{ee}eZ_w\boldsymbol{\nabla}\phi] = 0.
\end{equation}
Comparing the {previous} equation with Ohm's Law, the conductivity function can be written as, $\sigma(c) = M_{ee}eZ_w.$ Therefore, the potential field is calculated from the Laplace equation as
\begin{equation}
\boldsymbol{\nabla} \cdot [\sigma(c)\boldsymbol{\nabla}\phi] = 0, \label{eq:EMLaplaceeq}
\end{equation} 
where $\sigma(c)$\nomenclature{$\sigma$}{electrical conductivity dependent on order parameter $c$} is the electrical conductivity dependent on order parameter $c$, interpolated between the inclusion $\sigma_{\textrm{icl}}$\nomenclature{$\sigma_{\textrm{icl}}$}{conductivity of an inclusion} and the matrix $\sigma_{\textrm{mat}}$\nomenclature{$\sigma_{\textrm{mat}}$}{conductivity of the matrix} as, 
\begin{equation}
\sigma(c) = \sigma_{\textrm{icl}}h(c) + \sigma_{\textrm{mat}}[1-h(c)].
\end{equation}
 Any smooth function that satisfies $h(c)\vert_{c=0}$ = 0 and $h(c)\vert_{c=1}$ =1 can be employed. Some functions that satisfy these properties are $h(c) = c, h(c) = c^2(3-2c),$ and $h(c) = c^3(10-15c+6c^2)$. The linear interpolation function ($h(c) = c$) is considered in the present work for computational convenience and the form of interpolation functions does not alter the results as long as the interface width is small.

\section{Resemblance to sharp-interface description}
To facilitate comparison and physical significance, the relations of phase-field model parameters should correspond to sharp-interface theories. In this section, a linkage between the two methods is established.  

\subsection{Interfacial properties}

In the phase-field model, the physical properties of the phenomenon such as interfacial energy $\gamma_s$\nomenclature{$\gamma_s$}{interfacial energy} and interfacial width  $\delta_s$\nomenclature{$\delta_s$}{interface width} are related to the double-obstacle barrier height $X_A$ and the gradient-energy coefficient $\kappa$. 

The spatial gradient of the order parameter along the interface can be equated to the free energy density function to derive interfacial width, rearranging Euler-Lagrange Eq.~\eqref{eq:1DVariationalDerivativeSimplified}, of the form
\begin{equation}
{\textrm{d}x} = \sqrt{\frac{\kappa}{2f(c)}}{\textrm{d}c}, \label{eq:EMdcbydxAlongInterface}
\end{equation}
 Substituting $f(c)$ from Eq.~\eqref{fig:EMObstacleTypeFreeEnergy} into Eq.~\eqref{eq:EMdcbydxAlongInterface} and integrating along the interface from $c=0$ to $c=1$ $\Rightarrow$ $x=0$ to $x= \delta_s$. The interface width can be obtained as,
 \begin{equation}
 \delta_s = \pi \sqrt{\frac{\kappa}{2X_A}}. \label{eq:EMinterfaceWidth}
 \end{equation}
 \nomenclature{$\pi$}{Archimedes' constant}It is important to note that this relation is applicable to obstacle type free energy density, where the interface width can be strictly related to $\kappa$ and $X_A$. {However, for well-type free energy density, the diffuse interface reaches infinity. Therefore, to determine the interface width, an approximation should be considered by limiting the change in order parameter upto a few orders.}
 
 The system free energy expressed in Eq.~\eqref{eq:EMfreeEnergyFunctional} identically vanishes in the bulk phases where $c=0$ and 1, while it has finite value at the interface. In other words, since the energy of the system is solely dictated by the movement of the interface, the interfacial energy can be expressed of the form,
 \begin{equation}
 \gamma_s = \int_{V_\Omega} \Big\{ f^{\textrm{ob}}(c) + \frac{\kappa}{2} \vert \boldsymbol{\nabla} c \vert^2 \Big\} \textrm{d}{V_\Omega}.
 \end{equation}
 This expression can be calculated numerically to obtain the interfacial energy for a complex system of the free energy densities \cite{cahnhilliard1958free}. Alternatively, the derivation of an analytical relation is also plausible for the simple {systems considered} for the presented case. To derive the expression, assuming that the spatial dependency of the order parameter $c$ is restricted to X-direction, writing one-dimensional form of the {previous} equation, 
   \begin{equation}
 \gamma_s = \int_{-\infty}^{ \infty} \Bigg\{ f^{\textrm{ob}}(c) + \frac{\kappa}{2} \left( \frac{\textrm{d} c}{\textrm{d}x} \right)^2 \Bigg\} \textrm{d}x.
 \end{equation}
Substituting the Euler-Lagrange Eq.~\eqref{eq:EMdcbydxAlongInterface} into the above equation,
\begin{equation}
 \gamma_s = \sqrt{2\kappa} \int_{0}^{1}  \sqrt{f^{\textrm{ob}}(c)}  \textrm{d}c. \label{eq:EMinterfacialEnergyINtermediate}
\end{equation}
Substituting $f^{\textrm{ob}}(c)$ from Eq.~\eqref{fig:EMObstacleTypeFreeEnergy} into Eq.~\eqref{eq:EMinterfacialEnergyINtermediate} and integrating, the final form of the interfacial energy can be expressed as,
\begin{equation}
 \gamma_s = \frac{\pi}{4\sqrt{2}} \sqrt{\kappa X_A}. \label{eq:EMinterfacialEnergy}
\end{equation}
It is evident from these relations \eqref{eq:EMinterfaceWidth} and \eqref{eq:EMinterfacialEnergy} that the physical properties of the system, the interface width and the interfacial energy can be recovered in a phase-field model. Furthermore, by manipulating the energy barrier height $X_A = 4 \gamma_s/\delta_s$ and the gradient coefficient $\kappa = 8 \gamma_s \delta_s/\pi^2$ desired values of interface width $\delta_s$ and the interfacial energy $\gamma_s$ can be achieved.

\subsection{Determination of Mullins' constant}
\label{section:mullinsRelation}

\begin{figure}[t]
\begin{center}
\includegraphics[scale=0.70]{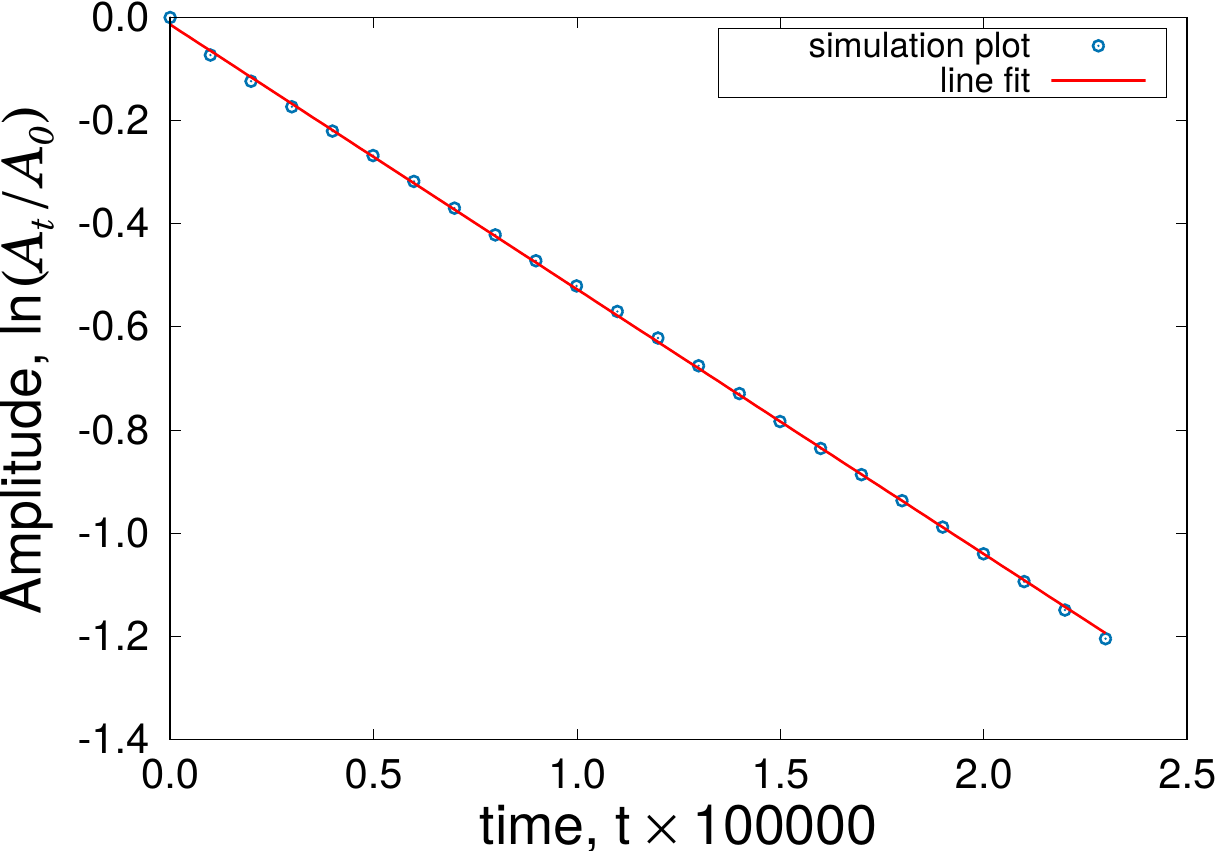}
\end{center}
\caption[Evolution of the sinusoidal amplitude with time.]{Evolution of the sinusoidal amplitude with time. $A_0$ {denotes} the initial amplitude of a perturbation with frequency $k$. At time $t$, $A_t$ denotes the temporal-dependent amplitude of the perturbations. The points are obtained from phase-field simulations, which are approximated by a linear line. The slope of the line indicates $-D_s \delta_s \Omega \gamma_s k^4/(k_BT)$.} \label{fig:EMMullinsRelation}
\end{figure}

\begin{figure}[t]
\begin{center}
\includegraphics[scale=0.40]{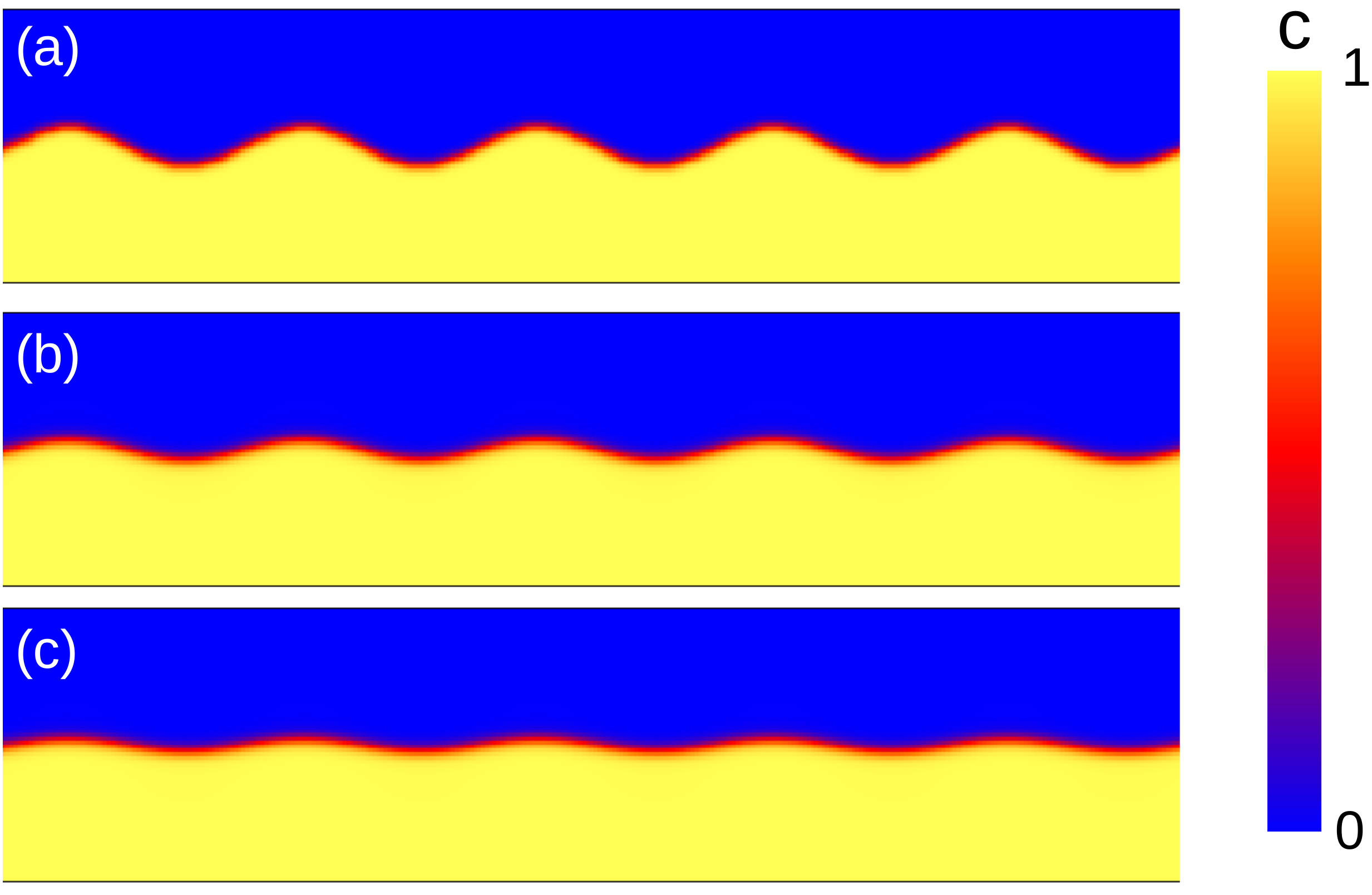}
\end{center}
\caption[Decay in the amplitude of the sinusoidal wave with time]{Decay in the amplitude of the sinusoidal wave with time sequence from (a) to (c).}\label{fig:EMMullinsSinusFigures}
\end{figure}
Surface diffusivity related to the Mullins' constant \nomenclature{$B$}{Mullins' constant}\nomenclature{$\Omega$}{atomic volume}$B (=D_s \delta_s \Omega \gamma_s/k_B T)$ is obtained by considering the dampening of a sinusoidal perturbation $ A_0$~sin($kx$) under surface diffusion \cite{joshi2017destabilisation}, as shown in Figure~\ref{fig:EMMullinsSinusFigures}. Here, $A_0$\nomenclature{$A_0$}{amplitude of the perturbation at time $t=0$} {represents} the amplitude of the perturbation at time $t=0$, and $k$\nomenclature{$k$}{frequency of the sinusoidal perturbation} {denotes} the frequency of the perturbation. 
 According to the Mullins' law for the curvature-driven surface diffusion \cite{mullins1959flattening}, the progressive dampening of the perturbation is expressed by
 \begin{equation}
 \frac{\partial y}{\partial t} = -\left( \frac{D_s \delta_s \Omega \gamma_s}{k_BT} \right) \frac{\partial^4 y}{\partial x^4}.
 \end{equation}
The analytical solution of this equation, which is valid for $A_0 k<1$, provides the expression for the amplitude of the perturbation:\nomenclature{$A_t$}{amplitude of the perturbation at time $t$}
\begin{equation}
A_t(t) = A_0 exp \left[ \left( -\frac{D_s \delta_s \Omega \gamma_s}{k_BT} \right) k^4 (t) \right].
\end{equation}
Thus, the value of the parameter $D_s/k_BT$ is obtained by the plot of $\textrm{ln}(A_t/A_0)$ vs. $t-t_0$ as shown in Figure~\ref{fig:EMMullinsRelation}, which provides the slope = $\left( -\frac{D_s \delta_s \Omega \gamma_s}{k_BT} \right) k^4$.

\section{Numerical procedures}
\label{sec:NumericalProcedures}

Eqs. \eqref{eq:EMCahnHilliardModified} and \eqref{eq:EMLaplaceeq} constitute the coupled partial differential equations to describe the inclusion dynamics under electromigration. It has been reported in Ref. \cite{mukherjee2019electric} via formal asymptotic analysis, that at the sharp-interface limit i.e. when the interface width tends to zero, these coupled equations recover the motion by {surface laplacian of the curvature and the electric potential}. In addition, these equations are normalized for the computational convenience \cite{santoki2019phase} with $\lambda'=210\textrm{ nm}$\nomenclature{$\lambda'$}{reference length scale for normalization} being the length scale and $\tau' =\lambda'^4 k_BT/ \gamma_s\Omega D_{s} \delta_s$\nomenclature{$\tau'$}{reference time scale for normalization} being the time scale. The physical values of the material properties and simulation conditions are listed in Table~\ref{tab:1}.\nomenclature{$\rho_{\textrm{mat}}$}{electrical resistivity of the matrix}
\begin{table}
\caption[SnAgCu material parameters and experimental conditions adopted in the model]{The following SnAgCu material parameters \cite{zhang2006effect} and experimental conditions are adopted in the model.}
\label{tab:1}       
\centering
\begin{tabular}{ l l }
\noalign{\smallskip}\hline\noalign{\smallskip}
 Atomic volume & $\Omega=2 \times 10^{-29}$ m$^3$\\
 Surface diffusivity & $D_s = 3.99 \times 10^{-10}$ m$^2$/s \\
 Thickness of surface layer & $\delta_s = 2.86 \times 10^{-10}$ m \\
 Surface energy & $\gamma_s = 1.4$ J/m$^2$ \\
 Temperature & $T$ = 419 K  \\
 Electron charge & $e$ = $1.6 \times 10^{-19}$ c  \\
 Boltzmann constant & $k_B$ = $1.38 \times 10^{-23}$ J/K  \\
 Electrical resistivity & $\rho_{\textrm{mat}}$ = $1.32 \times 10^{-7}$ $\Omega$m  \\
\noalign{\smallskip}\hline
\end{tabular}
\end{table}

An explicit finite-difference scheme is utilized for the implementation of the model. The spatial derivatives of the coupled Eqs.~\eqref{eq:EMCahnHilliardModified} and \eqref{eq:EMLaplaceeq} are discretized {utilizing} a combination of forward, backward, and central differences on a staggered grid, leading to second-order accuracy. 
A first-order explicit Euler technique is employed to discretize the temporal derivative. Isolate boundary conditions are imposed on the order parameter $c$, for all directions. The Dirichlet boundary conditions are set at the right and left ends, for the electric potential $\phi$ in the form of prescribed constant value, and are isolate at the top and bottom. The Laplace equation is solved iteratively {employing} the conjugate gradient scheme. 
The termination criterion is selected such that the maximum permissible difference in the electric potential is less than $1 \times 10^{-6}$ at a given position between the current and the previous iteration.

The model is implemented into the in-house software package PACE3D\nomenclature{PACE3D}{Parallel Algorithms for Crystal Evolution in 3D} version 2.2.0 \cite{hoetzer2018the}. 
{In this work}, the study is intended to track the complete temporal morphological transition of a circular inclusion. 
This may require a {extremely} large simulation domain of the order of $5000-6000$ grid cells in the direction of inclusion propagation, which increases the computation efforts dramatically.
In the present study, any appreciable change in the field variables $c$ and $\phi$ takes place only in the vicinity of the inclusion, due to the surface diffusion.   
The computational cost is restrained within reasonable limits by utilizing the so-called moving window approach \cite{vondrous2014parallel}.
Hence, a simulation starts with a fixed domain size.  
When any inclusion part reaches the center of the simulation domain, for each unit cell ($\Delta x$) migration of the inclusion, a grid point is removed from the rear {end and} is correspondingly added in front of the inclusion.

\section{Model applicability and limitations}

Some specific features of the developed model for a transgranular inclusion propagation are discussed as follows:
\begin{enumerate}
\item The presence of an electric field in the conductors can induce motion of the diffusing metal species in two ways: (1) the direct electrostatic force and (2) the electron-wind force. Firstly, the direct electrostatic force which drives the species towards the negative terminal (cathode). Second, the negatively charged electrons accelerated in the direction of the positive terminal (anode), which are colliding with the species, thus transferring momentum results in the advancement of the species in the direction of electron flow. This contribution is termed as electron wind.  At the operating current densities of {interconnects, that is of the order of $10^8-10^{10}$ A/m$^2$, electron-wind force is the dominant of the two. Therefore, within the framework of irreversible thermodynamics, the cross-effect arising due to the interaction between the conducting electrons and the diffusing species is considered}. This term has been added phenomenologically in the diffusion equation~\eqref{eq:EMCahnHilliardModified}. 

\item Although the inclusion nucleation, which is due to electromigration, is directly related to the reliability of the interconnects \cite{blech1969electromigration}, the process of inclusion nucleation is neglected. 
Instead, the studies are conducted to specific characteristics associated with the shape evolution of the inclusion.
\item The metallic conductor is assumed to be a single crystal, such that the inclusion-grain boundary interactions are neglected.  
\item The passage of an electric current, through a conductor, produces local heating near sharp corners and bends \cite{huang2006thermomigration}. 
Therefore, thermomigration may accompany electromigration. In the presented work, such effects of Joule heating are neglected. 
The inclusion shape changes are explored as a result of the competition between the relative magnitude of the electromigration force and the surface capillary force.
\item The surface diffusivity at the inclusion-matrix interface is very high in the operating temperature conditions $(T < 500 \textrm{ K})$ \cite{ho1970motion}. 
Hence, the surface diffusion is assumed to be the only atomic transport mechanism, and the bulk diffusion is neglected.
\item In general, the conductor materials might contain more than a single inclusion \cite{zeng2005kirkendall}. 
The local electric fields of different inclusions may interact with each other and may {alter} the overall dynamics of the inclusion evolution. 
In the presented dissertation, a study of morphological dynamics of an isolated inclusion is performed neglecting the effect of neighboring inclusions.
\end{enumerate}

\section{Conclusion}
 In this chapter, the phase-field method for morphological evolution of inclusion under the external electric field is described. The double-obstacle type of free energy density is considered to represent the temporal evolution of inclusions. The Laplace equation is employed to facilitate the distribution of the electric field in the simulation domain. {Besides}, numerical strategies considered for model implementation are discussed. The special provision in the mobility equation~\eqref{eq:EMmobilityFunction} of the model allows {simulating} isotropic as well as anisotropic mobilities of inclusions. The results obtained for isotropic inclusions are presented in Chapter~\ref{chapter:EM1}, which also compares sharp-interface theories with numerical results. Thereafter, results on anisotropic inclusions consisting of two-fold symmetry are described in Chapter~\ref{chapter:EM2}, followed by fourfold and sixfold symmetries in Chapter~\ref{chapter:4Fold6FoldEM}.

\afterpage{\blankpagewithoutnumberskip}
\clearpage

\newpage
\thispagestyle{empty}
\vspace*{8cm}
\phantomsection\addcontentsline{toc}{chapter}{III Results and Discussion: \\ Phase separation in lithium-ion batteries}
\begin{center}
 \Huge \textbf{Part III} \\
 \Huge \textbf{Results and Discussion: \\ Phase separation in lithium-ion batteries}
\end{center}
\afterpage{\blankpagewithoutnumberskip}
\clearpage

\chapter{Surface irregularities of a cathode particle}
\label{chapter:NeumannConstantFlux}

\section{Introduction}
Phase-field literature, which models the Li transport within a single particle \cite{dileo2014a, stein2016effects, christensen2006stress}, is ample. Previous models simulate particle geometry with various simplifications to perform the study in 1D \cite{huttin2012phase, walk2014comparison} or 2D \cite{xie2015phase, hong2016anisotropic}. There are few exceptional works simulated in 3D, which are limited to regularly shaped geometries, such as spherical particles \cite{welland2015miscibility, Klinsmann2016_1000050956}. The experimental results \cite{harris2012direct, guo2015direct, lampe2001benchmark, takahashi2015examination} reveal that the electrode consists of particles that show a shape polydispersity. The effect of particle geometry should be considered to obtain simulated results that are closer to the real electrode particle \cite{guo2017analytical}. Stein and Xu \cite{stein20143d} described the influence of particle geometry on the Li concentration profile in ellipsoidal particles. Guo et al. \cite{guo2017analytical} reported that the difference in maximum and minimum Li concentration is larger in elliptical particles than in spherical particles. Chakraborty et al. \cite{chakraborty2015combining} found that there are important differences between the plastic stretches in cylindrical particles and those in spherical particles, investigated earlier by Cui et al. \cite{cui2012finite}.

The presented work has two objectives. The first objective is to study the insertion dynamics of an irregularly shaped particle, to imitate a real particle in physical existence. The second objective is to consider the effect of various parameters influencing the charge dynamics. The foundation of the model was developed in chapter~\ref{chapter:phaseFieldModelLIB}. Based on that model, next section~\ref{section:NeumannFluxCondition} describes employed boundary conditions to study the surface irregularities presented in this chapter. In addition, the theory is validated with commercial multi-physical software COMSOL Multiphysics by considering a particular case from a benchmark study \cite{huttin2012phase}. Thereafter, the simulation results are presented in section \ref{sec:NeumannFluxResults} with plausible justification in section~\ref{sec:GalvanostaticBasicFeatures}. Afterward, the numerical study is extended in section~\ref{sec:GalvanostaticParameterStudy} to investigate the effect of various material parameters on the transportation mechanism. In section \ref{sec:NuemannFluxConclusions}, the chapter is concluded with some remarks on the results. Some parts of this chapter are published in the journal  \textit{Modelling and Simulation in Materials Science and Engineering} \cite{santoki2018phase}.

\section{Constant Neumann flux condition at the particle surface}
\label{section:NeumannFluxCondition}
 For a simplicity, a temporally independent lithium flux $J_n$ is considered at the particle surface of the form,
 \begin{equation}
J_n= \boldsymbol{J} \cdot \boldsymbol{n} =     \left\{ \begin{array}{ccl}
\frac{-C R_0}{N_{\textrm{ex}}d \times 3600 } & \mbox{  for}& c_s < 1, \\ 
      0  & \mbox{  for} & c_s = 1. 
\end{array}\right. \label{eq:LIBGalvanostaticFlux}
\end{equation}
The parameter $C$\nomenclature{$C$}{C-rate} refers to the C-rate, and  x C-rate measures the Li insertion or extraction rate at which complete charging or discharging takes place within 1/x hours~\cite{walk2014comparison}. Furthermore, $R_0$\nomenclature{$R_0$}{radius of the reference particle} is the radius of the reference particle, $c_s$\nomenclature{$c_s$}{concentration at the surface of the particle} is the concentration at the surface of the particle, $d$ defines the dimension of the simulation study, e.g., $d=3$ for a 3D simulation study and the characteristic extension\nomenclature{$N_{\textrm{ex}}$}{characteristic extension}
\begin{equation}\label{Eq:LIBGalvanostaticCurvature}
N_{\textrm{ex}} =     \left\{ \begin{array}{rcl}
\frac{\mathrm{Particle\, circumference \,} \times \mathrm{ \,Reference\, particle\, area}}{\mathrm{Particle \,area \,} \times \mathrm{\, Reference\, particle\, circumference}}& \mbox{  for the}& \mathrm{2D\, case,} \\ 
      \frac{\mathrm{Particle\, surface\, area \,} \times \mathrm{ \,Reference\, particle\, volume}}{\mathrm{Particle\, volume \,} \times \mathrm{\, Reference\, particle\, surface\, area}}  & \mbox{  for the} & \mathrm{3D \,case.} 
\end{array}\right.
\end{equation}

The constant Neumann flux boundary condition in Eq.~\eqref{eq:LIBGalvanostaticFlux} on the arbitrary geometrical shapes is implemented through the evolution equation~\eqref{eq:LIBNeumannFluxEvolutionEquation} for the Li diffusion. In addition, homogeneous nucleation from the surface is considered in the form of Eq.~\eqref{eq:LIBisotropicSurfaceEnergy}.  For the 2D~(or~3D) simulation study, a circular (or spherical) particle with radius R$_0$~=~1.0~$\mu$m is considered as a reference particle along with other parameter set is provided in Table \ref{tab:NeumannMaterialProperties}. A study is conducted to observe the Li diffusion from the surface of the particles, which are reported here.

\subsection{Simulation domain setup}

  The general boundary condition of arbitrary geometrical shapes is implemented through the smoothed boundary method \cite{hong2016anisotropic,yu2012extended}, in which a domain parameter, $\psi_1$, is registered to differentiate and interpolate between different phases. The simulation is initialized with the domain parameter $\psi_1$, continuously defined in the entire domain. The field quantity $\psi_1$ takes the value 1.0 for the bulk cathodic particle, 0.0 for the electrolyte and varies smoothly in the boundary between the particle and the electrolyte. The specific features of a diffuse interface between the particle and the electrolyte are responsible for the configuration of the surface normals, as shown in Figure~\ref{fig:LIBGalvanostaticSchematic}. As a consequence, the particle surface normal is exploited into the flux boundary condition given in Eq.~\eqref{eq:LIBGalvanostaticFlux}.

In the present chapter, particle morphology is considered to be constant during lithiation. Therefore, the simulations are performed in two independent steps. In the first step, the particle is formed from the Allen-Cahn equation presented in Appendix~\ref{appendix:singleParticleMicrostructure}. The resultant domain encompasses a smooth interface between the cathode particle and the electrolyte. In the second step, the smooth surface is exploited to implicitly incorporate the flux boundary condition in the Cahn-Hilliard equation~\eqref{eq:LIBNeumannFluxEvolutionEquation}, in which lithiation inside the particle obtained from the first step is considered.

The model simulation of the species flux infiltrates from the lithium abundant electrolyte to the cathode particle, through the exterior surface. A simulator is developed to obtain the numerical solution of the compound partial differential equation of the Cahn-Hilliard equation combined with a smoothed boundary method. By employing standard message passing interface (MPI)\nomenclature{MPI}{message passing interface} concepts, a parallel three-dimensional solver for large-scale computations on a rectangular mesh is realized as described in section~\ref{section:LIBimplementationStrategies}.

It is important to note that owing to the smoothed boundary method, the electrolyte is considered as LMO material as well. Even though the evolution equation~\eqref{eq:LIBNeumannFluxEvolutionEquation} is solved everywhere in the domain, i.e. in the region of the particle ($\psi_1=1$) as well as in the electrolyte ($\psi_1=0$), the specific manipulation in the smoothed boundary method motivates no concentration accumulation at $\psi_1=0.0$ (i.e., electrolyte). As a consequence, the influence of the electrolyte material parameter on the concentration accumulation in the electrolyte is negligible.

\begin{table}
\caption[Material properties of LiMn$_2$O$_4$ cathode and operational conditions]{Material properties of LiMn$_2$O$_4$ cathodic intercalation materials \cite{huttin2012phase} and operational conditions.\label{tab:NeumannMaterialProperties}}
\begin{center}
\begin{tabular}{ l c c c } 
 \hline
 Parameter & Symbol & Value  & Unit\\
 \hline 
 Diffusion coefficient & D$_1$ & 7.08 $\times$ 10$^{-15}$ & m$^2$/s \\
 Length scale & $L$ & 1 & $\mu$m \\ 
 Gradient energy coefficient & $\kappa$ & 7 $\times$ 10$^{-18}$ & m$^2$ \\
 Regular solution parameter & $\alpha_1'$ & -0.1 & - \\
  & $\alpha_1''$ & 2.6 & - \\
  Temperature & $T_{\mathrm{ref}}$ & 300 & K \\
 \hline
\end{tabular}
\end{center}
\end{table}

\subsection{Validation of the presented approach}
\label{section:NuemannFluxValidation}
\begin{figure}[t]
\begin{center}
\includegraphics[scale=0.95]{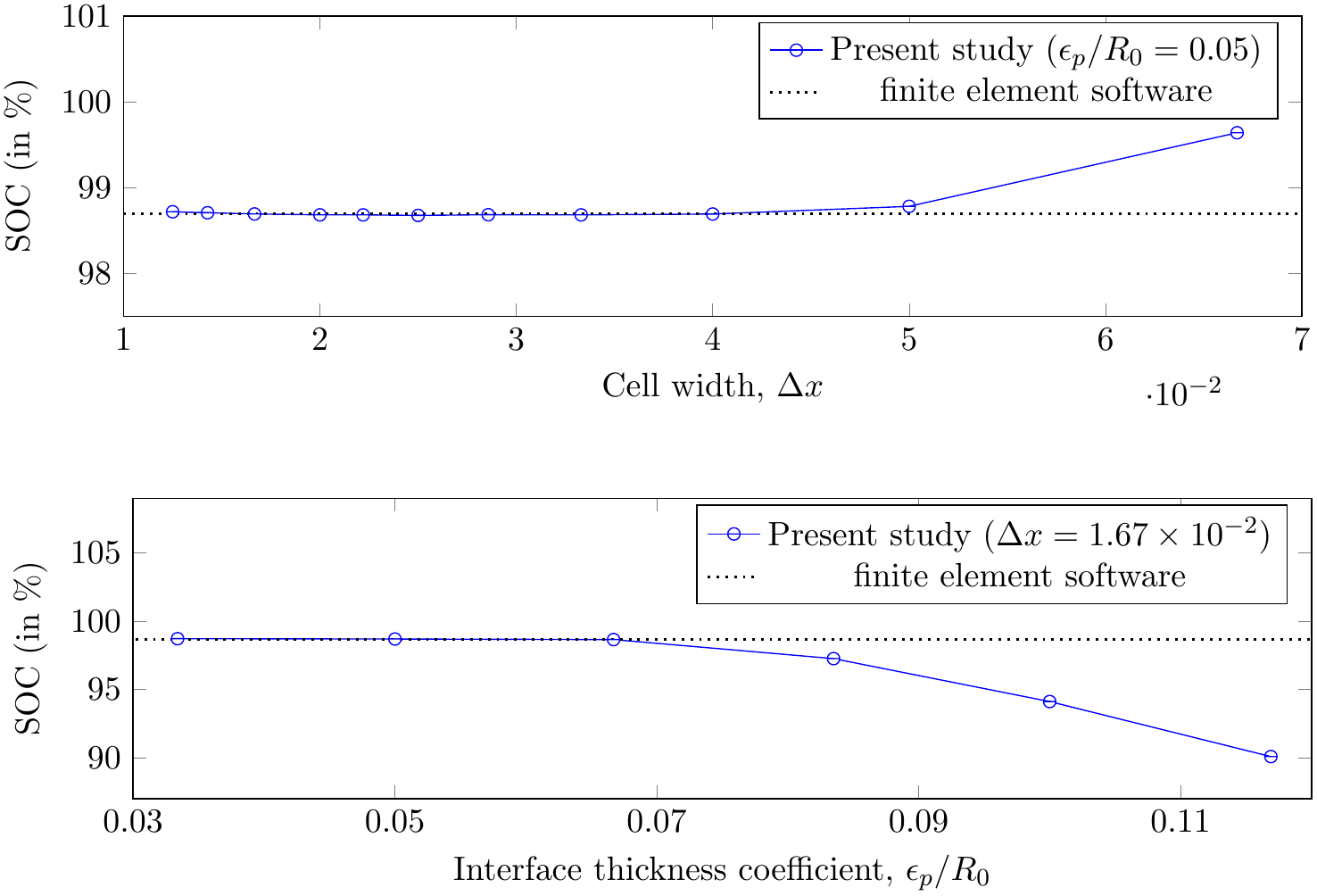}
\end{center}
\caption[Effect of the dimensionless cell width and the interface thickness coefficient on the SOC value]{Effect of the dimensionless cell width and the interface thickness coefficient on the SOC value: Comparison of the present approach with results obtained by the finite element software COMSOL Multiphysics\textsuperscript{\textregistered}.}
\label{fig:convergenceStudy}
\end{figure}

The simulation studies are performed for LMO\nomenclature{LMO}{lithium manganese oxide, LiMn$_2$O$_4$} cathodic particle at a lithiation of default 1 C-rate discharging condition. Therefore, the inflow of the lithium flux fills up the particle. The natural logarithmic terms $\mathrm{ln}({c})$ (or $\mathrm{ln}(1-{c})$) in the chemical-free energy formulation are singular at a concentration of ${c}=$ 0.0 (or 1.0). To avoid the uncertainty, the particles are filled with a concentration of a very small positive fraction, e.g., ${c_i}=$ 0.03, unless stated otherwise. Similarly, the simulations stop when any of the positions reaches ${c}=$ 0.999, instead of a fully lithiated state, inside the particle.

Yu et al. \cite{yu2012extended} presented a smoothed boundary method based on a finite difference scheme for various applications, such as the mechanical equilibrium equation, phase transformation with the presence of additional boundaries and the diffusion equation with Neumann and Dirichlet boundary conditions. In this section, the presented approach in section~\ref{sec:smoothedBoundaryMethod} is validated by considering a particular case in a phase-separating particle. During the insertion of the lithium species, a concentration gradient develops inside the particle from the higher concentration at the surface to the lower concentration at the center \cite{zhang2007numerical}. As a consequence, the surface reaches the threshold value of $c=0.999$ to stop the simulation before the bulk region of the particle. In turn, the simulations stop well before the particle attains a \nomenclature{SOC}{state of charge}state of charge (SOC = $100/{V_\Omega} \int_{V_\Omega} \psi {c})$ of 100\% for 1 C-rate species inflow. Therefore, the latest SOC attained by different values of simulation cell width of spherical particles are compared, with the 1D computations performed in Ref. \cite{huttin2014phase} by using the software COMSOL Multiphysics\textsuperscript{\textregistered}, which is based on the finite element method. The results obtained from the presented model are in agreement with the study of Huttin and Kamlah \cite{huttin2012phase}. The comparison suggests that at most a dimensionless cell width of $\Delta \boldsymbol{x} = 5 \times 10^{-2}$ is acceptable, to obtain a sufficient convergence of the simulation results, see Figure~\ref{fig:convergenceStudy}. Moreover, the calculated concentration value deviates further from the benchmark value as the interface thickness increases. The results ensure convergence for a thin interface thickness, which is in agreement with literature \cite{yu2012extended}. Therefore, the simulation study is carried out at $\Delta \boldsymbol{x} = 1.67 \times 10^{-2}$ for 3D\nomenclature{3D}{three-dimensional} and at $1 \times 10^{-2}$ for 2D\nomenclature{2D}{two-dimensional} studies with the interface thickness coefficient\nomenclature{$\epsilon$}{interface thickness coefficient} ${\epsilon_p}/R_0=0.05$, which are confined to the convergence of the smoothed boundary method.

\section{Concentration profiles}
\label{sec:NeumannFluxResults}
\begin{figure}[h]
\begin{center}
\includegraphics[scale=1.00]{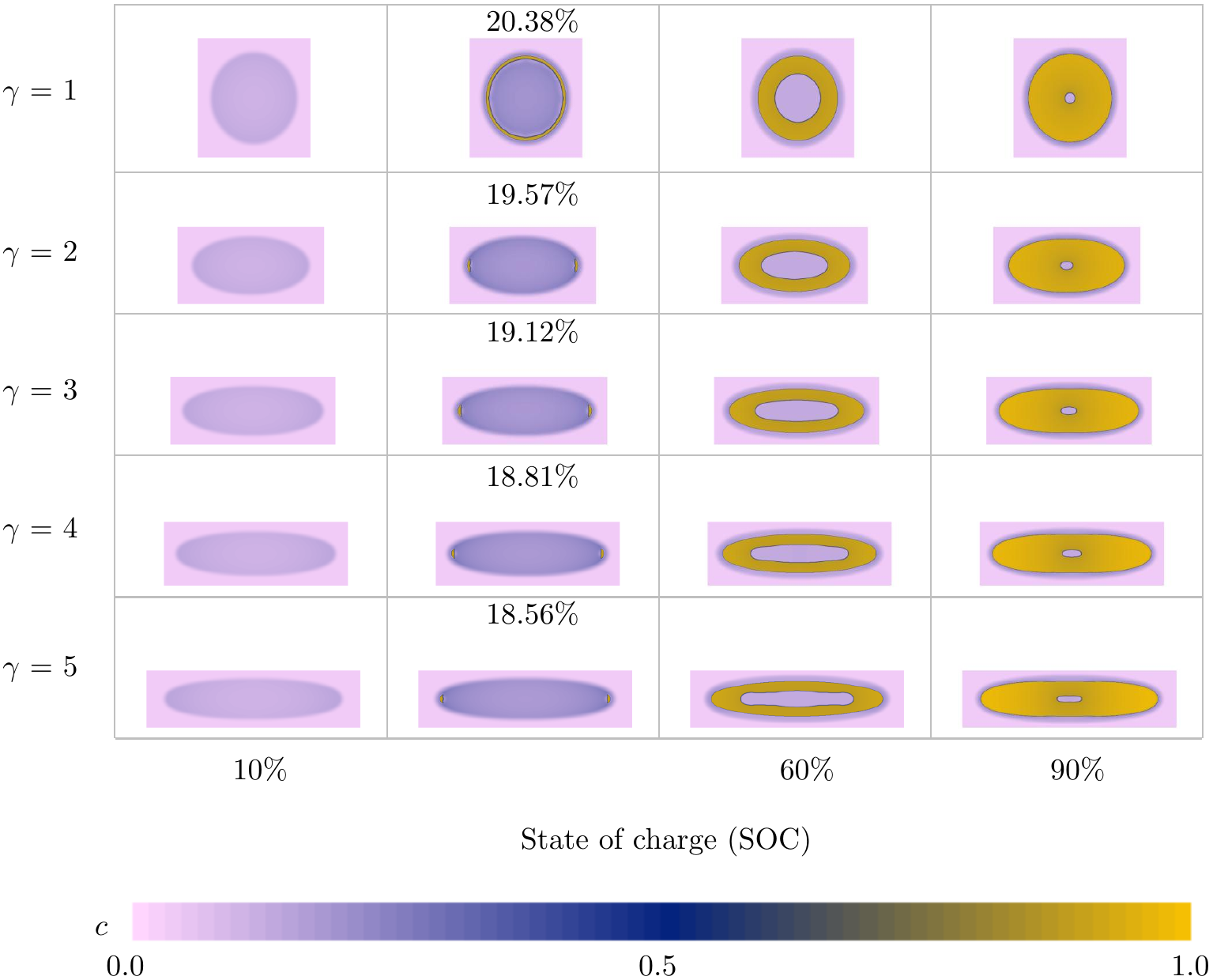}
\caption[Temporal evolution of isolated 2D elliptical particles with an initial concentration of ${c_i}=$ 0.03.]{Temporal evolution corresponding to a 10\%, 60\%, and 90\% state of charge, including an intermediate concentration profile at the initiation of a phase separation inside 2D elliptical particles, with aspect ratios $\gamma=$ 1, 2, 3, 4, and 5. The gold and magenta colors respectively correlate to the Li-rich and Li-poor phases. The particles are filled with an initial concentration of ${c_i}=$ 0.03.}
\label{fig:2DaspectRatiocrate1}
\end{center}
\end{figure}

\begin{figure}[h]
\begin{center}
\includegraphics[scale=1.0]{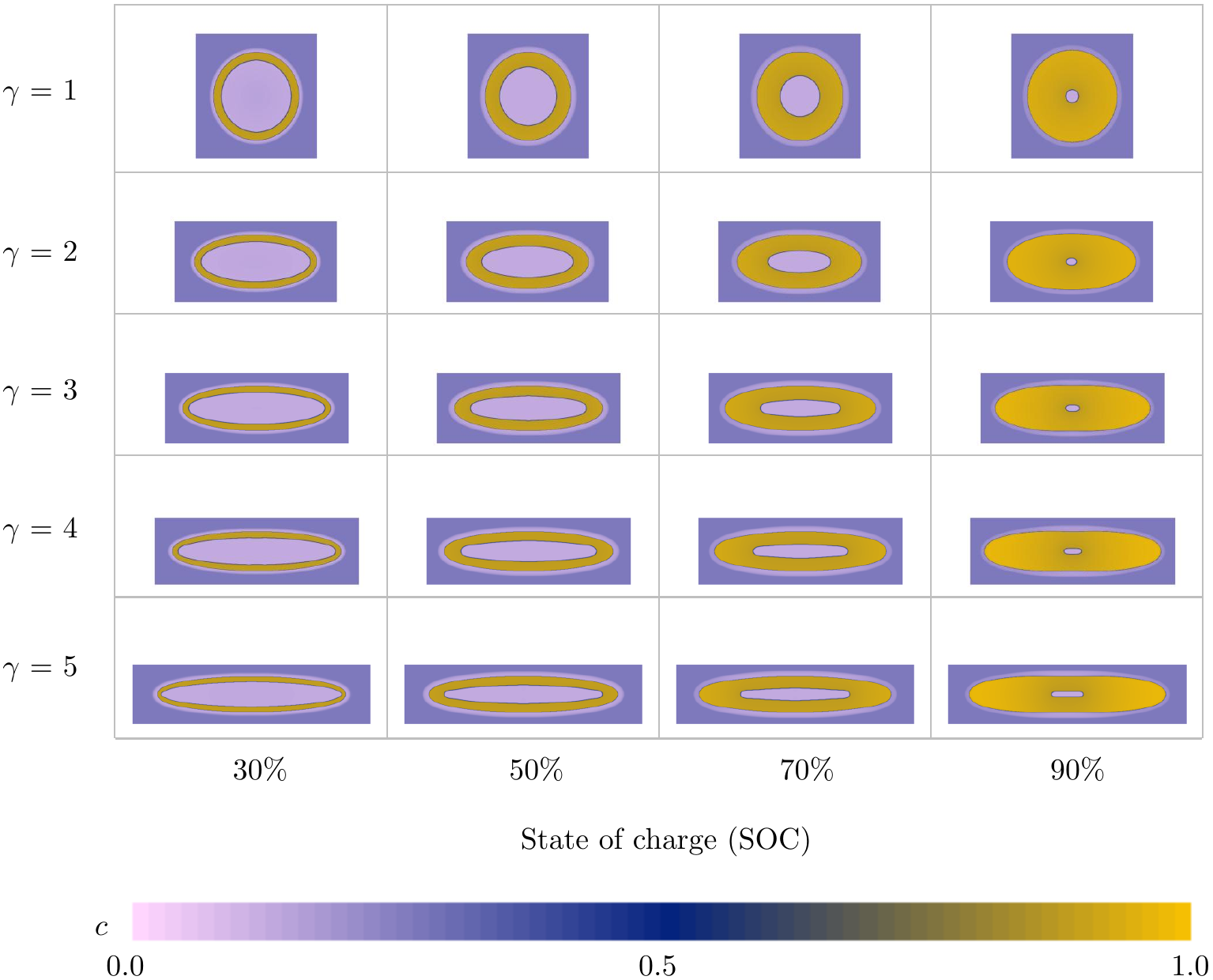}
\caption[Concentration profiles of isolated 2D elliptical particles with an initial concentration of ${c_i}=$ 0.26.]{Concentration profiles corresponding to 30\%, 50\%, 70\%, and 90\% state of charge inside 2D elliptical particles, with aspect ratios $\gamma=$ 1, 2, 3, 4, and 5. The gold and magenta colors correlate to the Li-rich and Li-poor phases, respectively. The particles are filled with an initial concentration of ${c_i}=$ 0.26.}
\label{fig:fiveParticle}
\end{center}
\end{figure}

To begin with, a 2D simulation study with a varying aspect ratio of ellipses of equal area is considered. The ellipses with the aspect ratios $\gamma$\nomenclature{$\gamma$}{aspect ratio of ellipse} = 1, 2, 3, 4, and 5 with major semiaxes are 1.00, 1.42, 1.74, 2.00, and 2.25 $\mu$m, respectively in Figure~\ref{fig:2DaspectRatiocrate1} with initial filling ${c_i} =0.03$\nomenclature{$c_i$}{initial concentration filling} and in Figure~\ref{fig:fiveParticle} with ${c_i} =0.26$. The first column in Figure~\ref{fig:2DaspectRatiocrate1} shows that a 10\% state of charge (SOC) corresponds to a homogeneous concentration profile. It is evident that the phase separation initiates in the particles with a higher aspect ratio, prior to the lower ones. For instance, phase segregation starts at 20.38\% SOC for $\gamma=1$, while at 18.56\% SOC for $\gamma=5$. Furthermore, the circular particle (i.e., $\gamma=1$) evolves with a continuous phase separation across the interface, while two Li-rich islands form at the maximum curvature points in the ellipses (i.e., $\gamma=2, 3, 4,$ and 5). The Li-rich islands continue to grow further at the particle surface. In Figure~\ref{fig:2DaspectRatiocrate1}, beyond 60\% SOC, the Li-poor phase is surrounded by the Li-rich phase in all particles, meaning they eventually follow the ``core-shell model", and the core shrinks as the shell enlarges with the evolution of time, which is called the ``shrinking-core model". Contrary to ${c_i} =0.03$ cases, there are no distinct Li-rich islands are observed for ${c_i} =0.26$, instead a continuous region of Li-rich phase completely surrounds the Li-poor phase for each particles as shown in Figure~\ref{fig:fiveParticle}.

\begin{figure}[h]
\begin{center}
\includegraphics[scale=1.02]{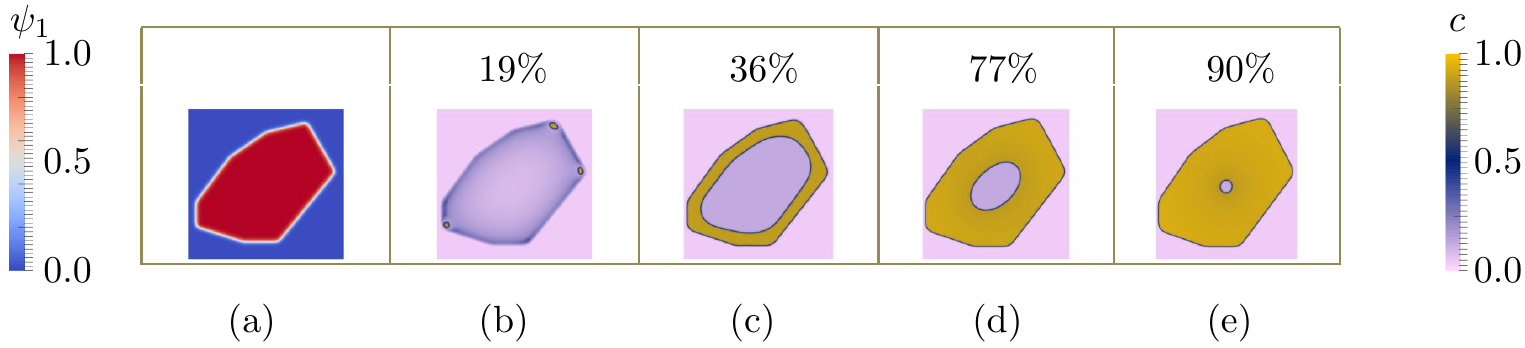}
\caption[Concentration profiles of 2D irregularly shaped particle]{ Concentration profiles of an irregularly shaped particle, when the SOC ranges from approximately 19\% to 90\%. The images correspond to 19\%, 36\%, 77\% and 90\%. The red color, $\psi_1$ = 1.0, corresponds to the cathodic particles, and the blue color, $\psi_1$ = 0.0, corresponds to the electrolyte in (a). The gold color represents the lithium-rich phase, and the magenta color inside the lithium-rich phase visualizes the lithium-poor phase from (b) onwards.}
\label{fig:2DirregularParticle}
\end{center}
\end{figure}

Figure~\ref{fig:2DirregularParticle} shows the evolution history of a 2D irregularly shaped particle. It is evident that the Li-rich phase initiates at the maximum curvature regions, as indicated by the three gold color islands in Figure~\ref{fig:2DirregularParticle}(b). These islands grow and, as a consequence, merge to form a single continent. Simultaneously, the Li-poor phase takes a curvaceous shape, followed by an ellipse and eventually transforms into a circle as it shrinks, which can be seen in Figure~\ref{fig:2DirregularParticle}(b), (c) and (d) respectively.

\begin{figure}[h]
\begin{center}
\includegraphics[scale=1.18]{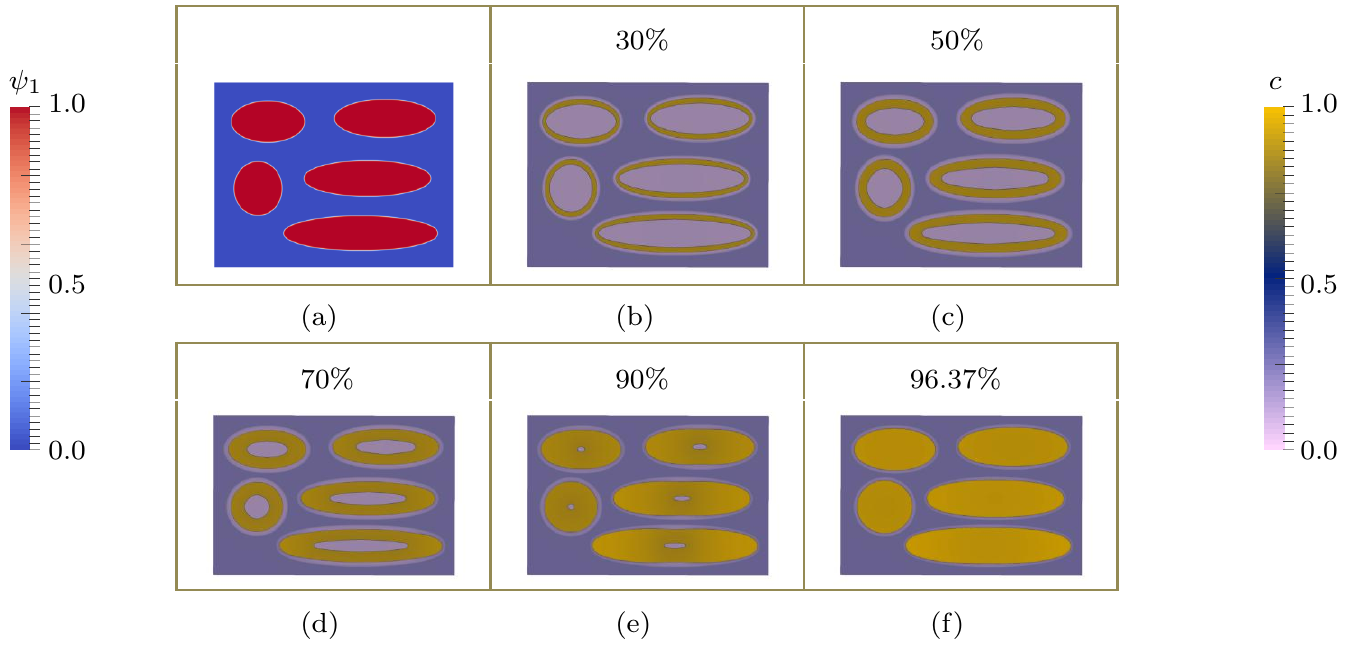}
\caption[Concentration profiles of five elliptical particles simultaneously immersed in an electrolyte with initial concentration $c_i = 0.26$]{Concentration profiles 
of five particles, simultaneously immersed in an electrolyte, when the SOC ranges from approximately 30\% to 90\%, displayed with an increment of 20\%, until it reaches 96.37\% in the last frame (f). In the entire domain, the domain parameter $\psi_1$ is continuously defined. The yellow color, $\psi_1$ = 1.0, corresponds to the cathodic particles and the blue color, $\psi_1$ = 0.0, corresponds to the electrolyte in (a). The gold color represents the lithium-rich phase, and the magenta color inside the lithium-rich phase visualizes to the lithium-poor phase from (b) onwards.}
\label{fig:fiveParticleSimultaneously}
\end{center}
\end{figure}

A 2D simulation study of five elliptical particles simultaneously immersed in the electrolyte, see Figure~\ref{fig:fiveParticleSimultaneously}. The particles are filled with an initial concentration of ${c_i}=$ 0.26. The particle aspect ratio varies from $\gamma = $ 1, 2, 3, 4 and 5 with 1.0, 1.54, 2.13, 2.72, and 3.35 $\mu$m major semiaxes and 1.00, 0.77, 0.71, 0.68, and 0.67 $\mu$m minor semiaxes ellipses, respectively. The circumference to area ratio of the particles is fixed. Therefore, $N_{\textrm{ex}}$ = 1 for all particles according to Eq.~\eqref{Eq:LIBGalvanostaticCurvature} and consequently, the particles are subject to identical applied flux and the particles evolve with identical SOC. The evolution pattern follows the widely known ``core-shell model." 

\begin{figure}[hbt!]
\begin{center}
\includegraphics[scale=1.4]{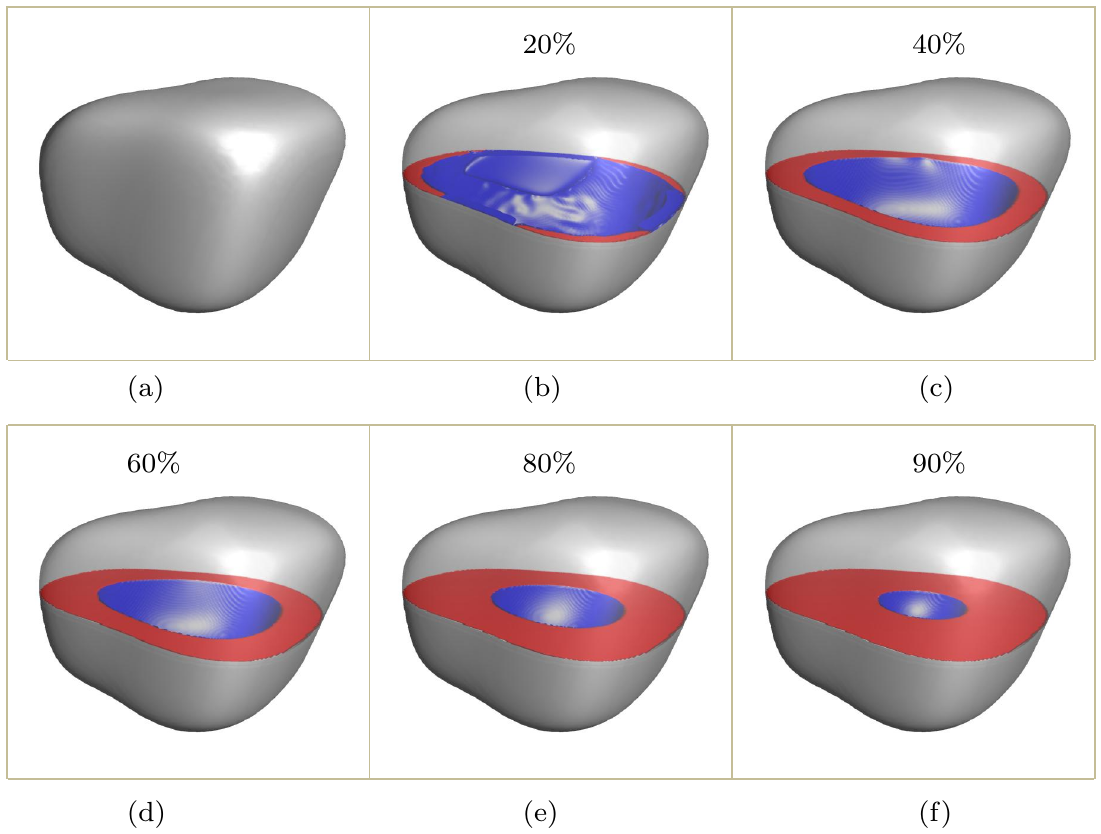}
\caption[Concentration profiles of 3D irregularly shaped particle]{Concentration profiles of an irregularly shaped particle, when the SOC ranges from approximately 20\% to 90\%. The images correspond to 20\%, 40\%, 60\%, 80\% and 90\%. The opaque silver colored figure in (a) indicates the contour of the particle surface at the value $\psi_1$ = 0.5 of the domain parameter, and the particle from the upper section is clipped for the sake of a better visual sight from (b) onwards. The red color is a contour of ${c}=0.80$ and it represents the lithium-rich phase. The blue color is a contour of ${c} = 0.22$ and the blue color cavity inside the lithium-rich phase corresponds to the lithium-poor phase.}\label{fig:3DParticle}
\end{center}
\end{figure}

Figure~\ref{fig:3DParticle} shows the evolution history of a 3D particle with an irregular shape. The Li-rich islands, which are marked in red color, develop at the convex surface, which corresponds to the maximum curvature of the particle. The islands progress to form a single continent at 40\% SOC and eventually, the core shrinks with time.

\begin{figure}[h]
\begin{center}
\includegraphics[scale=0.91]{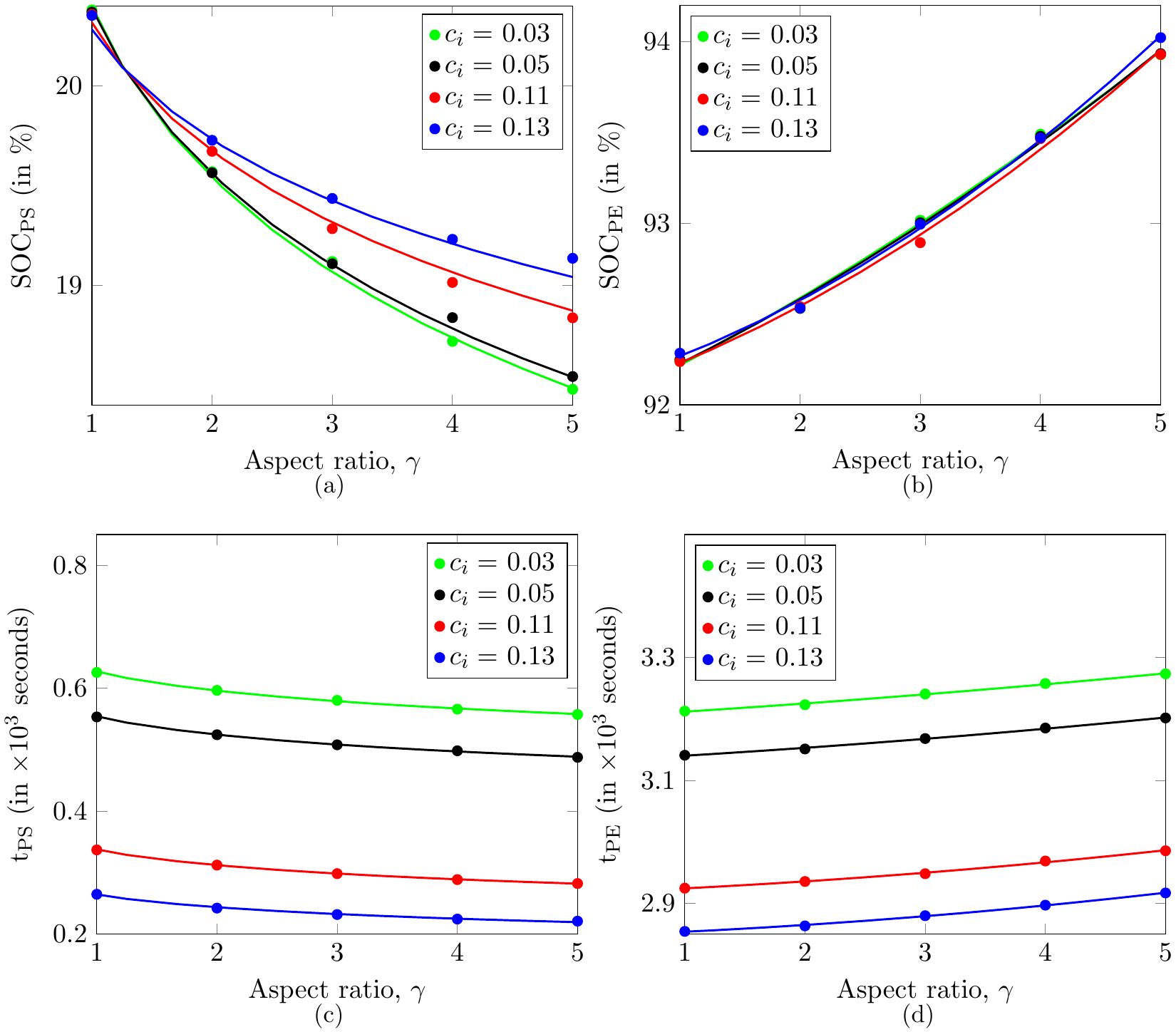}
\caption[Effect of initial concentration ${c}_i$ on the phase separation dynamics]{The SOC (or duration) at which phase separation initiates, SOC$_\mathrm{PS}$ in (a) (or $t_{\mathrm{PS}}$\nomenclature{$t_\mathrm{PS}$}{time at the onset of phase separation} in (c)) and completes\nomenclature{SOC$_\mathrm{PE}$}{state of charge attained before the end of phase separation} SOC$_\mathrm{PE}$ in (b) (or $t_{\mathrm{PE}}$\nomenclature{$t_\mathrm{PE}$}{time attained at the end of phase separation} in (d)) are plotted as a function of the aspect ratio of the 2D elliptical particles for the initial concentrations ${c_i} =$ 0.03, 0.05, 0.11 and 0.13. The dotted lines in (a) are plots of Eq. (\ref{eq:SOCvsGamma}), while in (b), (c) and (d) are line fits for respective values of ${c}_i$.}
\label{fig:2DSOCvsGamma}
\end{center}
\end{figure}

For various initial concentrations, a comprehensive study is performed as a function of aspect ratio $\gamma$ for a SOC at which phase separation is expected, \nomenclature{SOC$_\mathrm{PS}$}{state of charge at the onset of phase separation}$\mathrm{SOC}_\mathrm{PS}$, see Figure~\ref{fig:2DSOCvsGamma}(a). The curve fit suggests the empirical relation,
\begin{equation}\label{eq:SOCvsGamma}
\mathrm{SOC}_\mathrm{PS} = (-6.8  {c}_i^2 +20) \gamma^{(1.3  {c}_i^2-0.061)} \textrm{ for } 0.00<{c}_i<0.20,
\end{equation}
where, SOC$_\mathrm{PS}$ is the SOC at the onset  of phase separation. It is evident that increasing the initial concentration reduces the difference in the values of SOC$_\mathrm{PS}$ between the higher and the lower aspect ratio particles continuously. As a consequence, at a certain initial concentration, namely ${c_i} = 0.26$, phase separation starts in all particles simultaneously as seen in Figure \ref{fig:fiveParticle}.

The basic features of the charge dynamics and its dependence on the various model parameters are explained with feasible reasons in the following section.

\begin{figure}
\begin{center}
\includegraphics[scale=1.0]{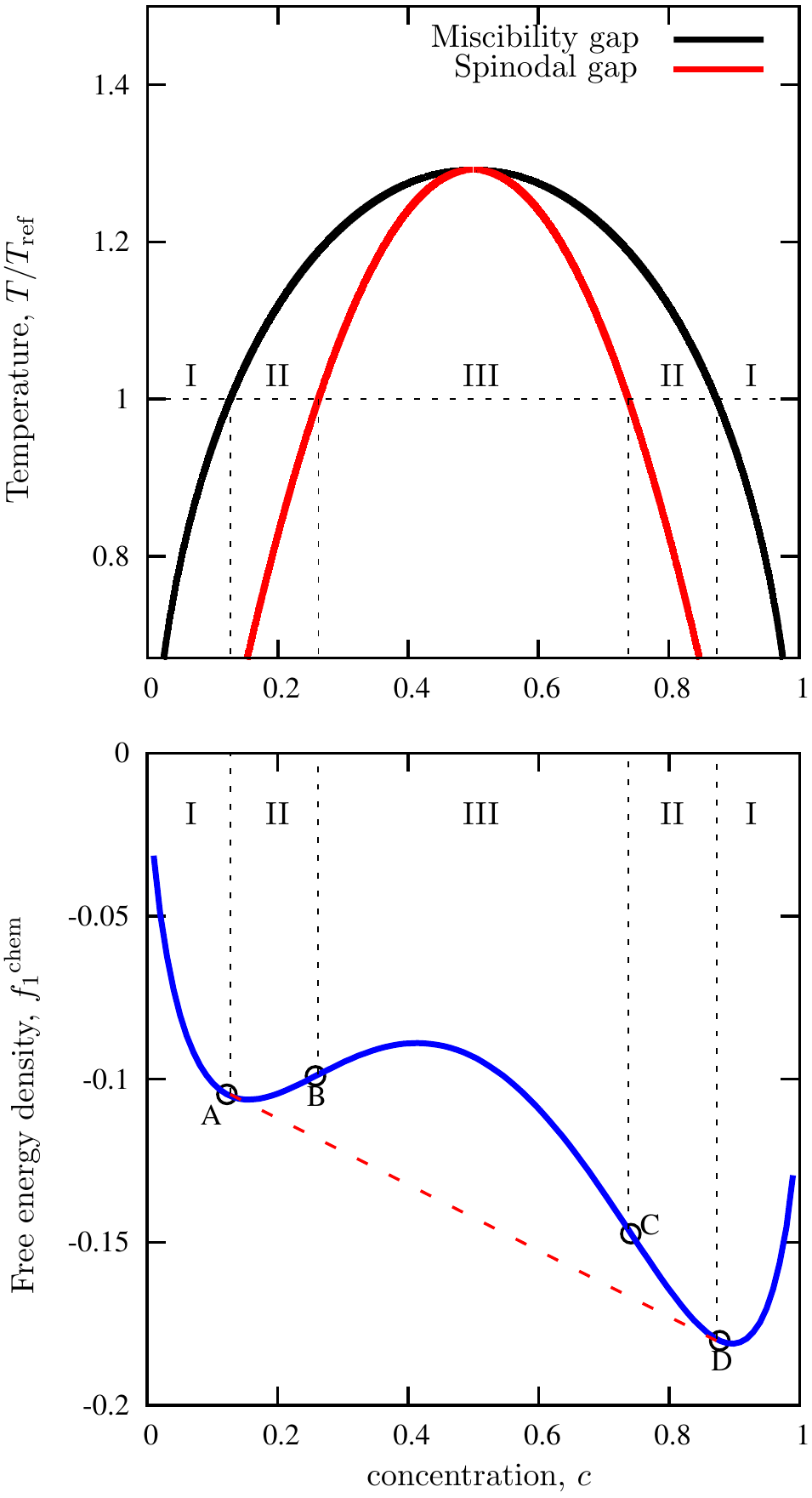}
\caption[Phase diagram indicating various zones characterized by the miscibility and the spinodal gaps]{The phase diagram shows various zones, which are characterized by the miscibility gap and the spinodal gap. Zones I, II and III respectively correspond to homogeneous states, nucleation states, and phase-separated states. The solid blue line in the bottom graph is a chemical-free energy density, $f_1^{\mathrm{chem}}$ at absolute temperature, $T=300$ K. The projections of the miscibility gap, on the free energy density curve, indicate the equilibrium conditions expressed in Eqs.~(\ref{Eq:potential}) and (\ref{Eq:grandPotential}). The projections of the spinodal gap, on the free energy density curve, indicate zero curvature points. The dotted red line in the bottom graph corresponds to the minimum energy path, widely known as Maxwell construction.}
\label{fig:phaseDiagram}
\end{center}
\end{figure}

\section{Miscibility gap}
\label{sec:GalvanostaticBasicFeatures}

The local concentration characterizes a condition of the particles, which corresponds to the zones I, II and III in Figure~\ref{fig:phaseDiagram}, which determine a phase-segregated state or a homogeneous state during intercalation. In the phase equilibrium diagram of the LMO particle, these zones are related to spinodal and miscibility gaps. In Figure~\ref{fig:phaseDiagram}, the spinodal gap is the locus of points where the curvature of the chemical free energy density changes its sign, i.e. from convex to concave, at point B and from concave to convex, at point C. Let $c^L$\nomenclature{$c^L$}{lower concentration level of a phase-separated state} and $c^H$\nomenclature{$c^H$}{higher concentration level of a phase-separated state} be the two concentration levels of a phase-segregated state for the Li-rich and the Li-poor phases, respectively. In electrode particle ($\psi_1=1$), the miscibility gap is obtained by respectively equating the chemical potential $\mu_1^{\mathrm{chem}}$ and the grand-chemical potential \nomenclature{$G_a^{\mathrm{chem}}$}{grand-chemical potential of phase $a$}$G_1^{\mathrm{chem}}=f_1^{\mathrm{chem}} - c \mu_1^{\mathrm{chem}}$ for both concentrations, $c^L$ and $c^H$, as
\begin{equation}
\mu_1^{\mathrm{chem}}(c^L) = \mu_1^{\mathrm{chem}}(c^H), \label{Eq:potential}
\end{equation}
\begin{equation}
f_1^{\mathrm{chem}}(c^L) - c^L \mu_1^{\mathrm{chem}}(c^L) =f_1^{\mathrm{chem}}(c^H) - c^H \mu_1^{\mathrm{chem}}(c^H).\label{Eq:grandPotential}
\end{equation}

The simulation study is carried out at an absolute temperature of $T=$ 300 K. The corresponding chemical free energy density plot in Figure~\ref{fig:phaseDiagram} represents the double-well structure, which entails the presence of a concavity between the two wells, and hence the presence of miscibility and spinodal gaps. The concentration of the miscibility gap fulfills the equilibrium conditions expressed in Eqs.~(\ref{Eq:potential}) and (\ref{Eq:grandPotential}), which are marked by the points A and D. Between these points, the free energy can be decreased by decomposing the homogeneous state of higher energy to a phase-segregated state of lower energy. The spinodal gap concentrations between the zero curvature points of the free energy curve, marked as points B and C, render the homogeneous distribution states unstable and force the particle into phase segregation. In Figure~\ref{fig:phaseDiagram}, the red dotted line refers to the common tangent construction and represents the minimum energy line.

With the application of flux, a concentration gradient is established from the surface to the center of the particle. Therefore, the phase segregation initiates from the surface of the particle. The generation and increase in the pre-existing phase boundary between the Li-rich and the Li-poor phases are energetically costly, and the system follows a path with the minimum phase boundary possible. The increase in the phase boundary portion leads to an energy penalty. Thus, the equilibrium states tend towards a minimum phase boundary inside the particle, following the evolution in the direction of the minimum free energy path. As a consequence, the inner phase shrinks by assuming an irregular shape with a curved boundary, at the beginning, which then transforms into an ellipse, and eventually into a circle, as shown in Figure~\ref{fig:2DirregularParticle}.

 In Figure~\ref{fig:2DaspectRatiocrate1}, the uniform curvature along the surface of the circular particle ($\gamma$~=~1) results in a continuous phase separation across the particle surface. In the case of $\gamma$~=~2, 3, 4 and 5, two Li-rich islands develop at the maximum curvature points. The likely reason behind the formation of the islands is that the local concentration at these positions reaches the spinodal zone first and causes a local phase separation. In an ellipse, the largest and shortest distances from the center of the ellipse to the surface correspond to the major and minor semiaxes. There, the maximum and minimum curvature points are also located. In the vicinity of the maximum curvature points, the particles have a larger circumference to the adjacent particle area, i.e., a larger local circumference to area ratio, which is in contrast to the minimum curvature points. As a consequence, a higher concentration builds up at the surroundings of the endpoints of the major semiaxis, in contrast to the minor semiaxis, which is in agreement with the literature~\cite{guo2017analytical, stein20143d, crank1975the}. This comparison suggests that more sites are available for the applied flux, compared to the sites that host lithium species near the major semiaxis. This triggers a phase separation process during the insertion as seen in Figure~\ref{fig:2DaspectRatiocrate1}. 

\begin{table}
\caption[Maximum and minimum concentration, along the surface of the particles, for the initial concentrations ${c}_i = 0.03$ and $0.13$ at 18\% SOC]{The maximum and minimum concentration, along the surface of the particles, for the initial concentrations ${c}_i = 0.03$ and $0.13$, at 18\% SOC. The maximum concentration observed in a region with maximum curvature, while the minimum concentration is observed in a region with minimum curvature in ellipses. \label{Table:maxMinConcentration}}
\begin{center}
\begin{tabular}{ c  c  c  c c}
 \hline
Aspect ratio, $\gamma$ & \multicolumn{2} {c} {Maximum concentration, ${c}$} & \multicolumn{2} {c} {Minimum concentration, ${c}$}   \\
 &${c_i}=0.03$ & ${c_i}=0.13$ & ${c_i}=0.03$& ${c_i}=0.13$\\
 \hline 
 1 & 0.200 & 0.201 & 0.200 & 0.201\\
 2 & 0.216 & 0.214 & 0.187 & 0.188\\
 3 & 0.224 & 0.220 & 0.180 & 0.182\\
 4 & 0.231 & 0.223 & 0.175 & 0.179\\
 5 & 0.239 & 0.225 & 0.172 & 0.177\\
 \hline
\end{tabular}
\end{center}
\end{table} 
 
 The difference between the maximum and minimum concentration across the surface of the particle increases with the aspect ratio, as can be seen in Table~\ref{Table:maxMinConcentration}. Therefore, the maximum concentration point reaches the spinodal gap with less SOC for particles with a higher aspect ratio. Hence, the SOC required to initiate a phase separation decreases with the aspect ratio. Similar to Figure~\ref{fig:3DParticle}, the higher curvature region accumulates more lithium, and arrives at the spinodal region, prior to the lower curvature region. Therefore, the phase separation starts at the corners and edges which correspond to the higher curvature region. In the meantime, it can be observed that the particle surfaces close to the minor semiaxis accumulate further species. The Li-rich islands continue to grow and eventually, the islands merge to form a single continent. 
 
 \section{Influence of various parameters on phase separation mechanism}
 \label{sec:GalvanostaticParameterStudy}
\subsection{Effect of initial concentration}
Figure \ref{fig:2DSOCvsGamma}(a) shows the increment in SOC$_{\mathrm{PS}}$ for higher aspect ratio particles with a rise in initial concentration ${c}_i$. An explanation is the decrease in the difference between a maximum and a minimum concentration at higher ${c}_i$, see Table \ref{Table:maxMinConcentration}. Furthermore, an increase in ${c}_i$ results in less concentration required locally to reach the spinodal zone, and drive phase separation. If the initial concentration ${c}_i$, in all particles of Figure~\ref{fig:fiveParticle} falls in the spinodal zone, phase separation is initiated everywhere at the surface of the particles simultaneously, regardless of the curvature variation across the particle surface. Furthermore, the particle with the higher ${c_i}$ requires less time to reach the same SOC compared to the lower. Therefore, the duration required to observe the onset of phase separation (i.e., $t_{\mathrm{PS}}$) decreases for higher initial concentration, as reported in Figure~\ref{fig:2DSOCvsGamma}(c).
 
 Although the initiation of the phase separation process is affected by the initial concentration of ${c_i}$, the completion of phase separation is independent, as depicted in Figure~\ref{fig:2DSOCvsGamma}(b). During the spinodal decomposition, the particle attains two concentration levels (i.e., Li-rich and Li-poor phases), irrespective of the initial concentrations. In other words, the phase separation process hinders the effect of initial concentration. Therefore, the effect of different initial concentrations on the SOC required to complete the phase separation ( i.e., SOC$_\mathrm{PE}$) is not observed. In addition, the effect of variation of ${c_i}$ on the duration until the phase separation observed $t_{\mathrm{PE}}$ is reported in Figure~\ref{fig:2DSOCvsGamma}(d). Obviously, the increase in ${c_i}$ shows additional leverage to SOC, which does not require to accumulate during operation. As a consequence, the increase in ${c_i}$ leads to a decrease in the time to complete the phase separation $t_{\mathrm{PE}}$.
 
\begin{figure}[h]
\includegraphics[scale=0.96]{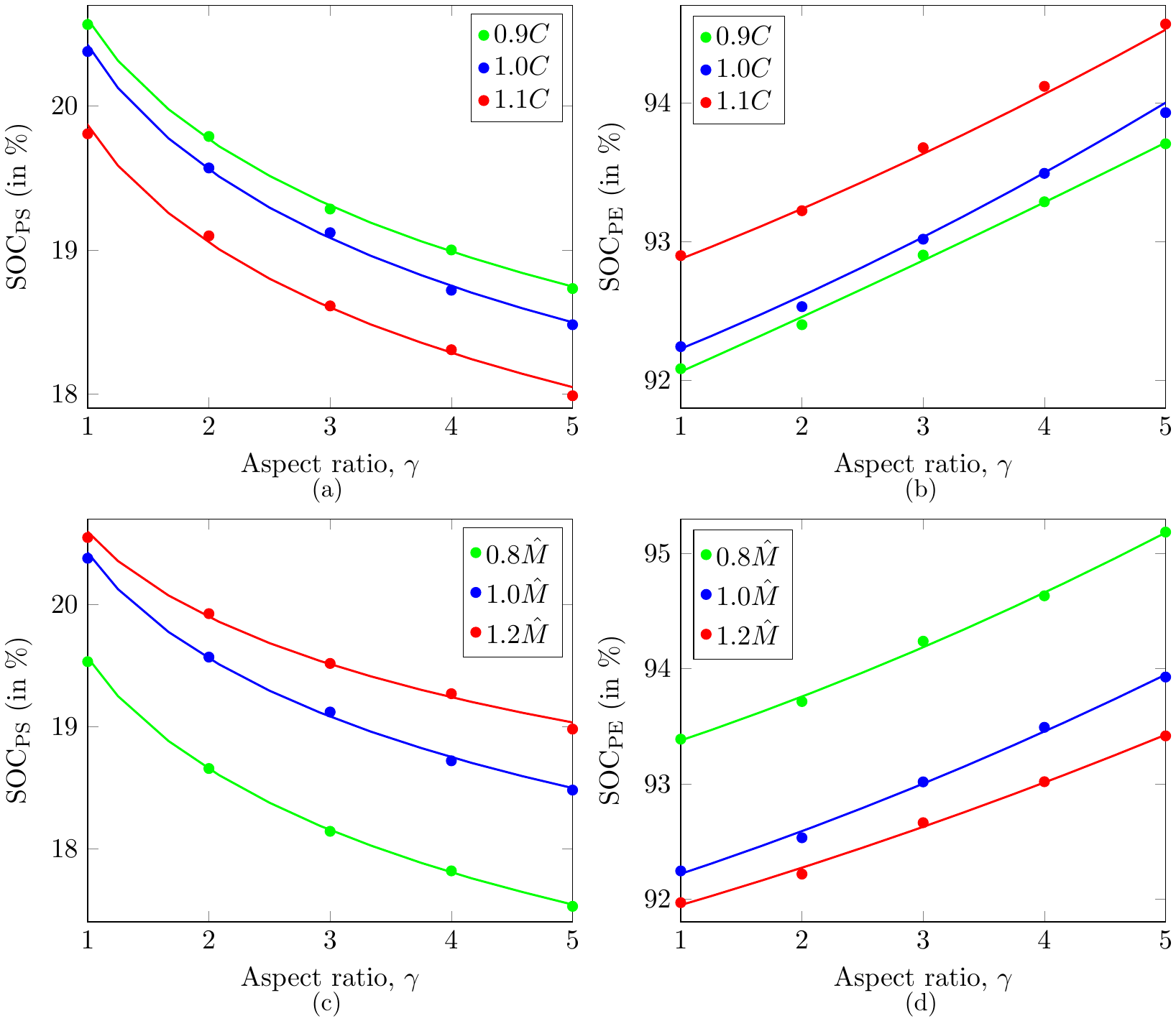}
\caption[Variation of C-rate and mobility as a function of the aspect ratio of the elliptical particles]{Variation of C-rate as a function of the aspect ratio of the elliptical particles for SOC at which the onset of phase separation starts  SOC$_{\mathrm{PS}}$  in (a) and ends SOC$_{\mathrm{PE}}$ in (b). The variation of mobility $\hat{M}$ as a function of SOC$_{\mathrm{PS}}$ in (c) and SOC$_{\mathrm{PE}}$ in (d).}\label{fig:variationCrateMobility}
\end{figure}

\subsection{Effect of C-rate and mobility}

Ideally in the quasi-static condition (i.e., C-rate $C\ll 1$) \cite{huttin2012phase}, the phase separation starts when the average concentration reaches the spinodal point B and ends outside the miscibility gap  (point D in Figure~\ref{fig:phaseDiagram}). In fact, the local concentrations at all positions are almost equal to the average concentration in quasi-static conditions. However, higher values of C-rate stimulate a steeper concentration gradient from the surface to the center. Therefore, a local concentration at the surface reaches the spinodal point before the average concentration. The difference between these two increases with flux. As a result, SOC required to observe the onset of phase separation SOC$_{\mathrm{PS}}$ decreases for higher flux rates, which is depicted in Figure~\ref{fig:variationCrateMobility}(a). Contrarily, the phase separation ends when the Li-poor phase is completely depleted by the Li-rich phase. The slope of the concentration gradient increases with the C-rate. As a consequence, the presence of gradient penalizes the SOC required to end the phase separation SOC$_{\mathrm{PE}}$ on a higher side, which can be seen in Figure~\ref{fig:variationCrateMobility}(b).

On the one hand, mobility can be related to the inherent ability of the system to homogenize the concentration distribution by transferring species in order to eliminate any gradients developed during operation. On the other hand, C-rate enhances the concentration gradient at the surface by depositing the flux of lithium species. The interplay between the deposition and transfer rates characterizes the equilibrium states. At the elevated mobility, the species transfer rate is more prominent compared to the species deposition. As a consequence, the species disperse immediately after the deposition at the surface. Thus, a particle requires higher SOC to observe phase separation at higher mobility as depicted in Figure~\ref{fig:variationCrateMobility}(c). Furthermore, the slope of concentration decreases at the agile mobility, which results in decreased SOC$_{\mathrm{PE}}$ (see Figure~\ref{fig:variationCrateMobility}(d)). Conversely, the lower mobility provides comparatively enough time for the accumulation of the species, which are deposited at the particle surface. Hence, SOC$_{\mathrm{PS}}$ decreases and SOC$_{\mathrm{PE}}$ increases for sluggish mobility as an effect from the increased slope of the concentration gradient.

A very high mobility (or very low C-rate) corresponds to the steady-state, in which phase separation starts at spinodal point B in Figure~\ref{fig:phaseDiagram}, for all particles irrespective of curvature. In other words, the curvature effects are suppressed in steady-state. As the mobility decreases (or C-rate increases), the curvature effects become prominent as a result of the development of concentration gradients. For instance, the increase in the difference between SOC$_{\mathrm{PS}}$ for the $1.0\hat{M}$ and $1.2\hat{M}$ with aspect ratio $\gamma$ is plausible due to the particle curvature.

\begin{figure}
\begin{center}
\includegraphics[scale=0.89]{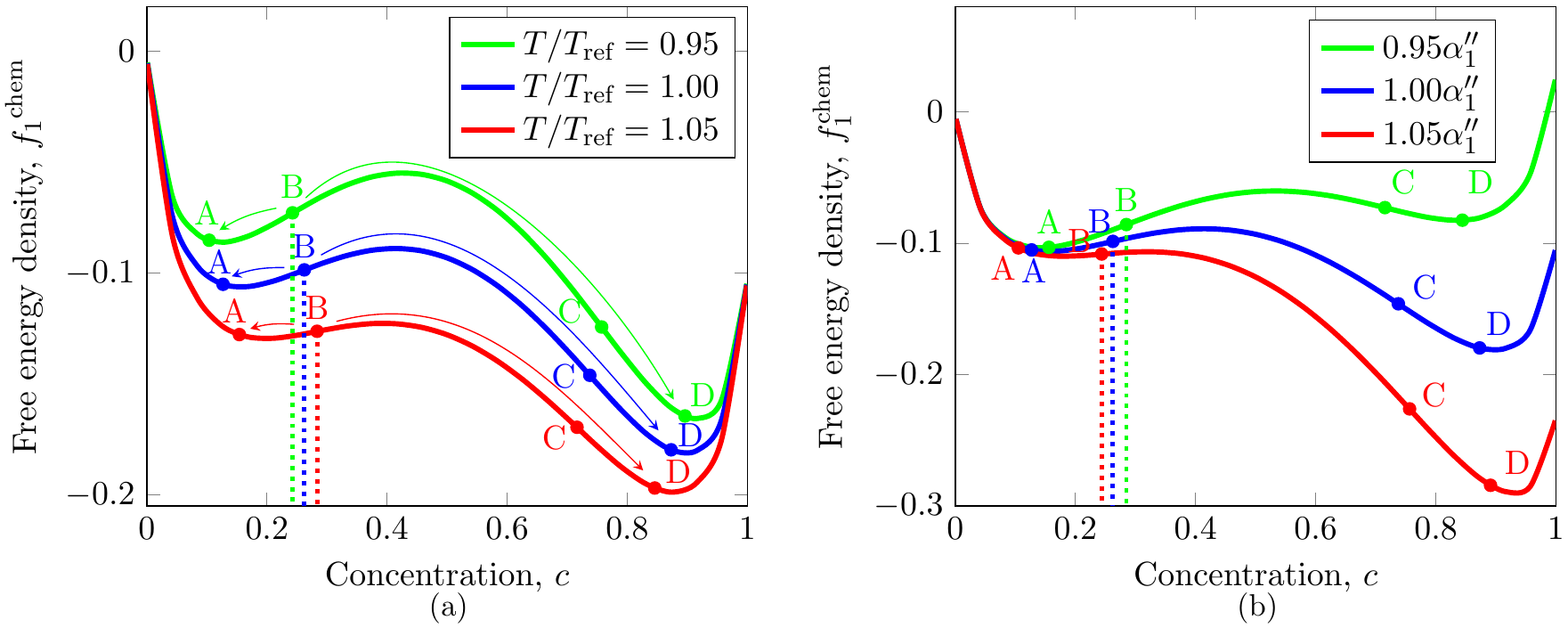}
\caption[Effect of temperature $T/T_{\mathrm{ref}}$ and free energy parameter $\alpha_1''$ on free energy density plots]{The effect of variation in (a) temperature $T/T_{\mathrm{ref}}$ and (b) free energy parameter $\alpha_1''$ on free energy density plots. The points A and D represent free energy local minima obtained from Eqs.~(\ref{Eq:potential}) and (\ref{Eq:grandPotential}), while B and C are spinodal points where the curvature of the plot changes its sign.}\label{fig:threeEnergyPlots}
\end{center}
\end{figure}

\subsection{Effect of temperature and free energy parameter $\alpha_1''$}

\begin{figure}[h]
\includegraphics[scale=0.96]{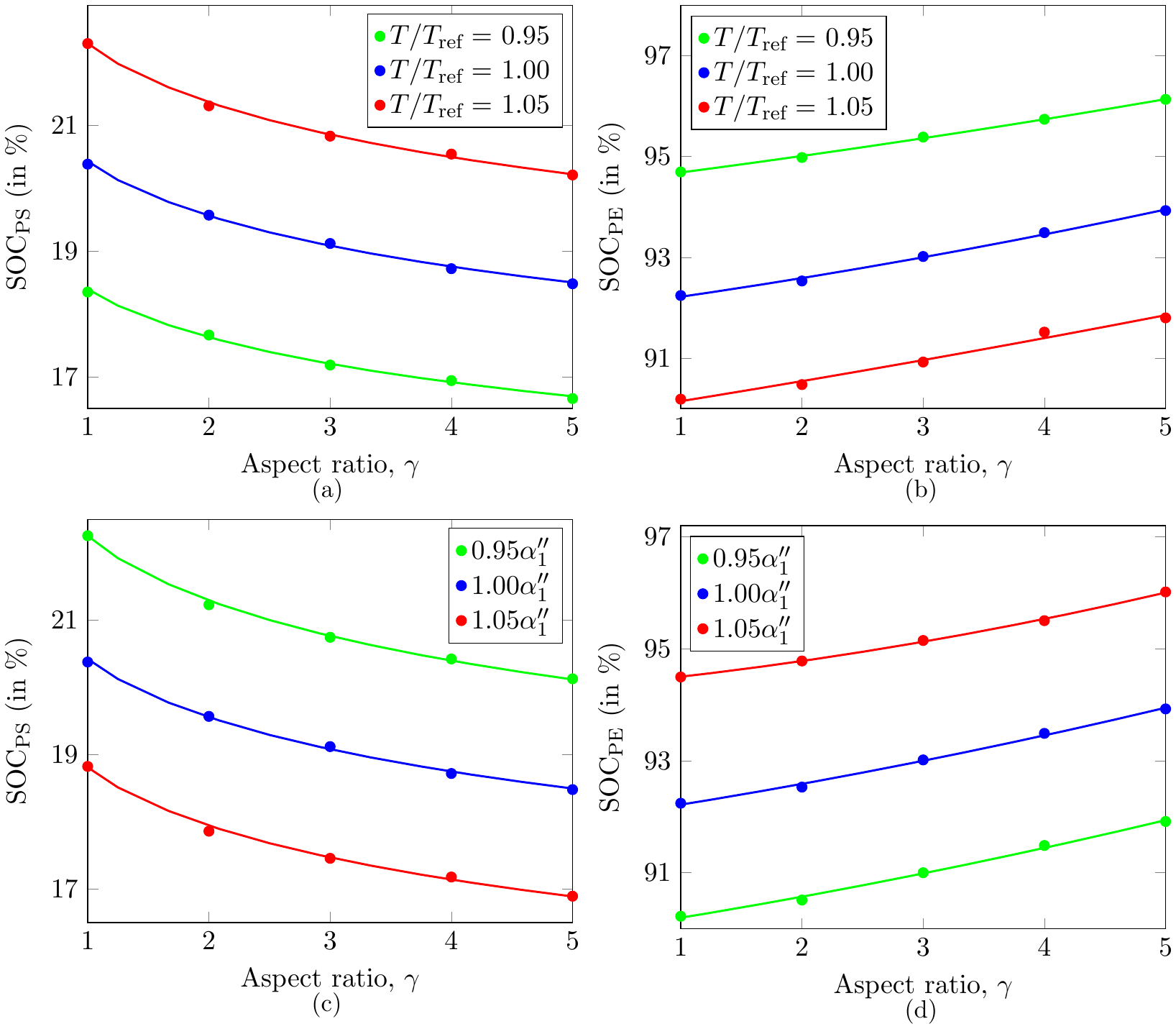}
\caption[Variation of the temperature and free energy parameter as a function of the aspect ratio of the elliptical particles]{Variation of the temperature $T/T_{\mathrm{ref}}$ as a function of the aspect ratio of the elliptical particles for SOC at which the onset of phase separation starts  SOC$_{\mathrm{PS}}$ in (a) and ends SOC$_{\mathrm{PE}}$ in (b). The variation of free energy parameter $\alpha_1''$ as a function of SOC$_{\mathrm{PS}}$ in (c) and SOC$_{\mathrm{PE}}$ in (d). }\label{fig:variationAlphaTemp}
\end{figure}

During operating conditions, the battery is seldom subjected to a constant temperature. Thus, understanding the effect of variation in operating temperature on the phase separation dynamics has technological implications. The free energy density for three different values of the temperature $T/T_{\mathrm{ref}}$ is plotted in Figure~\ref{fig:threeEnergyPlots}(a). The particle, which is initially filled with concentration value of ${c_i}=0.03$, starts accumulating species during insertion and consequently increases in local concentration. Note that, if concentration at a particular position of the particle reaches spinodal zone (i.e., point B), it decomposes into two concentration levels (i.e., lower concentration point A and higher concentration point D). The rise in temperature shifts the spinodal point B on the higher side (see Figure~\ref{fig:threeEnergyPlots}(a)). In other words, the system requires a higher value of concentration so that any of the positions lie in the spinodal zone at the elevated temperature. Therefore, SOC required to observe the phase separation SOC$_\mathrm{PS}$ increases with temperature $T / T_{\mathrm{ref}}$, which can be seen in Figure~\ref{fig:variationAlphaTemp}(a). On the contrary, the rise in temperature shortens the miscibility gap, i.e., a shift in point D towards the lower concentration values (see Figure~\ref{fig:threeEnergyPlots}(a)). As the point D is closely related to SOC$_\mathrm{PE}$, the effect of a variation of the temperature $T/T_{\mathrm{ref}}$ is justified in Figure~\ref{fig:variationAlphaTemp}(b).

Figures~\ref{fig:variationAlphaTemp}(c) and (d) show the effect of variation of the free energy parameter $\alpha_1''$. In Figure~\ref{fig:threeEnergyPlots}(b), it can be seen that the miscibility and spinodal gaps enlarge with an increase in $\alpha_1''$, the behavior is opposite to the temperature $T/T_{\mathrm{ref}}$.  Therefore, SOC$_\mathrm{PS}$ decreases and SOC$_\mathrm{PE}$ increases for higher $\alpha_1''$, which shows an opposite tendency compared to the temperature $T/T_{\mathrm{ref}}$.
 
 Finally, Figure~\ref{fig:fiveParticleSimultaneously} and Figure~\ref{fig:3DParticle} exhibits the capability of the method to simulate multiple particles simultaneously and three-dimensional particles with an irregular shape immersed in an electrolyte. As the particle boundary condition is implicitly defined in the smoothed boundary method, this method can be applied to particles with almost any geometry. Hence, this technique is very powerful and convenient to solve differential equations in complex geometries with complex boundary conditions that are often difficult to mesh.

\section{Conclusion}
\label{sec:NuemannFluxConclusions}
A phase-field study is performed to simulate phase separation during the insertion process. The presented model employs two coexisting phases, which is validated with a benchmark. In addition, the increase in mesh resolution conveys the convergence of the presented method. In the finite-difference framework, the smoothed boundary method enables the modeling of particles with almost any geometry. The effects of the particle geometry on the concentration evolution are explored numerically. For a constant flux boundary condition at the particle surface, the free energy density and the chemical potential are discussed in detail. Based on that, the spinodal and miscibility gaps are estimated. 

The simulation results show that the growth of the phase boundary between the Li-poor and Li-rich phases, inside the particle, leads to an energy penalty. As a consequence, the morphological evolution of the concentration profile suggested a minimum phase boundary pathway. With the application of flux, the particle surface with a higher curvature preferentially accumulates lithium species. Therefore, phase segregation starts in the vicinity of the regions with a higher curvature. The reason for the inhomogeneous phase separation across the particle surface is explained in detail, by means of miscibility and spinodal gaps. Furthermore, the elliptical particle with a higher aspect ratio is subjected to the onset of the phase separation, prior to the lower ones.

Finally, it is evident from Figure~\ref{fig:fiveParticleSimultaneously} that the smoothed boundary method can be applicable to multiple particles simultaneously. However, the significance of the simulation involving multiple particles is restricted due to the boundary conditions are applied on the particle surfaces. Therefore, the influence of neighboring particles is limited. In fact, the transportation of species in any particle is independent of the other particles and therefore could be simulated separately of the others, as shown in Figure~\ref{fig:fiveParticle}. Hence, the concentration profiles are identical for the densely packed particles and the sparsely packed, which should not be the case generally. This provokes an extension of the simulation study to identify the significance of the microstructural properties of the electrodes consists of many particles. Therefore, these issues are addressed in the next chapter.

\chapter{Morphological descriptors in multiple particle porous electrodes}
\label{chapter:Potentiostatic}

High-performance batteries necessitate electrodes of superior active material and more importantly, optimized porous microstructures \cite{dreyer2011behavior,arico2011nanostructured}. Recent advancements in manufacturing facilities provide great control over the particle size and the complex-tortuous microstructure of electrodes \cite{ebner2014tortuosity}. Thus, electrode engineering allows producing high-density electrodes without compromising its rate capability. The physical microstructure of the electrode has a direct impact on battery performance measures such as cyclic stability, intercalation rates, and power density amongst others \cite{marks2011a}. Therefore, it is important to quantify the parameters that control electrode microstructure, such as particle shape, size, porosity, and tortuosity. For instance, an experimental study \cite{buqa2005high} reported that the rate performance of the graphite electrode to be a function of particle size and porosity of the electrode. In addition, images from X-ray tomography suggest that the microstructural inhomogeneities largely influence the direction of lithium transport \cite{harris2010direct, harris2013effects}.

Systematic numerical studies of porous electrodes may complement the scientific understanding of the hierarchical microstructures and help to optimize the process parameters efficiently and economically \cite{santhanagopalan2006review, newman1975porous‐electrode, lai2011mathematical}. Despite some theoretical and numerical studies are performed to account essential features of phase-separating porous electrodes \cite{bai2013statistical, li2014current, smith2017multiphase, chueh2013intercalation}, the literature providing exclusive information on the microstructural properties of the electrodes is scarce. For instance, Ferguson and Bazant \cite{ferguson2012nonequilibrium} showed a change in effective diffusivity relates to the system porosity and tortuosity. Besides, Orvananos et al. \cite{orvananos2014architecture} demonstrated the impact of electrode architecture in terms of ionic and electronic connectivities by considering interactions between active particles. Furthermore, Vasileiadis et al. \cite{vasileiadis2018toward} investigated that the tortuosity, particle sizes, porosity, and electrode thickness influence the capacity of Li$_4$Ti$_5$O$_{12}$ (LTO)\nomenclature{LTO}{lithium titanate, Li$_4$Ti$_5$O$_{12}$} electrodes as a function of C-rate. However, most of the studies utilize an approximation by the Bruggeman relation \cite{bruggeman1935berechnung} for the calculation of tortuosity, instead of estimating the tortuosity based on the electrode microstructure under investigation \cite{joos2011reconstruction, ender2012quantitative, ender2011three}. In addition, these porous electrode phase-field models are focused on LiFePO$_4$ or Li$_4$Ti$_5$O$_{12}$ electrode materials. Relatively few models have been developed for LMO materials \cite{huttin2012phase, zhang2018nonlocal, santoki2018phase, walk2014comparison}, which are limited to single particles. Therefore, this chapter aims to provide crucial measures to define the electrode microstructures and the relation of those attributes to the performance of LMO electrodes. Additionally, the microstructural properties, specifically the tortuosity, are estimated numerically to provide a better understanding of the transport mechanism \cite{joos2011reconstruction}. 

In the present chapter, the transportation rates of porous electrode structures with the two-phase coexistence in LMO particles is investigated. In this initial effort, neglecting the reaction kinetics at the electrode-electrolyte interface, the focus is based on the diffusional properties of the electrode and the electrolyte. In addition, an explicit treatment of mechanical effects due to misfit strains at the phase boundary is avoided. Instead, the emphasis is placed on the influence of various microstructural properties such as particle sizes, porosity, and tortuosity of electrode consists of ellipsoid-like particles with a focus on the transport mechanism. The chapter organized as follows: Section \ref{section:simulationSetup} illustrates considered simulation setup to investigate the electrode microstructures. Few typical cases of insertion dynamics are discussed in section~\ref{sec:LIB2BasicFeaturesModel} with highlighting the underlying mechanism. Also, the obtained results are analyzed and compared with known relations. Thereafter, a numerical study of various morphology descriptors is performed in section~\ref{sec:LIB2MorphologyDescriptors}. Finally, the chapter is concluded by a brief discussion on the applicability of the presented results in Section~\ref{sec:remarks}. Some parts of this chapter are submitted for publication as a journal article.

\section{Simulation setup}
\label{section:simulationSetup}

The generic model of phase separation for multiphase systems of LMO material is described in chapter~\ref{chapter:phaseFieldModelLIB}. In the present chapter, the evolution equation~\eqref{eq:LIBCahnHilliard} is solved numerically with considering a constant concentration boundary condition~\eqref{eq:boundaryConditionSeparator}, as shown in Figure~\ref{fig:boundaryConditionsSeparator}.  As a demonstration of the model, the results are obtained for considering only two phases in the present chapter, the electrode and the electrolyte with highlighting the broad applicability. Particularly, the simulation results are presented for propylene carbonate +1$M$ LiClO$_4$ (lithium perchlorate) as an electrolyte and Li$_x$Mn$_2$O$_4$ as an electrode. The utilized values of parameters for the simulation study are expressed in Table \ref{tab:simulationParameters}. The characterization of the electrode microstructure consists of several particles is described in Appendix~\ref{sec:electrodeFormation}.

The simulation setup consists of cathode particles and the electrolyte. The respective regions are filled with their equilibrium concentration values to avoid any self-generated driving force to the diffusing species, which may hinder the effect of external lithium driving force (constant Dirichlet concentration condition). In the present work, the study is intended to focus the insertion of lithium species. Therefore, the electrode region is filled with lower equilibrium value ${c_i}(i, j, k, {t}=0)=0.12$, while the electrolyte region with its respective equilibrium ${c_i}(i, j, k, {t}=0)=0.50$. A higher concentration value is prescribed at the boundary ${c}_{ps}(i={N}_x, j, k,{t})$(= 0.57 in the present study) to provide a driving force to the diffusing species. The simulation domain is observed by monitoring the state of charge (SOC) frequently, once there are no changes observed for a reasonable time then the simulations can be stopped. Note that the same model can be utilized to study the lithium extraction process. Then, the minute changes are need to be done in terms of electrode regions filled with higher equilibrium value and the prescribed constant concentration boundary condition should be lower than 0.5.
\begin{table}
\caption[Simulation conditions and material properties of propylene carbonate +1$M$ LiClO$_4$ electrolyte and Li$_x$Mn$_2$O$_4$ electrode.]{Simulation conditions and material properties of propylene carbonate +1$M$ LiClO$_4$ electrolyte and Li$_x$Mn$_2$O$_4$ electrode.}\label{tab:simulationParameters}
\centering
\begin{tabular}{lccc}
\hline
Parameter&Symbol&Value&Unit\\
\hline
Reference Temperature&T$_{\textrm{ref}}$&300\cite{santoki2018phase}&K\\
Absolute Temperature&T&300\cite{santoki2018phase}&K\\
Reference length scale&$L$&55.5 &nm\\
Electrode Parameters:&&&\\
Diffusion coefficient&D$_1$&$1\times 10^{-13}$\cite{fuller1994simulation}&m$^2$\,\,s$^{-1}$\\
Gradient energy coefficient&$\kappa_1$&$7\times 10^{-18}$\cite{huttin2012phase}&m$^2$\\
Regular solution parameters&$\alpha'_{1}$&2.5\cite{huttin2012phase}&-\\
&$\alpha''_{1}$&-5.2\cite{huttin2012phase}&-\\
Electrolyte Parameters:&&&\\
Diffusion coefficient&D$_2$&$2.58\times 10^{-10}$\cite{fuller1994simulation}&m$^2$\,\,s$^{-1}$\\
Gradient energy coefficient&$\kappa_2$&$0.0$ (Appendix \ref{subsection:ficksLawElectrolyte})&m$^2$\\
Regular solution parameter&$\alpha''_{2}$&0.0 (Appendix \ref{subsection:ficksLawElectrolyte})&-\\
\hline
\end{tabular}
\end{table}

\section{Basic features of the model}
\label{sec:LIB2BasicFeaturesModel}
\subsection{Concentration profiles for porous and planar electrodes}
\begin{figure}[hbt!]
\begin{center}
\includegraphics[scale=0.65]{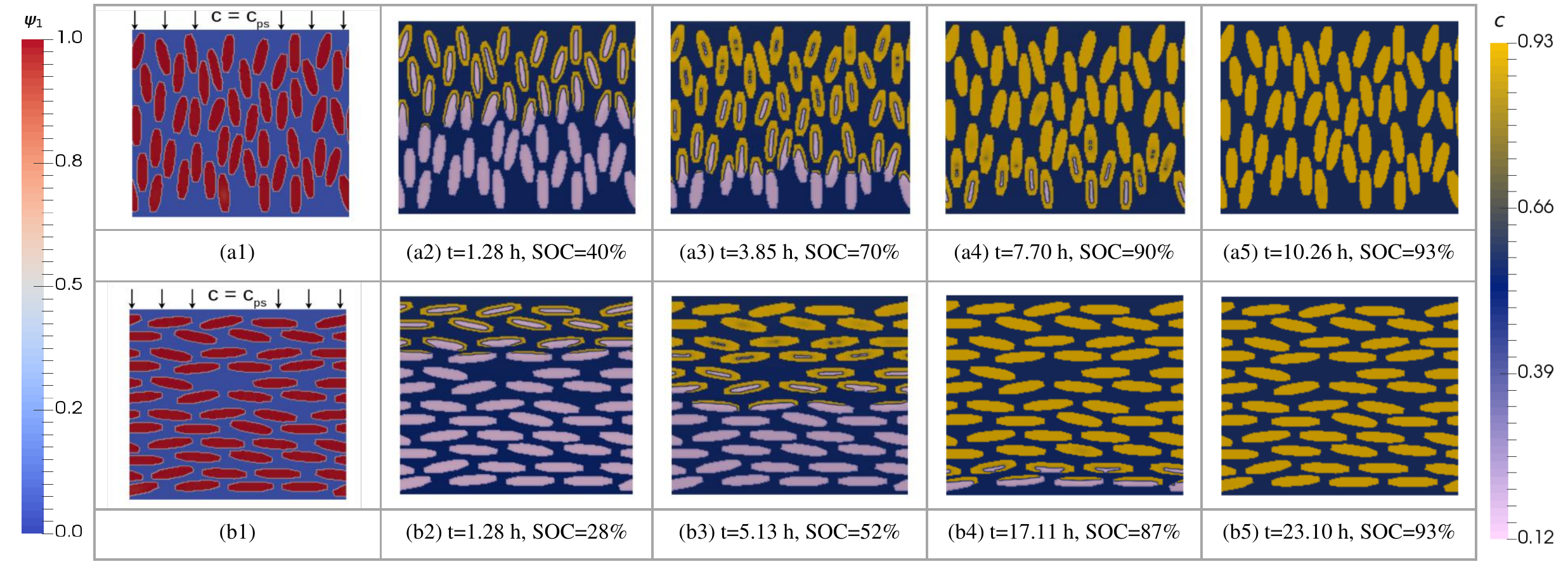}
\end{center}
\caption[Representative cases of insertion of species in 2D porous electrode under a constant concentration boundary condition at the separator]{Representative cases of insertion of species in a porous electrode with $\unit[11.1]{\mu m}$ depth along with current collector under a constant concentration boundary condition at the separator in two-dimensional space. (a) shows insertion in particles aligned to the species flow, i.e. perpendicular to the current collector, while (b) shows particle parallel to the current collector. The leftmost column demonstrates simulation geometry, while the other four columns depict phase-separated morphologies for various time-steps and SOCs.}\label{fig:2Dintercalation}
\end{figure}

Figure~\ref{fig:2Dintercalation} shows insertion in particles aligned perpendicular to the current collector in (a) and parallel to the current collector in (b) for equal porosity and particle size. Firstly, the onset of the development of the Li-rich phase is observed near the separator initially, where the constant concentration boundary condition is employed as evident in (a2) and (b2). On the one hand, the coexistence of two-phases can be observed for a greater depth of electrode in the particles perpendicular to the current collector in (a3). On the other hand, even after a longer period, the coexistence of two-phases is observed for a shorter length span of the electrode in (b3), along with completely intercalated and deintercalated particles near the separator and the current collector respectively. The former case corresponds to more concurrent intercalation, while the latter case rather follows the particle-by-particle intercalation \cite{li2014current}. 

Within a single particle, due to the diffusing species conveniently cover the depth of almost the complete electrode region in the former case, phase separation initiates nearly from the whole surface of a particle, which is in contrast to the latter where phase separation observed mainly at the regions oriented toward the separator as shown in (b4) compared to (a4). Therefore, the core-shell type phase separation within a single particle is more prominent in the former case as shown in (a4), while a phase-separated traveling front from the separator to the current collector is apparent in the latter case in (b4) for the porous electrode.
\begin{figure}[ht]
\centering
\includegraphics[scale=0.58]{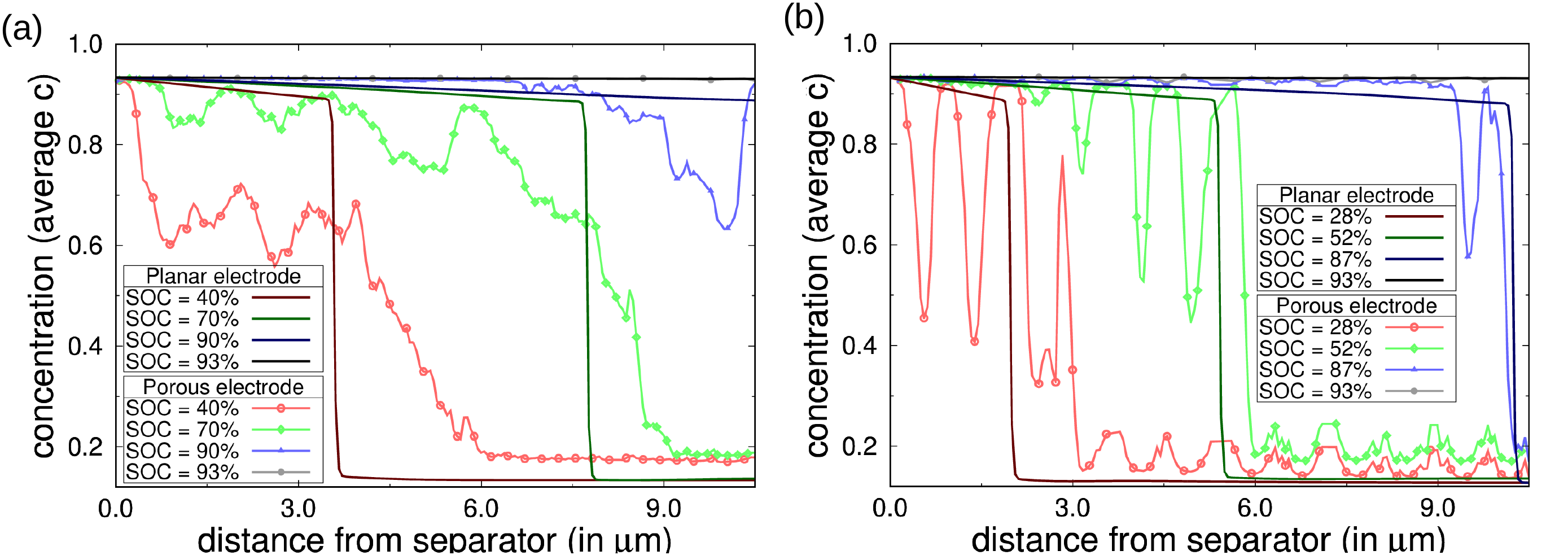}
\caption[Average of concentration ${c}$ in porous and planar electrodes as a function of the depth of the electrode from the separator]{Average of concentration ${c}$ as a function of the depth of the electrode from the separator for various SOC values. The concentration profiles of the porous electrode (Figure~\ref{fig:2Dintercalation}(a) and (b) correspond to (a) and (b) of the present Figure respectively) is compared with the planar single electrode particle model.}\label{fig:depthStudy}
\end{figure}

The porous structure of an electrode is of interest in battery applications due to high surface area yet minimum energy loss \cite{minakshi2010anodic}. Another type of structure that is widely under scrutiny is the planar electrode, as utilized in solid-state thin-film batteries \cite{lai2011mathematical}. The thin film electrode layer can be viewed as a solid foil, which is a continuous film of active electrode material directly deposited on the current collector. Therefore, the electrolyte can not infiltrate the electrode, rather it is in contact with the planar surface. The lithium species have to travel through this film via diffusion after being deposited at the interface. 

The numerical results of the planar electrode and the porous electrode are compared in Figure~\ref{fig:depthStudy}. The concentration evolution of the planar electrode shows a clearly defined region of interface with a lithium-rich phase near the separator and lithium-poor phase near the current collector. However, even though more fraction of region near the separator corresponds to the Li-rich phase during the concentration evolution in the porous electrode, no clearly defined interface is detectable along the depth of the electrode. In addition, Figure~\ref{fig:depthStudy}(a) shows concentration profiles of the porous electrode with electrode particles aligned perpendicular to the separator, while (b) shows concentration profiles of electrode particles aligned parallel to the separator. The concentration profiles in the latter case oscillate more compared to the former case (red and green curves in Figure~\ref{fig:depthStudy}(a) and (b)). This can be justified from the fact that the particles parallel to the separator apparently are stacked layer-wise as shown in Figure~\ref{fig:2Dintercalation}(b1). As the phase separation in a particle starts from the section towards the separator, several particles at the same distance in an affected layer initiate the phase separation simultaneously. Due to the enhanced diffusivity of the electrolyte, the lithium flux traverses through the surrounding of the particles. This initiates the phase separation in the next layer prior to the end of phase separation in the previous layer as shown in Figure~\ref{fig:2Dintercalation}(b3). Therefore, the wave-like pattern can be observed in Figure~\ref{fig:depthStudy}(b).  

\subsection{Comparison with bulk-transport and surface-reaction limited models}
\begin{figure}[hbt!]
\begin{center}
\includegraphics[scale=0.77]{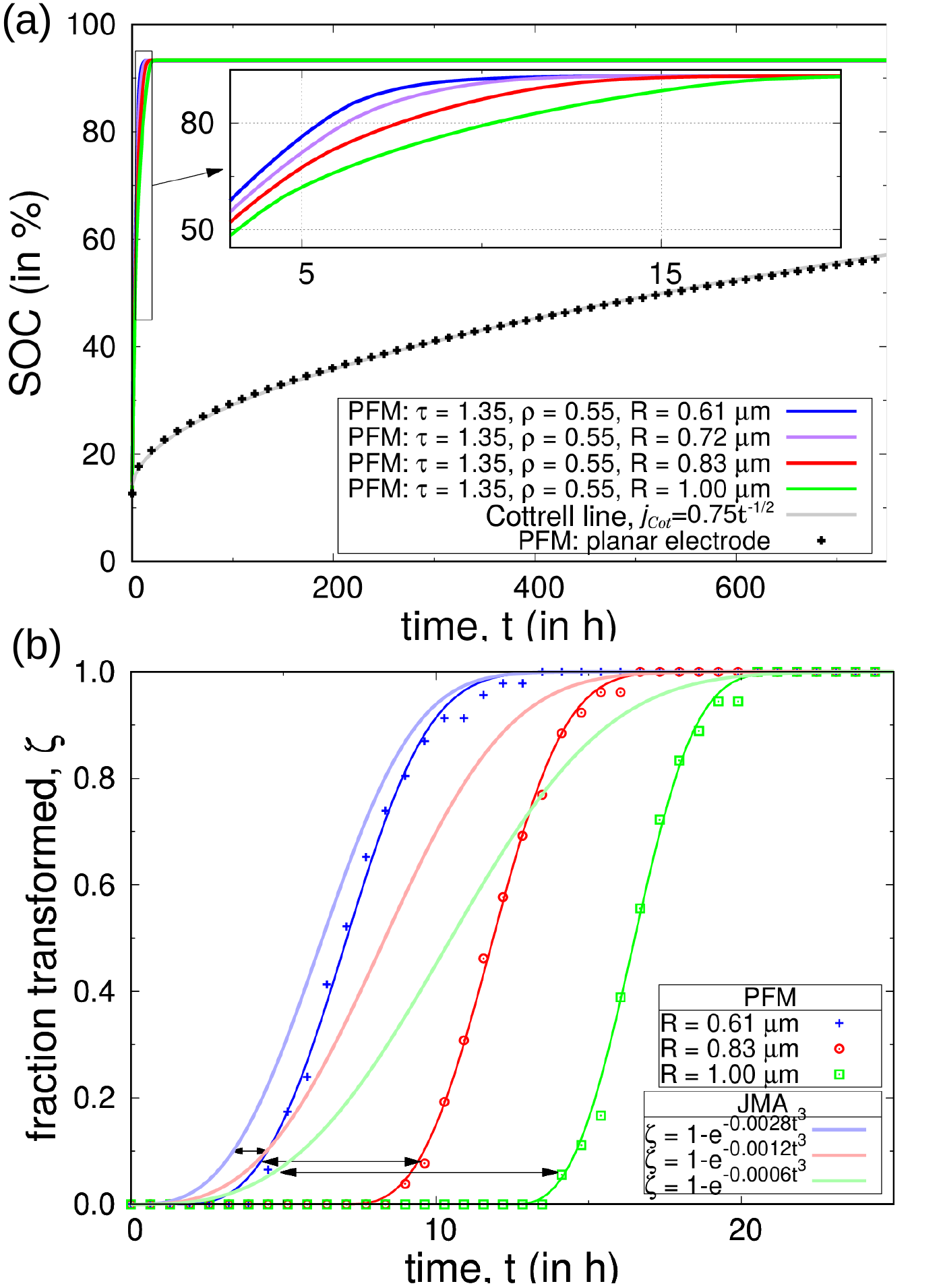}
\end{center}
\caption[Evolution of SOC and fraction transformed from PFM compared with analytical relations for bulk-transport and surface-reaction limited theories]{(a) Change in SOC of the electrode and (b) fraction of total particles transformed to Li-rich with time for various particle sizes obtained from the phase-field model (PFM)\nomenclature{PFM}{phase-field model}. Here $\tau$\nomenclature{$\tau$}{tortuosity} denotes tortuosity, $\rho$\nomenclature{$\rho$}{porosity} indicates porosity of the system, and $R$\nomenclature{$R$}{particle whose area is equivalent to the area of the circular particle of radius $R$} denotes particle size whose area is equivalent to the area of the circular particle of radius $R$. Along side, the evolution of SOC for the planar electrode from PFM and the analytical relation for bulk transport limited theory $j_{\textrm{Cot}} \propto t^{-1/2}$ (Cottrell \cite{cottrell1903residual} line) are displayed for the comparison in (a). The solid dark-colored lines in (b) are guide to eye for the results from PFM, while light-colored lines are the relations obtained from Johnson-Mehl-Avrami (JMA)\nomenclature{JMA}{Johnson-Mehl-Avrami} equation \eqref{eq:JMAequation}.}\label{fig:particleSize}
\end{figure}

Simplified kinetics of the system is studied by measuring the net species insertion response of the material under a prescribed concentration at the simulation boundary. The SOC of the porous electrode increases rapidly during the initial phase and shows a plateau subsequently as shown in Figure~\ref{fig:particleSize}(a), which indicates that the flux (rate of change of SOC) of diffusing species of the presented model decays with time. The obtained results are compared with the analytical relations from bulk-transport and surface-reaction limited diffusion \cite{singh2008intercalation}, which are the standard techniques employed to measure the change in species current in a controlled potential environment in electrochemistry.

\begin{figure}
\begin{center}
\includegraphics[scale=0.50]{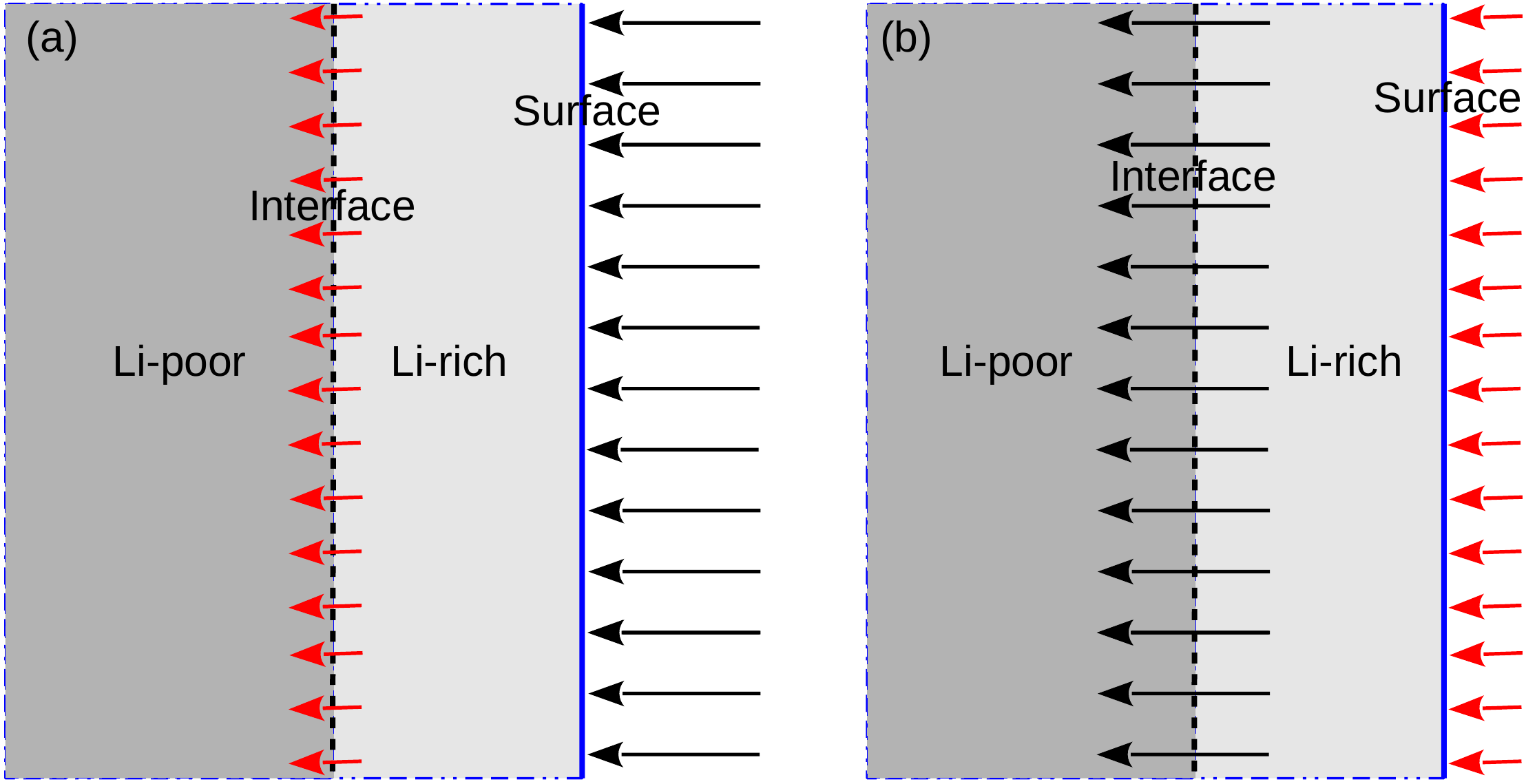}
\end{center}
\caption[Schematic of bulk diffusion and surface reaction limited transportation dynamics.]{Schematic of (a) bulk diffusion and (b) surface reaction limited transportation dynamics. The black arrows indicate dominant driving force, while red arrows illustrate sluggish response. The thick blue line represents electrode surface, while dotted-black line indicates interface between the Li-rich and Li-poor regions.}\label{fig:schematicBulkSurface}
\end{figure}

The bulk-transport limited theory assumes the diffusion of species in the bulk electrode being slower than the deposition at the surface, as illustrated in Figure~\ref{fig:schematicBulkSurface}(a). Therefore, the transportation rate (or species flux) is controlled by bulk diffusion. A simplistic relation of species flux with time in a bulk-transport limited electrode under a potentiostatic condition is described by Cottrell equation \cite{cottrell1903residual},
\begin{equation}
j_{\textrm{Cot}} = k_{\textrm{Cot}} t^{-1/2}, \label{eq:CottrellEquation}
\end{equation}
where $j_{\textrm{Cot}}$\nomenclature{$j_{\textrm{Cot}}$}{Cottrell flux}  (in h$^{-1}$ units) is the flux of lithium species measured as the rate of change of SOC and $k_{\textrm{Cot}}$\nomenclature{$k_{\textrm{Cot}}$}{Cottrell flux-time proportionality constant} (in h$^{-1/2}$ units) is the proportionality constant associated with operating conditions and electrode properties. The Cottrell equation is commonly considered for planar electrodes in Potentiostatic Intermittent Titration Technique (PITT)\nomenclature{PITT}{Potentiostatic Intermittent Titration Technique} to measure the diffusivity \cite{wen1979thermodynamic}. Therefore, the Cottrell equation \eqref{eq:CottrellEquation} for $k_{\textrm{Cot}}=\unit[0.75]{h^{-1/2}}$ is plotted with the results obtained from planar electrode for comparison. After the deposition at the particle surface, the species transport through the bulk of the electrode. Due to the lower diffusivity of the electrode, species transport through the bulk of the electrode is the transportation rate-limiting factor, which correlates to the Cottrell equation \eqref{eq:CottrellEquation}, as shown in Figure~\ref{fig:particleSize}(a). In addition, the results obtained from the porous electrode of various particle sizes are presented. Compared to electrodes of smaller particles, it is evident that the response from the electrode of larger particles ($R=\unit[1.00]{\mu m}$) tends towards the planar electrode and the Cottrell line. Therefore, it can be inferred that the bulk diffusion-limited transportation is more prominent in larger particles compared to the smaller ones.

Another widely recognized theory of the species diffusion is the surface-reaction limited transportation, which considers the bulk diffusion to be more favorable compared to the species deposition at the surface, as shown in Figure~\ref{fig:schematicBulkSurface}(b). Therefore, the transportation rate is controlled by the surface reaction, in which the nucleation events are a primary focus. Due to dominant bulk diffusion, ideally, the nucleated particles can be regarded as a completely transformed to the Li-rich phase, otherwise consisting of Li-poor phase. An analytical relation of the transformation rate with time can be obtained by invoking few assumptions such as, the nucleation is assumed to occur randomly over the entire untransformed portion and the previously nucleated particle does not influence the likelihood of nucleation around that particle. With these conditions, Johnson-Mehl-Avrami (JMA) equation \cite{balluffi2005kinetics} relates the transformed fraction in two-dimensional systems as,
\begin{equation}
\zeta = 1-e^{N_ct^3}, \label{eq:JMAequation}
\end{equation}  
where $\zeta$\nomenclature{$\zeta$}{transformed fraction} denotes transformed fraction and $N_c$\nomenclature{$N_c$}{nucleation rate coefficient} is the constant associated with nucleation rate. The JMA equation defines the transformation of the particles to follow a sigmoidal shape. Figure~\ref{fig:particleSize}(b) shows results obtained from PFM for three different particle sizes compared with the relations from the JMA equation. As the particle size increases, the transformation rate deviates from the sigmoidal shape considerably. Therefore, the electrode with smaller particles prefers surface reaction limited transportation.

Ultimately, these two theories, utilized in Ref. \cite{singh2008intercalation} for intercalation compounds, can be viewed as special cases of the presented numerical model. As an inference from the comparison, the transition of SOC evolution between two different regimes, the bulk-transport and the surface-reaction limited theories can be induced by changing the electrode characteristics such as the size of the particles.

\section{Influence of morphology descriptors on transportation rate}
\label{sec:LIB2MorphologyDescriptors}
\subsection{Effect of tortuosity}
\label{sec:EffectOfTortuosity}
\begin{figure}[ht]
\centering
\includegraphics[scale=1.0]{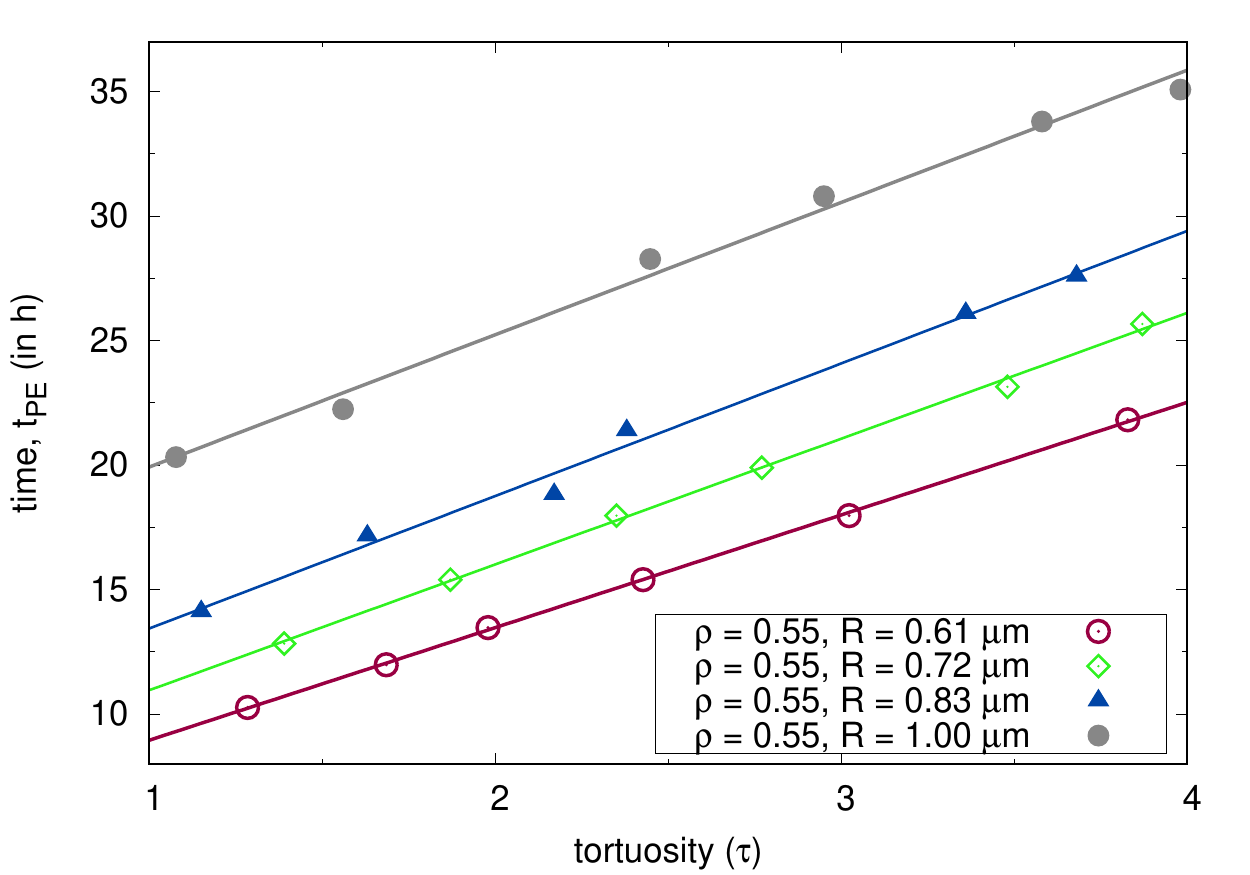}
\caption[Effect of tortuosity $\tau$ on transportation dynamics for different particle sizes $R$]{Effect of tortuosity $\tau$ for different particle sizes $R$ as a function of the time required to complete the phase separation process $t_{PE}$, where the SOC stagnates.}\label{fig:variationInTortuosity}
\end{figure}
The complete process of phase separation is displayed in Figure~\ref{fig:2Dintercalation} for two representative systems: (a) electrode particles aligned perpendicular and (b) parallel to the separator. System (a) consumes $t_{PE}=\unit[10.26]{h}$ for a phase separation process to complete, while system (b) requires $t_{PE}= \unit[23.10]{h}$. As the systems correspond to equal particle sizes and porosity, the distribution of particles can be considered immensely influential on the $t_{PE}$. One of the parameters that quantify the geometrical particle distribution is the tortuosity of the system $\tau$. Based on studies of Refs. \cite{joos2011reconstruction, ender2012quantitative, ender2011three}, the technique to measure the tortuosity is described in Appendix \ref{sec:calculationOfTortuosity}. For the provided representative cases in Figure~\ref{fig:2Dintercalation}, tortuosity $\tau = 1.28$ attained $t_{PE} = \unit[10.26]{h}$, while $\tau = 2.43$ necessitates $t_{PE} = \unit[23.10]{h}$. Furthermore, for different $\tau$ and maintaining the porosity and the mono-disperse particle sizes of the systems consistent, $t_{PE}$ is  linearly related to tortuosity $\tau$, which is plotted as the red curve in Figure~\ref{fig:variationInTortuosity}. Therefore, the increase in tortuosity of the system linearly increases the time required to end the phase separation, $t_{PE}$.

\subsection{Effect of particle size}

 The rate of change of SOC and fraction transformed relate to the particle sizes $R$, as displayed in Figure~\ref{fig:particleSize}(a) and (b) respectively, which shows the numerical result of systems containing several particles of a given size with maintaining equal tortuosity and porosity, where $R$ denotes particle size whose area is equivalent to the area of the circular particle of radius $R$. The system of smaller particles (blue curve) tends towards the surface-reaction limited line in (b), while that of the larger particles (green curve) is situated close to the bulk-transport limited line (black curve) in (a). The discrepancy can be explained by considering the distance diffusing species have to travel after being deposited at the surface of the particle. The diffusing species avail more time to migrate from the surface to reach the center in the larger particles. Therefore, bulk transport is the rate-limiting factor in the larger particles, which is analogous to the bulk-transport limited model. This effect diminishes with the reduction in particle size. Therefore, smaller particles observe higher species flux (rate of change of SOC) in the initial stage and saturate quickly, similar to the surface-reaction limited model.

 Figure~\ref{fig:variationInTortuosity} shows the effect of different mono-disperse particle sizes as a function of tortuosity $\tau$ and $t_{PE}$ with maintaining porosity. For different particle sizes, as shown in Figure~\ref{fig:particleSize}, the system of smaller particles reaches saturation earlier compared to larger particles. Apart from that, linearity in the $\tau$ and $t_{PE}$ relation is observed for all systems of various particle sizes. Furthermore, $\tau$ vs $t_{PE}$ lines are parallel for different particle sizes with the same porosity. However, the porosity certainly affects the time of $t_{PE}$, which is described in the next section.
\subsection{Effect of porosity}
\begin{figure}[ht]
\centering
\includegraphics[scale=1.0]{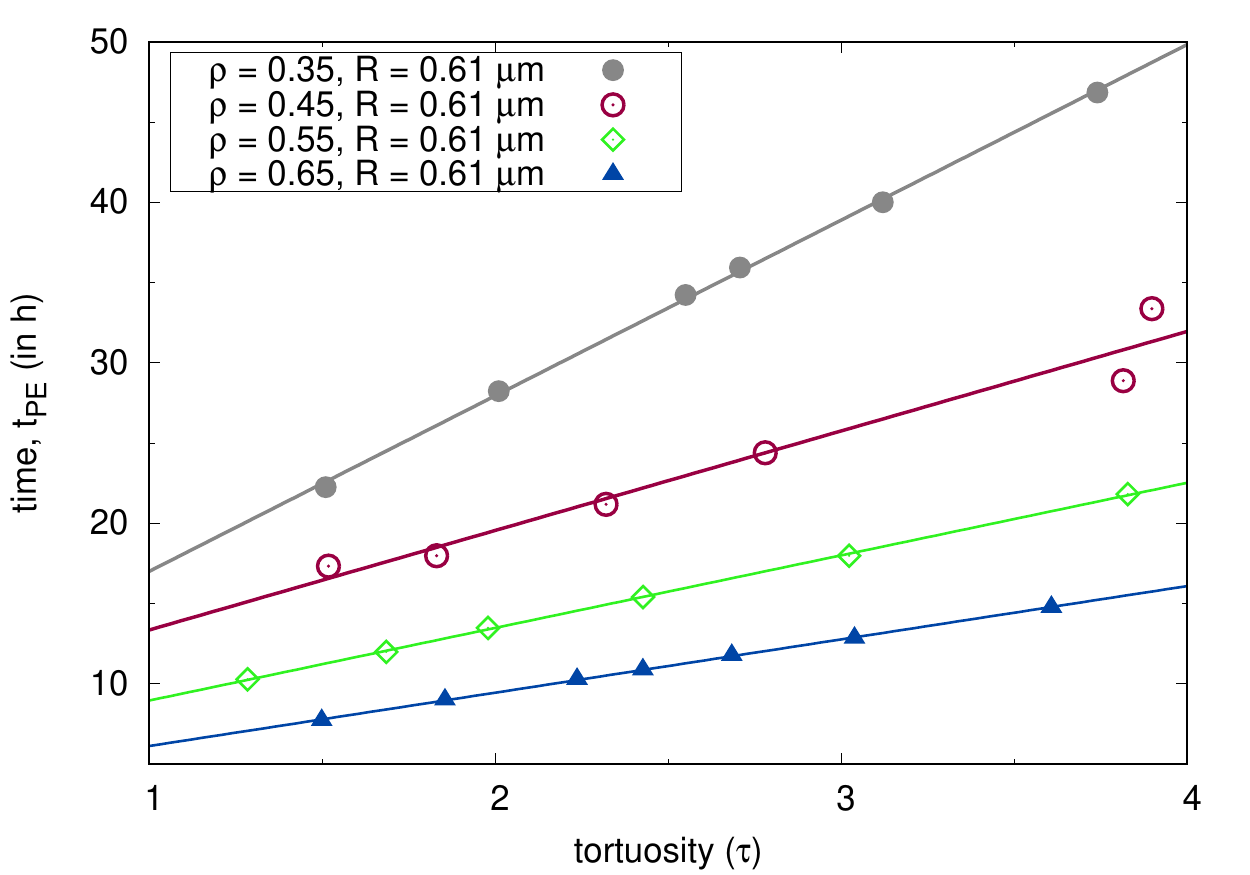}
\caption[Effect of porosity $\rho$ on transportation dynamics for different tortuosity $\tau$]{Effect of porosity $\rho$ for different tortuosity $\tau$ as a function of the time required to complete the phase separation process $t_{PE}$, where the SOC stagnates.}\label{fig:variationInPorosity}
\end{figure}
During the formation of electrode packing, the selection of an optimum mass fraction of host material has gathered much attention due to its implications on energy density and performance of the battery. The mass fraction of the host material is directly linked to the porosity of the electrode. Therefore, understanding the effect of variation in porosity on the phase separation dynamics has technological implications. Figure~\ref{fig:variationInPorosity} shows the relation of $\tau$ and $t_{PE}$ with the porosity $\rho$ for a consistent particle size $R$. A linear relation of $t_{PE}$ with $\tau$ can be observed for respective values of porosity. 

\begin{figure}[ht]
\centering
\includegraphics[scale=0.65]{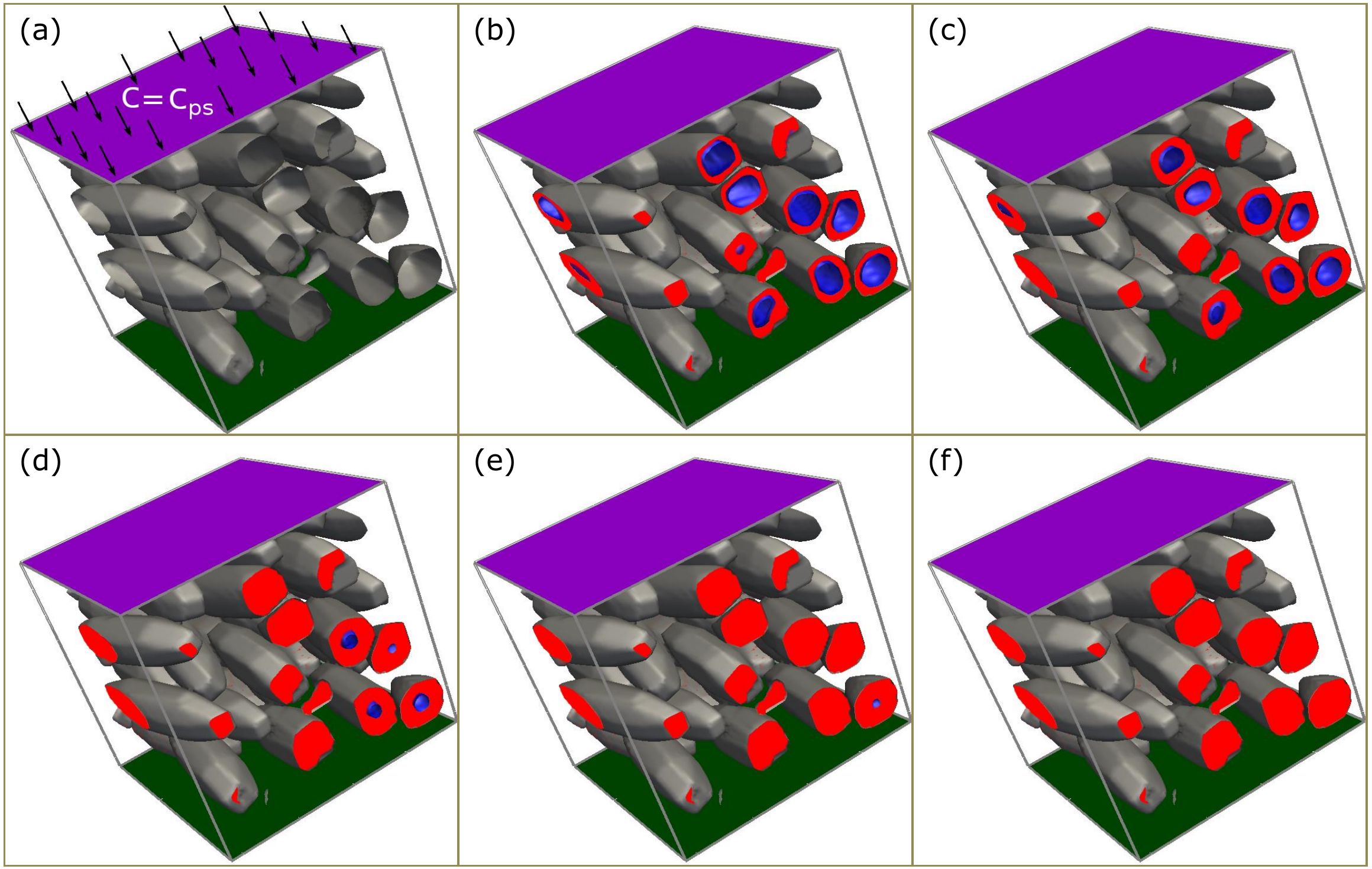}
\caption[Insertion of species in a 3D cathode electrode under a constant concentration boundary condition at the separator]{Insertion of species in a cathode electrode with $\unit[4.6]{\mu m}$ depth along with current collector under a constant concentration boundary condition at the separator in three-dimensional space. (a) demonstrates simulation geometry, while others depict phase-separated morphologies for various time-steps. The images correspond to SOC (b) 58\% ($t= \unit[0.5]{h}$), (c) 70\% ($t= \unit[1.0]{h}$), (d) 85\% ($t= \unit[2.6]{h}$), (e) 90\% ($t= \unit[3.6]{h}$), and (f) 93\% ($t= \unit[4.9]{h}$). The red color represents the lithium-rich phase. The blue color cavity inside the lithium-rich phase corresponds to the lithium-poor phase.}\label{fig:3Dparticle}
\end{figure}

The decrease in porosity results in the increased mass fraction of host material and decreased ion carriers in the form of electrolytes. Hence, the system of lower porosity requires more time $t_{PE}$ to obtain the SOC plateau. Also, the slope of the linear relationship ($\tau$ with $t_{PE}$) is not constant, instead increases with decreasing porosity. This implies that the system of higher $\tau$ exhibits appreciable increment in $t_{PE}$ compared to the lower ones for the same porosity difference. 

Finally, Figure~\ref{fig:3Dparticle} demonstrates the capability of the employed model that the method can be readily extended to simulate three-dimensional systems. The 3D electrode is considered for the numerical experiment to study the transport mechanism. The constant concentration at the separator initiates the formation of the Li-rich phase from the outer surface of the particles, otherwise consists of the Li-poor phase throughout, as shown in (b). The Li-rich phase continues to grow further at the surface of the particles in (c), (d), and (e). Eventually, the lithium-poor phase has been eliminated by the lithium-rich phase in (f). Note that the 3D case shows a significantly different pattern compared 2D case during transportation in the particles. Even though particles are aligned parallel, it shows a transportation mechanism more like the perpendicular 2D case than the parallel 2D case. This is due to the fact that the tortuosity here ($\tau=1.23$) is more like the perpendicular 2D one. Furthermore, the depth of the electrode in 3D is much less than the 2D case in order to minimize the computational efforts. According to the preliminary findings, the depth of the electrode certainly influences the SOC evolution, which will be subject of future investigations.

\section{Conclusion}
\label{sec:remarks}
A numerical result of the porous electrode is presented to understand the effect of various microstructural properties such as particle size, porosity, and tortuosity on the lithium transport mechanism. In this chapter, the phase separation mechanism in LMO particles has been successfully demonstrated in a multiple particle model system. Ellipsoid-like particles are considered as an example, however, the model can be readily applicable to particles of complicated geometries.  According to the diffusional properties of electrode and electrolyte, a study is conducted on transportation rate dependence with various morphological descriptors of electrode microstructures.

The obtained results suggest that the transportation rate of the system is strongly related to the tortuous pathways formed by the particle orientation, which can be quantified by the tortuosity parameter $\tau$. When controlling the lithium concentration at the separator, the lithium transportation rate is observed to be linearly related to the tortuosity. Furthermore, the slope of this linear relation is independent of the particle size, while the slope alters with a change in the porosity of the electrode. Therefore, the tortuosity, the porosity, and the particle size can be suitable descriptors for the characterization of electrode morphologies. 

 Furthermore, the evolution of the state of charge of the obtained results for the porous structures of mono-disperse particles are compared with the bulk-transport and surface-reaction limited theories. The results suggest that systems consisting of smaller particles are limited by surface reaction, while larger particles tend towards the bulk-transport limited theory derived for planar electrodes. In order to identify the promising hierarchically structured electrodes, the presented simulation results could be utilized to optimize the experimental efforts.






\afterpage{\blankpagewithoutnumberskip}
\clearpage

\newpage
\thispagestyle{empty}
\vspace*{8cm}
\phantomsection\addcontentsline{toc}{chapter}{IV Results and Discussion: \\ Electromigration in metallic conductors}
\begin{center}
 \Huge \textbf{Part IV} \\
 \Huge \textbf{Results and Discussion: \\ Electromigration in metallic conductors}
\end{center}

\afterpage{\blankpagewithoutnumberskip}
\clearpage
\chapter{Motion of isotropic inclusions}
\label{chapter:EM1}

Phase-field methods are extensively applied to study isotropic inclusion propagations under electromigration. In the stimulating work by Mahadevan and Bradley \cite{mahadevan1999simulations}, the development of a slit-like feature from an edge perturbation is studied by the phase-field model due to electromigration. The work focused on the physical mechanism of slit growth at the location of a preexisting notch and subsequent propagation transverse to the line. As a result, the change in volume of the inclusion depends strongly on the applied current. Contrarily, experimental evidence also corroborates that the motion of volume-preserving inclusions along the conductor line \cite{joo1993evolution, sanchez1992slit}, which instigated interest in modeling community in recent years. Bhate et al. \cite{bhate2000diffuse} presented the morphological evolution of a geometrical shape under the influence of surface energy, electric and stress fields on single isolated inclusions. Thereafter, Barrett et al. \cite{barrett2007a} studied the migration, splitting, and coalescence of inclusions along the metallic conductors. Furthermore, Ba{\v{n}}as et al. \cite{banas2009phase} presented results of geometrical shape evolution in three dimensions space. There are instances where propagation of a finger-like slit with shape and volume preserved is observed in SnAgCu solder bumps \cite{zhang2006effect}, which has not received attention from a modeling perspective until now. Furthermore, a critical comparison between the solutions obtained from phase-field and sharp-interface methods is lacking.

Therefore, the emphasis of the presented chapter is to compare the results from the phase-field model derived in Chapter~\ref{chapter:phaseFieldModelElectromigration}  with the sharp-interface analysis to investigate the selection of slit width and velocity of an initially circular inclusion at an applied electric field. More importantly, the second objective of the study is to investigate the geometrical morphologies of inclusions from the phase-field model, which is otherwise unfeasible with sharp-interface theories. The sharp-interface analysis is considered in section~\ref{section:LinearStabilityAnalysisCircularIsland} to derive a critical limit of circular inclusions stability with assuming equal conductivities to the matrix. Afterward, the assumption of equal conductivities is relieved in section~\ref{section:linearStabilityAnalysis}. Thereafter, results from the phase-field numerical study are compared with linear stability analysis in section~\ref{section:chi0andBetaInitial}. Subsequently, the description of the finger-like slits obtained from the numerical study is presented in section~\ref{sec:EM1SpeceficFeaturesSlits}. Also, the characteristics associated with the slit shapes are derived from sharp-interface model 
and a critical comparison with numerical results are also presented.
Finally, the chapter is concluded by its practical implications in section~\ref{sec:EM1Discussion} with subsequent discussion of the important results in section~\ref{sec:IsotropicInclusionsConclusions}. Some parts of this chapter are published in the \textit{Journal of Electronic Materials} \cite{santoki2019phase}.

\section{Sharp-interface analysis for circular islands}
\label{section:LinearStabilityAnalysisCircularIsland}
\begin{figure}
\begin{center}
\includegraphics[scale=0.55]{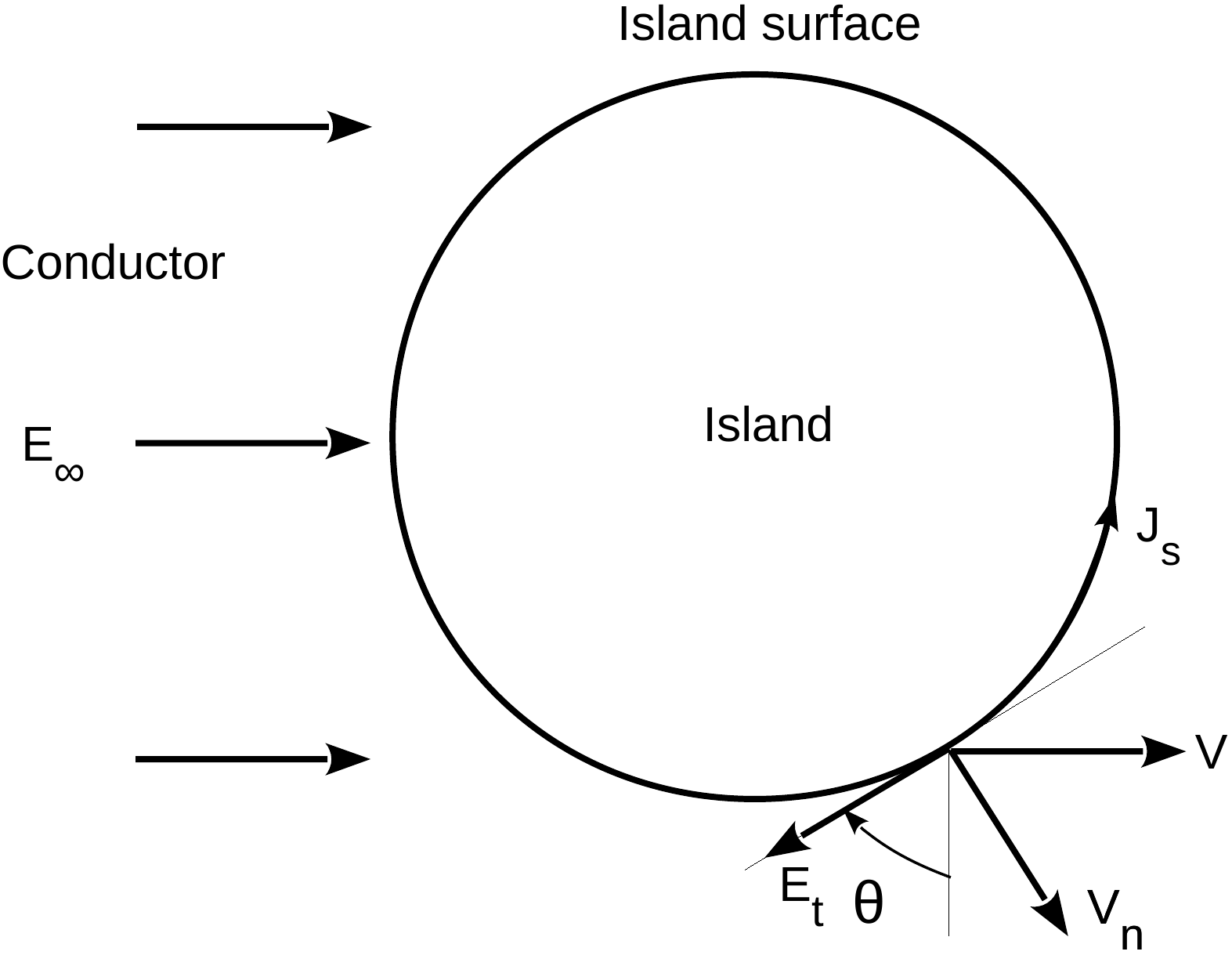}
\end{center}
\caption[Schematic of a circular island, subjected to an external electric field $E_\infty$, in an infinite conductor]{Schematic of a circular island, subjected to an external electric field $E_\infty$, in an infinite conductor domain.}\label{fig:EM1sharpInterfaceCircularIsland}
\end{figure}

The stabilty analysis of a circular dislocation loop propagation, under the assumption of a steady-state, is derived by Yang et. al \cite{yang1994cavity}. For brevity, the basic steps of the derivations are casted here for islands migrating in a conductor. In the present context, the term, island describes a cluster of material entrapped in the metallic conductor consists of equivalent conductivity to that of the conductor.

Consider an island, translating along the length of the conductor under an external electric field, as shown in Figure~\ref{fig:EM1sharpInterfaceCircularIsland}. The migration of the island is due to the result of a mass transport flux, which is induced by surface electromigration, and by the capillarity given by the Nernst-Einstein relation:
\begin{equation}
J_s = \frac{D_s \delta_s}{\Omega k_B T} \Big( -eZ_sE_t + \Omega \gamma_s \frac{\textrm{d} \kappa_s}{\textrm{d} s} \Big), \label{eq:EM1SharpInterfaceFlux}
\end{equation}
where $J_s$\nomenclature{$J_s$}{surface atomic flux} represents the number of species passing per length of island surface per unit time, $D_s$ is the surface diffusivity, $\delta_s$ is the thickness of the surface layer, $\Omega$ denotes the atomic volume, $e$ is the electron charge, $Z_s$\nomenclature{$Z_s$}{effective valence} represents the effective valence at the surface, $k_B$ is the Boltzmann constant, $T$ denotes the absolute temperature, $\gamma_s$ is the surface energy, $\kappa_s$ represents the local curvature of the surface, and $s$\nomenclature{$s$}{arc length along the island/inclusion surface} denotes the arc length along the island surface. 
The capillary-mediated flux is driven by the gradients of curvature, along the island surface, while the electromigration flux is dictated by the tangential component of the electric field along the island surface $E_t$\nomenclature{$E_t$}{tangential component of the electric field along the island/inclusion surface}, given by
\begin{equation}
E_t = - E_{\infty} \textrm{ sin } \theta, \label{eq:EM1sharpInterfaceElectricalPotentialIsland}
\end{equation}
which is responsible for the island transmission \cite{wang1996a}. $E_{\infty}$\nomenclature{$E_{\infty}$}{applied electric field} is the applied electric field, $\theta$ is the tangent angle of the island surface. 

Mass conservation relates the surface divergence of the flux to the normal velocity\nomenclature{$V_n$}{velocity along the surface normal} as follows:
\begin{equation}
\frac{\textrm{d}J_s}{\textrm{d}s} = \frac{V_n}{\Omega}. \label{eq:EM1MassConservationNormalVelocity}
\end{equation}
By assuming a steady state, the normal velocity can be related to the velocity\nomenclature{$V$}{velocity along the external electric field} along applied electric field as
\begin{equation}
V_n = V \textrm{ cos } \theta. \label{eq:EM1normalVelocity}
\end{equation}
The \nomenclature{$\kappa_s$}{curvature along the island/inclusion surface}curvature along the island surface is the fraction of change in tangent angle to the length of line segment, which can be expressed as,
\begin{equation}
\kappa_s = \frac{\textrm{d}\theta}{\textrm{d}s}.\label{eq:EM1surfaceCurvatureIsland}
\end{equation}
Substituting Eqs.~\eqref{eq:EM1SharpInterfaceFlux}, \eqref{eq:EM1sharpInterfaceElectricalPotentialIsland}, \eqref{eq:EM1normalVelocity}, and \eqref{eq:EM1surfaceCurvatureIsland} into the mass conservation equation \eqref{eq:EM1MassConservationNormalVelocity}, the resultant expression can be written as,
\begin{equation}
V\textrm{ cos } \theta  = \frac{D_s \delta_s}{ k_B T} \left( e Z_s E_\infty \textrm{ cos } \theta \kappa_s + \Omega \gamma_s \kappa_s \frac{\textrm{d} }{\textrm{d} \theta} \left( \kappa_s \frac{\textrm{d}  \kappa_s}{\textrm{d} \theta} \right) \right). \label{eq:EM1resultantExpression}
\end{equation}
This equation can be non-dimensionalized using the relation,
\begin{equation}
\left[1/\kappa_s \right] = \left[ 1/\hat{\kappa}_s \right] \frac{D_s \delta_s}{k_B T V} eZ_s E_\infty \label{eq:EM1nonDimensional}
\end{equation}
Here $\hat{\kappa}_s$\nomenclature{$\hat{\kappa}_s$}{non-dimensional curvature of the island surface} denote the non-dimensional curvature of the island. Substituting Eq.~\eqref{eq:EM1nonDimensional} into Eq.~\eqref{eq:EM1resultantExpression}, the shape equation can be expressed in a non-dimensional form as,
\begin{equation}
\frac{\textrm{d} }{\textrm{d} \theta} \left( \hat{\kappa}_s \frac{\textrm{d}  \hat{\kappa}_s}{\textrm{d} \theta} \right) + \chi \left( 1 - \frac{1}{\hat{\kappa_s}} \right) \textrm{ cos } \theta =0. \label{eq:EM1differentShapes}
\end{equation}
where the dimensionless number $\chi$\nomenclature{$\chi$}{dimensionless number associated with the steady state shapes of an island} is defined as,
\begin{equation}
\chi = \left( \frac{D_s \delta_s}{k_B T V} eZ_s E_\infty \right)^2 \frac{eZ_sE_\infty}{\Omega \gamma_s}. \label{eq:EM1ChiDifferentShapes}
\end{equation}
The non-linear ordinary differential equation~\eqref{eq:EM1differentShapes} determines the steady-state shapes. Note that, heretofore, the expressions are equally applicable to non-circular islands. Moving on to circular shaped islands, their stability can be analyzed by introducing perturbations to the surface in the form, 
\begin{equation}
\hat{\kappa}_s = 1 + \varepsilon \varkappa (\theta) \label{eq:EM1islandPerturbation}
\end{equation}
where $\varepsilon$\nomenclature{$\varepsilon$}{perturbation parameter} is the small perturbation parameter and $\varkappa (\theta)$\nomenclature{$\varkappa$}{perturbation function} denotes the perturbation function. Substituting Eq.~\eqref{eq:EM1islandPerturbation} into Eq.~\eqref{eq:EM1differentShapes}, the growth of perturbations can be expressed as,
\begin{equation}
\frac{\textrm{d}^2 \varkappa}{\textrm{d}\theta^2} + \chi \varkappa \textrm{ cos } \theta =0. \label{eq:EM1MathieuEquation}
\end{equation}
This non-trivial expression is similar to Mathieu equation. A MATLAB based subroutine is developed with \textit{bpv4c}  algorithm to seek eigenvalue of Eq.~\eqref{eq:EM1MathieuEquation}. The second order ordinary differential equation splits into two first order equations with initial guess function and boundary conditions are employed in the form of a cosine function. The lowest eigenvalue is estimated as $\chi=10.65$. 

If the island does not collapse, and keeps its circular shape of radius $R_i$\nomenclature{$R_i$}{radius of circular island/inclusion}, the steady-state velocity\nomenclature{$V_0$}{steady-state velocity of an island/inclusion} \cite{arzt1994electromigration} can be determined by using $\kappa_s =1/R_i$ and $\textrm{d} \kappa_s / \textrm{d}s =0$. Therefore, substituting these relations in Eq.~\eqref{eq:EM1resultantExpression}, the velocity of a circular island can be expressed as,
\begin{equation}
V_0 = - \frac{D_s \delta_s}{R_ik_BT} eZ_s  E_{\infty} \label{eq:EM1steadyStateCircularVelocity}
\end{equation}
Therefore, the dimensionless number in Eq.~\eqref{eq:EM1ChiDifferentShapes} can be expressed for steady-state circular islands of the form,
\begin{equation}
\chi_0 = \frac{eZ_sE_\infty R_i^2}{\Omega \gamma_s}. \label{eq:EM1chi0}
\end{equation}
This equation represent relative strength of the capillarity $\Omega \gamma_s/R_i^2$ and the electromigration force $eZ_s E_\infty$. The capillarity attempts to reduce any curvature gradient in an ultimate aim to form a circular shape, while the atomic transfer due to electromigration in the direction of electron wind alters the preexisting circular shape of the island. Therefore, the island shape is controlled by the parameter $\chi_0$\nomenclature{$\chi_0$}{dimensionless number associated with the stability of circular island/inclusion}, which represents the competition between the electromigration force and the capillarity. A critical limit of the electric field, at which a circular island maintains its shape is defined by $\chi_{0\textrm{ct}}=$\nomenclature{$\chi_{0\textrm{ct}}$}{critical value of $\chi_0$, above which circular island/inclusion is unstable} 10.65. Above the critical value, i.e., when the electron wind force prevails over the capillary force, a circular island is unstable and undergoes shape bifurcation.

Heretofore, a study of the homogeneous inclusion is considered. Meaning, the conductivity of the circular inclusion is equivalent to the metallic conductor. To differentiate, the homogeneous inclusions are termed as the islands. However, the presented theory needs to be extended in order to capture the behavior of insulating voids and different conductivity material entrapped into metallic conductors, which has technological relevance. Therefore, in the next section, a theory is developed to determine the stability limit of circular heterogeneous inclusions.

\section{Linear stability analysis of circular inclusion} 
\label{section:linearStabilityAnalysis} 
 
 A direct projection of external electric field at the infinity is considered on the surface of inclusion in previous section \ref{section:LinearStabilityAnalysisCircularIsland}(see Eq.~\eqref{eq:EM1sharpInterfaceElectricalPotentialIsland}). This can be justified for homogeneous inclusions. However, for heterogeneous inclusions, a general expression of electrical potential in a matrix containing a circular inclusion is determined. Then, this expression is extended to a circular inclusion with small perturbations on the surface. Finally, a stability limit of a circular inclusion is determined from the basic species transport relation. The presented theory is in coherence with the work of Ref. \cite{hao1998linear}. For the brevity and completeness, the approach with necessary assumptions is described.
 
 \subsection{General expression of electrical potential in a matrix containing a circular inclusion}
 
 Consider a circular inclusion of radius $R_i$ in an infinite matrix subject to an external electric field, $E_{\infty}$. In the present study, the plane polar coordinate system $(r, \vartheta)$\nomenclature{$r, \vartheta$}{polar coordinates of a point} is considered with the origin at the center of the inclusion. $r$ represents a distance of a point in the system (i.e., matrix + inclusion) and $\vartheta$ corresponds to an angle formed by the point under consideration at the origin with the direction of electric field. The electrical potential at any far point from the inclusion is stated as,
 
 \begin{equation} \label{eq:linearStabilityElectricPotential}
 \phi (r,\vartheta) \rvert_{r \rightarrow \infty} = - E_{\infty} r \textrm{ cos } \vartheta.
 \end{equation}
 
 It is easy to perceive that the inclusion of conductivity $\sigma_{\textrm{icl}}$ disrupts the electrical potential distribution in the matrix of different conductivity, $\sigma_{\textrm{mat}}$, inside and nearby the inclusion. The spatial distribution of the electrical potential can be obtained by Laplace's equation, in cylindrical coordinates with azimuthal symmetry,
 
\begin{equation}
\frac{1}{r}\frac{\partial}{\partial r} \left( r \frac{\partial \phi (r,\vartheta)}{\partial r}\right) + \frac{1}{r^2}  \frac{\partial }{\partial \vartheta}\left( \frac{\partial \phi (r,\vartheta)}{\partial \vartheta} \right) = 0.
\end{equation} 
The general solution of the above equation can be derived by the variable separation method,

\begin{equation}
\phi (r,\vartheta) = \sum^{\infty}_{l=1} \left( A_l r^l +\frac{B_l}{r^l} \right) (C_l \textrm{ cos }l\vartheta + D_l \textrm{ sin }l\vartheta).
\end{equation}
$A_l, B_l, C_l,$ and $D_l$ are the real coefficients, which need to be determined by known conditions. At the far distance, above equation of the electrical potential should correspond to Eq.~\eqref{eq:linearStabilityElectricPotential}. Thus the equation reduced to the form, 
 
\begin{equation}
\phi(r,\vartheta) = A r\textrm{ cos } \vartheta + \frac{B }{r} \textrm{ cos } \vartheta,
\end{equation} 
 containing only cos $\vartheta$ term. Furthermore, as the origin is at the center of the inclusion, the second term diverges at the origin. Therefore, electrical potential outside and inside of the sphere is expressed separately as,
 
 \begin{eqnarray}
 \phi_{\textrm{in}}(r,\vartheta) &=& A_{\textrm{in}} r \textrm{ cos } \vartheta,\\
 \phi_{\textrm{out}}(r,\vartheta) &=& A_{\textrm{out}} r \textrm{ cos } \vartheta +  \frac{B_{\textrm{out}}} {r} \textrm{ cos } \vartheta.
 \end{eqnarray}
 Comparing above equation with the asymptotic limit at the far distances, the expression $A_{\textrm{out}} = - E_{\infty}$ can be obtained. It is easy to notice that, from the continuity equation, on the inclusion surface,
 
 \begin{equation} \label{eq:linearStabilitycontinuityConditions}
{ \left. \frac{\partial \phi_{\textrm{in}}}{\partial r}\right|}_{r=R_i} = \beta
{ \left. \frac{\partial \phi_{\textrm{out}}}{\partial r}\right|}_{r=R_i} \hspace{0.5cm} \textrm{and} \hspace{0.5cm} { \left. \frac{\partial \phi_{\textrm{in}}}{\partial \vartheta}\right|}_{r=R_i} = 
{ \left. \frac{\partial \phi_{\textrm{out}}}{\partial \vartheta}\right|}_{r=R_i} .
 \end{equation}
 Here $\beta = \sigma_{\textrm{mat}}/\sigma_{\textrm{icl}}$\nomenclature{$\beta$}{conductivity ratio, $\sigma_{\textrm{mat}}/\sigma_{\textrm{icl}}$}. The resultant electrical potential reads,
\begin{equation}
\phi(r,\vartheta) =     \left\{ \begin{array}{ccl}
\frac{-2\beta}{(1+\beta) }E_{\infty} r \textrm{ cos }\vartheta, & \textrm{  for}& r \leq R_i, \\ 
      -E_{\infty} r \textrm{ cos } \vartheta + \frac{E_{\infty} R_i^2}{r} \frac{(1-\beta)}{(1+\beta)} \textrm{ cos } \vartheta & \textrm{  for} & r \geq R_i. 
\end{array}\right. \label{eq:linearStabilityflux}
\end{equation}
Note that both expressions are satisfied at the surface of the inclusion.

\subsection{Electrical potential in a matrix containing a small perturbed circular inclusion} 

Using the conformal mapping to express complex potential (Eq. \eqref{eq:linearStabilityflux}) for the circular inclusion at point $z = x + iy$\nomenclature{$i$}{complex number} in complex form, 
\begin{eqnarray}
\Phi_{in}(z)&=& \frac{-2\beta }{(1+\beta)} E_{\infty} z,\\
\Phi_{out}(z)&=& - E_{\infty} z + \frac{(1-\beta)}{(1+\beta)} E_{\infty} \frac{R_i^2}{z}.
\end{eqnarray}
Here $\Phi(z) = \phi + i \psi$, where $\psi$\nomenclature{$\psi$}{orthogonal function to the electrical potential} is the orthogonal function to the electrical potential. Heretofore, the linear stability of a perfectly rounded inclusion is considered. Now, a small perturbation $(\varepsilon \ll 1)$ is introduced to the surface of inclusion. In the present part, an expression of electric field is derived for the matrix containing inclusion of circular shape with small perturbations around the surface. The mapping of inclusion surface, 

\begin{equation}
z(R_i, \vartheta) = x + iy = R_i(\textrm{cos }\vartheta + i \textrm{ sin }\vartheta ) + \varepsilon \sum^{\infty}_{n=0} a_n (\textrm{cos } n \vartheta - i \textrm{ sin }n \vartheta ),
\end{equation}
here $a_n$ are real constants\nomenclature{$a_n$}{surface perturbation coefficient corresponds to n$^{\textrm{th}}$-term}, which associated with the strength of surface perturbations. The expression for the electrical potential needs to be updated, as the small perturbations on the inclusion surface modify the electrical potential distribution. From the arguments based on a convergence of the electrical potential inside and outside of the inclusion, the expression of electrical potential can be given by,

\begin{eqnarray}
\Phi_{in}(z)&=& \frac{-2\beta }{(1+\beta)} E_{\infty} z + \varepsilon \sum^{\infty}_{n=0} C_n z^n,\\
\Phi_{out}(z)&=& - E_{\infty} z + \frac{(1-\beta)}{(1+\beta)} E_{\infty} \frac{R_i^2}{z} + \varepsilon \sum^{\infty}_{n=0} D_n z^{-n}.
\end{eqnarray}
Here $C_n$ and $D_n$ are coefficients, which are determined by substituting $z$ in the above equation and applying continuity conditions, Eq. \eqref{eq:linearStabilitycontinuityConditions}, on the surface of the inclusion, after transforming, from $\partial \phi_{in}/\partial r = \beta (\partial \phi_{out}/\partial r) $ to $\psi_{in} = \beta \psi_{out}$ by the Cauchy-Riemann condition (i.e., $\partial \phi / \partial r= \partial \psi / \partial \vartheta$) and from $\partial \phi_{in}/\partial \vartheta = \partial \phi_{out}/\partial \vartheta $ to $\phi_{in} = \phi_{out}$. Therefore, the resultant electrical potential on the surface of inclusion,

\begin{eqnarray}
\Phi_{in} &=& \Bigg{\lbrace}-\frac{2\beta}{(1+\beta)} E_{\infty} R_i \, e^{i\vartheta}  - \varepsilon \frac{2\beta}{(1+\beta)} E_{\infty} \sum^{\infty}_{n=0} a_n e^{-in\vartheta}\\
&&\hspace{5.5cm} - \varepsilon \frac{2 \beta(1-\beta)}{(1+\beta)^2}  E_{\infty} \sum^{\infty}_{n=0} a_n e^{in\vartheta}\Bigg{\rbrace}, \nonumber\\
\Phi_{out} &=& -E_{\infty} R_i \, e^{i \vartheta} + \frac{(1-\beta)}{(1+\beta)} E_{\infty} R_i \, e^{-i \vartheta} - \varepsilon \frac{4 \beta}{(1+\beta)^2}E_{\infty}\sum^{\infty}_{n=0} a_n e^{-in\vartheta}.\label{eq:linearStabilityelectricalPotential}
\end{eqnarray}
 Here $e^{i\vartheta} = \textrm{ cos }\vartheta +i \textrm{ sin }\vartheta$.
 
\subsection{Linear stability analysis of transient inclusion} 
 
 The mapping function $z$ can be expressed for a transient inclusion, which drifts in the matrix. The mapping function of the inclusion, whose position changes with time can be given by,
 
 \begin{equation}\label{eq:linearStabilitysurfaceMapping}
 z(R_i,\vartheta, t) = R_i\, e^{i\vartheta} + S(t) + \varepsilon \sum^{\infty}_{n=0} a_n (t) \, e^{-in\vartheta}.
 \end{equation}
where $S(t)$ is the spatial displacement function related to path traveled by the inclusion\nomenclature{$S$}{displacement function}. In the present case, the coefficients $a_n$ vary with time as the inclusion shape transforms as it evolves.

\begin{figure}
\begin{center}
\includegraphics[scale=0.3]{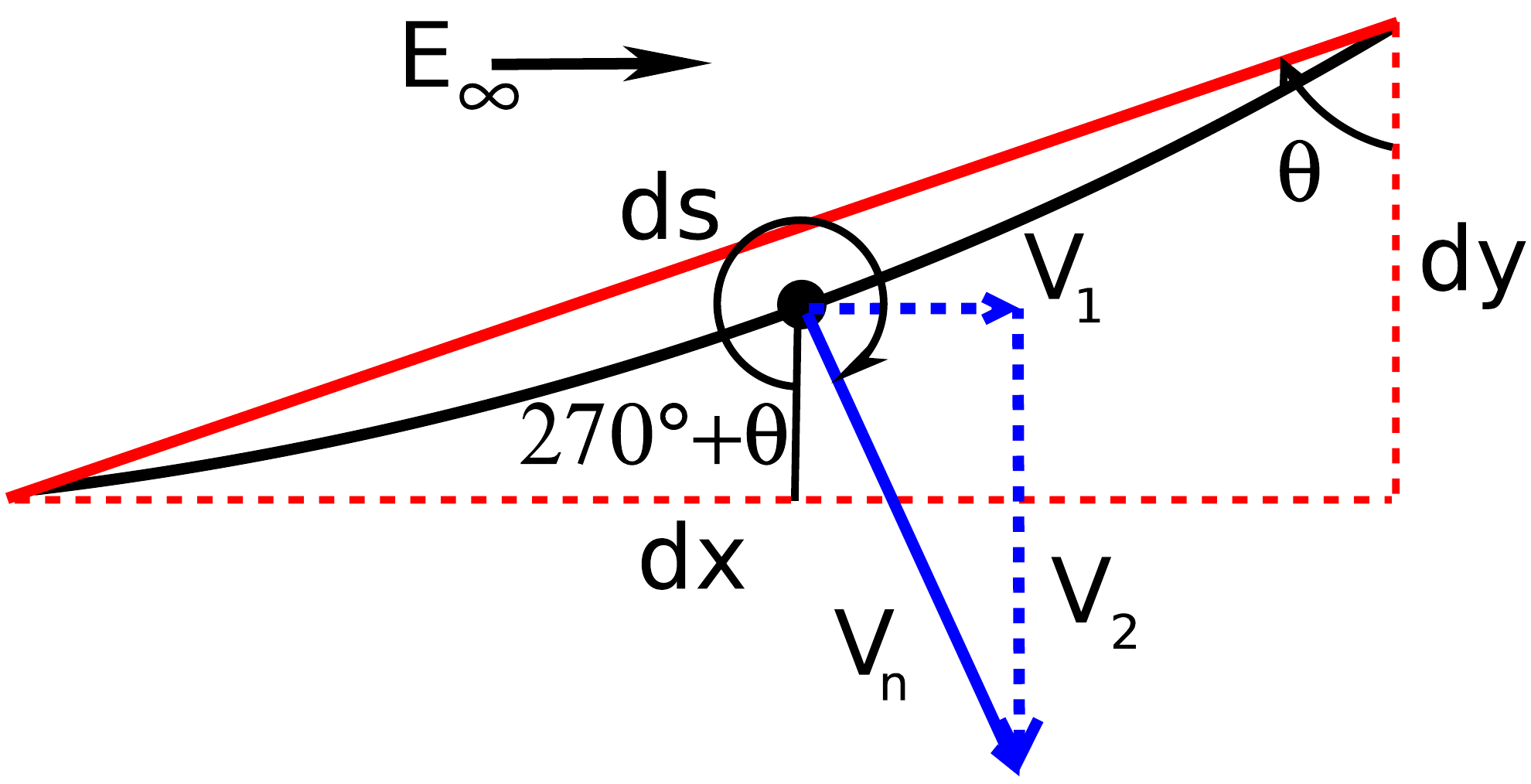}
\end{center}
\caption[Schematic of a line segment of a surface of inclusion.]{Schematic of a line segment of a surface of inclusion.}\label{fig:schematicSegment}
\end{figure}

Consider a small segment on the inclusion surface as shown in Figure~\ref{fig:schematicSegment}. Length of the segment can be expressed as,
\begin{equation}\label{eq:linearStabilitysegmentLength}
\textrm{d}s = \vert \textrm{d}z\vert =\left\vert \frac{\textrm{d}z}{\textrm{d} \vartheta} \frac{\textrm{d}\bar{z}}{d\vartheta} \right\vert^{1/2} d\vartheta = R_i\left( 1 -   \frac{\varepsilon}{R_i} \sum^{\infty}_{n=0} na_n \textrm{ cos }(n+1) \vartheta \right) d\vartheta.
\end{equation}
 Here, $\bar{\bullet}$ is a complex conjugate of the respective function. The curvature of the segment can be expressed as,
 \begin{equation}
 \kappa_s = \frac{\textrm{d} \theta}{ \textrm{d} s} = \frac{\textrm{d}}{\textrm{d}s}\left(\textrm {tan}^{-1} \left(- \frac{\textrm{d}x/\textrm{d}s}{\textrm{d}y/\textrm{d}s} \right) \right).
 \end{equation}
where $\theta$ is an angle formed by line segment with the perpendicular to external electric field. Simplifying above equation by substituting the expressions for $\textrm{d}y/\textrm{d}s$, $\textrm{d}x/\textrm{d}s$ and $\textrm{d}s/\textrm{d}\vartheta$ from the Eqs. \eqref{eq:linearStabilitysurfaceMapping} and \eqref{eq:linearStabilitysegmentLength}, one obtains
\begin{equation}\label{eq:linearStabilitycurvatureMapping}
\kappa_s = \frac{1}{R_i} \left( 1 + \frac{\varepsilon}{R_i} \sum^{\infty}_{n=0} n(n+2) a_n \textrm{ cos }(n+1)\vartheta \right).
\end{equation}
Now, the electric field can be evaluated from Eqs. \eqref{eq:linearStabilityelectricalPotential} and \eqref{eq:linearStabilitysegmentLength} as,
\begin{equation}\label{eq:linearStabilityElectricFieldOnPerturbedVoid}
\begin{aligned}
  E_t = - \frac{\textrm{d} \phi_{out}}{\textrm{d}s} = \Bigg\lbrace- \frac{2\beta}{(1+\beta)}&E_{\infty} \textrm{ sin } \vartheta - \varepsilon \frac{\beta (3-\beta)}{(1+\beta)^2} \frac{E_{\infty}}{R_i} \sum^{\infty}_{n=0} n a_n \textrm{ sin }n \vartheta \\
  &- \varepsilon \frac{\beta}{(1+\beta)} \frac{E_{\infty}}{R_i} \sum^{\infty}_{n=0} a_{n-2} (n-2) \textrm{ sin } n\vartheta\Bigg\rbrace, 
\end{aligned}
\end{equation}
It is easy to verify that by putting $\varepsilon=0$ in the above equation, the expression of the electric field on a perfectly circular inclusion is obtained, which is widely accepted form in the sharp-interface analysis \cite{santoki2019phase, wang1996simulation}.

The normal velocity at any point on the line segment is given by,
\begin{equation}
V_n = \Omega \frac{\textrm{d}J_n}{\textrm{d}s} = V_2 \textrm{ cos }(270^\circ + \theta) +V_1 \textrm{ sin }(270^\circ + \theta) = \frac{\textrm{d}y}{\textrm{d}t} \frac{\textrm{d}x}{\textrm{d}s} - \frac{\textrm{d}x}{\textrm{d}t} \frac{\textrm{d}y}{\textrm{d}s} 
\end{equation} 
Substituting Eqs. \eqref{eq:linearStabilitysurfaceMapping} and \eqref{eq:linearStabilitysegmentLength} in the above equation and integrating, the resultant equation can be given by,
\begin{equation}\label{eq:linearStabilityOmegaJ}
\Omega J_s = - \dot{S}R_i\textrm{ sin }\vartheta - \varepsilon R_i \sum^{\infty}_{n=0} \frac{\dot{a_n}}{(n+1)} \textrm{ sin }(n+1)\vartheta + \varepsilon \dot{S} \sum^{\infty}_{n=0} a_n \textrm{ sin } n\vartheta .
\end{equation}
Substituting Eqs. \eqref{eq:linearStabilitycurvatureMapping}, \eqref{eq:linearStabilityElectricFieldOnPerturbedVoid}, and \eqref{eq:linearStabilityOmegaJ} into the basic mass transport equation,
\begin{equation}
\Omega J_s = \frac{D_s \delta_s}{k_B T}\left( -eZ_s E_t +\Omega \gamma_s \frac{\textrm{d}\kappa_s}{\textrm{d}s} \right).
\end{equation}
In addition, the obtained equation is equivalently applicable to the perfectly circular inclusion. Therefore, by substituting $\varepsilon=0$, one finds a relation of a velocity of a perfectly circular inclusion,
\begin{equation}
\dot{S} = V_0= -\frac{D_s \delta_s}{k_B T} \frac{2\beta}{(1+\beta)}\frac{eZ_sE_{\infty}}{R_i}. \label{eq:linearStabilityinclusionVelocity}
\end{equation}
and the equation of evolution of inclusion surface reduced to,
\begin{equation}
\begin{aligned}
  -\dot{a}_{n-1}= \frac{D_s \delta_s}{k_BT} \frac{\beta}{(1+\beta)} \frac{eZ_sE_{\infty}}{R_i^2}n \Bigg( (n-2) a_{n-2} &-\frac{(1+\beta)}{\chi_0 \beta} n(n^2-1) a_{n-1}\\
  &+ \left(\frac{(3-\beta)}{(1+\beta)}n+2 \right)a_n \Bigg), 
\end{aligned} \label{eq:linearStabilityChi}
\end{equation}
where, a dimensionless parameter is defined as $\chi_0 = eZ_sE_{\infty}R_i^2/(\Omega \gamma_s).$ At the equilibrium of inclusion shape, i.e. $\dot{a}_{n-1}=0$, a set of an infinite number of homogeneous linear equations are obtained of the form,
\begin{figure}[!htbp]
\begin{center}
\includegraphics[scale=0.8]{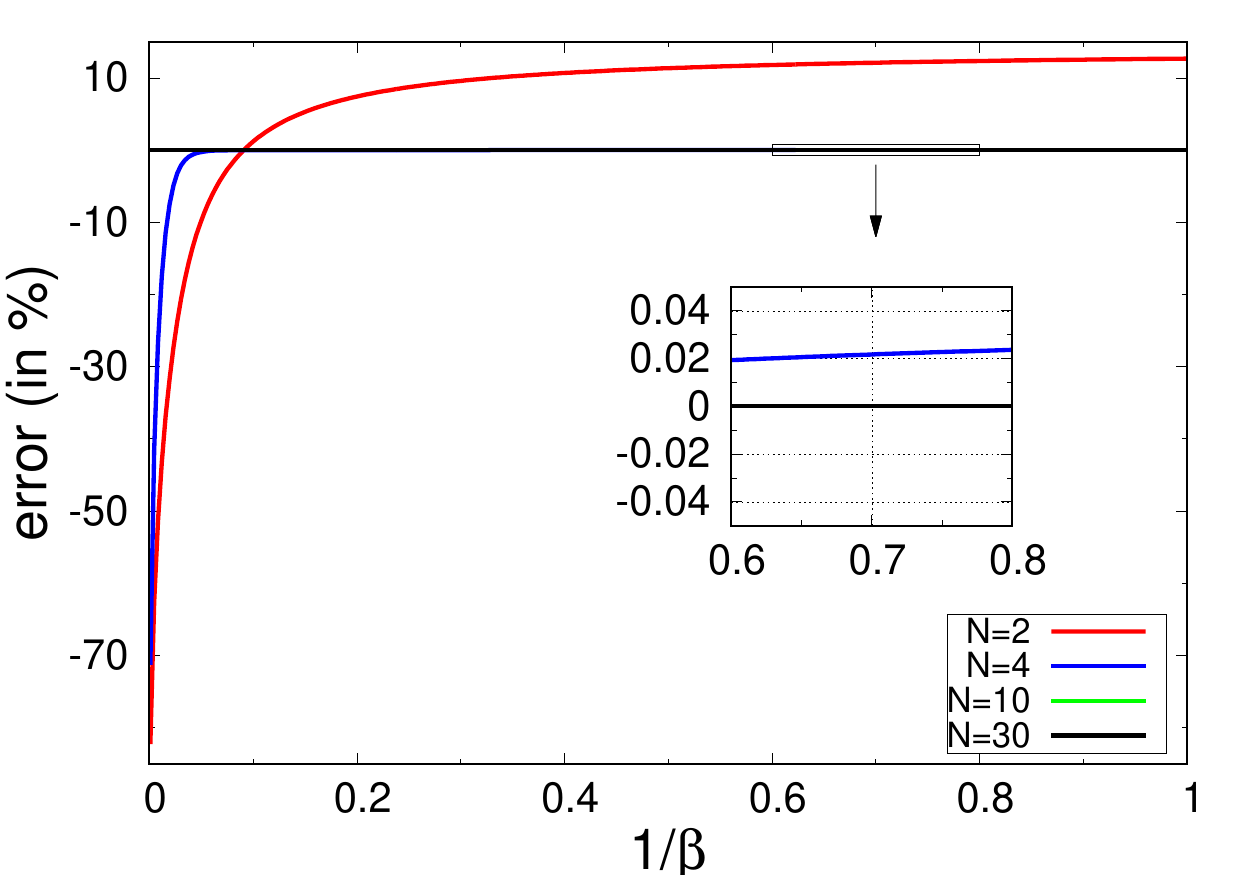}
\caption[Convergence of the numerical solution for linear stability analysis]{Convergence of the numerical solution for linear stability analysis. $N$ indicates the number of equations considered for the study. The error is calculated on the basis of relation obtained from 36 equations.}\label{fig:linearStabilityconvergenceStudy}
\end{center}
\end{figure}
\begin{equation}
(n-2)a_{n-2} - \frac{(1+\beta)}{\beta \chi_{0ct}} n (n^2-1)a_{n-1} + \left( n\frac{3-\beta}{1+\beta} +2 \right) a_n =0,\hspace{1cm} n\geq2.\label{eq:linearStabilitySharpInterface}
\end{equation}
For all integer values of $n\geq2$, a homogeneous linear equation can be obtained in the form of coefficients, $a_{n-2}$, $a_{n-1}$, $a_{n}$, $\beta$, and $\chi_{0ct}$. The purpose of these equations is to find a relation between $\beta$ and $\chi_{0ct}$ for non-trivial coefficients, $a_n$'s. Only a subset of an infinite number of equations is considered due to restrictions on the computational and analytical approaches. In other words, finite set of equations are considered with assumption that coefficients beyond $a_N$ (i.e., $a_{N}$, $a_{N+1}$,...) are zero, where $N$ indicates the number of equations considered for the study. The analytical solution can be obtained by substitution of variables up to five equations. Beyond that, the difficulty to obtain analytical relation increases exponentially with an increase in the number of equations. Therefore, a numerical tool is developed on MATLAB platform by using \textit{vpasolve} algorithm. The convergence study of the resultant relation is plotted in the Figure~\ref{fig:linearStabilityconvergenceStudy}. The derived relation between $\beta$ and $\chi_{0ct}$ in the present section (Eq. \eqref{eq:linearStabilitySharpInterface}) is considered further for the comparison with the results obtained from the phase-field method in Figure~\ref{fig:EMLSA}.

\section{Phase-field results of initially-circular inclusion migrating under isotropic diffusion}
\label{section:chi0andBetaInitial}

\begin{figure}[t]
\begin{center}
\includegraphics[scale=0.9]{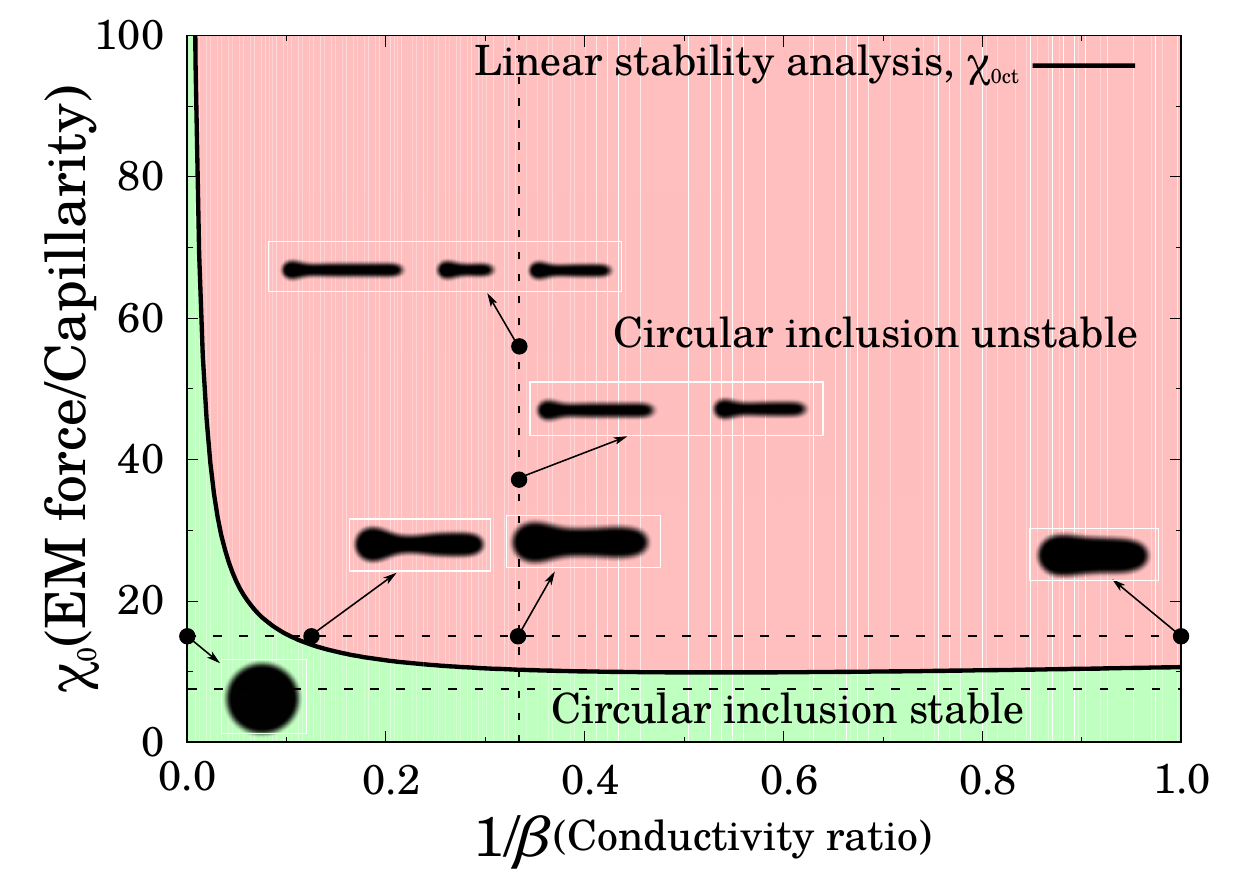}
\end{center}
\caption[Dependence of an inclusion equilibrium shape on the dimensionless parameter, $\chi_0$ and the conductivity ratio $\beta$]{Dependence of an inclusion equilibrium shape on the dimensionless parameter, $\chi_0$ and the conductivity ratio $\beta$. The black solid curve corresponds to $\chi_{0ct}$, which is obtained by linear stability analysis. This line identifies whether an inclusion to retain its circular shape or evolves to subsequent collapse. The black dashed line indicates lines of action on which the phase-field simulations are obtained. The inset of images are the snapshots from the phase-field model.}\label{fig:EMLSA}
\end{figure}

The competition between the EM wind force and the capillary force, defined by the non-dimensional parameter $\chi_0$, determines the stability of the circular inclusion. Accordingly, there exists a critical value $\chi_{0ct}$, above which the circular inclusion can no longer be stable and loose its circular shape. The theoretical value of the dimensionless parameter, $\chi_{0ct}=10.65$ , for homogeneous island ($\beta=1$) is obtained by computing the eigenvalue of Mathieu-type equation~\eqref{eq:EM1MathieuEquation}, obtained in section~\ref{section:LinearStabilityAnalysisCircularIsland}. The derived value is in agreement with linear stability analysis presented in section~\ref{section:linearStabilityAnalysis}. Furthermore, as shown in Figure~\ref{fig:EMLSA}, $\chi_{0ct}$ shows a strong dependency with conductivity ratio, $\beta$.
Even though, this theory provides information regarding the stability of circular inclusion. However, it serves no information about the shape of the inclusion after the collapse from the circular shape. These inferences can be understood by the analysis of the phase-field simulations.

  Inset of images from phase-field simulations in Figure~\ref{fig:EMLSA} shows the influence of the non-dimensional parameters $\chi_0$ and $\beta$ on inclusion equilibrium shapes. On the one hand, the competition between electromigration wind force and the capillary force decides the consequences on the inclusion shape. The non-dimensional parameter $\chi_0$ determines the relative strength of the electromigration force compared to the capillarity. On the other hand, $\beta$ identifies maximum and minimum values of $E_x$\nomenclature{$E_x, E_y$}{local electric field components in x and y-directions} and $E_y$ for a known external electric field $E_{\infty}$, which is responsible for the change of local electric field distribution. The significance of these two parameters is discussed in the following paragraphs. The simulation parameter set considered for this study is summarized in Table~\ref{tab:EM1SignificanceChiBeta}.
  
\begin{table}
\centering
\caption{Values of parameters considered for the simulation sets of an isotropic inclusions to study significance of $\chi_0$ and $\beta$.}\label{tab:EM1SignificanceChiBeta}
\begin{tabular}{ l c c }
\hline
&$\chi_0$&$\beta$\\
\hline
Set: 1&$5.0$ to $60.0$&$3$\\
Set: 2&$7.5$&$1$ to $10000$\\
Set: 3&$15.0$&$1$ to $10000$\\
\hline
\end{tabular}
\end{table}

\subsection{Significance of $\chi_0$}
\label{subsection:significanceOfChi0}
\begin{figure}[t]
\begin{center}
  \includegraphics[scale=0.93]{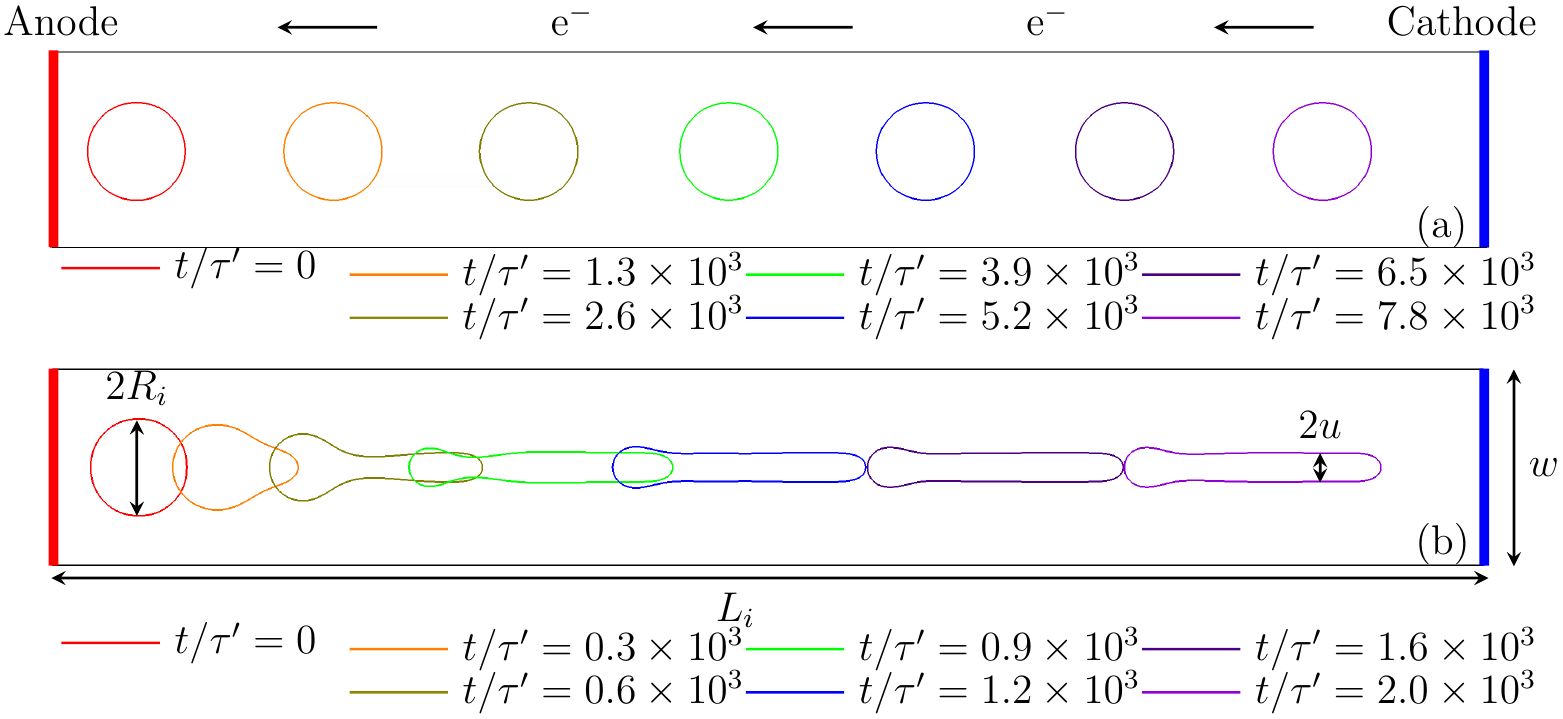}
\end{center}
\caption[Representative case of steady-state circular inclusion and circular inclusion to slit transition]{(a) A circular inclusion migrates by preserving its shape, and (b) evolves its shape to the finger-like slit. The results correspond to the dimensionless parameter $\chi_0=5.39$ for (a), and $\chi_0=26.99$ for (b) keeping the conductivity contrast constant, $\beta=3$. The arrows show the direction of the electron wind, from the cathode to the anode. Here $u$ is the half-slit width\nomenclature{$u$}{half-slit width} and $w$ denotes the line width of the conductor\nomenclature{$w$}{line width of the conductor}.}
\label{fig:SchameticVoidEvolution}     
\end{figure}
The shape of the inclusion is governed according to the competition between the electromigration wind force and the capillary force, as it is evident from the non-dimensional parameter $\chi_0$. 
The capillary force seeks a uniform curvature and consequently diffuses species to reduce any curvature gradient along the inclusion surface.
 Electromigration, on the other hand, instigates an atomic transport in the direction of the electron wind force, leading to its shape alterations. 
 In the present case, this corresponds to the diffusion of the species, from the right towards the left surface.

The migration of the circular inclusion, at a low electromigration force, is shown in Figure~\ref{fig:SchameticVoidEvolution}(a). 
From the phase-field simulations, the inclusion surface is extracted as an isoline of value $c=0.5$. 
The diffusion of the atomic species, in the direction of the wind force, induces the drift of the inclusion, in the direction of the electric field. Inherently, the capillary force suppresses the presence of any curvature gradient along the inclusion surface. The dominant capillary force, however, encourages the inclusion to preserve the circular shape during the course of the drift as shown in Figure~\ref{fig:SchameticVoidEvolution}(a). Any non-uniformity of the curvature is almost instantaneously eliminated by the capillarity. Therefore, the circular inclusion is the most preferred shape by the capillarity. 

A dominant electromigration force, on the other hand, leads to a biased mass transport towards the anode side, thereby producing a sharp protrusion on the cathode side of the inclusion (see Figure~\ref{fig:SchameticVoidEvolution}(b)).
 The differential curvature induces a counter mass transport towards the cathode end of the inclusion.
  The curvature-mediated mass transport is evident in the shortening rear end of the inclusion.
   However, owing to the significantly higher electromigration force, the protrusion on the inclusion surface, at the cathode end, increases in length, and ultimately forms a narrow finger-like slit. 

\subsection{Significance of $\beta$ and the current crowding}
\label{subsection:currentCrowding}

\begin{figure}[hbt!]
\begin{center}
\includegraphics[scale=1.0]{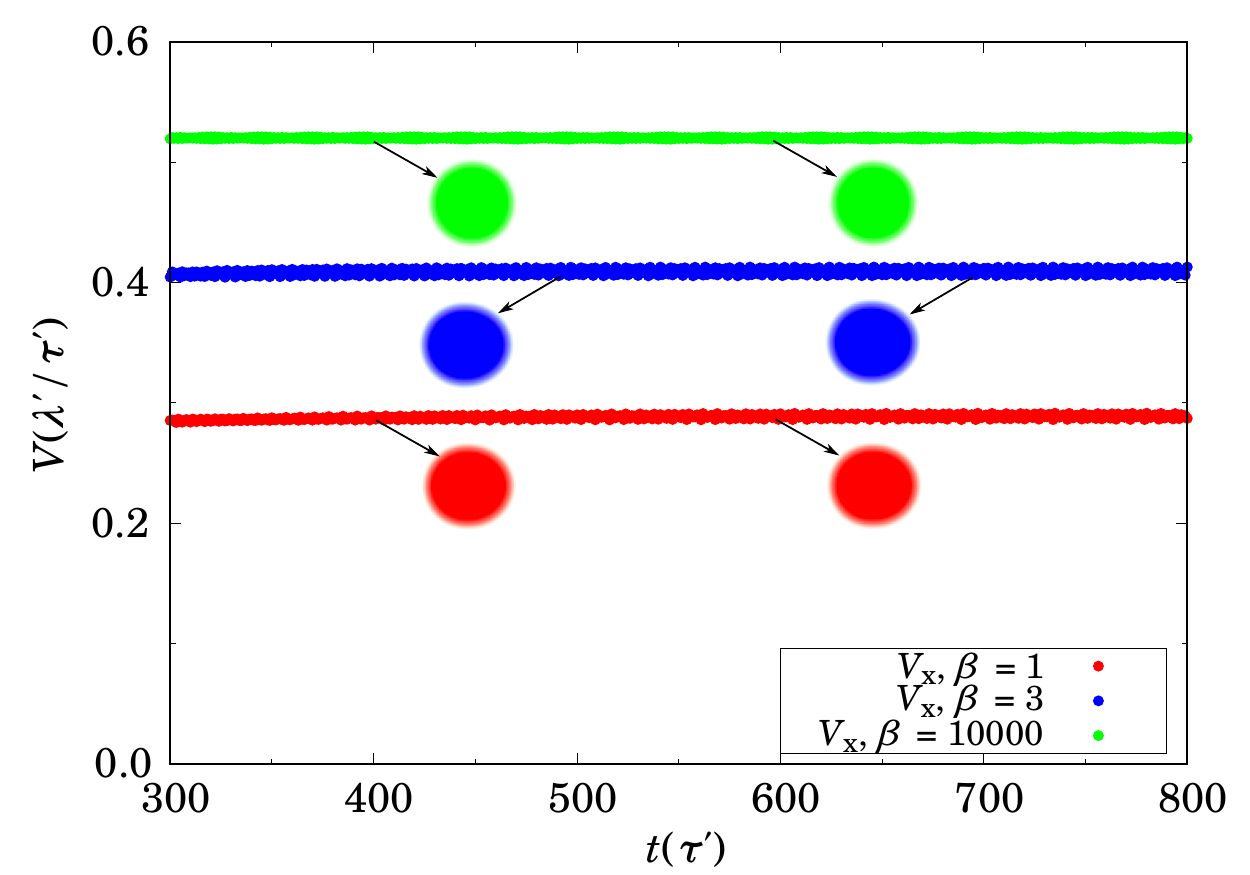}
\end{center}
\caption[Velocity of the centroid of the inclusions along the direction of external electric field, $V_0$ for conductivity ratios, $\beta=1, 3,$ and 10000]{Velocity of the centroid of the inclusions along the direction of external electric field, $V_0$ for conductivity ratios, $\beta=1, 3,$ and 10000, and the dimensionless parameter $\chi_0=7.50$.  The inset of images are the snapshots from the phase-field model in equilibrium.}\label{fig:EMcircularisland}
\end{figure}

 To understand the significance of $\beta$, simulations along two horizontal dotted lines shown in Figure~\ref{fig:EMLSA} are considered. According to linear stability analysis, all inclusions on the lower line should migrate maintaining their circular shapes, while shape alterations are expected on the right part of the upper line.
 
  From the phase-field simulations, the migration of the initially-circular inclusions for lower electromigration force ($\chi_0=7.5$) are shown in Figure \ref{fig:EMcircularisland}. Due to capillarity mediated diffusion, circular inclusions are unchanged from its initial configuration irrespective of conductivity ratio $\beta$. However, the velocities of the inclusions are modified by $\beta$, which is in agreement with the literature \cite{santoki2019phase, wang1996simulation} and linear stability analysis (see Eq. \eqref{eq:linearStabilityinclusionVelocity}). In fact, the adjustment of the inclusion velocity approximately corresponds to the factor of $2\beta/(1+\beta)$. This infers that the $\beta$ increases velocity of the circular inclusions, as inclusions of heterogeneous conductivities ($\beta>1$) travels faster compared to its homogeneous ($\beta=1$) counterpart. Therefore, the increase in conductivity ratio improves propagation speed, up to two times for a circular inclusion. 
 
\begin{figure}[hbt!]
\begin{center}
\includegraphics[scale=0.1]{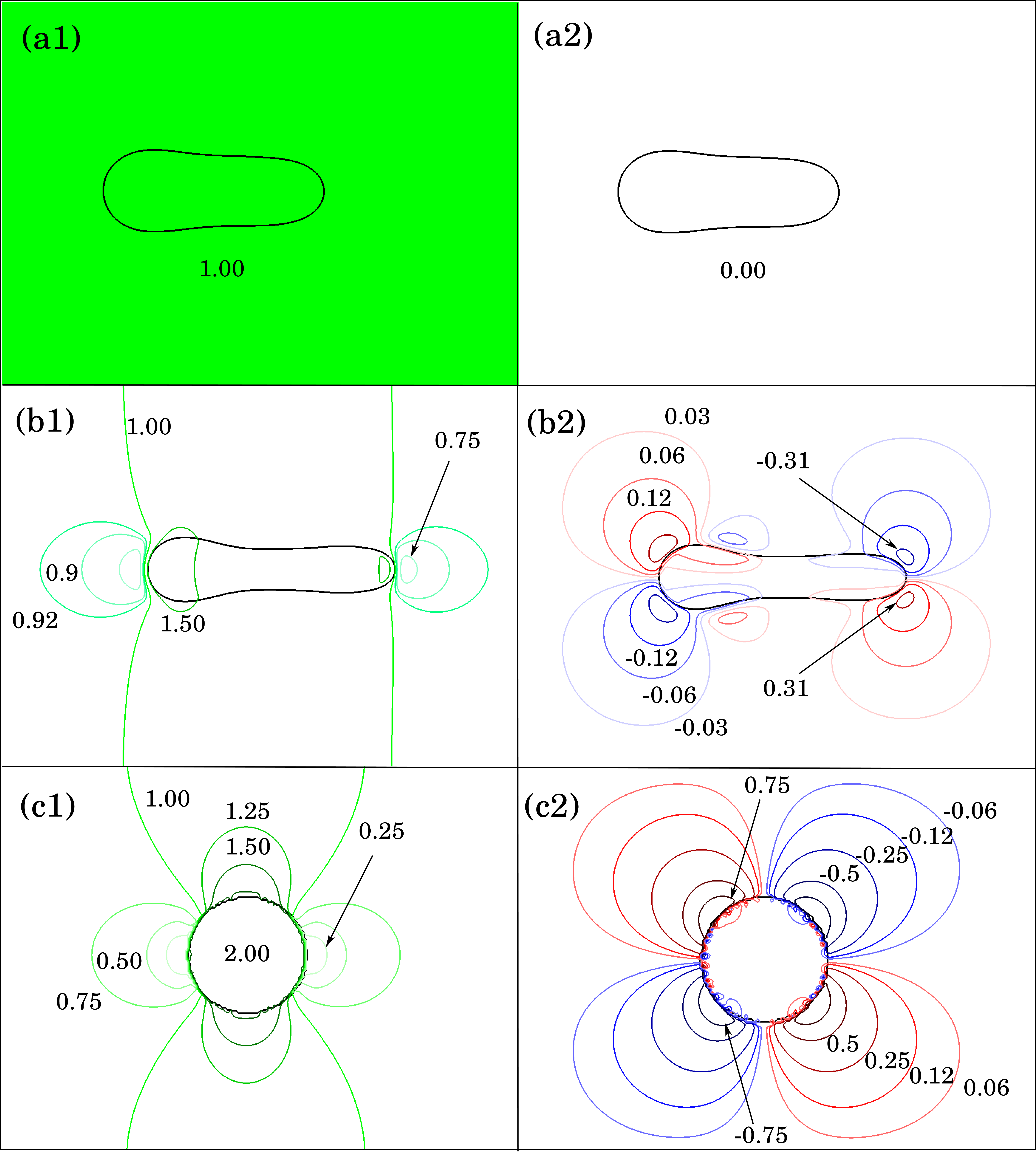}
\end{center}
\caption[Change in local electric field distribution due to current crowding near inclusion surface]{The change in local electric field distribution during the evolution of the inclusion. The distribution of electric field components $E_x$ in first column and $E_y$ in second column are shown for $\beta=1, 3,$ and $10000$ from top to bottom respectively. The electric field components are normalized by the external electric field $E_{\infty}$. The electric field distribution remains unchanged for $\beta=1$. On the contrarily, although it remains nearly constant at the distance from the inclusion, while inside and around the inclusion, electric field changes effectively for $\beta=3$ and $10000$. The contour of maximum value of $E_x$(=1.5) is divided into two regions inside the inclusion for $\beta=3$. In addition, $E_x$ enhances local electric field up to two times, while $E_y$ forms a quadrupole pattern around the inclusion of $\beta=10000$.} \label{fig:EMcurrentCrowding}
\end{figure}

 During the inclusion migration at higher electromigration force ($\chi_0=15$), when electron wind passes from the periphery, it interacts with the surface. Therefore, the electromigration force perceives the length of the inclusion along the traverse direction to the electric field as a resistance in the path of electron wind. The higher electromigration force reduces the resistance in the path by altering inclusion shape to the formation of a finger-like slit (see inset at the right side of Figure. \ref{fig:EMLSA}). However, these arguments are only valid for the homogeneous systems, meaning, of similar conductivities ($\beta=1$), because of uniform electric field distribution as shown in Figure \ref{fig:EMcurrentCrowding}(a1) and (a2). For heterogeneous systems due to different conductivities in the matrix and the inclusion, the distribution of electric field near the inclusion disrupted, which significantly contributes to its shape. Figure \ref{fig:EMcurrentCrowding}(b) and (c) show the local change in electric field distribution in the vicinity of the inclusion with help of the electric field components, $E_x$ and $E_y$, in the longitudinal and traverse directions respectively.

 The distribution of $E_x$ in Figure \ref{fig:EMcurrentCrowding}(c1) describes an increase in the local electric field near the inclusion in the vertical direction, while a decrease in the horizontal direction. The asymmetric allocation of the electric field promotes local species diffusion from the lower electric field to the higher one. Therefore, the local atomic flux for $\beta=10000$ asserts the preservation of inclusion shape during the drift. For $\beta=3$ in Figure \ref{fig:EMcurrentCrowding}(b1), due to relatively weak intensity of change in electric field and the separated regions of maximum $E_x$ (i.e., $E_x/E_{\infty}=1.5$ contour) at the rear and the front ends motivates slit of thicker portions at both ends, while thinner at the middle portion. As shown in Figure \ref{fig:EMLSA}, the intermediate steady states between the finger-like slit and the circular inclusion, i.e., from $\beta=1$ to 10000, reveals smooth transition of inclusions by continuous aggregation of the matter at the rear end. These arguments from phase-field simulations facilitate reasoning for a continuous increase in $\chi_{0ct}$ with $\beta$, obtained by linear stability analysis in Figure~\ref{fig:EMLSA}. The stiff gradient in $\chi_{0ct}$ on the left side of Figure~\ref{fig:EMLSA} can be explained by the relative strength of the external electric field and the local change in the electric field. The higher electron wind force due to further increase in the dimensionless parameter $\chi_0$ for the constant $\beta$ establishes a higher velocity of inclusions. This promotes protrusion to the surface of the inclusion with the same conductivity ratios for higher $\chi_0$.

As evident, the finger-like slits are the frequently observed morphology in isotropic inclusions. Therefore, to analyze the characteristics of the finger-like slits, only the occurrence of single slits is considered for a detailed analysis in the following section. The phase-field simulations are performed for parameter set $\beta=3.0$ and $\chi_0=13.0$ to $28.0$. The specific features of the finger-like slits obtained from the numerical study are discussed in subsection~\ref{subsection:EM1SFFLS}. In addition, the sharp-interface theory is derived for these slits in subsection~\ref{section:sharpInterfaceAnalysisSlitProfile}. Subsequently, the findings from phase-field simulations are critically compared with sharp-interface analysis in subsection~\ref{section:SlitComparisonWithSharpInterface}. Furthermore, the influence of various factors influencing this slit selection are discussed in subsections~\ref{subsection:conductorWidth} and \ref{subsection:EM1SelectionSlitWidthVelocity}.

\section{Finger-like slit propagation} 
\label{sec:EM1SpeceficFeaturesSlits}
 
  Gan et al. \cite{gan2002electromigration} observed that inclusions nucleate and grow at the intermetallic compound (IMC)\nomenclature{IMC}{intermetallic compound} and the grain boundary interface. Afterward, the inclusions migrate into the conductor lines in a transgranular manner, which might be subjected to shape alterations \cite{shingubara1991electromigration}.
 Kraft et. al. \cite{kraft1993observation} reported that a slight change in the void shape can be amplified by the electron wind leading to a slit. 
   Perhaps, the most pervasive form of inclusion migration is the transition to a slit-like morphology.
The presence of finger-shaped (also known as pancake-type) inclusions have been reported in Refs. \cite{miyazaki2006electromigration, gan2002electromigration}. 




\subsection{Specific features of finger-like slits}
\label{subsection:EM1SFFLS}
\begin{figure}[t]
\begin{center}
  \includegraphics[scale=0.9]{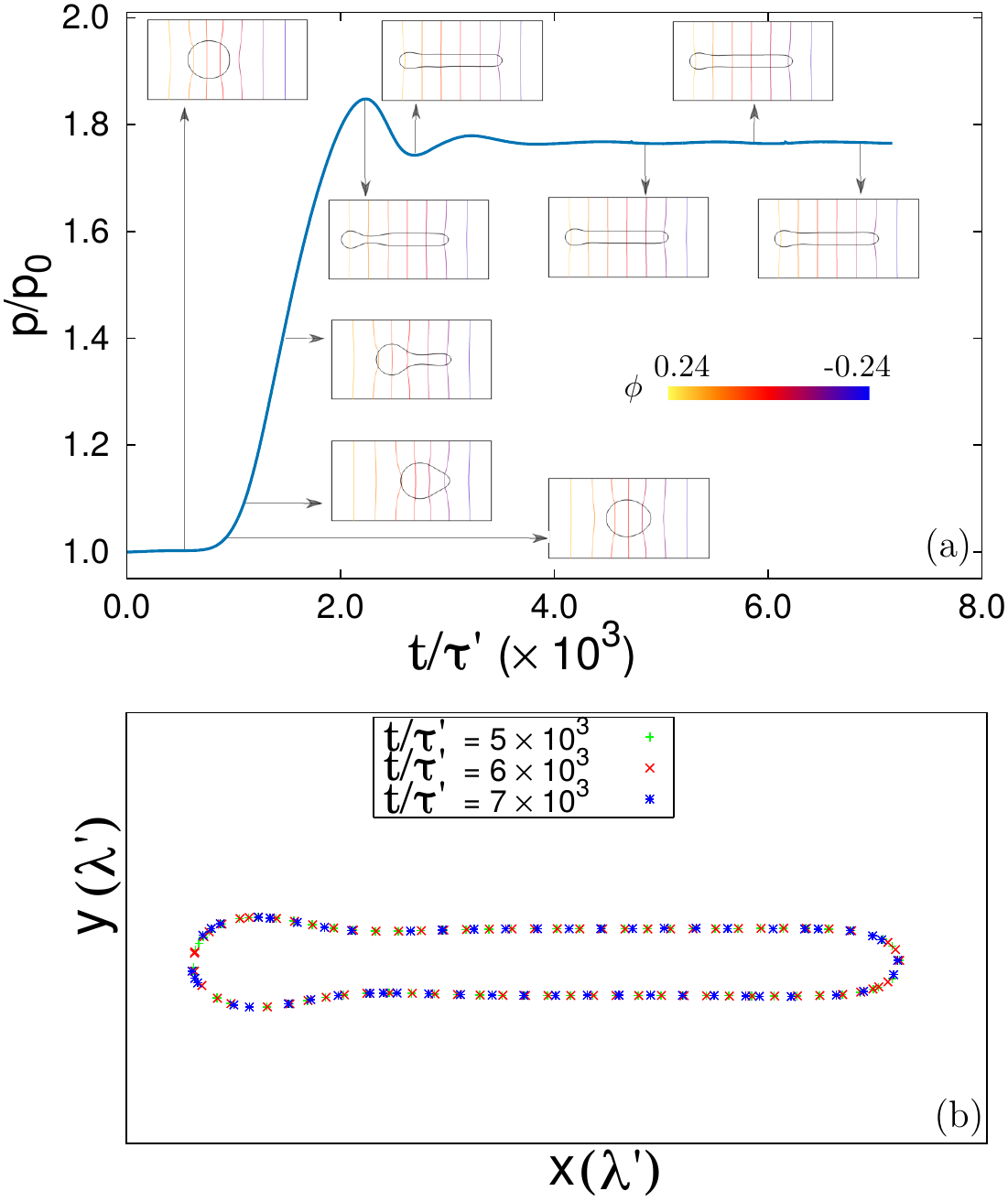}
  \end{center}
\caption[Temporal evolution of the perimeter as a function of time from the initial circular inclusion to the finger-like slit and contour plots of the steady-state slit at three different times]{(a) Temporal evolution of the perimeter as a function of time, from the initial circular inclusion to the finger-like slit. (b) Contour plot of the slit, at three different times indicating that the slit propagates with an invariant shape. The results are shown for the dimensionless parameter $\chi_0=27.27$.}
\label{fig:perimeter}     
\end{figure}
\begin{figure}[t]
\begin{center}
\includegraphics[scale=1.0]{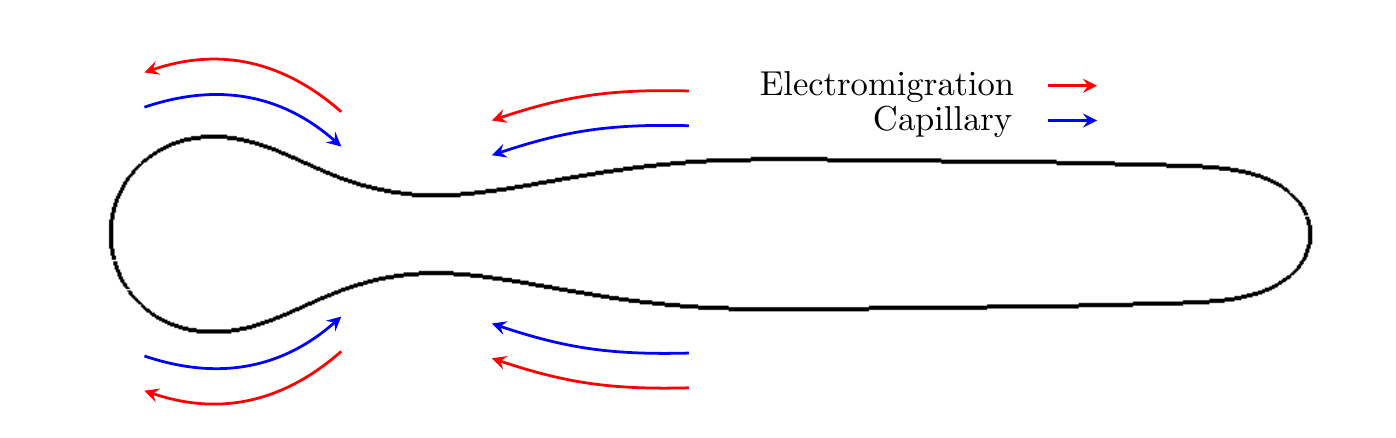}
\end{center}
\caption[Schematic diagram of constricted neck region with EM force and capillarity]{A schematic diagram of constricted neck region at $t/\tau'=2 \times 10^3$ in Figure \ref{fig:perimeter}. The blue arrows show atomic mass transport due to the capillary force, while the red arrows are the flux due to the electron wind.}
\label{fig:EMCapillaryComp}    
\end{figure}

The shape changes during the transition of the circular inclusion to a stationary slit can be captured effectively by tracking the temporal inclusion perimeter of the phase-field simulation, as shown in Figure~\ref{fig:perimeter}(a). 
The perimeter $p$\nomenclature{$p$}{transitory perimeter of the inclusion} is calculated from the slit profile of the isoline with the value $c=0.5$, and is normalized by the initial perimeter of the circular inclusion $p_0$\nomenclature{$p_0$}{initial perimeter of the inclusion}.
  As the protrusion initiates, the inclusion perimeter increases with time, until a maximum at $t/\tau'=2 \times 10^3$.
   At this stage, the junction at the end of the parallel region of the slit and its circular end become constricted as shown in Figure~\ref{fig:EMCapillaryComp}.
    The surface is concave, with respect to the neighboring regions on either side. 
    Thus, the flux from either side, which is induced by the curvature gradient, in addition to the electromigration-induced flux from the cathode side, leads to the momentary decrease in the inclusion perimeter.
     At a later stage ($t/\tau'>4 \times 10^3$), the inclusion perimeter saturates to a constant value, which indicates the time-invariant migration of the inclusion.
      The time-invariant drift of the inclusion is the manifestation of the equilibrium of the capillary-induced and electromigration-induced flux at every point on the surface.
       The invariance of the inclusion shape is also evident from the complete overlap of the slit profile, at the late stages, as shown in Figure~\ref{fig:perimeter}(b).
        The slit profiles have been translated by a factor $V t$, where $V$ is the steady-state velocity.

\begin{figure}[t]
\begin{center}
  \includegraphics[scale=1.0]{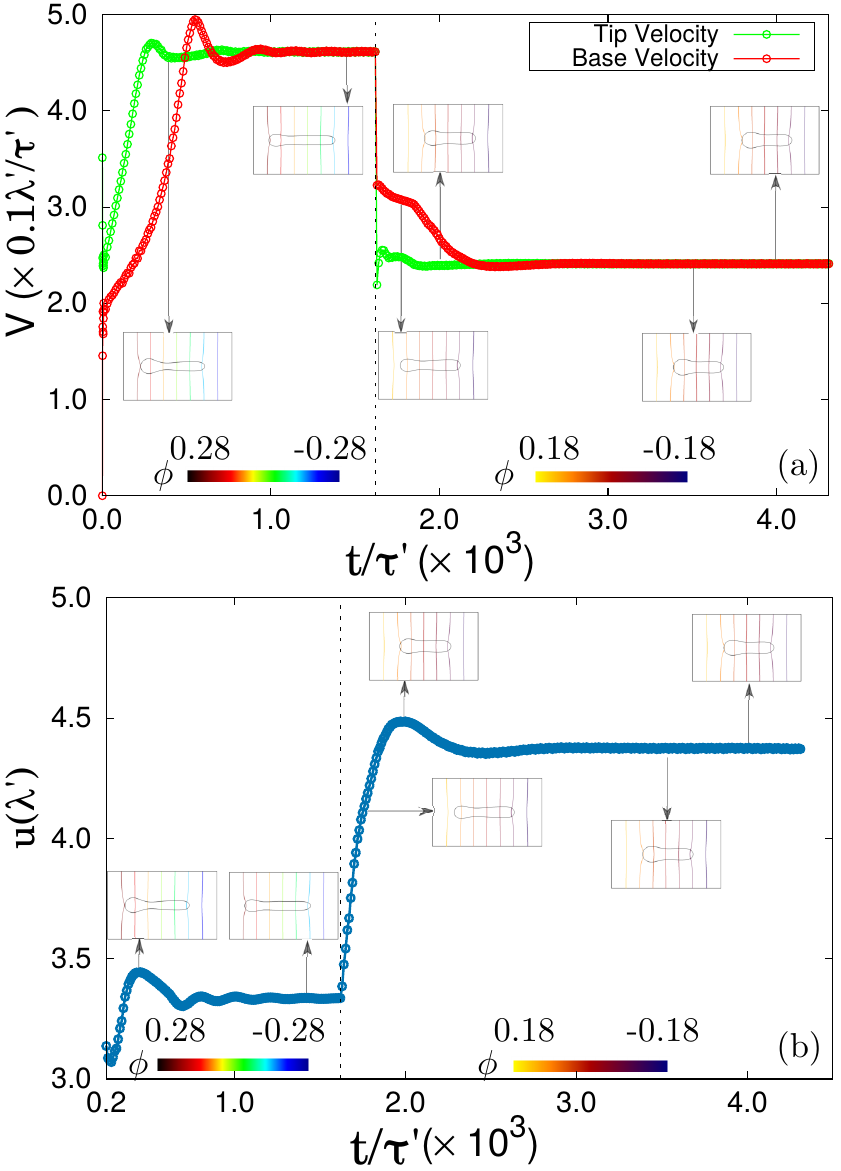}
  \end{center}
\caption[Effect of change in the electrical potential on propagation velocity and width of the slit]{The change in the electrical potential at time $t/\tau' =1.6 \times 10^3$ results in the alteration of (a) the base velocity, the tip velocity, and (b) the width of the slit propagation. The dotted-black line in the graph represents the change in the electrical potential, from $\phi_\infty$ = 0.28, -0.28 ($\chi_0=20.72$) to $\phi_\infty$ = 0.18, -0.18 ($\chi_0=13.81$). The slit adjusts the width and the velocity according to the external electric field, and evolves to a steady-state finger-like shape.}
\label{fig:potentialChange}     
\end{figure}

The interplay between the capillary force and the electromigration wind force is further reflected by changing the applied potential, after the equilibration of the slit profile. 
During operational conditions, the interconnect lines are seldom subjected to a constant electric field.
 Thus, understanding the effect on the inclusion shape, which is due to a change in the electric field, has technological implications. 
 Figure~\ref{fig:potentialChange}(a) depicts the change of the inclusion shape, when the applied potential is abruptly changed from $\phi_\infty=-0.28,0.28$ ($\chi_0=20.72$) to $\phi_\infty=-0.18,0.18$ ($\chi_0=13.81$), after the slit attains its equilibrium shape in the steady state conditions.

To ascertain that every point on the slit propagates with the same velocity, the drift speed is traced at the slit tip, and at the rear end (base). 
Although the front and the rear ends drift with different velocities initially, during the first electric field cycle, the equilibrium of the electromigration-induced and capillary-induced flux leads to both, an invariant shape as well as a steady-state velocity at each point along the slit surface. 
The decrease in the electrical potential reduces the electromigration force, and hence disrupts the already equilibrated electromigration and capillary forces. 
With the decrease in electromigration flux, the capillary-mediated flux from the rear end towards the cathode end becomes prominent, leading to a decrease in the inclusion perimeter. 
In the second cycle, the species transport faster from the rear end, due to the momentarily steep capillary force, which leads to an initially faster base velocity, in contrast to the former. 
The transport, which is mediated by the capillary, leads to an increase in the slit width, in the attempt to minimize the curvature gradient. 
This process can be understood as follows: The curvature at the tip of the slit is approximately $\kappa_s \approx 1/u$, and the curvature gradient $d \kappa_s/d s \approx 1/u^2$. 
Since the decrease in $E_{\infty}$ leads to a decrease in the electromigration flux (first part in Eq.~\eqref{eq:EM1SharpInterfaceFlux}), the corresponding decrease in the capillary part is achieved by an increase in the width of the slit.
 The temporal evolution of the slit width $u$, with the change in the electric field cycle is illustrated in Figure~\ref{fig:potentialChange}(b).
 The slit width is calculated from the averaged values of the difference between the ordinates of the slit profile.
 The width increases until a new equilibrium is established between the two counteracting forces, after which the entire surface again drifts with a uniform steady-state velocity.

The above results imply a strong dependence of the selection of the slit width and velocity on the electric field, which is discussed in the last part of this section (in subsection~\ref{subsection:EM1SelectionSlitWidthVelocity}).

\subsection{Sharp-interface analysis for slit profile}
\label{section:sharpInterfaceAnalysisSlitProfile}

\begin{figure}[h]
\centering
  \includegraphics[scale=1.0]{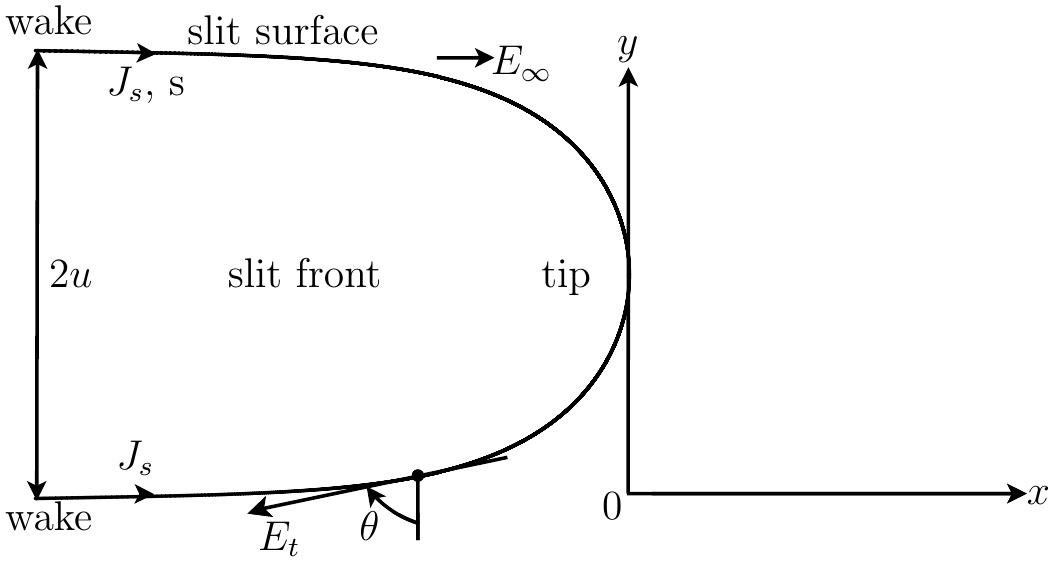}
\caption[Schematic diagram of the finger-like slit front]{Schematic diagram of the finger-like slit front, subjected to an external electric field, in an infinite interconnect domain. The origin lies at the point of intersection between the tangent to the slit tip and a straight line through the flat wake.}
\label{fig:slitProfileSchametic}     
\end{figure}
Transgranular slit propagation, under the assumption of a steady-state, is derived by Suo et. al. \cite{suo1994electromigration}.
 For brevity, the basic steps of the derivations are described. A complete description can be found in Refs. \cite{suo1994electromigration, yao2015an}.

Consider a slit front, translating along the length of the conductor, with a time-invariant shape.
 The coordinate system is assumed to move in the frame of reference, which is attached to the slit, with the origin coinciding with the flat region, as shown in Figure~\ref{fig:slitProfileSchametic}. 
 The slit migrates as a result of a mass transport flux, which is induced by the surface electromigration and the capillarity. Therefore, the Nernst-Einstein relation~\eqref{eq:EM1SharpInterfaceFlux}, mass conservation equation~\eqref{eq:EM1MassConservationNormalVelocity}, and the normal velocity~\eqref{eq:EM1normalVelocity} are equally applicable to finger-like slit propagation. However, to generalize the tangential electric field component Eq.~\eqref{eq:EM1sharpInterfaceElectricalPotentialIsland}, which includes the effect due to distinct conductivities in the matrix and the inclusion, the electric field along the slit surface can be expressed as,
\begin{equation}
E_t =  - \frac{2\beta}{1+\beta} E_{\infty} \textrm{ sin } \theta, \label{eq:SIelectricalPotentialComponent}
\end{equation}
which is responsible for the slit transmission \cite{wang1996a}. 

Substituting Eq.~(\ref{eq:EM1normalVelocity}) into Eq.~(\ref{eq:EM1MassConservationNormalVelocity}), and utilizing the relation $\textrm{d}y=\textrm{d}s\textrm{ cos } \theta $, the resultant expression can be written as,
\begin{equation}
\frac{\textrm{d}J_s}{\textrm{d}y} = \frac{V}{\Omega}. \label{eq:MassConservationSlitPropagation}
\end{equation}
Integrating once, 
\begin{equation}
J_s = \frac{V}{\Omega}y + C_{st}. \label{eq:MassConservationWithConstant}
\end{equation}
Here $C_{st}$ denotes the integration constant, which is determined from the fact that in the flat wake, i.e., $\textrm{d} \kappa_s/\textrm{d}s =0$ at $y=0$, substituting this relation in the Nernst-Einstein relation~\eqref{eq:EM1SharpInterfaceFlux}, the flux in the flat wake can be expressed as,
\begin{equation}
J_s\vert_{y=0}=\frac{D_s \delta_s}{\Omega k_BT}e Z_s\frac{2\beta}{1+\beta} E_{\infty} \label{eq:fluxFlatWake}
\end{equation}
Equating Eqs.~\eqref{eq:MassConservationWithConstant} and \eqref{eq:fluxFlatWake}, the resultant flux,
\begin{equation}
J_s = \frac{V}{\Omega} y +\frac{D_s \delta_s}{\Omega k_B T} e Z_s  \frac{2\beta}{1+\beta} E_{\infty}. \label{eq:fluxwithVelocitySI}
\end{equation}

For a circular inclusion of radius $u$, which is migrating maintaining its shape, the velocity relation can be given by Eq.~\eqref{eq:EM1steadyStateCircularVelocity},
\begin{equation}
V_0 = - \frac{D_s \delta_s}{uk_BT}eZ_s \frac{2\beta}{1+\beta} E_{\infty} \label{eq:voidVelocity}
\end{equation}
 Henceforth, the derivation digresses from the work of Suo et al.~\cite{suo1994electromigration}. 
According to Yao et. al.~\cite{yao2015an}, the shape evolution changes the velocity simultaneously. 
The parameter $\xi$\nomenclature{$\xi$}{discrepancy coefficient} can then be introduced to reflect the discrepancy between the velocities of the circular inclusion with the radius $u$, and the finger-like slit with the same half slit width. 
Thus, the slit propagation velocity relative to the circular one is defined by
\begin{equation}
V = \xi V_0. \label{eq:slitVelocity}
\end{equation}
The value $\xi>1$ implies that the slit moves faster, after collapsing from a circular inclusion, whereas the value $0<\xi<1$ denotes that the slit moves more slowly. 
 
A lower half of the slit profile is calculated by the following description, which considers symmetry at $y=u$.
 Consider a small segment $\textrm{d}s$, whose tangent makes an angle $\theta$ with the y-axis. Let $q=\textrm{ sin } \theta = (-\textrm{d}x/\textrm{d}y)/\sqrt{1+(\textrm{d}x/\textrm{d}y)^2} $.
  Note that $\textrm{d}y/\textrm{d}s = (1-q^2)^{1/2}$ and $\kappa_s = \textrm{d} \theta/\textrm{d} s =  \textrm{d}q/\textrm{d}y$.
   The curvature gradient writes as
\begin{equation}
\frac{\textrm{d} \kappa_s}{\textrm{d}s} = \left( \frac{\textrm{d} \kappa_s}{\textrm{d}y} \right) \left( \frac{\textrm{d} y}{\textrm{d}s} \right) = (1-q^2)^{1/2} \frac{\textrm{d}^2 q}{\textrm{d}y^2}. \label{eq:curvatureGradient}
\end{equation}
Substituting Eqs.~\eqref{eq:EM1SharpInterfaceFlux},\eqref{eq:SIelectricalPotentialComponent},\eqref{eq:voidVelocity},\eqref{eq:slitVelocity},and \eqref{eq:curvatureGradient} into Eq.~\eqref{eq:fluxwithVelocitySI} results in
\begin{equation}
\frac{1}{\eta} (1-q^2)^{1/2} \frac{\textrm{d}^2 q}{\textrm{d}Y^2} + q + \xi Y = 1, \label{eq:SIMainEquation}
\end{equation} 
where Y = y/u and a dimensionless number, $\eta$\nomenclature{$\eta$}{slit characteristic parameter} is expressed as
\begin{equation}
\eta = \frac{eZ_s {2\beta} E_{\infty} u^2}{\Omega \gamma_s{(1+\beta)}}\label{eq:etaEquation}.
\end{equation}
The dimensionless group $\eta$ reflects the relative strength between the electromigration wind force and the capillary force. It is worth noticing that even though both dimensionless parameters, $\chi_0$ and $\eta$ determines relative strength between electromigration force and capillarity, the former provides condition for the stability of the circular inclusion, while letter helps to find physical characteristics associated to the slit, which is elaborated further in the forthcoming paragraphs.

For the flat wake, the vanishing slope and curvature provide the boundary conditions:
\begin{equation}
q=1, \quad \textrm{d}q/\textrm{d}Y=0 \quad \textrm{ at } Y=0. \label{eq:SIBoundaryCondition}
\end{equation}
The sharp-interface equation~(\ref{eq:SIMainEquation}) is solved under the boundary conditions in Eq.~(\ref{eq:SIBoundaryCondition}), by using a combined shooting and Runge-Kutta methods. 
The value of $\eta$ is held fixed and shooting for the value of $\xi$, such that the angle at the slit tip is zero (i.e., $q=0$ at $Y=1$). 
Finally, by combining Eqs.~(\ref{eq:voidVelocity}), (\ref{eq:slitVelocity}), and (\ref{eq:etaEquation}), the resultant expressions can be obtained as,
\begin{equation}
u=\left(\frac{\eta \Omega \gamma_s{(1+\beta)}}{Z_s e {2\beta}  E_{\infty}}\right)^{1/2}, \label{eq:SISlitWidthFinal}
\end{equation}
\begin{equation}
V= \xi \frac{D_s \delta_s}{k_BT} \frac{(Z_s e {2\beta}  E_{\infty})^{3/2}}{(\eta \Omega \gamma_s)^{1/2}{(1+\beta)}^{3/2}}. \label{eq:SISlitVelocityFinal}
\end{equation}

\subsection{Comparison with the sharp-interface description}
\label{section:SlitComparisonWithSharpInterface}
\begin{figure}
\begin{center}
\includegraphics[scale=1.0]{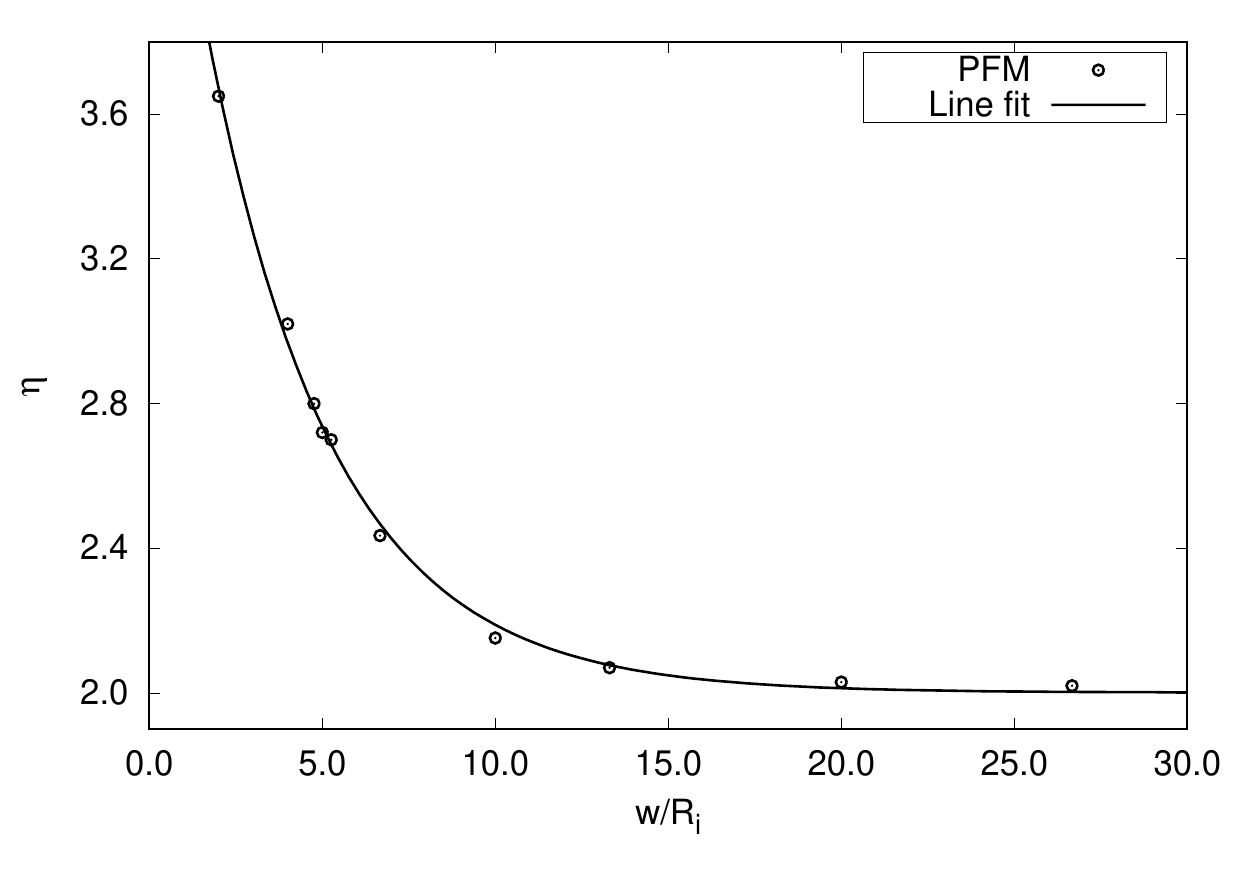}
\end{center}
\caption[Non-dimensional parameter $\eta$ as a function of conductor line width to inclusion radius ratio $w/R_i$]{The values of $\eta$, obtained from the phase-field model, together with a line fit according to Eq.~(\ref{eq:functionFit}) are plotted as a function of $w/R_i$. A smaller values of $w/R_i$ result in wider slit width for same applied electric field.}
\label{fig:etaVsWbyU}    
\end{figure}
The parameter $\chi_0$ (Eq.~\eqref{eq:EM1chi0}) is introduced to provide a condition for the circular inclusion stability. 
If $\chi_0$ exceeds a critical value $\chi_{0ct}$, an initially circular inclusion collapses into a slit. 
However, the parameter contains no information about the dimensions of the newly formed slit, and about the propagation velocity after the breakdown \cite{suo1994electromigration}.
By incorporating the value $\eta$ into the sharp-interface model, all
  corresponding slit characteristics can be determined.
$\eta$ is an input parameter in sharp-interface analysis. While, in the phase-field model, the surface energy, the electric field and the ratio of the initial inclusion to the line width are given as the input, and the slit selects its width and velocity accordingly.
  Hence, $\eta$ is \textit{a priori} unknown quantity. 
   To facilitate a comparison of the slit characteristics, namely the slit profile, the width, and the velocity obtained from the sharp-interface and diffuse-interface formalism, the value of $\eta$ needs to be extracted from the phase-field simulation. 
 Furthermore, from Figure~\ref{fig:potentialChange}, it is evident that an increase (or decrease) in the electric field leads to a finer (or wider) slit. 
 The same is also apparent from Eq.~(\ref{eq:SISlitWidthFinal}). 
 However, it is worth noting that $E_{\infty}$ and $u^2$ appear as a product in the numerator, in the expression of $\eta$. 
 Since the cause ($E_{\infty}$) and the effect ($u^2$) act in the opposite way, it is inexplicable beforehand whether an increase (or decrease) in $E_{\infty}$ entails a corresponding increase (or decrease) in $\eta$.

Additionally, it is to be noted that the sharp-interface analysis does not consider the effect of the line width on the selection of the final width and the velocity of the slit. 
However, the phase-field simulations reveal a strong dependency of the line width on the final slit characteristics. 
In Figure~\ref{fig:etaVsWbyU}, $\eta$ is shown as a function of the interconnect width $w$ to the inclusion radius $R_i$. 
$\eta$ exhibits a steep increase, with a corresponding decrease in the values of $w/R_i$. 
This implies that the slit originating from the higher values of $w/R_i$ result in narrower slit width for the same value of the applied electric field.     
The relationship between the two parameters $\eta$ and $w/R_i$ is fitted by using an exponential function:
\begin{equation}\label{eq:functionFit}
\eta = 2.88 { \, \,} e^{-0.27w/R_i} +2.00.
\end{equation}  
The values of $\eta$ in the present work are obtained from simulations where $E_{\infty}$ and $w/R_i$ are simultaneously changed.
Moreover, such a procedure further illuminates the effect of the line width.

\subsection{Effect of conductor line width, $w$}
\label{subsection:conductorWidth}

\begin{figure}[hbt!]
\begin{center}
\includegraphics[scale=1.0]{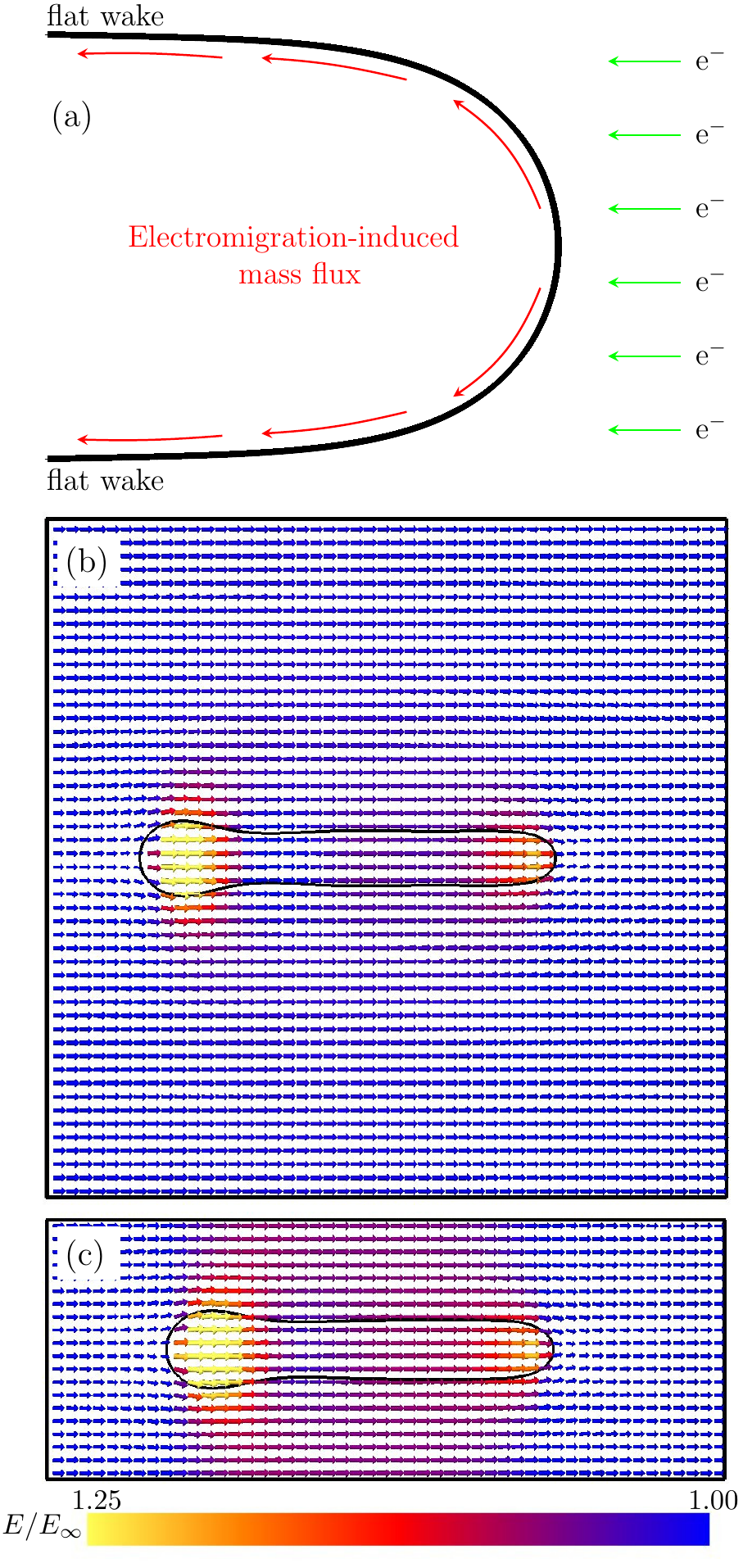}
\caption[Schematic diagram and simulation results with electric field vectors]{Schematic diagram of the interaction between the slit surface and the electron wind in (a), normalized electric field vectors are superimposed on the steady-state slit profile for $w/R_i=8$, $\eta=2.55$ in (b) and $w/R_i=3$, $\eta=2.95$ in (c) under the dimensionless parameter $\chi_0=20.60$.}
\label{fig:MassFluxWidthChange}
\end{center}
\end{figure}

The most interesting result of the present study is the dependence of line width on the final slit characteristics. Further, the line width has a significant impact on the inclusion shape and, as a consequence, the performance of the interconnect material. Xia et al.~\cite{xia1997a} computed critical values of $\chi_0$, at which a circular inclusion collapses, as a function of line width analytically. Kraft~et~al~\cite{kraft1995shape} reported shape changes of inclusions with line width experimentally. In the present study, the dependence of line width on the slit characteristics can be justified by the analysis of the electric field. On one hand, at elevated electric field collisions of electrons on the inclusion surface may overcome the surface energy and consequently result in the occurrence of shape changes in Figure \ref{fig:SchameticVoidEvolution}(b). On the other hand, electrons flow at lower electric field colliding on the inclusion may not overcome the surface energy and, therefore, circular inclusions maintain their shape as seen in Figure \ref{fig:SchameticVoidEvolution}(a). Furthermore, decreasing the line width of the interconnect motivates enhanced interaction between the electric field vectors and the inclusion surface locally, as depicted in Figure\ref{fig:MassFluxWidthChange}(b) and (c). Therefore, the enhanced electric field in the vicinity of the inclusion surface, in Figure \ref{fig:MassFluxWidthChange}(a), provokes electromigration-induced mass flux from the inclusion tip to the inclusion wake. As a consequence, the slit width, and accordingly $\eta$, increases, while the velocity decreases, for the shorter line width.

\begin{figure}[hbt!]
\begin{center}
  \includegraphics[scale=0.9]{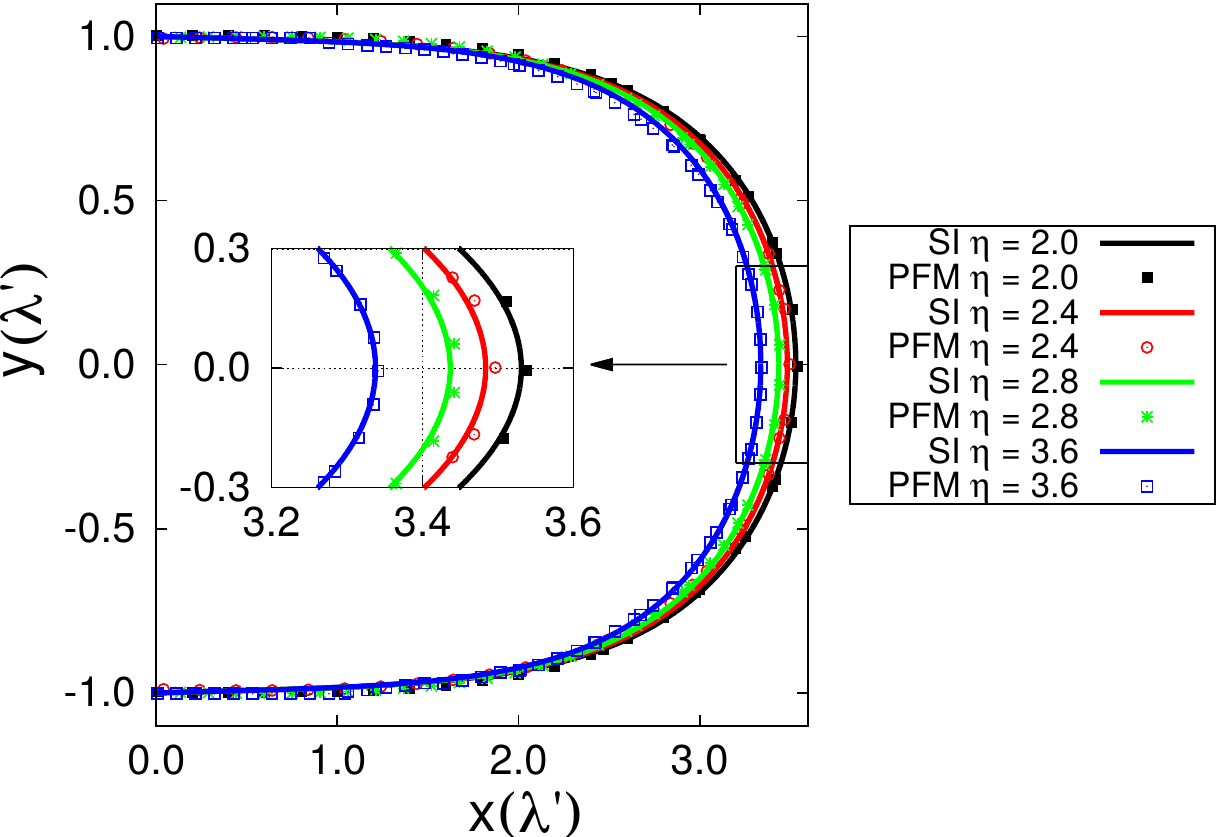}
\end{center}
\caption[Comparison of the front part of the slit profile from the phase-field model with the sharp-interface analysis]{A comparison of the front part of the slit profile from the phase-field model with the sharp-interface analysis for different values of $\eta$. All dimensions obtained from the phase-field model are normalized by the slit width $u$.} 
\label{fig:slitProfileZoomIn}     
\end{figure}
A comparison of the slit profiles at the tip, obtained from the numerical solution of Eq.~(\ref{eq:SIMainEquation}), and from the the phase-field simulations for different $\eta$ are presented in Figure~\ref{fig:slitProfileZoomIn}. 
The results reveal a good agreement with the sharp-interface analysis across all values of $\eta$.
 Another point to be noted is that the asymmetry in the profile, relative to the circular inclusion, decreases with an increasing $\eta$.

\subsection{Selection of slit width and velocity}
\label{subsection:EM1SelectionSlitWidthVelocity}
\begin{figure}[hbt!]
\begin{center}
  \includegraphics[scale=1]{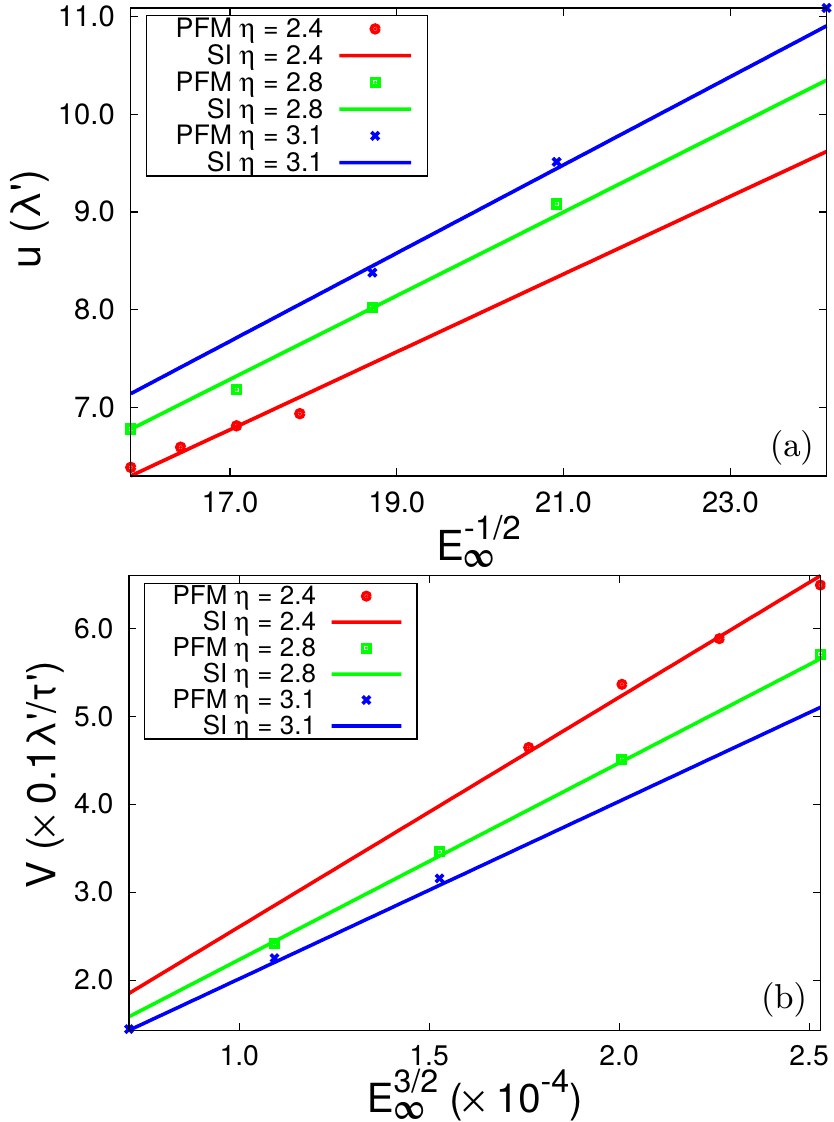}
\end{center}
\caption[Comparison of the phase-field results with the sharp-interface analysis for the half slit width and the velocity of the slit as a function of the external electric field]{A comparison of the phase-field simulation results with the sharp-interface analysis for (a) the half slit width and (b) the velocity of the slit, as a function of the external electric field, for different values of $\eta$. The plots suggest the power-law dependence $u \propto E_{\infty}^{-1/2}$ and $V  \propto E_{\infty}^{3/2}$. The half slit width and the velocity are calculated in the steady-state part of the motion.}
\label{fig:width_velocity}     
\end{figure}
Next, the dependence of the electric field strength on the selection of the slit width and the velocity is addressed.
 The slit width, which corresponds to different values of $\eta$ and the dependency on the field strength, is plotted in Figure~\ref{fig:width_velocity}(a).
  An excellent agreement between the sharp interface relation Eq.~(\ref{eq:SISlitWidthFinal}) and the phase-field simulations is observed.
   The slit width scales as $E_{\infty}^{-1/2}$, implying a narrower width at higher field strengths.
\begin{figure}[hbt!]
\begin{center}
\includegraphics[scale=0.8]{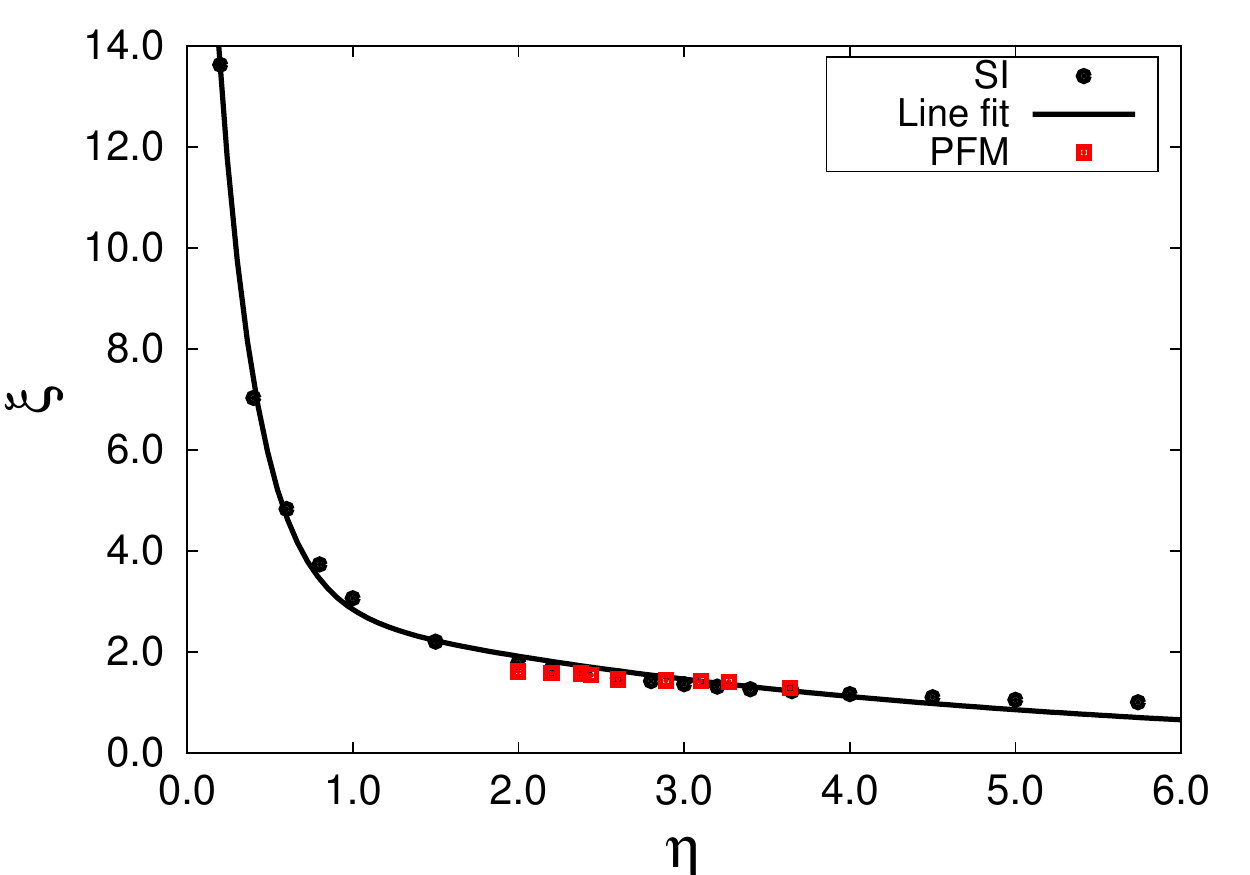}
\end{center}
\caption[Dependence of $\xi$ on $\eta$ obtained from sharp interface, and the phase-field model]{The dependence of $\xi$ on $\eta$ obtained from sharp interface, and the phase-field model. The solid line represents exponential fit to the sharp-interface data.}
\label{fig:etaVsEPS}    
\end{figure}

The slit velocity is measured after it attains the equilibrium.
 Note that in the steady-state, every point on the slit moves at a constant velocity $V$. 
  In Figure~\ref{fig:width_velocity}(b), the velocity is plotted as a function of the electric field strength, for different values of $\eta$. 
 The phase-field results show good agreement with the sharp-interface relation in Eq.~(\ref{eq:SISlitVelocityFinal}).
  The slit velocity scales as $E_{\infty}^{3/2}$, which indicates that the slit migrates faster at higher electric field strengths.

Yao et. al. \cite{yao2015an} reported that the inclusion propagation velocity changes simultaneously if the inclusion collapses to a finger-like slit.
 They introduced a parameter $\xi$, as the ratio of the velocities of a finger-like slit to the circular inclusion of radius equal to half slit width. 
 The value $\xi>1$ implies that the slit moves faster after collapsing from a circular inclusion, whereas the value $0<\xi<1$ denotes that the inclusion moves more slowly.
  The phase-field results suggest a strong alliance between the parameters $\eta$ and $\xi$, which is nothing but the relationship between the slit width and the velocity.
   For each simulation result, the parameter $\xi$ can be associated with a unique value of $\eta$.
    The relation between the two parameters $\eta$ and $\xi$, as displayed in Figure~\ref{fig:etaVsEPS} is fitted with

\begin{equation}
\xi = 24.49 e^{-4.26\eta}+3.26e^{-0.26\eta}.
\end{equation}
The slit velocity changes slowly for $\eta>2$, while it increases rapidly for the lower values.
 The parameter $\xi$ approaches the value of unity, at $\eta=5.74$.
This implies that at $\eta=5.74$ the finger-like slit and the corresponding circular inclusion of the radius equal to half slit width, drift with the same velocity. 
  Furthermore, Figure~\ref{fig:etaVsEPS} reveals that the narrower slit propagates faster.

\begin{table}[hbt!]
\caption[Comparison of the values of the void size and the velocity, for eutectic SnAgCu solder bumps obtained from phase-field, and experiments]{A comparison of the values of the void size and the velocity, for eutectic SnAgCu solder bumps obtained from phase-field, and experiments. Only the experimental results from Zhang et al. \cite{zhang2006effect} are considered for the comparison with presented phase-field and sharp-interface methods.}
\label{tab:2}      
\begin{center}
\begin{tabular}{cccc}
\hline\noalign{\smallskip}
& PFM & Experiment \cite{zhang2006effect}  \\
\noalign{\smallskip}\hline\noalign{\smallskip}
void size ($2u$) & $0.80-2.49$ $\mu$m & 2.44 $\mu$m\\
velocity ($V$)  & $1.38-4.93$ $\mu$m/h & 4.40 $\mu$m/h\\
\noalign{\smallskip}\hline
\end{tabular}
\end{center}
\end{table}

\section{Discussion and experimental correlation}
\label{sec:EM1Discussion}

The movement of voids, as in cavities, is one of the applications of species diffusion due to electromigration in the interconnects. The evidence is reported that the metallic lines often observe a dramatic change in the resistance and even the performance of the conductors is being affected due to sudden alteration of rounded voids to complex shapes \cite{lin2017electromigration}. Although the results presented in this chapter examined the case of the transgranular voids migrating along the metal line, which previously was considered uncritical in terms of failure \cite{shingubara1991electromigration, yang1994cavity}, the present study is applicable in the flip-chip solder bumps. 
The nucleated voids at the current crowding zone propagate in the direction of the electron wind, along the interface between the solder and the intermetallic compound. 

In order to make a comparison between the magnitude of the void size and the velocity observed in the solder bumps and in the phase-field model, the experimental data at the current density $1.1 \times 10^{8}$ A/m$^2$ from \cite{yao2015an} as shown in Table~\ref{tab:2} are utilized. 
Since the descriptions of the phase-field, in Chapter~\ref{chapter:phaseFieldModelElectromigration} correspond to a single-component system, the material parameters for Sn are assumed to be the dominant diffusing species.
 Furthermore, it is important to emphasize that the effective valence $Z_s$ of Sn is reported to be 17.
 However, to compare this with the simulation, the results are obtained at $Z_s=5$ in this chapter. 
 This should not alter the order of magnitude of the void size and the velocity, as these should only differ by a factor 0.54 and 6, as evident from Eqs.~(\ref{eq:SISlitWidthFinal})~and~(\ref{eq:SISlitVelocityFinal}), respectively. 
 According to this argument, the phase-field results are scaled by these factors, in Table~\ref{tab:2}, where the comparisons between the phase field and the experimental observation are presented. 
 In the solution of the void size and the velocity, the parity in the order of magnitude can be appreciated. 
 Secondly, even though a constant electric field for comparison is considered, the effect of a variable field cycle on the selection of the width and the velocity has been highlighted in Figure~\ref{fig:potentialChange}, which has physical significance.

Although the present study considered a propagation of the transgranular voids, the scaling laws can be extrapolated to the voids migrating in the transverse direction to the electric field. Since the void velocity scales as $E_{\infty}^{3/2}$ (from Eq.~(\ref{eq:SISlitVelocityFinal})), the same law is valid for the latter case on the dimensional ground. As the velocity is inversely proportional to the failure time of the interconnect material. Therefore, the exponent in the Black's law is expected to be n~$=3/2$ for a slit propagating in a steady-state.

The present work does not take into account the presence of a passivated dielectric interface. However, void nucleation and migration along the passivated interface has been observed frequently \cite{zaporozhets2005three, choi2008effects}. Since the motion of the voids is governed by the flux divergence, the diffusivity along the void surface in addition to that of the passivated interface dictates the void velocity. However, in most cases, the void surface diffusivity is much greater than the interface diffusivity of the passivated layer due to which the result presented here is expected to agree considerably in the presence of the dielectric interface. 

In addition to the technological implications in the efficient design of interconnects, the results of the present study can be exploited in the fabrication of wires with a high aspect ratio, ranging from the nano- to microscale. 
Since the width of the slit scales as $E_{\infty}^{-1/2}$, the features of the desired dimension can be achieved by tuning the inclusion radius and $E_{\infty}$. 
The pattern formation, which is induced by the electric field, already enjoys a great deal of success in modulating the morphology in block copolymers \cite{mukherjee2016influence, mukherjee2016electric}, metal conductors  \cite{du2004electrostatic, gill2008electric}, fluids \cite{morariu2003hierarchical}, and more recently in an electromigration-induced flow in liquid metals \cite{dutta2009electric}.

\section{Conclusion}
\label{sec:IsotropicInclusionsConclusions}
Sharp-interface models, both analytical as well as numerical approaches, have been employed extensively in the past to simulate electromigration-induced inclusion migration \cite{gungor1999theoretical, cho2007theoretical, cho2006current, kraft1993observation, ho1970motion, yang1994cavity, suo1994electromigration, xia1997a, gungor1998electromigration}.
 Analytical methods (section~\ref{section:LinearStabilityAnalysisCircularIsland} and \ref{section:linearStabilityAnalysis}) provide information with regard to the onset of the shape bifurcation of circular inclusions.
  Subsequent changes in the shape, however, are inexplicable. 
 In the present context of slit propagation, the derivation in Section~\ref{section:sharpInterfaceAnalysisSlitProfile} assumes a slit shape and quantifies information regarding the selection of the steady-state width and velocity for a given electric field.
   It does not reveal the dependence of the initial inclusion radius to the line width on the selection of final slit width (and hence velocity). 
   The presented diffuse-interface approach, on the other hand,  allows one to track the entire temporal events elegantly, leading from circular inclusion to slit transition without requiring to track the interface explicitly. 
   Thus, in addition to the electric field, the effect of the initial inclusion to the line width ratio is elucidated through the phase-field simulations, as shown in Figure~\ref{fig:etaVsWbyU}, which otherwise is obscure in the sharp-interface analysis.

Numerical sharp-interface methods have experienced a great deal of success in simulating inclusion migration \cite{wang1996a}, inclusion to slit transition, and concurrent faceting \cite{gungor1999theoretical} and void coalescence \cite{wang2017surface, dasgupta2013surface}, amongst others. 
The studies hitherto have been limited to two dimensions, owing to the onerous task of interface tracking.
 In addition, the electric field on the inclusion/slit surface is approximated by the local tangential component $-(2\beta/(1+\beta))E_{\infty}\textrm{ sin } \theta$. 
This limitation of sharp-interface theories is eluded in the presented results by solving the Laplace equation to obtain the electric field distribution at every spatial coordinate. 
  Moreover, the effect of the current crowding at sharp corners and bends, if any, is inherently captured (section~\ref{subsection:currentCrowding}). Besides, the effect due to distinct conductivities in the matrix and the inclusion is incorporated (section~\ref{subsection:significanceOfChi0}). Furthermore, the model presented here can be readily extended to three dimensions, without any further complexity, albeit with an increased computational cost.
  The ability of the phase-field method to predict damage in polycrystalline interconnect in three dimensions has been exemplified recently \cite{mukherjee2018electromigration}.

Insights of phase-field simulations on the morphological evolution of initially circular inclusions are considered in the present chapter. 
A morphological map is constructed in the plan of $\beta$ and $\chi_0$, as shown in Figure~\ref{fig:EMLSA}.
Circular inclusions and finger-like slits are observed for the isotropic inclusions. 
To this end, the effect of diffusional anisotropy, between the inclusion and the matrix, can further be utilized towards efficient guidance of the morphological patterns \cite{kumar2016current}. Therefore, anisotropic inclusions are considered in the next chapters. Particularly,  anisotropic inclusions of two-fold symmetry are investigated in Chapter~\ref{chapter:EM2}, followed by fourfold and sixfold symmetries in Chapter~\ref{chapter:4Fold6FoldEM}. 

\afterpage{\blankpagewithoutnumberskip}
\clearpage
\chapter{Motion of  anisotropic inclusions consist of 2-fold symmetry}
\label{chapter:EM2}

\section{Introduction}


The phase-field literature focused on the development of inclusion morphology is limited \cite{santoki2019phase, hauser2010the, banas2009phase, li2012the, bhate2000diffuse, mahadevan1999simulations}. The majority of the works are established based on finger-like slit formation \cite{santoki2019phase, hauser2010the, banas2009phase}, wedge-shaped inclusion \cite{li2012the}, crack-type feature \cite{li2012the, bhate2000diffuse, mahadevan1999simulations}, and/or cardioid-type morphology \cite{bhate2000diffuse, banas2009phase}.  However, the literature is devoid of a variety of inclusion morphology. This limitation is likely due to assumptions of isotropic diffusivity. More importantly, the morphological influence by differing the conductivities for inclusions and matrices is absent. In addition, the inclusions have been known to undergo complex topological transitions such as splitting \cite{schimschak1998electromigration}, coalescence \cite{cho2006current} and coarsening \cite{ma1993precipitate}. Faceting is another common feature that has been observed experimentally \cite{arzt1994electromigration, arzt1994facet}, which can not be explained by diffusional isotropy. 

These observations can be explained by the consideration of either surface energy anisotropy or adatom mobility anisotropy. Most interconnect materials are face-centered-cubic (FCC) metals with a strong crystallographic direction-dependent surface energy \cite{sun1997a} and adatom mobility \cite{kraft1997electromigration}. In the present work, diffusional anisotropy and isotropic surface energy are considered, due to the fact that the intensity of anisotropy in adatom mobility for the facet development during electromigration-induced surface diffusion is much higher than that of the surface energy \cite{giesen2004step, kuhn2005complex}. Moreover, the faceted inclusions are known to revert back to circular shapes when the electric field is switched off, further highlighting the dominance of anisotropy in species diffusion over surface energy \cite{arzt1994facet}.

\subsection{Parameter space for inclusion morphologies}

The common finding of the previous studies is that the morphological evolution of the inclusion can be characterized by six dimensionless parameters. 

\begin{enumerate}
\item $\chi_0 = eZ_sE_{\infty}R_i^2/\Omega \gamma_s$, denotes the ratio of surface electromigration force to capillary force. Lower values of $\chi_0$ promote rounded shape while higher values distort the shape into the slit. The effect of this parameter is discussed in chapter~\ref{chapter:EM1}.

\item $\Lambda_L = w/R_i$ defines the ratio of line width $w$ to the initial inclusion radius $R_i$. Larger $\Lambda_L$ leads to finer and faster propagating slit and vice versa \cite{santoki2019phase}. The significance of this parameter is addressed in subsection~\ref{subsection:conductorWidth}.

\item $A$ characterizes the strength of anisotropy in species diffusion.

\item $m$ denotes the grain symmetry parameter.

\item $\varpi$ related to misorientation of the fast surface diffusion direction with respect to a perpendicular to the external electric field, $E_{\infty}$.

\item Finally, $\beta = \sigma_{\textrm{mat}}/\sigma_{\textrm{icl}}$ depicts the ratio of conductivity of the matrix $(\sigma_{\textrm{mat}})$ to that of the inclusion $(\sigma_{\textrm{icl}})$.
\end{enumerate} 

A complete understanding of the inclusion dynamics requires an exploration of the complete six-dimensional parameter space of $\chi_0$, $\Lambda_L$, $A$, $m$, $\varpi$, and $\beta$. Such an exercise is certainly computationally expensive and does not help in unraveling the effect of individual parameters. A considerable simplification of the problem is then to explore the two or three-dimensional parametric space. Kuhn et al. \cite{kuhn2005complex} studied the effect of $\chi_0$ and $A$ on inclusion dynamics on the substrate using a continuum-based numerical approach. They proposed a morphological map bifurcating the regions of steady-state, oscillatory-state, zigzag motion, and breakup amongst others. Dasgupta et al. \cite{dasgupta2013surface, dasgupta2018analysis} further extended the work to explore $\chi_0$, $m$, and $\varpi$ space. Maroudas and coworkers \cite{gungor1999theoretical, gungor1998electromigration, gungor1998non} have performed a series of systematic study using front tracking dynamical simulations to unravel the dynamics of edge inclusions in $\chi_0$, $\Lambda_L$\nomenclature{$\Lambda_L$}{ratio of line width $w$ to the initial inclusion radius $R_i$}, $A$, $m$, and $\varpi$ five-dimensional space. They highlighted the stability of wedge-shaped inclusions \cite{maroudas1995dynamics}, facet-selection mechanisms \cite{gungor1999theoretical} and propagation of soliton-like feature \cite{gungor2000current, koronaki2007current} on the inclusion surface.

\subsection{Conductivity ratio, $\beta$}
While the above-mentioned parameters have received much attention until now, the study on the role of $\beta$ is relatively scarce. The conductivity contrast, $\beta$ can significantly influence inclusion dynamics and morphology. For instance, careful consideration of the conductivity contrast between the inclusion and matrix is important to ascertain the contribution of bulk and surface electromigration using marker motion technique \cite{ho1970motion}. In interconnect lines, conductive species might be trapped inside the inclusions leading to a finite conductivity \cite{wang1996simulation,hao1998linear}. Therefore, understanding the importance of $\beta$ is of scientific interest. Hao and coworkers \cite{wang1996simulation, hao1998linear} studied the stability of a circular inclusion analytically in $\chi_0$-$\beta$ space using linear stability analysis. In addition, Li and Chen \cite{li2008electromigration} studied the effect of conductivity on the electromigration-driven motion of an elliptical inclusion analytically. Furthermore, Li et al. \cite{li2012the} focused on the role of $\chi_0$-$\beta$ on elliptical and wedge-shaped inclusions using the phase-field method. However, these studies have one or more of the following limitations: Although the linear stability analysis sheds light on the stability of the circular inclusion, it does not provide further information about the topological transition. Moreover, sharp-interface based front tracking methods usually approximate the electric field by a local surface projection instead of solving the Laplace equation thereby neglecting the effect of current crowding. Furthermore, previous phase-field studies have only considered the effect of inclusion size \cite{liang2015morphological, liang2017numerical}. The only study \cite{li2012the} which considered the effect of conductivity contrast is applicable to edge inclusions migrating on $\{111\}$ planes of FCC metals. 

The aim of the present chapter is thus to systematically investigate the unexplored topological changes in $\beta$-$\varpi$ space. In addition, the emphasis is laid on inclusion dynamics on $\{110\}$ textured single crystal. The phase-field model developed in chapter~\ref{chapter:phaseFieldModelElectromigration} illustrates the employed numerical method in the form of the phase-field model. Few typical cases of morphological evolution of the inclusion highlighting the underlying mechanism are discussed in section \ref{sec:TwoFoldEMResults}. The presented results suggest that while the shape of the inclusion strongly depends on the conductivity contrast, the stability is dictated by the misorientation. The chapter is summarized by a brief discussion on the applicability of the presented results in section~\ref{sec:twoFoldEMDiscussion}, followed by discussion in section~\ref{sec:twoFoldEMConclusion}. Some parts of this chapter are published in the \textit{Journal of Applied Physics} \cite{santoki2019role}.

\section{Results}
 \label{sec:TwoFoldEMResults}

\begin{figure}[h]
\centering
\includegraphics[scale=0.40]{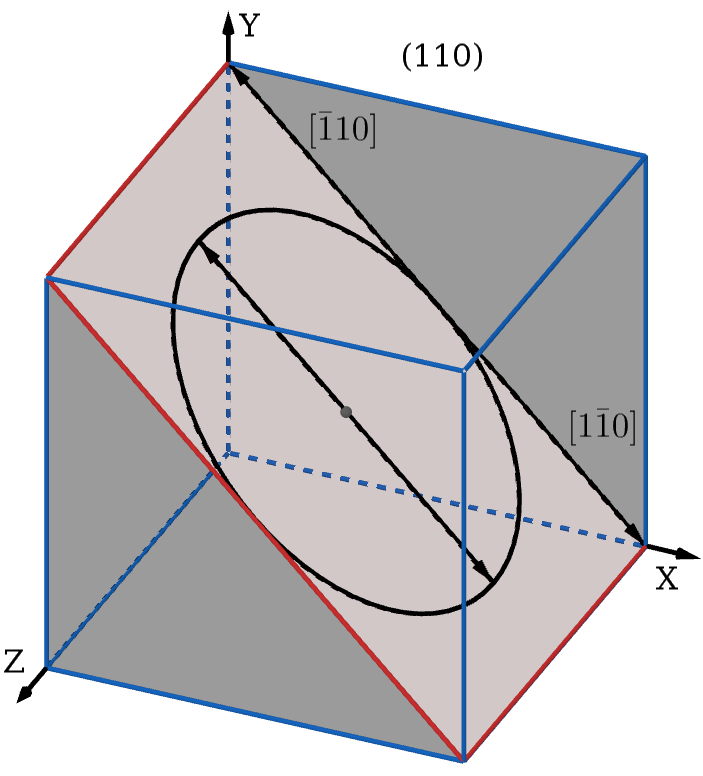}
\caption[Schematic of an inclusion in (110) crystallographic plane]{ Schematic of an inclusion in (110) crystallographic plane. The directions of dominant atomic diffusivity are represented by black arrows. The red color plane is the surface of the film perpendicular to the circular inclusion. The double-headed arrows define positions of maximum surface diffusivity.}\label{fig:EM1twoFoldSymmetry}
\end{figure}


The twofold symmetry in the inclusion is a result of the two fast diffusion locations existing in the plane. For instance, in FCC crystals, the symmetry axes corresponding to $\langle110\rangle$ crystallographic directions have higher surface diffusivity compared to $\langle111\rangle$ and $\langle100\rangle$ directions \cite{kraft1997electromigration}. Therefore, the number of $\langle110\rangle$ axes in a plane decides the number of fast diffusion axes in the inclusion. As shown in Figure~\ref{fig:EM1twoFoldSymmetry}, (110) plane contains two crystallographic directions $[ \bar{1}10]$ and $[1\bar{1}0]$, which are responsible for the fast diffusion. Similarly, each plane in $\{110\}$ family ($m=1$) contains two of $\langle110\rangle$ directions. Note that the fast diffusion sites are always opposite to each other in the same axis for all possible planes of twofold symmetry.

 A systematic study of the dynamics of a circular inclusion as a function of $\beta$ and $\varpi$ is performed. The parameters $\chi_0$, $\Lambda_L$, $A$, and $m$ are held constant, and the numerical values utilized in this chapter are summarized in Table \ref{tab:simulationParametersTwoFold}. Due to the symmetry in the shape of the preexisting inclusion, the parameter space for $\varpi$ can be reduced to a range of $0^{\circ}-90^{\circ}(=180^{\circ}/2m)$, without compromising the features of the inclusion morphology. Few typical migration modes are discussed next.
\begin{table}[h]
\begin{center}
\caption[Values of parameters considered for the simulation study for an inclusion in \{110\} crystallographic plane]{Values of parameters considered for the simulation study for an inclusion in \{110\} crystallographic plane.}\label{tab:simulationParametersTwoFold}
\begin{tabular}{ c c }
\hline
Parameter&Value \\
\hline
$\chi_0$&15\\
$\Lambda_L$&6\\
$A$&10\\
$m$&1\\
$\varpi$&0$^\circ$ to 90$^\circ$\\
$\beta$&1 to 10000\\
\hline
\end{tabular}
\end{center}
\end{table}

\subsection{Steady-State Inclusion Dynamics}
\begin{figure}[h]
\begin{center}
\includegraphics[scale=1.2]{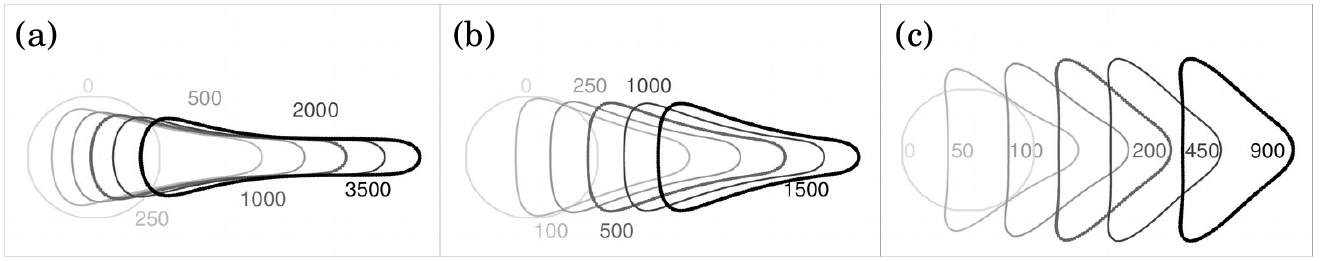}
\end{center}
\caption[Inclusion morphology for conductivity ratio, $\beta=1$, $3$, and $10000$ at the misorientation angle, $\varpi=90^\circ$]{Inclusion morphology for conductivity ratio, $\beta=1$ in (a), $3$ in (b) and $10000$ in (c) at the misorientation angle, $\varpi=90^\circ$. The inclusions with surface contours are presented with time, t($\tau'$). The increasing darkness of the contour indicates inclusion evolution. The inclusions are displaced in space for better visual inspection. The inclusion for $\beta=1$ shows a nanowire morphology, while wedge shape is observed for $10000$.}\label{fig:r0}
\end{figure}
When the electric field aligns along the fast diffusion direction (i.e., $\varpi=90^\circ$), the species at the inclusion front are displaced faster leading to the formation of a protrusion as shown in Figure~\ref{fig:r0}(a). With time, the protrusion progressively assumes a narrow slit-like shape, which eventually migrates with an invariant shape with a high aspect ratio elongated in the direction of the electric field. The slit formation can be attributed to the high value of the parameter $\chi_0$ due to which the capillary force is unable to compete with the surface electromigration force.

Higher $\beta$ leads to the slit with a lower aspect ratio in Figure~\ref{fig:r0}(b). On increasing $\beta$, the tangential component of the electric field along the inclusion surface becomes more prominent due to current crowding, as shown in Figure~\ref{fig:EMcurrentCrowding}. Hence, the species transported relatively faster from the front end accumulate at the diametrically opposite ends, which are perpendicular to the applied field corresponding to the slowest species mobility regions. As a result, the rear end becomes progressively thicker and flatter. On further increasing of $\beta = 10000$, the inclusion assumes a faceted triangular shape as shown in Figure~\ref{fig:r0}(c). While the rear edge is perpendicular, the other two are tilted with respect to the applied field. Once the wedge shape develops, the inclusion migrates along the line without further change in shape. The last darkest contours of the inclusion profiles in Figure~\ref{fig:r0} are corresponding to steady-state shapes.

\begin{figure}[t]
\centering
\includegraphics[scale=1.2]{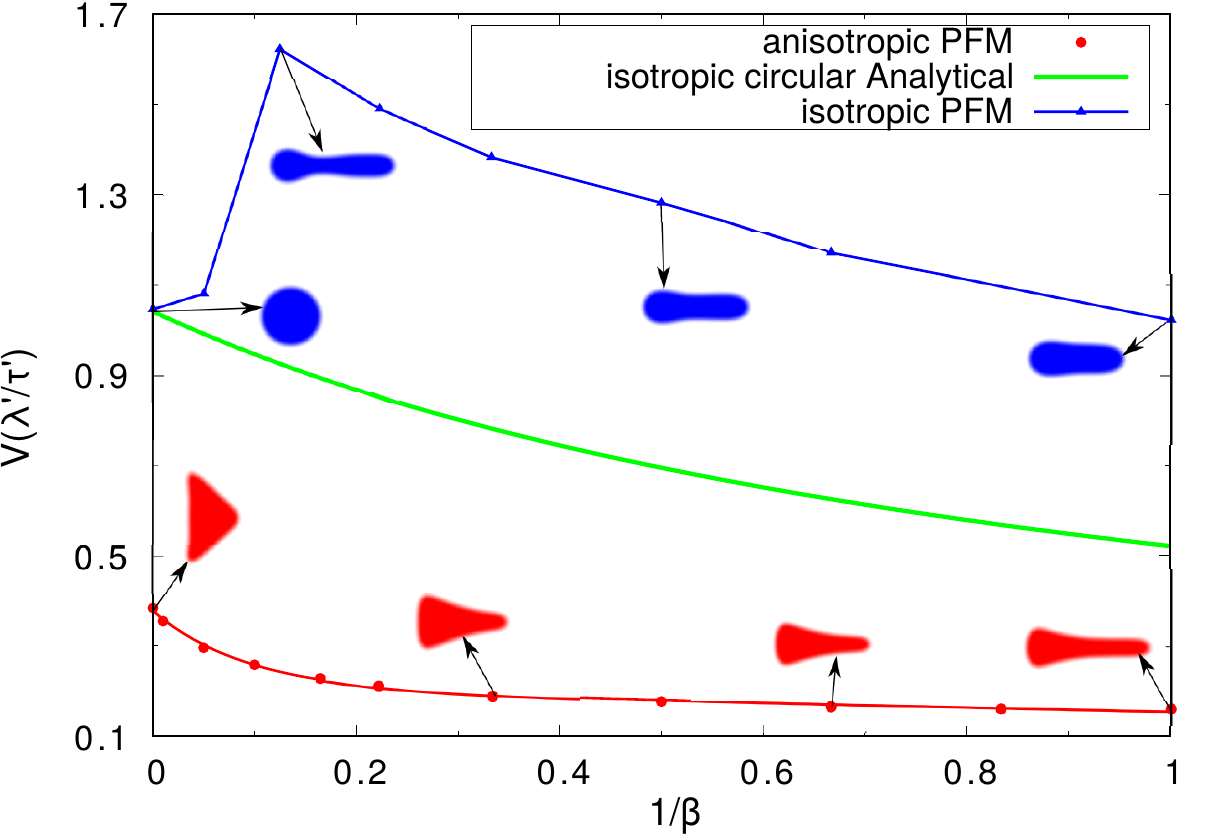}
\caption[Velocities of the centroid of steady-state inclusions for two-fold anisotropy ($m=1$), $\varpi=90^\circ$ and for isotropic ($m=0$) surface diffusion (numerical and analytical) as a function of the conductivity ratio $\beta$.]{The velocities of the centroid of steady-state inclusions for two-fold anisotropy ($m=1$), $\varpi=90^\circ$ (in red graph) and for isotropic ($m=0$) surface diffusion (in blue graph) as a function of the conductivity ratio $\beta$. The green curve corresponds to a steady-state velocity of the circular inclusion of isotropic surface diffusivity, derived from Eq.~\eqref{eq:voidVelocity} considering the diffusion coefficient $D_{s}$. The inset of images show inclusion morphology in steady state. $\beta$ governs the inclusion shapes and the steady-state velocities.}\label{r0BetaVelocityGraph}
\end{figure}

The steady state velocity of the centroid of the inclusion obtained from phase-field simulations, after it attains an invariant shape for different values of $\beta$ is shown in Figure~\ref{r0BetaVelocityGraph} (red curve). For comparison, the velocity of a shape-preserving circular inclusion obtained from Ho's analytical expression similar to Eq.~\eqref{eq:voidVelocity} (green curve) and that migrating under isotropic diffusion (blue curve) obtained from the phase-field simulations ($m=0$ in Eq.~\eqref{eq:anisotropy}) are also plotted. The velocity is dependent on both, the shape and the conductivity contrast. This can be understood by examining the tangential component of the electric field in the sharp-interface limit which can be expressed by a projection on the surface of inclusion as,
\begin{equation} \label{eq:2foldElectricFIeld}
E_t=E_\infty g^{\theta \beta}(\theta, \beta),
\end{equation}
where $g^{\theta \beta}(\theta, \beta)$\nomenclature{$g^{\theta \beta}$}{function of shape and conductivity contrast} denotes a function of shape ($\theta$) and conductivity contrast ($\beta$). The surface mass flux expressed for isotropic inclusions in Eq.~\eqref{eq:EM1SharpInterfaceFlux} is modified for the anisotropic inclusion of the form,
\begin{equation}
J_s = \frac{D_s f^{\theta}(\theta) \delta_s}{\Omega k_B T} \Big( -eZ_sE_t + \Omega \gamma_s \frac{\textrm{d} \kappa_s}{\textrm{d} s} \Big), \label{eq:2foldSharpInterfaceFlux}
\end{equation}
 The electric field projection Eq.~\eqref{eq:2foldElectricFIeld} is substituted in the flux equation~\eqref{eq:2foldSharpInterfaceFlux} and subsequently into the mass conservation equation~\eqref{eq:EM1MassConservationNormalVelocity}, to obtain normal velocity of the form,
\begin{equation}\label{velocityexpanded}
 V_n =  \frac{D_s \delta_s}{k_B T} \frac{\text{d}}{\text{d}s} \Bigg[ f^{\theta}(\theta) \Bigg\{ \Omega \gamma_s \frac{\text{d}{\kappa_s}}{\text{d}s} - q_sE_{\infty}g^{\theta \beta}(\theta,\beta)\Bigg\} \Bigg ],
\end{equation}
The normal velocity in the phase-field governing Eqs. \eqref{eq:EMCahnHilliardModified} and \eqref{eq:EMLaplaceeq} reverts to the above expression as the interface thickness approaches zero, has already been proved in Refs. \cite{bhate2000diffuse, mahadevan1999phase, mukherjee2019electric} via formal asymptotic analysis. The first term in the curly parentheses represents the contribution due to capillarity or curvature-gradient, while the second term denotes the effect of electromigration. The electric field along the surface Eq.~\eqref{eq:2foldElectricFIeld} is a function of both, the shape (given by $\theta$) and the conductivity ratio ($\beta$). In general, the function $g^{\theta \beta}(\theta,\beta)$ cannot be written down as an analytically closed form solution for any arbitrary inclusion geometry. For a shape-preserving circular inclusion ($\text{d}\kappa_s/\text{d}{s} = 0$ and $g^{\theta \beta}(\theta, \beta) = -2\beta \cos\theta/(1+\beta)$) migrating due to isotropic diffusion ($m=0$ and $f^{\theta}(\theta) = 1$) Eq.~(\ref{velocityexpanded}) simplifies to Eq.~\eqref{eq:linearStabilityinclusionVelocity}, which is the same as the solution of Ho \cite{ho1970motion} and is plotted in Figure~\ref{r0BetaVelocityGraph} (green color). The non-linear decrease in the velocity with $1/\beta$ is evident from the above expression. For the same magnitude of applied electric field and initial radius of the inclusion, a shape-preserving circular inclusion due to isotropic diffusion (green curve) will migrate faster than a triangular or slit-shaped inclusion induced as a result of two-fold anisotropic diffusion (red curve). Although a similar non-linear decrease with $1/\beta$ is observed due to anisotropic diffusivity (red curve), the velocity is lowered roughly by a factor of three. 

In absence of diffusional anisotropy (blue curve), however, a slit has a faster propagation velocity than a circular one (green curve) indicating the amplification of the factor $g^{\theta \beta}(\theta, \beta)$. At the lowest $1/\beta$ (i.e., $\beta=10000$) which leads to a shape-preserving circular inclusion, the velocity obtained from phase-field simulation is in accordance with Eq.~\eqref{eq:linearStabilityinclusionVelocity}. It is to be noted that intermediate values of $1/\beta$ ($\approx 0.12-0.20$) lead to finer slits which propagate faster than the wider slits originating at higher $1/\beta$. In addition, on comparison of the red and the blue curve, it appears that the effect of anisotropy $f^{\theta}(\theta)$ is to further slow down the velocity of inclusion migration.

\subsection{Effect of misorientation}

\begin{figure}[h]
\centering
\includegraphics[scale=1.6]{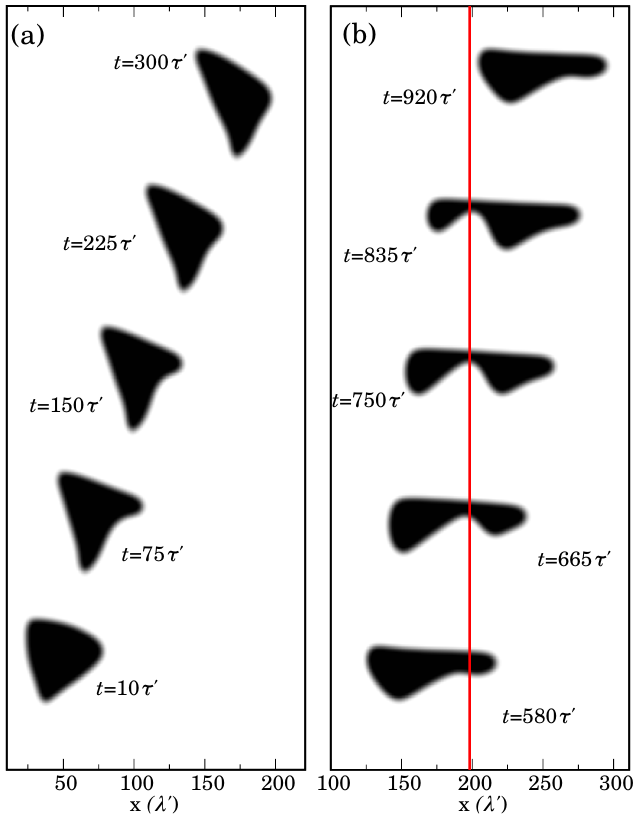}
\caption[Complex shape dynamics of inclusions at misorientation angle $\varpi = 75^{\circ}$ for the conductivity ratio $\beta=10000$ and $1$]{The complex shape dynamics of inclusions at misorientation angle $\varpi = 75^{\circ}$ during the morphological evolution for the conductivity ratio $\beta=10000$ in (a) and $1$ in (b). The snapshots of the inclusions are shifted upwards in time. The solid red line represents a position of the valley during evolution. The inclusion for $\beta=1$ shows time-periodic oscillations with rounded hills and valleys. While a steady-state inclusion morphology is observed for $\beta=10000$.}\label{fig:r15}
\end{figure}

With the incorporation of misalignment of the fast diffusion sites to the electric field, $\varpi = 75^{\circ}$ (keeping $\beta = 10000$ fixed), similar to the previous case, a triangular shape ensues with a slightly tilted rear edge with respect to the direction of the applied field at $t = 10\tau^{\prime}$ (Figure~\ref{fig:r15}(a)). The upper edge develops a boomerang appearance at $t = 75\tau^{\prime}$, with the upper vertex growing perpendicular to the field direction. The upper edge of the advancing inclusion reverts to the flat surface at $t = 300\tau^{\prime}$ and, consequently, the inclusion propagates preserving the shape.

\begin{figure}[hbt!]
\centering
\includegraphics[scale=1.2]{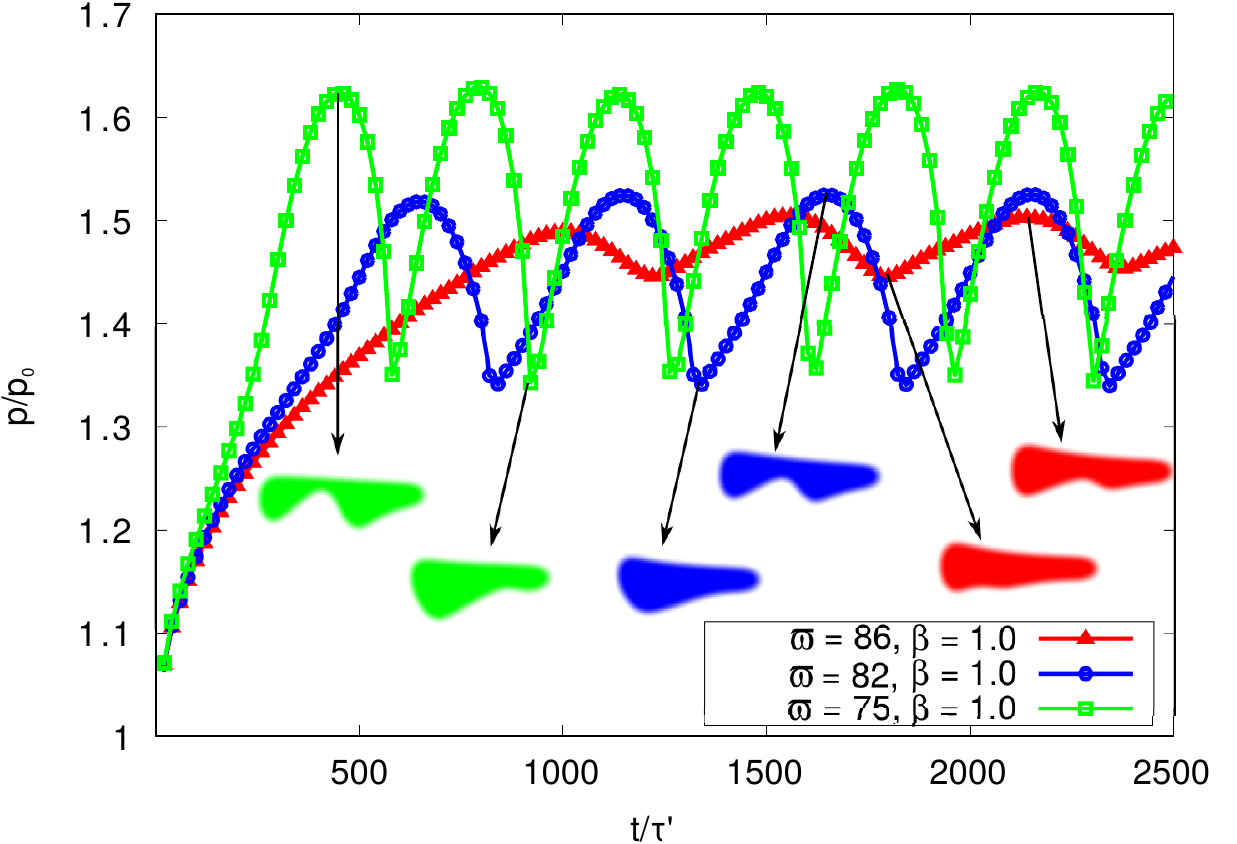}
\caption[Evolution of normalized perimeter of inclusions with misorientation angles $\varpi =86^{\circ}$, $82^{\circ}$, and $75^\circ$ for $\beta=1$]{The evolution of normalized perimeter of inclusions with misorientation angles $\varpi =86^{\circ}$, $82^{\circ}$, and $75^\circ$ for $\beta=1$. The inclusions undergo time-periodic oscillations. The period of oscillations decreases  while the amplitude increases with a decrease in $\varpi$. The normalized inclusion perimeter is defined as the instantaneous perimeter (p) over the initial perimeter (p$_0$).}\label{fig:2fChangeRotation}
\end{figure}

Another interesting case arises for $\varpi = 75^{\circ}$ and $\beta = 1$ as shown in Figure~\ref{fig:r15}(b). The initial circular inclusion evolves to a slit with a straight lower edge and a protrusion almost perpendicular to the field direction at the rear end of the upper edge ($t = 580\tau^{\prime}$). A new protrusion emerges at the forepart at $t = 665\tau^{\prime}$, the amplitude of which increases with time, and shifts towards the left. The amplitude of the protrusion at the rear end decreases simultaneously. At $t = 920\tau^{\prime}$, the inclusion reverts back to the shape that was observed at $t = 580\tau^{\prime}$, implying the completion of a period of wave-like motion on the surface of the inclusion. The cycle is repeated indefinitely thereafter. The time-periodic dynamics can be inferred by studying the temporal evolution of the normalized inclusion perimeter as shown in Figure~\ref{fig:2fChangeRotation} (green curve). After the initial transient regime, which is characterized by an increase of the perimeter, the inclusion undergoes a time-periodic oscillatory state with constant amplitude. 

\begin{figure}[hbt!]
\centering
\includegraphics[scale=1.1]{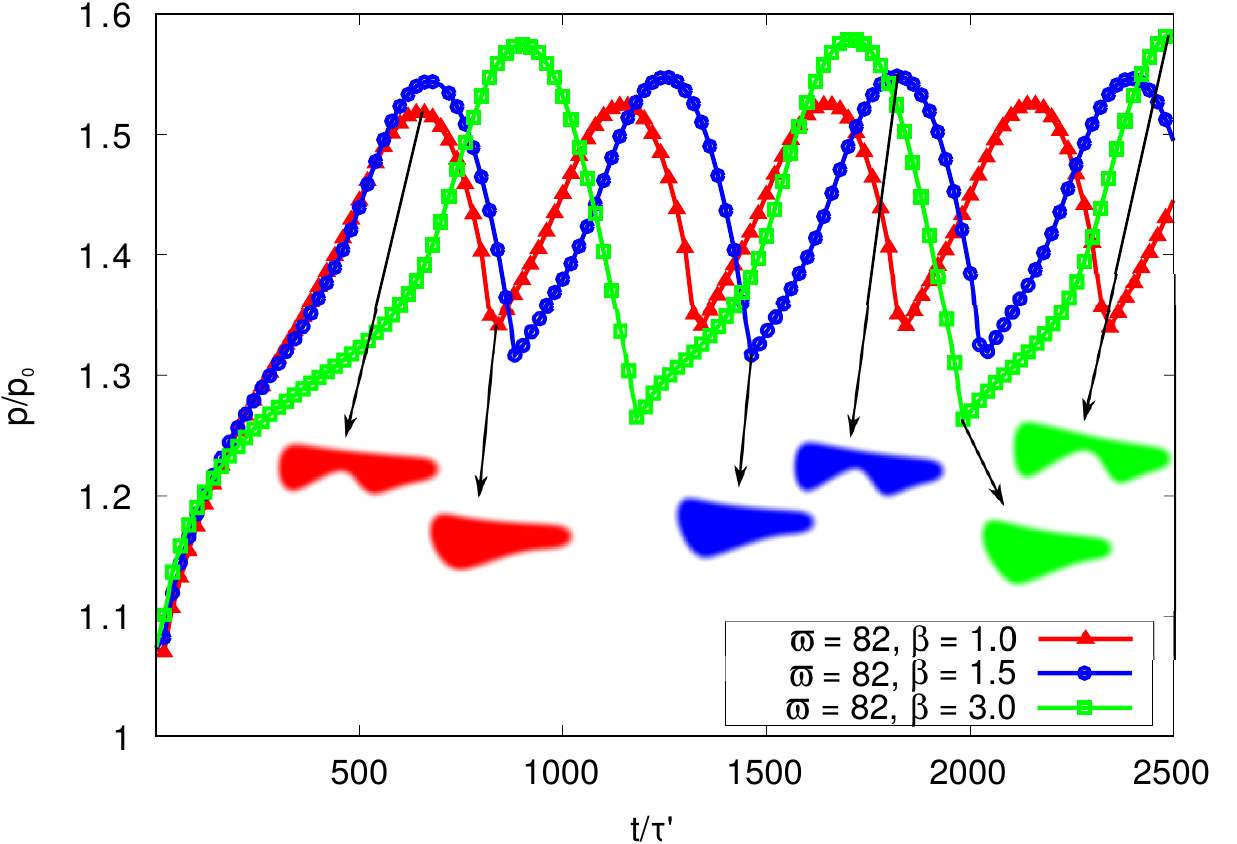}
\caption[Evolution of perimeter of inclusions with misorientation angle $\varpi = 82^{\circ}$ for $\beta=1$, 1.5, and 3.]{The evolution of perimeter of inclusions with misorientation angle $\varpi = 82^{\circ}$ for $\beta=1$, 1.5, and 3. The inclusions observe time-periodic oscillations. The period of oscillations increases with an increase in $\beta$, in addition, to increase in the amplitude.}\label{fig:2fChangeConductivity}
\end{figure}

The oscillatory dynamics was found to be dependent on both, $\varpi$ and $\beta$. This can be understood considering the following two cases. Firstly, for the same $\beta$, a decrease in the $\varpi$ decreases the period as shown in Figure~\ref{fig:2fChangeRotation}, which implies accelerated wave propagation on the inclusion surface. In addition, the amplitude of the oscillations, which signifies the limit of shape variation, increases with a decrease of $\varpi$. The state with higher amplitude entails greater shape deviations than that of the lower. The highest perimeter corresponds to shape with the valley at the center and two peaks on either side, while the lowest perimeter relates to a single peak at the rear end. Furthermore, it is to be noted that the time elapsed in reaching the periodic state decreases with a decrease in $\varpi$. Secondly, as presented in Figure~\ref{fig:2fChangeConductivity}, increasing the conductivity ratio $\beta$, for the same $\varpi$ increases the amplitude, the period of oscillation and the time taken to reach the periodic state simultaneously.

\begin{figure}[hbt!]
\centering
\includegraphics[scale=1.4]{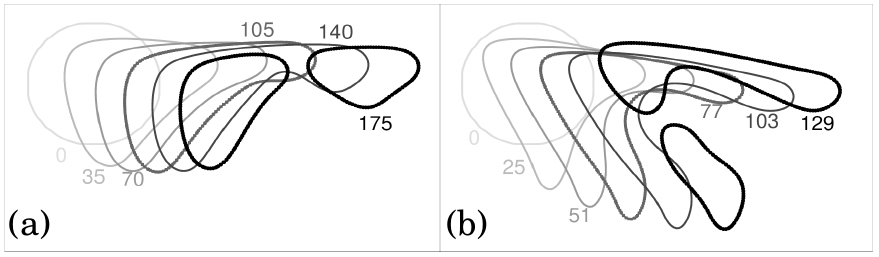}
\caption[Inclusion morphology for conductivity ratio, $\beta=1$ and $10000$ at the misorientation angle $\varpi=60^\circ$.]{Inclusion morphology for conductivity ratio, $\beta=1$ in (a) and $10000$ in (b) at the misorientation angle $\varpi=60^\circ$. The inclusions with surface contours are presented with time, t($\tau'$). The increasing darkness of the contour indicates inclusion evolution. The inclusions are displaced in space for better visual inspection. The preferential elongation in traverse direction for $\beta=10000$ ruptures lower edge, while the enlargement along the electric field for $\beta=1$ break at the upper.}\label{fig:r30}
\end{figure}

All inclusions breakup for misorientation $\varpi = 60^\circ$ (Figure~\ref{fig:r30}). Owing to the misaligned fast diffusivity sites, both species flux directions became prominent. This provokes species flow in two separate directions, which ultimately leads to a necking instability. However, the morphological evolution strongly depends upon the conductivity ratio, $\beta$. For instance, some typical cases for $\varpi = 60^{\circ}$ are shown in Figure~\ref{fig:r30}, where inclusions breach uniquely depending on the conductivity ratio. The preferential elongation in the traverse direction to the electric field for $\beta = 10000$ (Figure~\ref{fig:r30}(b)) ruptures the lower edge. This is in contrast to $\beta=1$ (Figure~\ref{fig:r30}(a)), which shows elongation in the longitudinal direction and hence the inclusion breaks at the upper edge.

\begin{figure}[hbt!]
\centering
\includegraphics[scale=1.6]{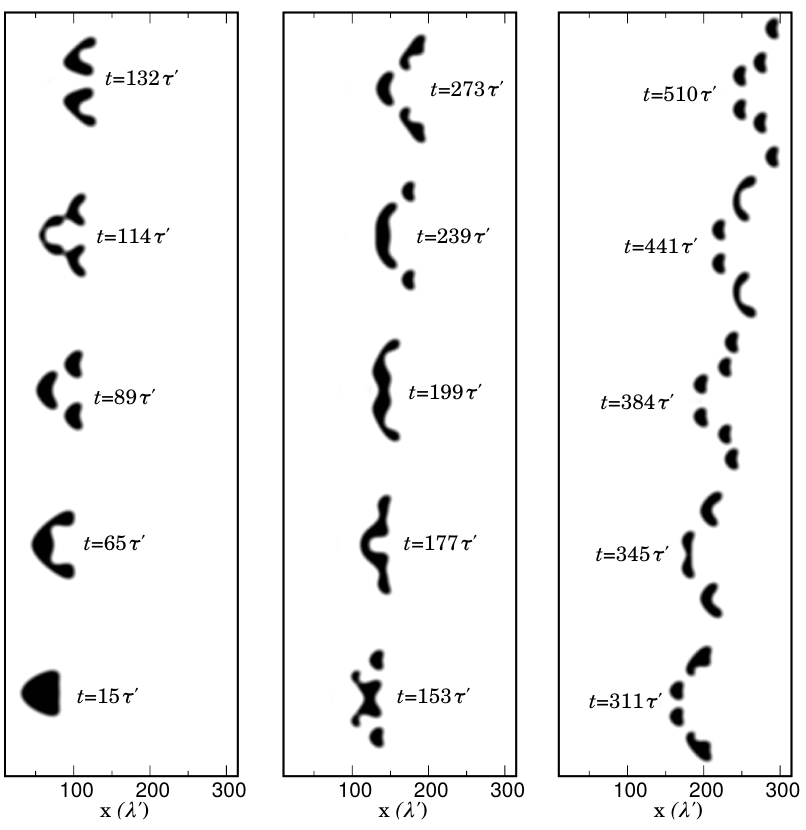}
\caption[Complex shape dynamics of inclusions at misorientation angle $\varpi = 0^{\circ}$ for the conductivity ratio $\beta=1$]{The complex shape dynamics of inclusions at misorientation angle $\varpi = 0^{\circ}$ during the morphological evolution for the conductivity ratio $\beta=1$. The snapshots of the inclusions are shifted upwards in time.}\label{fig:r90}
\end{figure}

For lower $\varpi = 0^{\circ}$ (Figure~\ref{fig:r90}), the high diffusivity regions at the two diametrically opposite ends are perpendicular to the applied field direction, which migrates faster relative to the rest of the inclusion. This results in a triangular shape with the apex at the rear end ($t=15\tau^{\prime}$). The flat front undergoes a transition to a boomerang shape ($t=65\tau^{\prime}$), followed by a pinch-off at the two ends and ultimately breaks up into three smaller inclusions ($t=89\tau^{\prime}$). All three inclusions experience a similar boomerang morphological transition. It has been shown by Ho \cite{ho1970motion} (similar to the expression in Eq.~\eqref{eq:linearStabilityinclusionVelocity}) that under isotropic diffusion a smaller inclusion has a faster migration velocity than a larger one. Although the parent inclusion is larger than the two daughter inclusions, the faster-growing slit tips of the parent (due to anisotropic diffusion) relative to the rear end of the daughters, lead to a coalescence as evident in Figure~\ref{fig:r90} ($t=114\tau^{\prime}$). The present results are in agreement with the self-consistent numerical simulations of Cho et al. \cite{cho2004electromigration}, who showed that the inverse dependence of velocity on radius is violated due to diffusional anisotropy. A complex interaction between the broken parts of the inclusions follows thereafter engendering successive breakup and coalescence. At $t = 510\tau^{\prime}$ in Figure~\ref{fig:r90} the initially circular inclusion has split into six smaller ones. In addition, it is evident that the position of the inclusion parts are symmetrical with respect to the longitudinal (migration) direction.

\subsection{Morphological Map}

\begin{figure}[hbt!]
\centering
\includegraphics[scale=1.3]{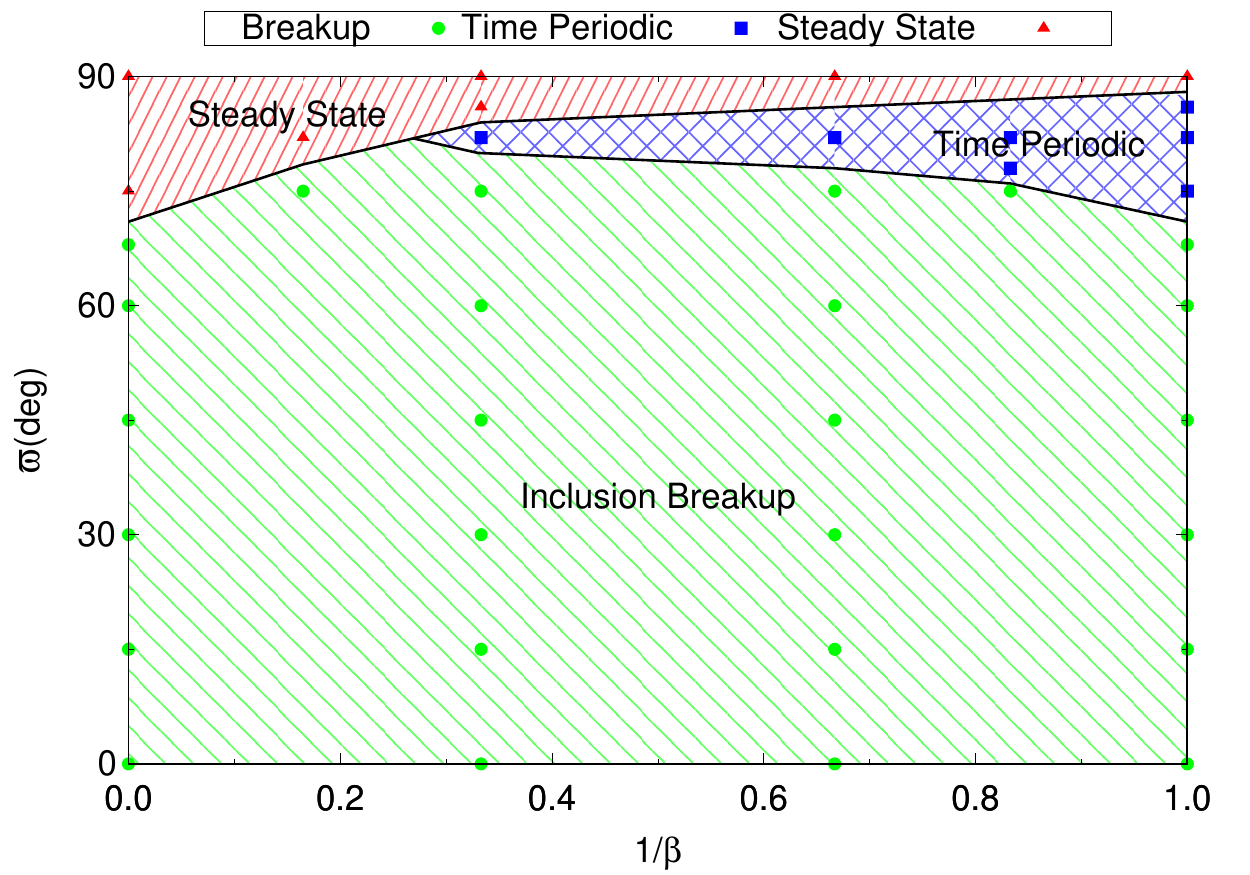}
\caption[Morphological map of inclusion migration modes as a function of $\beta$ and $\varpi$ for anisotropic twofold symmetry.]{Morphological map of inclusion migration modes as a function of $\beta$ and $\varpi$ for anisotropic twofold symmetry. The triangular, square and circular points correspond to steady-state, time-periodic, and breakup morphology respectively. The solid-black colored line is a guide to the eye.}\label{fig:morphologyMap}
\end{figure}

The effect of $\varpi$ and $\beta$ on the inclusion evolution can be summarized in the form of a morphological map as presented in Figure~\ref{fig:morphologyMap}. Higher $\varpi$ favors a shape-preserving steady-state drift. For the chosen values of $\chi_0 = 15$ and $A = 10$, the dominant electromigration force promotes a slit or triangular-like shape depending on the conductivity contrast. The steady-state region diminishes as the conductivity of the inclusion approaches that of the matrix. For intermediate $\varpi$ values and conductivity ratio ($1/\beta$) greater than $0.3$, the inclusion undergoes a time-periodic dynamics. At $\varpi$ less than about $70^{\circ}$, the inclusions become unstable and experience a complex cycle of break up and coalescence for all conductivity ratios.

\section{Discussion}
\label{sec:twoFoldEMDiscussion}

 For steady-state migration, the slit forming propensity of the inclusion changes from being along the line to perpendicular to the line as $\beta$ increases (Figure~\ref{fig:r0}). Such perpendicular growth of slits transverse to the line is expected to prove fatal to the lifetime of the line; especially for the case of smaller $\Lambda_L$ when the inclusion radius is comparable to the line width which enhances the current crowding effect. The results on the steady-state migration (Figure~\ref{r0BetaVelocityGraph}) are an important extension to the analytical theory of Ho \cite{ho1970motion}, which was developed for inclusion motion under isotropic diffusion. Compared to a circular inclusion, the triangular and slit-shaped ones engendered due to anisotropic diffusion have about three times lower velocity. Moreover, the two-fold anisotropy in diffusivity lowers the steady-state velocity approximately by a factor of seven in comparison to the slits originating due to isotropic diffusion.

\begin{figure}[hbt!]
\centering
\includegraphics[scale=0.8]{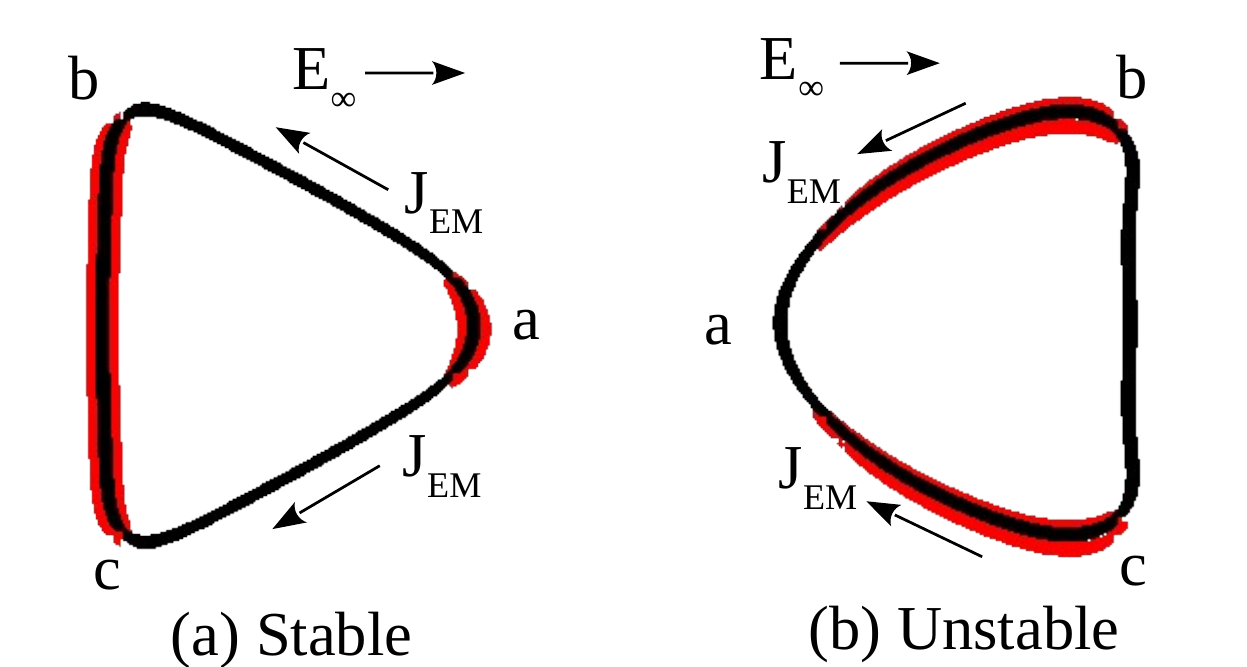}
\caption[Schematic of electromigration-induced atomic flux from the apex to the base, and from the base to the apex on the surface of the wedge-type inclusions]{Schematic of electromigration-induced atomic flux (a) from the apex to the base, and (b) from the base to the apex on the surface of the wedge-type inclusions corresponding to Figure~\ref{fig:r0}(c) and Figure~\ref{fig:r90} ($t=15\tau^{\prime}$). The former migration mode leads to steady-state propagation, while the latter instigates inclusion breakup. The red region highlights the locality of high diffusion.}\label{fig:apexPlots}
\end{figure}

Two kinds of triangular-shaped inclusions were found during the course of evolution depending on the value of the misorientation angle as shown in Figure~\ref{fig:apexPlots}. First, as in Figure~\ref{fig:r0}(c), where the apex is located at the migrating front while the rear end is perpendicular to the direction of the electric field. Second, in which the shape of the triangular inclusion is exactly reversed with the apex forming at the rear end as in Figure~\ref{fig:r90} ($t=15\tau^{\prime}$). While the former shape is found to be stable, the latter subsequently disintegrate into a number of daughter inclusions. The stability of the triangular-shaped inclusions can be understood in terms of the mass flux along the surface (Figure~\ref{fig:apexPlots}). In the former case, electromigration induces mass transport from a to b and a to c, while no mass transport takes place along the edge bc as it is perpendicular to the field direction, as shown in Figure~\ref{fig:apexPlots}(a). As a result, the inclusion migrates along the line preserving the shape. In contrast, in the latter case (Figure~\ref{fig:apexPlots}(b)) mass transport from b to a and c to a leads to slit-like growth towards line edges. The slit subsequently pinch-off at the thinnest section due to necking instability.

The importance of the misorientation angle on the stability of the inclusion as evident in Figure~\ref{fig:morphologyMap} needs to be emphasized. The assumed two-fold anisotropic atomic mobility results in two fast diffusing directions. However, the starting condition of a pre-existing circular inclusion renders the high diffusivity rear end effectively insignificant due to the unidirectional motion of species from the cathode to the anode end. Beyond a critical misorientation angle, both the fast diffusivity sites become operative. As a consequence, these ends migrate faster relative to the rest of the surface leading to the onset of instability.

The focus of the chapter has been understanding the dynamics of inclusion under electromigration in thin-film metallic conductors. The case of $\beta=10000$ corresponds to an insulating void as in vacancy which is most commonly encountered in interconnect lines. In addition, certain cases are equally applicable to the related phenomenon of directed assembly of single layer islands on crystalline substrates \cite{yamanaka1989surface, yongsunthon2011surface}. The case of $\beta = 1$ would correspond to homoepitaxial islands i.e. the monoatomic layer of atoms of a given metal deposited on the substrate of the same metal. Tuning the orientation of the substrate can then be used to precisely control the island morphology to fabricate nanowires (Figure~\ref{fig:r0}(a)), triangular (Figure~\ref{fig:r0}(c)), boomerang (Figure~\ref{fig:r30}(b) and \ref{fig:r90}) and L-shaped (Figure~\ref{fig:r30}(a)) nanostructures. 

A prior study on current-driven nanowire formation by Kumar et al. \cite{kumar2016current} investigated single layer island dynamics on conducting \{110\}, \{100\}, and \{111\}-oriented substrates, with a slightly different parameterization. While the present chapter is reported for a parameter $\chi_0$ which denotes the ratio of wind force to capillary force and has been held fixed, Kumar et al.\cite{kumar2016current} defined the length scale as the radius of the island for which the capillary force balances the electron wind force. The island radius measured in units of the defined length scale was varied. In the present work, this implies varying $\chi_0$ by varying the island radius while holding the applied electric field constant. The nanowire formation in the present study obtained from parameter set $\varpi = 0^\circ$, $\beta = 1$, $m = 1$ and $\chi_0 = 15$, thus corresponds to $\phi$ = 0$^\circ$, $m = 1$ and $R_E = \sqrt{15 \pi}l_E$ of their article, where $l_E = \sqrt{\gamma_s \Omega/q_s E_\infty}$ and $R_E$ is the square root of island area.

\section{Conclusion} \label{sec:twoFoldEMConclusion}

The numerical results are presented on the migration of circular inclusion in a \{110\}-oriented single crystal of face-centered-cubic metals under the action of an external electric field. The simulations predict a rich variety of morphologies, which include steady-state, time-periodic, and inclusion breakup. The amplitude and the frequency of time-periodic oscillations are strongly dependent on the values of conductivity contrast $\beta$ and misorientation angle $\varpi$. Furthermore, higher $\beta$ promotes a transverse elongation, while similar conductivities lead to a slit-like feature along the direction of the electric field. Finally, a morphological map is constructed delineating the dependence of various migration modes on conductivity contrast and misorientation. Results presented here have important implications on void dynamics in interconnects and fabrication of nanostructures of desired features and dimensions.

 The focus of the present work has been on inclusion migration along $\{110\}$ planes. A straightforward extension would be to address the dynamics and morphologies of inclusion, migrating along $\{100\}$ and $\{111\}$ planes i.e. four and six-fold diffusional anisotropy respectively. 
  This study is performed in the Chapter~\ref{chapter:4Fold6FoldEM}.

\afterpage{\blankpagewithoutnumberskip}
\clearpage
\chapter{Motion of anisotropic inclusions consist of 4-fold and 6-fold symmetries}
\label{chapter:4Fold6FoldEM}

Even though {numerical approaches} employed for the electromigration-driven inclusion morphology {are} ample, there are only a few works focused on the conductivity contrast. For instance, Li et. al. \cite{li2012the} studied the effect on the stability of the elliptical {inclusion,} and further highlighted the morphological evolution of crack and wedge-shaped inclusions influenced by various conductivity contrasts. However, {this} literature is devoid of {a} systematic guideline of {a} conductivity contrast, in terms of a morphological map. In an attempt, results presented in chapter~\ref{chapter:EM2} {obtain} the morphological map for mobility anisotropy with two-fold symmetrical inclusions, as shown in Figure~\ref{fig:morphologyMap}. Therefore, in the present chapter, the morphological evolution of the higher-order (fourfold and sixfold) symmetrical inclusions are considered. It has two objectives: (i) to study the effect of change in {the} conductivity contrasts between the inclusion and matrix and (ii) to critically compare the results obtained by fourfold and sixfold symmetrical inclusions with those of isotropic and twofold inclusions.

This chapter demonstrates a detailed description and the analysis of the numerical results obtained from phase-field simulations presented in chapter~\ref{chapter:phaseFieldModelElectromigration} on the electromigration-induced dynamics of fourfold and sixfold inclusions at a constant volume considering $m=2$ and 3 respectively. For different parameter sets provided in Table~\ref{tab:parameters}, a study is conducted to observe the morphological changes and dynamical evolution of initially circular inclusions, which are reported here. The morphological map is constructed in section~\ref{sec:EM3MorphologicalMap} by highlighting different modes of migration. Followed by, description of various migration modes are presented in sections~\ref{sec:EM3InclsutionBreakup}, \ref{sec:EM3TimePeriodic}, \ref{sec:EM3SteadyState}, and \ref{sec:EM3Zigzag}. Finally, the chapter is concluded by outlining significant results in section~\ref{sec:4Fold6FoldConclusions}. Some parts of this chapter are submitted for publication as a journal article.

\begin{table}[hbt!]
\begin{center}
\caption[Values of parameters considered for the simulation study inclusions in \{100\} and \{111\} crystallographic planes]{Values of parameters considered for the simulation study inclusions in \{100\} and \{111\} crystallographic planes}\label{tab:parameters}
\begin{tabular}{ l c c }
\hline
Name of the parameter&symbol&value\\
\hline
Electric field to capillary force ratio&$\chi_0$&15\\
Conductor width to inclusion radius ratio&$\Lambda_L$&6\\
Anisotropy strength&$A$&10\\
Symmetry parameter&$m$&fourfold: 2\\
&&sixfold: 3\\
Misorientation angle&$\varpi$&fourfold: $0^\circ$ to $45^\circ$\\
&&sixfold: $0^\circ$ to $30^\circ$\\
Conductivity ratio&$\beta$& 1 to 10000\\
\hline
\end{tabular}
\end{center}
\end{table}

\begin{figure}[hbt!]
\centering
\includegraphics[scale=0.30]{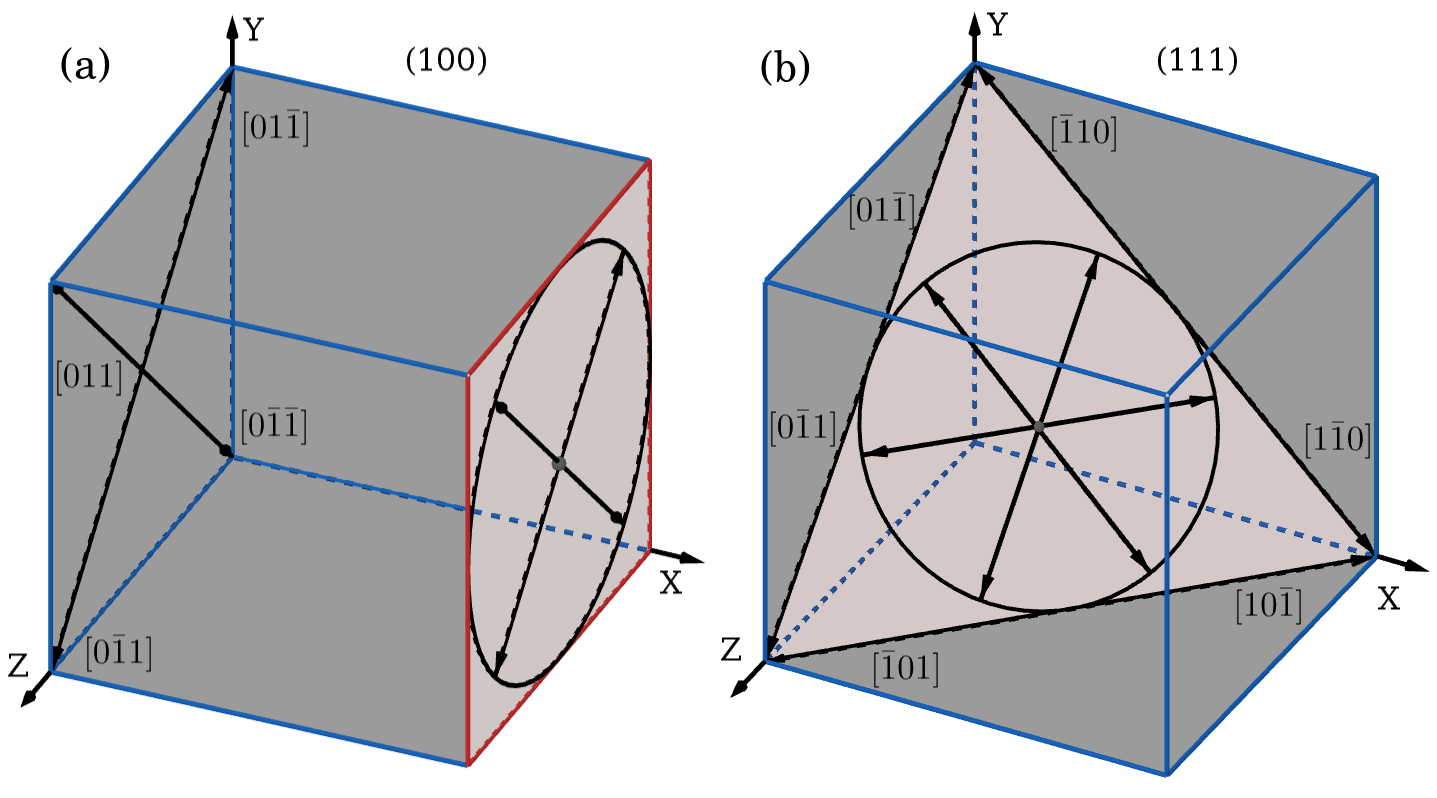}
\caption[Schematic of inclusions in (100) and (111) crystallographic planes]{Schematic of an inclusion in (100) crystallographic plane in (a) and (111) plane in (b). The directions of dominant species diffusivity are represented by black arrows. The red color plane is the surface of film perpendicular to the circular inclusions.}  \label{fig:millerPlanes}
\end{figure}
\section{Morphological map}
\label{sec:EM3MorphologicalMap}
The diffusional anisotropy in {an} inclusion is a result of the distinct fast diffusion locations in the plane. For instance, see Figure \ref{fig:millerPlanes}(a), (100) plane contains four crystallographic directions $[0\bar{1}1]$, $[01\bar{1}]$, $[ 011]$ and $[0\bar{1}\bar{1}]$, which are responsible for {fast} diffusion. Similarly, (111) plane consists of six fast diffusion directions $[1\bar{1}0]$, $[\bar{1}10]$, $[01\bar{1}]$, $[0\bar{1}1]$, $[\bar{1}01]$ and $[10\bar{1}]$, as shown in Figure~\ref{fig:millerPlanes}(b). In fact, each planes in $\{100\}$ and $\{111\}$ families contain four and six of $\langle110\rangle$ directions and referred as fourfold and sixfold respectively.
Due to the symmetry of initial inclusion shape, the parameter space of the misorientation $\varpi$ can be reduced to a range from $0^{\circ}$ to $45^{\circ}(=180^{\circ}/2m)$ for fourfold symmetry and from $0^{\circ}$ to $30^{\circ}$ for sixfold symmetry, without compromising the features of the inclusion morphology. In the systematic study of fourfold and sixfold symmetrical inclusions, the morphologically complex dynamics of inclusion shape is obtained as a function of the misorientation angle $\varpi$ and the conductivity ratio $\beta = \sigma_{\textrm{mat}}/\sigma_{\textrm{icl}}${, where} $\sigma_{\textrm{mat}}$ denotes conductivity of matrix and $\sigma_{\textrm{icl}}$ represents the conductivity of the inclusion.

\begin{figure}[hbt!]
\centering
\includegraphics[scale=1.0]{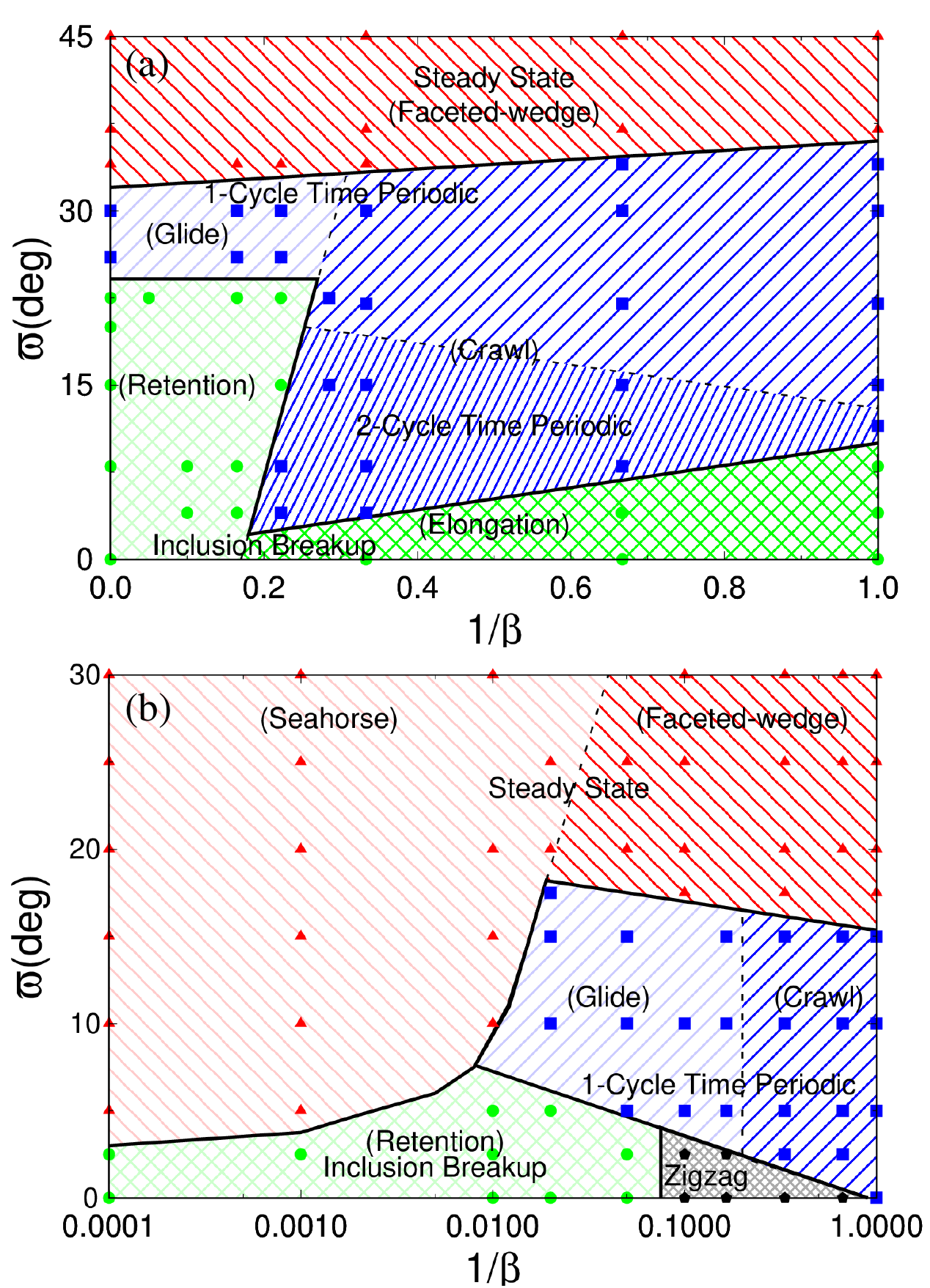}
\caption[Morphological map of inclusion migration modes as a function of $\beta$ and $\varpi$ for anisotropic fourfold ($m=2$) and sixfold ($m=3$) symmetries]{Morphological map of inclusion migration modes as a function of $\beta$ and $\varpi$ for anisotropic fourfold ($m=2$) symmetry in (a) and sixfold ($m=3$) symmetry in (b). The triangular, square,  pentagonal, and circular points correspond to steady-state, time-periodic, Zigzag oscillations and breakup morphology respectively. The solid-black colored lines are a guide to the eye to differentiate between different modes of migration, while the dotted lines separate distinct characteristics of a particular migration mode. Semi-log plot is employed for the morphological map of the sixfold symmetry for the better visual representation.}\label{fig:PhaseDiagram}
\end{figure}

Several inclusion morphologies can be observed during its propagation under electromigration-induced surface diffusion. Figure \ref{fig:PhaseDiagram}(a) and (b) present the morphological map of different migration modes in $\beta$ and $\varpi$ plane obtained by the phase-field simulations for anisotropic fourfold and sixfold symmetries of species diffusion respectively. The inclusions {assume} various morphologies such as a breakup, steady-state, time-periodic, and zigzag oscillations for sixfold symmetry, while only the former three migration modes can be identified for the fourfold symmetrical inclusions. Dynamics of the different regions within a particular migration mode in the morphological map can be recognized by the distinct characteristics during propagation. These aspects of the morphological map are described extensively in the forthcoming paragraphs.

\section{Inclusion breakup}
\label{sec:EM3InclsutionBreakup}
\begin{figure}[hbt!]
\centering
\includegraphics[scale=1.5]{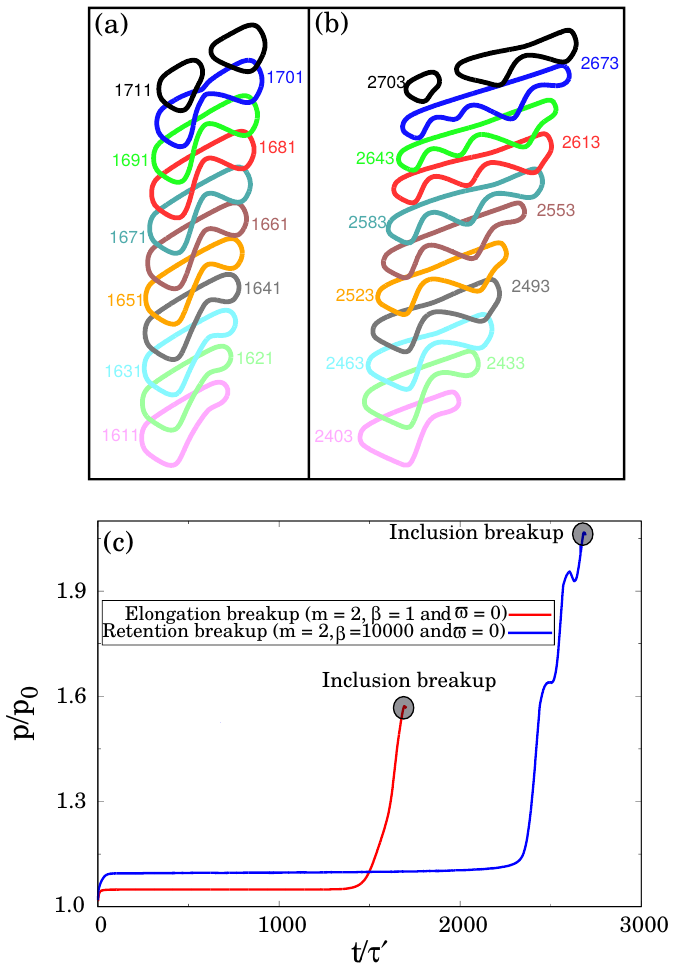}
\caption[Representative dynamics of inclusion breakup due to elongation and retention]{Representative dynamics of inclusion breakup due to elongation in (a) and retention in (b). The inclusion breakups are shown for fourfold inclusion ($m=2$) at misorientation angle $\varpi=0^\circ$ during the morphological evolution at conductivity ratio $\beta = 1$ and 10000 respectively. The inclusions with surface contours are presented with time, $t(\tau ')$. The inclusions are displaced in space for better visual inspection. (c) represents evolution of the inclusions perimeter (normalized by initial inclusion perimeter) until breakup.}\label{fig:inclusionBreakup}
\end{figure}

At low $\varpi$, {inclusions} breakup can be observed for all values of conductivity contrast in fourfold symmetry, as compared to only for conductivity contrast for sixfold symmetry. An inclusion breaks up due to two specific ways: retention and elongation. Both cases can be observed for fourfold symmetry, while only the former is prominent for sixfold symmetry.

 As a representative case, the characteristics of an inclusion migrating for fourfold symmetry at $\varpi=0^\circ$ is described. For all $\beta$, the inclusions of $\varpi =0^\circ$ and $m=2$ breaks while migrating along the electric field, which is a result of spontaneously broken symmetry of the initially circular inclusions with the external electric field. This is because of a slight change in the alignment of the axis instigates one of the fast diffusion sites in the traverse direction becomes more prominent compared to the other site. Although the inclusions eventually split into parts, two specific characteristics are observed during the course of evolution, as shown in Figure \ref{fig:inclusionBreakup}. Firstly, a monotonously increasing perimeter describes breakup due to elongation, where {an} inclusion of lower conductivity ratios ($\beta<6$) separates at the first valley formation (Figure \ref{fig:inclusionBreakup}(a)) due to insufficient species transport between two hills. Secondly, the wavy increase in perimeter (Figure \ref{fig:inclusionBreakup}(c)) demonstrates breakup due to retention, where {an} inclusion of higher conductivity ratio ($\beta>6$) surpasses the first valley without breakage (Figure \ref{fig:inclusionBreakup}(b) at $t=2523 \tau'$). Consequently, the inclusion attends several hills and valleys during its propagation before breakage. On increasing $\beta$, the tangential component of the electric field along the inclusion surface becomes more prominent, which encourages to maintain its traverse elongation. Due to that, the species diffusivity reduces at the last hill of the rear end of the inclusion. Also, the inclusion continues to form new hills and valleys, with maintaining last hill intact. Which leads to break up induced by the retention of species at the last hill (in Figure \ref{fig:inclusionBreakup}(b) at $t=2703 \tau'$).
\begin{figure}[hbt!]
\centering
\includegraphics[scale=1.3]{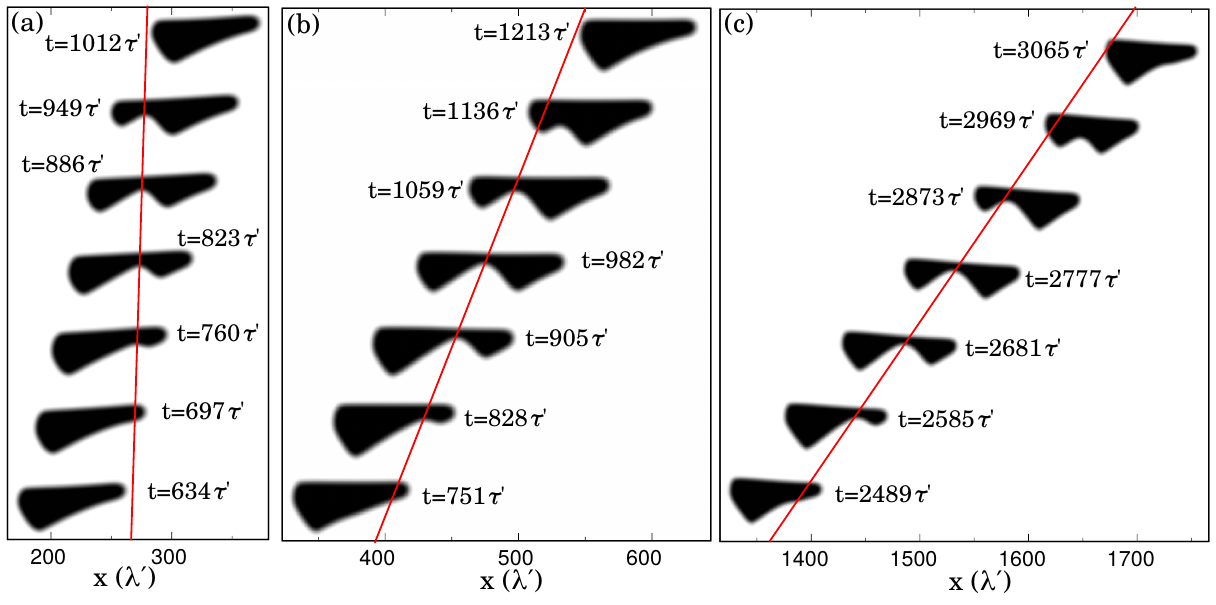}
\caption[Time-periodic shape dynamics of inclusions from crawling to gliding motion]{ The time-periodic shape dynamics of inclusions from crawling in (a) to gliding in (b) and (c). The inclusions are shown for fourfold symmetry ($m=2$) at misorientation angle $\varpi=30^\circ$ during the morphological evolution of an inclusion of conductivity ratio $\beta = 1$ in (a), 6 in (b), and 10000 in (c). The snapshots of the inclusions are shifted upwards in time. The solid red line is an attempt to recognize inclusion gliding by locating the position of the valley during evolution. The inclusion with equal conductivity to the matrix shows no appreciable gliding. The inclusion of higher conductivity contrast shows relatively more gliding compared to the lower one.}\label{fig:glideAndCrawl}
\end{figure}

It is important to note that, when the fast diffusion sites of {the} 2-fold ($\varpi=90^\circ$) and 6-fold ($\varpi=30^\circ$) inclusions align {to the} electric field, then a steady-state morphology can be observed. Contrarily, inclusions of the 4-fold undergo inclusion breakups when the fast diffusion sites align the electric field ($\varpi=0^\circ$). {Reason} for this can be explained by the position of the other fast diffusion sites which are not aligned to the electric field. There are two fast diffusion sites in the 4-fold inclusions exactly perpendicular to the electric field, which are not the case for 2-fold and 6-fold inclusions. Small deviations in species flux might lead to {broken} symmetry with the axis of the external electric field spontaneously. Furthermore, this slight change in the alignment of the axis instigates one of the fast diffusion sites in the perpendicular direction to the electric field slightly ahead compared to its diametrical counterpart. As a consequence, this fast diffusion site became favored for species diffusion. Which in turn alters the direction of {the} inclusion propagation altogether and eventually leads to breakups. In fact, when the fast diffusion sites are perpendicular to the electric field ($\varpi=0^\circ$ for all cases), then most likely the inclusions {exhibit} breakup. 

\section{Time-periodic oscillations}
\label{sec:EM3TimePeriodic}

The intermediate values of $\varpi$ promotes time-periodic oscillations for both the symmetries (blue regions in Figure \ref{fig:PhaseDiagram}). Only low conductivity contrast ($\beta<100$) shows time-periodic oscillations for sixfold symmetry, in contrary to all the conductivity contrast for fourfold symmetry. The time-periodic oscillations can be categorized based on various characteristics such as the type of motion or the type of periodic cycle. Firstly, the relative slippage during the propagation differentiates {crawling} and {gliding} motion. Secondly, a transitional number of valleys (negative curvature) points during a cycle of time-periodic oscillations differentiate 1-cycle and 2-cycle.

\subsection{Types of motion during propagation}
\begin{figure}[hbt!]
\centering
\includegraphics[scale=1.50]{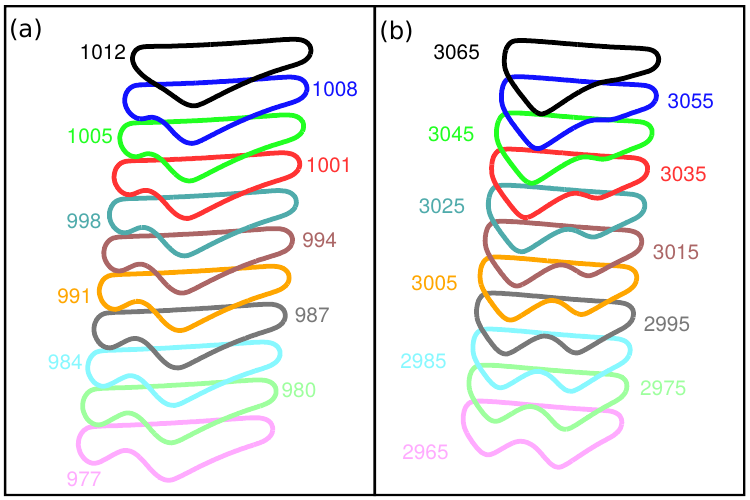}
\caption[Crawl (for $\beta = 1$) and Glide (for $\beta = 10000$) time-periodic shape dynamics of fourfold inclusions at misorientation angle $\varpi=30^\circ$]{(a) Crawl and (b) Glide time-periodic shape dynamics of fourfold inclusions at misorientation angle $\varpi=30^\circ$ during the morphological evolution at conductivity ratios $\beta = 1$ and 10000 respectively. The inclusions with surface contours are presented with time, $t(\tau ')$. The inclusions are displaced in space for better visual inspection. }\label{fig:glideAndCrawlEndPartContour}
\end{figure}

{Change} of position of a particular valley (red lines in Figure \ref{fig:glideAndCrawl}) can be utilized as a measure of gliding during evolution. The inclusion {with} no conductivity contrast ($\beta=1$) {reflects} crawling motion due to a negligible change {in} position of a particular valley, while $\beta=10000$ inclusion observes gliding motion due to appreciable change in {position}. 
Even though both inclusions attain their initial configurations after the completion of cycles as shown at $t=1012 \tau'$ in Figure~\ref{fig:glideAndCrawl}(a) and $t=3065\tau'$ in Figure~\ref{fig:glideAndCrawl}(c), {different} evolutionary pathways are observed during the motion. For that purpose,
the inclusion evolution for both types of time-periodic oscillations at the end of their respective cycles are shown in Figure \ref{fig:glideAndCrawlEndPartContour}. The complete elimination of the last hills (for $\beta=1$ in Figure \ref{fig:glideAndCrawlEndPartContour}(a)) and formation of new hills at the front part of the inclusion is observed during the propagation of the lower conductivity inclusions. While the tendency to maintain the traverse elongation at the last hill due to reduced diffusivity prohibits complete elimination of the last hill at $\beta=10000$ in Figure \ref{fig:glideAndCrawlEndPartContour}(b). Instead, inclusion propagates by contraction and stretching without any new hill formation. 

\begin{figure}[hbt!]
\centering
\includegraphics[scale=1.25]{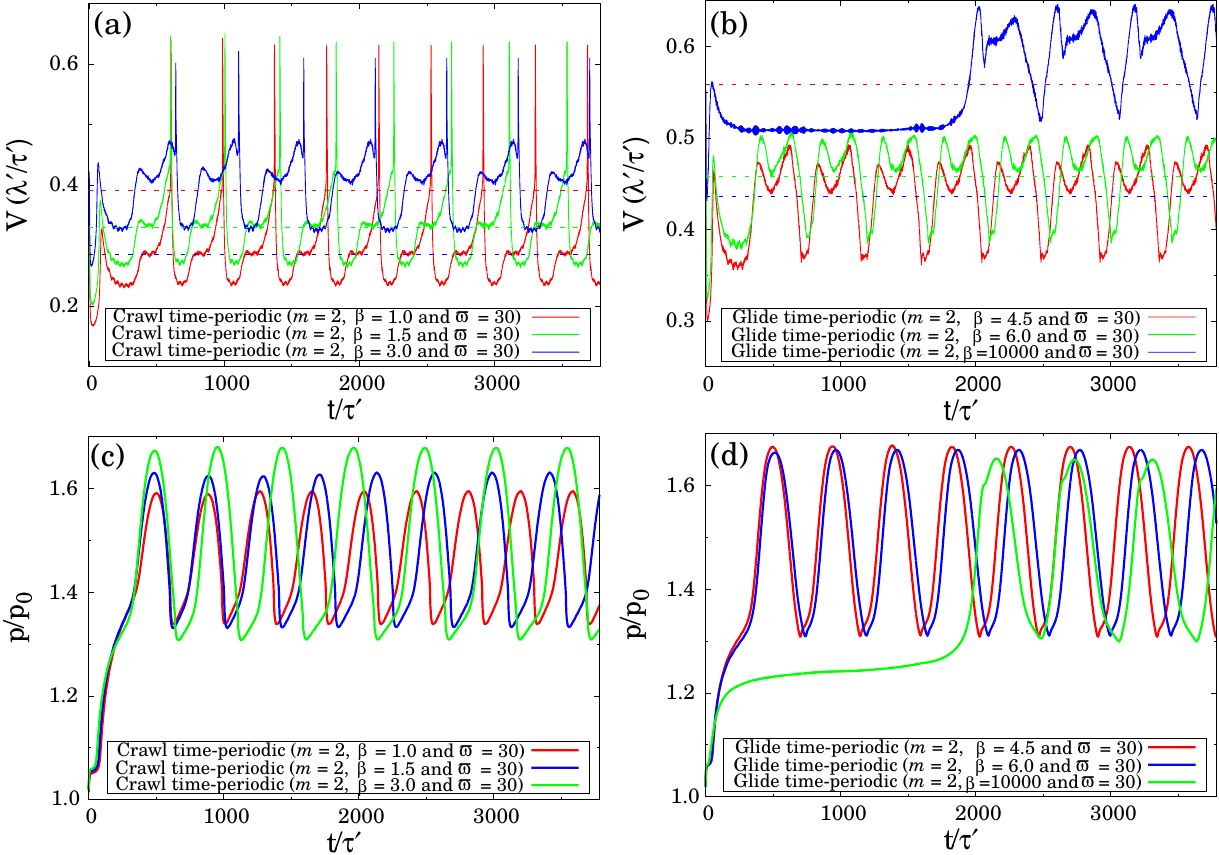}
\caption[Effect of conductivity contrast during the morphological evolution of crawling and gliding time-periodic oscillations]{Effect of conductivity contrast during the morphological evolution of crawling and gliding time-periodic oscillations. The inclusions are shown for misorientation angle $\varpi=30^\circ$ and various values of conductivity contrast $\beta$. The top row corresponds to the centroid velocity of the inclusions, while the bottom row shows normalized perimeter as a function of the crawling motion in the left column and the gliding dynamics in the right column. The dotted lines in the upper panel of the graph are mean centroid velocity of the respective solid curves.}\label{fig:glideAndCrawlPerimeterVelocity}
\end{figure}

The distinction between the crawling and gliding time-periodic oscillations can be further supported by the plots of centroid velocity. The velocity plots of crawling and gliding inclusions are demonstrated in Figure \ref{fig:glideAndCrawlPerimeterVelocity}(a) and (b) respectively. The inclusion velocities at the lower conductivity ratio intensify significantly in the region where time-periodic cycle ends, i.e., the elimination of the older hill is about to complete, for instance, between $t=1008$ to $1012 \tau'$ in Figure \ref{fig:glideAndCrawlEndPartContour}(a). This behavior is demonstrated for $\beta=1.0, 1.5$, and 3.0 in Figure \ref{fig:glideAndCrawlPerimeterVelocity}(a). However, no appreciable deviation is observed around the average centroid velocities in the case of gliding inclusions, as shown in Figure \ref{fig:glideAndCrawlPerimeterVelocity}(b) for $\beta=4.5,6.0,$ and 10000. In addition, the mean centroid velocity of the inclusions is continuously increasing with the increase in $\beta$ irrespective of the type of motion.

Furthermore, the evolution of perimeters of crawling and gliding inclusions are shown in Figure \ref{fig:glideAndCrawlPerimeterVelocity}(c) and (d) respectively. On the one hand, the increase in {the} conductivity contrast increases the amplitude and the period of {crawl}. On the other hand, the increase in the conductivity contrast increases the period, while decreases the amplitude of {glide}. In addition, time to reach the time-periodic cycle is approximately equal for various $\beta$ in the crawling dynamics, while it increases with $\beta$ for the gliding motion. A similar trend is observed for various $\beta$ of the sixfold symmetrical inclusions for both types of motion.

\begin{figure}[h]
\centering
\includegraphics[scale=1.25]{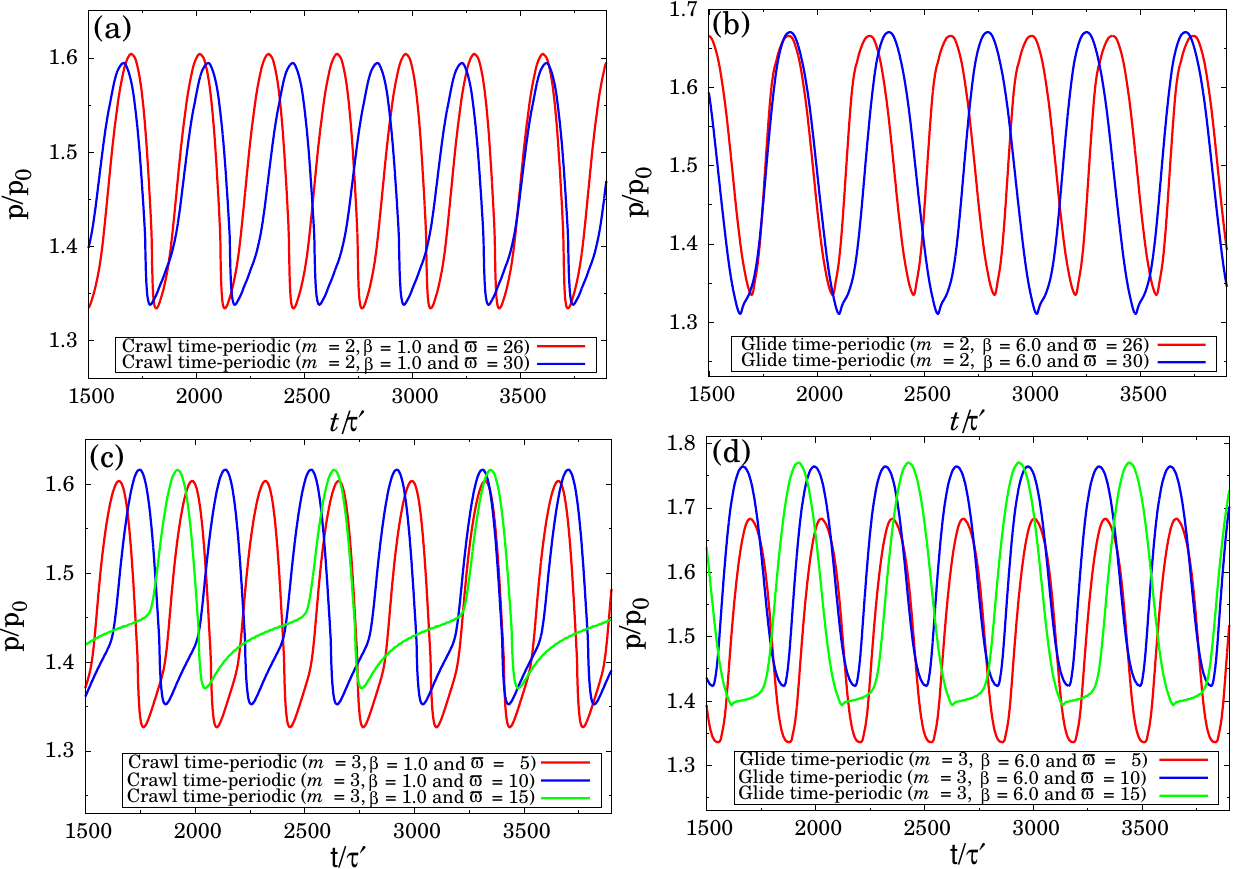}
\caption[Evolution of the perimeter as a function of $\varpi$  during time-periodic oscillations]{Evolution of the perimeter as a function of $\varpi$. The top row corresponds to fourfold, while the bottom row represents sixfold symmetrical inclusions. The left column shows results for crawling, while right column associates gliding during time-periodic oscillations.}\label{fig:varpiFourFoldSixFold}
\end{figure}

Next, the effect of misorientation in {the} fourfold and sixfold symmetrical inclusions on the different types of motion are highlighted in Figure~\ref{fig:varpiFourFoldSixFold}. A completely analogous behavior is observed between {the} fourfold and sixfold inclusions. For instance, {amplitude} of crawl motion of fourfold inclusion depreciates and period enhances with $\varpi$, which is similar to {the} sixfold inclusions. Similarly, the akin trend is observed between the fourfold and sixfold inclusions during gliding dynamics. It is important to note that the inclusions at highest $\varpi$ in fourfold (blue curves in Figure~\ref{fig:varpiFourFoldSixFold}(a) and (b)) and in sixfold symmetry (green curves in Figure~\ref{fig:varpiFourFoldSixFold}(c) and (d)) show oscillations of highest periods compared to their counterparts. This behavior can be rationalized from the location of the time-periodic region in the morphological map. Meaning, steady-state morphology is situated above the time-periodic region as shown in Figure~\ref{fig:PhaseDiagram}. {Steady}-state morphology can be perceived as time-periodic oscillation with an infinite period.

\begin{figure}[h]
\centering
\includegraphics[scale=0.58]{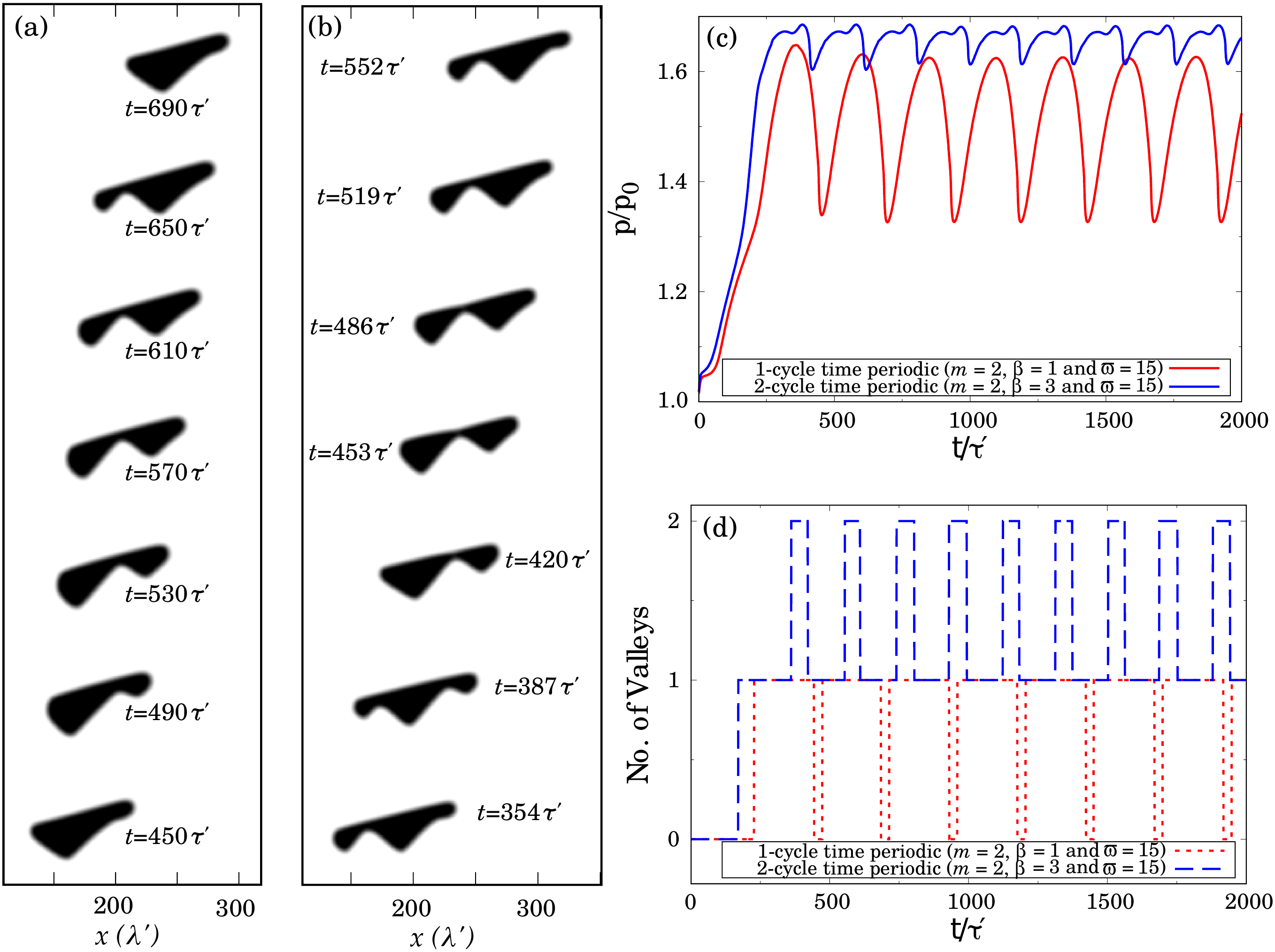}
\caption[Inclusion dynamics of 1-cycle and 2-cycle time-periodic oscillations]{Inclusion dynamics of 1-cycle time-periodic in (a) and 2-cycle time-periodic in (b).  The inclusion morphologies are shown for fourfold inclusion at misorientation angle $\varpi=15^\circ$ at the conductivity ratios $\beta = 1$ and 3 respectively.  The snapshots of the inclusions are shifted upwards in time. (c) represents the evolution of the normalized inclusions perimeter and (d) shows the number of valleys (locations of negative curvature) during the inclusion propagation.}\label{fig:1cycle2cycle}
\end{figure}

\subsection{Types of periodic cycle}
Based on {number} of valleys during the evolution, the time-periodic oscillations can be discriminated between the 1-cycle or 2-cycle as shown in Figure~\ref{fig:1cycle2cycle}. During {evolution, a} 1-cycle inclusion commutes between no valley point in one part of the cycle (near $t=450\tau'$ in Figure~\ref{fig:1cycle2cycle}(a)) to one valley in the other part, while between one to two valleys (near $t=387\tau'$ in Figure~\ref{fig:1cycle2cycle}(b)) in 2-cycle. 
{Evolution} of perimeter and {number} of valley points are displayed in Figure~\ref{fig:1cycle2cycle}(c) and (d) respectively. From the comparison, it is evident that {number} of valleys is directly associated with the value of the perimeter. Therefore, {presence} of two valley points during oscillations certainly increases the perimeter as shown in Figure~\ref{fig:1cycle2cycle}(c).  

\begin{figure}[h]
\centering
\includegraphics[scale=1.3]{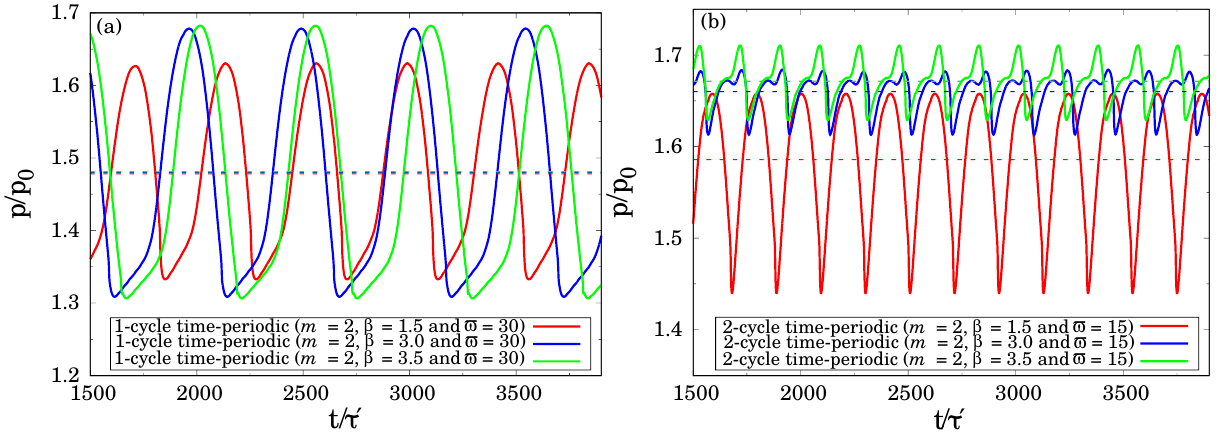}
\caption[Evolution of perimeter for 1-cycle and 2-cycle time-periodic oscillations as function of $\beta$.]{Evolution of perimeter for (a) 1-cycle and (b) 2-cycle time-periodic oscillations as function of $\beta$. The inclusion morphologies are shown for fourfold inclusion at misorientation angle $\varpi=30^\circ$ and $15^\circ$ respectively. The dotted lines in the graph are mean perimeter of a complete period of a respective solid curves.} \label{fig:1cycle2cyclePerimeter}
\end{figure}

{Difference} in dynamics of 1-cycle and 2-cycle can be further clarified from the influence of $\beta$. The evolution of perimeter for 1-cycle and 2-cycle time-periodic oscillations are displayed in Figure~\ref{fig:1cycle2cyclePerimeter}(a) and (b) respectively. Increasing $\beta$ enhances the amplitude of {perimeter} alongside the increment in the period for 1-cycle oscillations, which is contrary to 2-cycle. Furthermore, a significant disparity exists at the mean perimeter. The mean perimeter is approximately equal for all cases of 1-cycle, while the value of mean perimeter increases with $\beta$ for 2-cycles.

\section{Steady-state morphology}
\label{sec:EM3SteadyState}

\begin{figure}[hbt!]
\begin{center}
\includegraphics[scale=1.5]{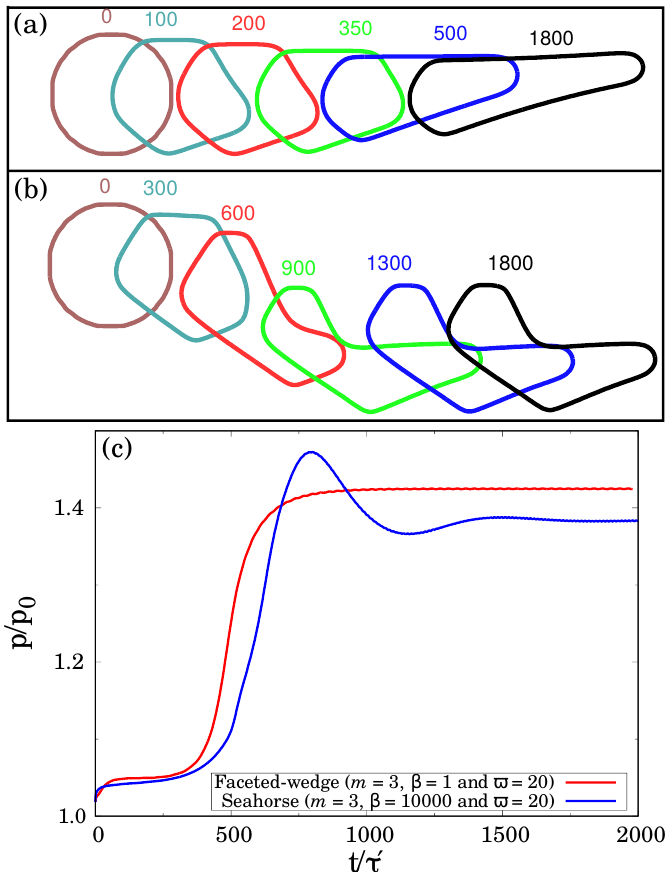}
\caption[Steady-state morphological evolutions of faceted-wedge and seahorse pattern]{Steady-state morphological evolutions of faceted-wedge in (a) and seahorse pattern in (b). The inclusion dynamics are shown for sixfold symmetry at misorientation angle $\varpi=20^\circ$ and conductivity ratios $\beta = 1$ and 10000 respectively. The inclusions with surface contours are presented with time, $t(\tau ')$. The inclusions are displaced in space for better visual inspection. (c) represents the evolution of the normalized inclusions perimeter until it attains equilibrium shape.}\label{fig:facetedWedgeSeahorse}
\end{center}
\end{figure}

For all values of conductivity contrast, the inclusions of higher misorientation in fourfold symmetry and sixfold symmetry (red regions in Figure \ref{fig:PhaseDiagram}) obtain steady-state morphologies after initial adjustments as shown in Figure \ref{fig:facetedWedgeSeahorse}. Faceted-wedge and seahorse morphologies are observed for sixfold symmetry, while only a former case is observed at fourfold symmetry. A representative morphological evolution of faceted-wedge and seahorse pattern is shown in Figure~\ref{fig:facetedWedgeSeahorse}(a) and (b) respectively. The inclusion perimeter graph (in Figure~\ref{fig:facetedWedgeSeahorse}(c)) shows a monotonously increasing perimeter for faceted-wedge shape, while a wavy behavior is observed for seahorse-shaped inclusions.

\begin{figure}[h]
\centering
\includegraphics[scale=1.3]{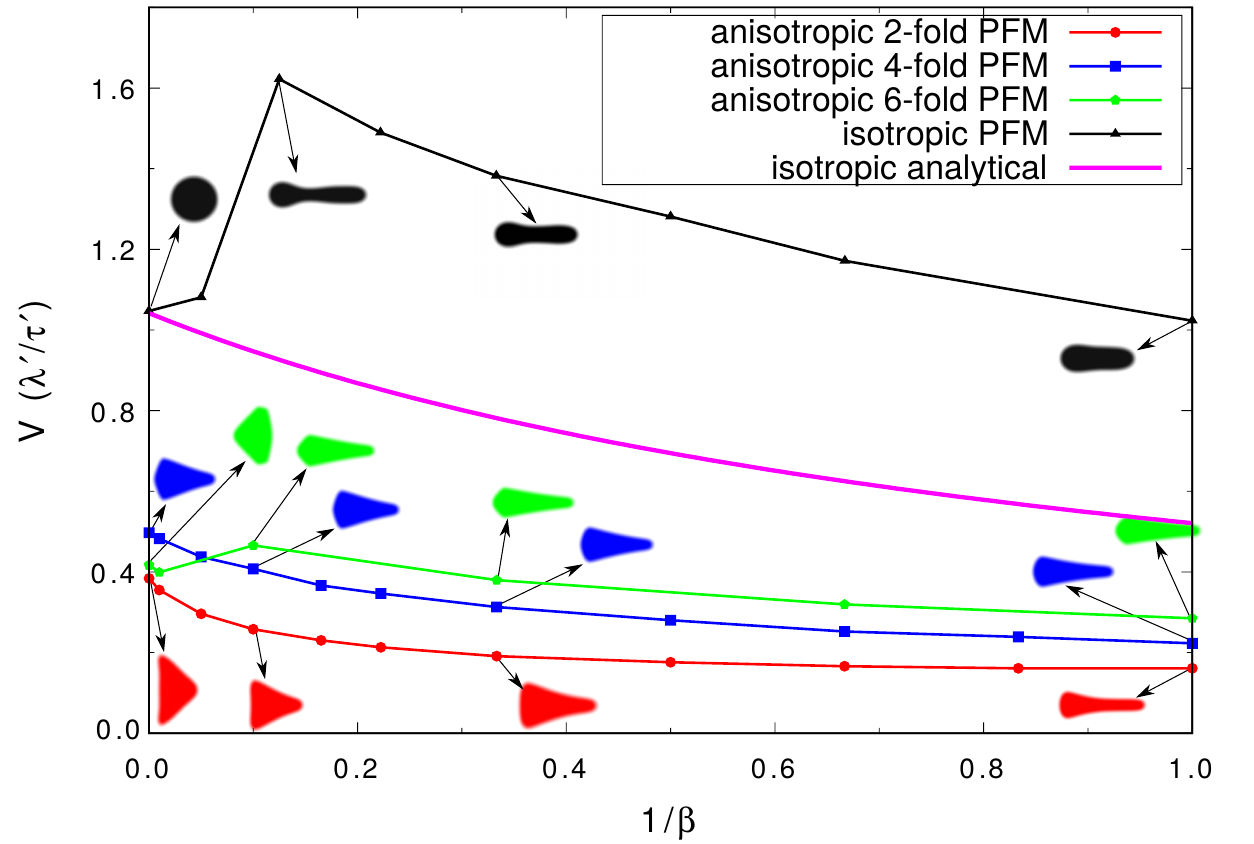}
\caption[Velocities of the centroid of steady-state inclusions for twofold, fourfold, and sixfold anisotropies, compared with isotropic surface diffusion (analytical and numerical) as a function of the conductivity ratio $\beta$]{ The velocities of the centroid of steady-state inclusions for twofold anisotropy ($m = 1$), $ \varpi = 90^\circ$ (in red graph), fourfold anisotropy ($m = 2$), $ \varpi = 45^\circ$ (in blue graph), sixfold anisotropy ($m = 3$), $ \varpi = 30^\circ$ (in green graph), and for isotropic ($m = 0$) surface diffusion (in the black graph) as a function of the conductivity ratio $\beta$. The inset of images correspond to steady-state morphologies of respective $m$ and $\beta$. The magenta curve corresponds to a steady-state velocity of the circular inclusion of isotropic surface diffusivity, derived from \cite{ho1970motion} considering the diffusion coefficient D$_s$. Evidently, $\beta$ governs the inclusion shapes and the steady-state velocities.}\label{fig:4foldSteadyVelocities}
\end{figure}

{Comparison} of the steady-state {morphologies} across the isotropic and anisotropic (twofold, fourfold, and sixfold) cases are conducted on the inclusions propagating with maintaining symmetricity along the electric field during migration. After the inclusions attain an invariant shape, the steady-state velocities of the centroid for fourfold and sixfold symmetries are obtained from phase-field simulations for different values of $\beta$ as shown in Figure~\ref{fig:4foldSteadyVelocities} (blue and green curves respectively). For comparison, the velocity of a shape-preserving circular inclusion (pink curve) is derived from Ho's analytical formula similar to Eq.~\eqref{eq:linearStabilityinclusionVelocity}. Along with, velocities of inclusions migrating under isotropic diffusion ($m=0$) and twofold symmetry ($m=1$) obtained from phase-field simulations of chapters~\ref{chapter:EM1} and \ref{chapter:EM2} are presented.

On the one hand, the morphologies at fourfold symmetry reveal progressively increment in the width of the inclusion and decrement in the length with the increase in the conductivity contrast $\beta$, in addition to the increase in the centroid velocities, as shown in the blue graph. On the other hand, a sudden shrinkage in the length (along with expansion in width) and decrement in the velocity of the inclusion from conductivity contrast $\beta=10$ to $\beta=10000$ are observed for sixfold symmetry, which is analogous to the isotropic case (black curve).

Even though nearly all inclusions attain a finger-like slit with thinner front and wider back for lower conductivity contrast ($\beta=1$), different morphologies are observed for the highest conductivity contrast ($\beta=10000$). Twofold symmetry encourages triangular shape with the apex at the front, while sixfold symmetry promotes nearly a triangular shape with the apex at the back. From {velocity} graph, it is evident from {green} curve (for sixfold) that the flatter front with the apex at the back significantly restricts inclusion velocity. Contrarily, the apex at the front in the direction of motion with the flatter back (red inset at $\beta=10000$ for twofold) assists the velocity of inclusion migration.

\section{Zigzag oscillations}
\label{sec:EM3Zigzag}

\begin{figure}[hbt!]
\centering
\includegraphics[scale=1.5]{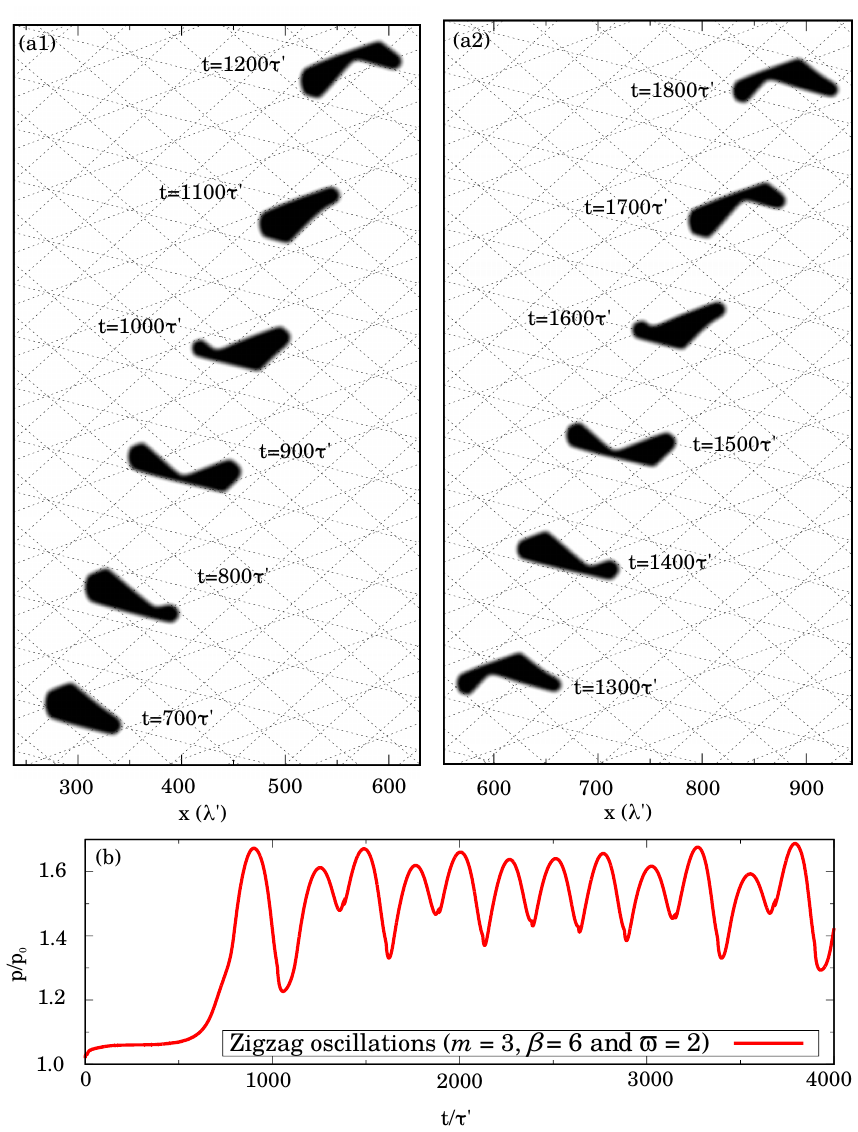}
\caption[Zigzag oscillations of the sixfold symmetrical inclusion at misorientation angle $\varpi=2^\circ$ and conductivity ratio $\beta=6$]{(a) Zigzag oscillations of the sixfold symmetrical inclusion at misorientation angle $\varpi=2^\circ$ and conductivity ratio $\beta=6$. The snapshots of the inclusions are shifted upwards in time. The gray dotted lines represent preferential facet orientations predicted by the linear stability theory. (b) represents the evolution of the normalized inclusions perimeter.} \label{fig:zigzag}
\end{figure}

{Zig-zag} oscillations are observed only for sixfold symmetry with lower misorientation and lower conductivity ratio ($1/\beta \approx$ 0.1 to 1.0). A representative morphological evolution is displayed in Figure~\ref{fig:zigzag} for $m=3$, $\beta=6$, and $\varpi=2^\circ$. In contrary to the time-periodic oscillations, where a straight edge on one side remains permanently impervious, while {hills and valleys} on the other side undergo shape changes during propagation. {Zigzag} oscillations are a result of the facet associated with the straight edge changes from the upper to the lower facet and reverses during evolution. The species diffusion from rear-end prioritize the formation of a rounded end (Figure~\ref{fig:zigzag}(a2) from $t=1400 \tau'$ to $1500 \tau'$) before further elongation of {straight} edge. The straight edge undergoes an alteration after an undefined time period. {Variation} of the perimeter is demonstrated in Figure~\ref{fig:zigzag}(b). The zig-zag pattern was reported previously in the homogeneous substrate in Ref.~\cite{kuhn2005complex}.


\begin{figure}[h]
\centering
\includegraphics[scale=0.5]{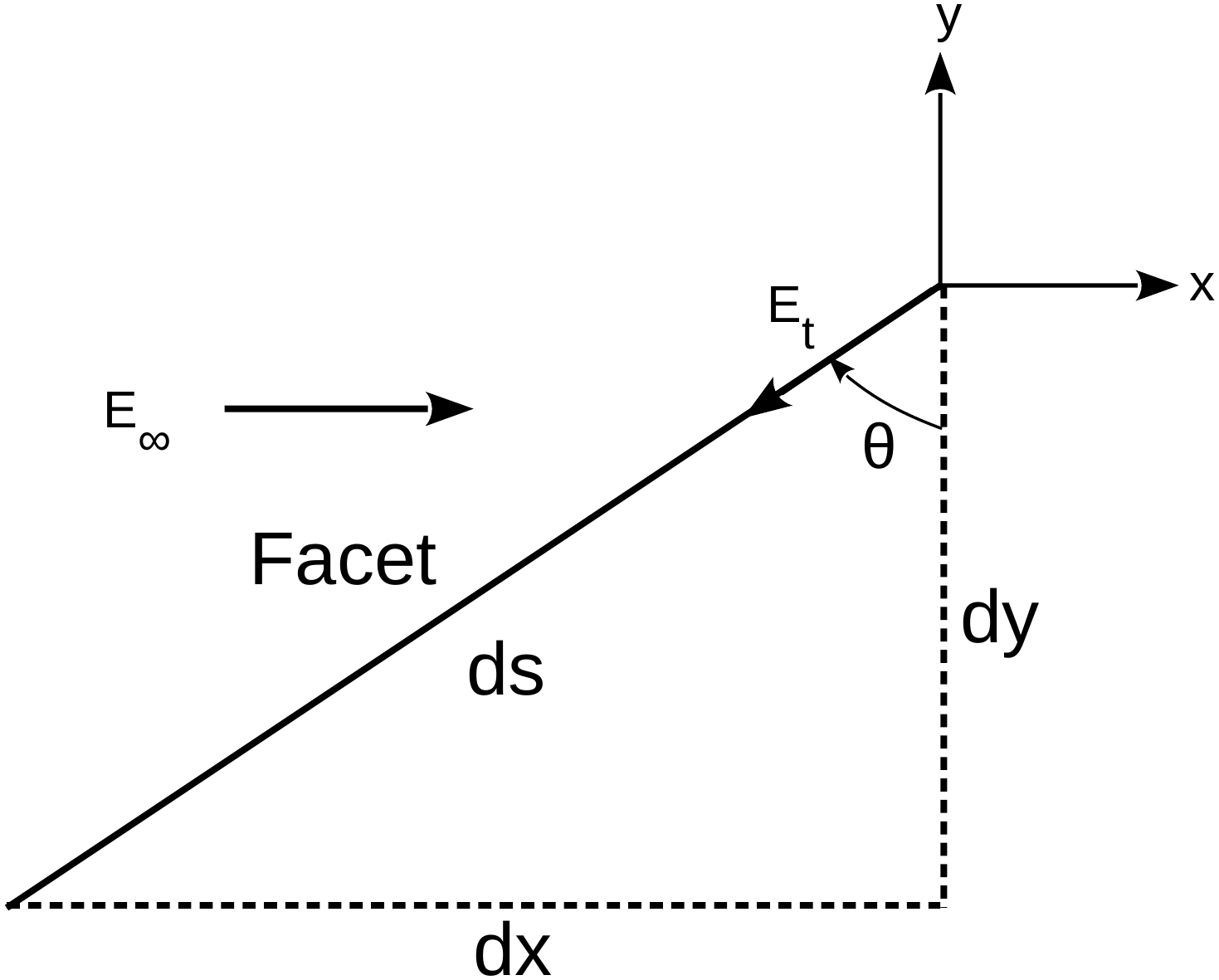}
\caption[Schematic of a facet of an inclusion.]{Schematic of a facet of an inclusion. Thick black line indicates the facet, which is subjected to an external electric field $E_\infty$. The projection of electric field on the surface of facet is $E_t$.} \label{fig:facetSchematic}
\end{figure}

{Dotted} lines in Figure~\ref{fig:zigzag}(a) are the preferential facet directions obtained from the continuum theory \cite{gungor1998electromigration, krug1994current, maroudas1995dynamics}. A facet $y=mx$, oriented at an angle $\theta$ with the perpendicular direction to the electric field is considered for the stability analysis, as shown in Figure~\ref{fig:facetSchematic}. Using the relation from Eq.~\eqref{eq:EM1MassConservationNormalVelocity} and $\textrm{d}x=-\textrm{d}s \textrm{ sin } \theta$, the mass conservation of surface flux can be expressed as,
\begin{equation}
\frac{\textrm{d}y}{\textrm{d}t} = -\Omega \frac{\textrm{d} J_s}{\textrm{d}x}. \label{eq:sharpInterfaceContinuity}
\end{equation}
This governing equation relates the height of the facet during inclusion propagation. Substituting the Nernst-Einstein relation~\eqref{eq:2foldSharpInterfaceFlux} into the governing equation \ref{eq:sharpInterfaceContinuity}, the resultant expression can be written as,
\begin{equation}
\frac{\textrm{d}y}{\textrm{d}t}=\frac{D_s \delta_s E_\infty e Z_s}{ k_B T} \Bigg\{ \frac{2\beta}{1+\beta} \frac{f^{\theta}(\theta) \textrm{ cot}\theta + \frac{\textrm{d}f^{\theta}(\theta)}{\textrm{d}\theta}}{(1+\textrm{ cot}^2\theta)^{3/2}} \frac{\textrm{d}^2 y}{\textrm{d}x^2} - \frac{\Omega \gamma_s}{e Z_s E_\infty}\frac{1}{(1+\textrm{ cot}^2\theta)^2} \frac{\textrm{d}^4 y}{\textrm{d}x^4}\Bigg\}. \label{eq:sharpInterfaceFacetResultant}
\end{equation}
where $\textrm{cot}\theta = -\textrm{d}y/\textrm{d}x$. A dispersion relation is derived from Eq.~\eqref{eq:sharpInterfaceFacetResultant} to predict the growth or decay rate of the small perturbations on the facet expressed as,
\begin{equation}
\omega_{d}(k_d)=\frac{D_s \delta_s E_\infty e Z_s}{ k_B T} \Bigg\{- \frac{2\beta}{1+\beta} \frac{f^{\theta}(\theta) \textrm{ cot}\theta + \frac{df(\theta)}{d\theta}}{(1+\textrm{ cot}^2\theta)^{3/2}} k_d^2  - \frac{\Omega \gamma_s}{e Z_s E_\infty}\frac{1}{(1+\textrm{ cot}^2\theta)^2} k_d^4\Bigg\}, \label{eq:sharpInterfaceFrequency}
\end{equation}
For a facet to be unconditionally stable, the perturbation frequency $\omega_d(k_d)$ should be negative. As the second term in Equation~\ref{eq:sharpInterfaceFrequency} is always negative, therefore the sufficient condition for the facet stability is
\begin{equation}
 f^{\theta}(\theta) \textrm{ cot}\theta + \frac{\textrm{d}f^{\theta}(\theta)}{\textrm{d}\theta}>0.
\end{equation}
Using this expression, the preferential facet orientations $\theta^* = \theta$ is derived for the required anisotropy parameters. For instance, these orientations are displayed as the dotted lines in Figure~\ref{fig:zigzag}(a). It is evident that the inclusion morphology precisely trails the preferential facet directions. 

\section{Conclusion}
\label{sec:4Fold6FoldConclusions}

The numerical results are presented on the migration of circular inclusion in \{100\} and \{111\}-oriented crystallographic planes under the action of the external electric field. Morphological maps are constructed in the plane of misorientation angle and conductivity contrast based on numerical results. The steady-state, time-periodic oscillations, zig-zag, and inclusion breakups are observed. In addition, these various migration modes can be further classified as stable faceted or seahorse morphology, crawling-based or gliding-based time-periodic oscillations, and elongation or retention breakups. Furthermore, the influence of variation in conductivity contrast and misorientation on the dynamics of the time-periodic oscillations are discussed. Finally, the steady-state dynamics obtained with the fourfold and sixfold symmetrical inclusions in the present study are critically compared with the twofold symmetry, isotropic analytical and numerical results.

The simulations predict reach a variety of morphologies during propagation. Some of the morphologies are already reported in the literature, such as steady-state faceted inclusions \cite{dasgupta2018analysis}, time-periodic crawls by formation and destruction of hills \cite{dasgupta2018analysis}, zigzag-type motion \cite{kuhn2005complex}, and inclusion breakups due to elongation \cite{schimschak1998electromigration}. Other morphologies are introduced to the scientific community for the first time in the present chapter, such as time-periodic glides without formation and destruction of hills, the breakup of inclusions due to species retention at the last valley, and seahorse-like inclusions.
\afterpage{\blankpagewithoutnumberskip}
\clearpage

\newpage
\thispagestyle{empty}
\vspace*{8cm}
\phantomsection\addcontentsline{toc}{chapter}{V Conclusions and Future Directions}
\begin{center}
 \Huge \textbf{Part V} \\
 \Huge \textbf{Conclusions and Future Directions}
\end{center}

\afterpage{\blankpagewithoutnumberskip}
\clearpage
\chapter{Conclusions and future directions}\label{chapter:conclusion}

Diffusion-driven processes during energy conversion and transmission of {electric} field are highlighted. Particularly, two-phase coexistence in cathode particles of lithium-ion batteries and electric-field induced inclusion migration in metallic conductors are investigated in the presented work. {Phase}-field models are formulated to obtain an enhanced understanding of the physical phenomena. {Numerical} results show that {phase}-field methods can efficiently apprehend the essential thermodynamics in inclusion migration in metallic conductors and two-phase coexistence in cathode particles. In addition, the results enhance the current understanding of the phenomena. In the following paragraphs, findings from {numerical} studies and their applicability are highlighted along with possible future directions. 

\section{Phase separation in lithium-ion batteries}

The presented method in Chapter~\ref{chapter:phaseFieldModelLIB} based on classical Cahn-Hilliard is capable to model the evolution of two-phase coexistence. In {finite}-difference framework, the derived model is parallelized through message passing interface (MPI) algorithms. The increase in mesh resolution conveys the convergence of the presented method, which is validated with a benchmark study. The reason for the two-phase coexistence, the presence of the miscibility gap is discussed in detail with the help of free energy density and the chemical potential. 

\subsection{Single particle study}
{Smoothed} boundary method enables the modeling of particles with almost any geometry. A simulation study is demonstrated on a cathode particle during discharge in Chapter~\ref{chapter:NeumannConstantFlux}, which is applicable to a constant flux boundary condition at the particle surface.  {Effects} of the particle geometry on {concentration} evolution are explored numerically. Numerical results show that the phase segregation starts in the vicinity of the regions with higher curvature. This is due to the fact that particle surface with a higher curvature preferentially accumulates lithium species and exhibit the spinodal decomposition earlier to other regions of the surface, with the application of constant flux at the particle boundary. Therefore, the elliptical particle with a higher aspect ratio is subjected to the onset of the phase separation, prior to the lower particle.

Microstructructure modeling on cathode particles of lithium-ion batteries can be helpful for better understanding and tailoring the material properties influencing the mechanisms underneath the actual process. Specifically, numerical studies are performed on C-rate, mobility, temperature and the energy parameter. It has been depicted that the C-rate and the species mobility related to deposition and transportation rates respectively, while the operating temperature and free energy parameters are regulating the miscibility gap. The results from the performed study can be utilized to accelerate or decelerate by tuning the process of phase separation. For instance, higher values of energy parameter and C-rate accelerate the phase separation, while higher values of temperature and mobility suppress the process.

While {numerical} study in Chapter~\ref{chapter:NeumannConstantFlux} is focused on constant lithium flux at the particle surface, the model is certainly not restricted to it. A straightforward extension is to consider time, concentration, or spatial-dependent flux at the particle surface in the form Neumann flux boundary condition expressed in Eq.~\eqref{eq:LIBNeumannFluxEvolutionEquation}. In addition, {a} study can be extended to numerically investigate {phase} separation mechanism under the Dirichlet concentration boundary condition (Eq.~\eqref{eq:LIBSmoothedBoundaryMethodDirichlet}) at the particle surface \cite{cheng2009evolution}. Finally, the described method can be generalized to incorporate mechanical effects. The development of stresses, due to diffusion gradients in the particle pose a threat to the life expectancy of the battery and may ultimately lead to crack formation \cite{Chen2012_1000028805} and permanent degradation of the material.   

\subsection{Multiple particles study}

A numerical study performed in Chapter~\ref{chapter:Potentiostatic} can be employed for the performance prediction of tortuous structures based on diffusional properties. For instance, the SOC evolution of the obtained results for the porous structures of mono-disperse particles are compared with the bulk-transport and surface-reaction limited theories. {Results} show that the smaller mono-disperse particles tend to {surface}-reaction limited theory, while the larger ones tend towards {bulk}-transport limited theory of the planar electrode. Thus, the lithium transport rate can be efficiently controlled through an appropriate selection of electrode particle sizes. 

Similarly, an extension of {numerical} findings presented here can be considered to prepare a guideline which tunes the transportation rates by selecting the appropriate values of structural descriptors, such as mono-disperse particle size $R$, porosity~$\rho$, and tortuosity~$\tau$. In addition, the porous structures are characterized numerically in the current investigations as described in Appendix \ref{sec:electrodeFormation}, which allows to control the desired geometrical properties of electrode microstructure. This procedure in combination with the presented model can be employed as a numeric prototype for preliminary investigations and testing, which might reduce the experimental efforts.

A number of interesting directions can be pursued hereafter. 
For instance, {focus} of the present study is on the effect of various geometrical parameters such as monodisperse particle size, porosity, and tortuosity on the charge dynamics. A straightforward extension is to investigate other crucial parameters such as electrode overpotential (value of $c_{ps}$), electrode thickness, size polydispersity, and shape of the particles. These parameters have physical significance, which should be the subject of future investigations. In addition, a constant concentration condition is employed at the separator in the form of Eq.~\eqref{eq:boundaryConditionSeparator}. This can be extended to be a time, local concentration, and/or spatially-dependent function. Furthermore, numerical investigations of Neumann flux boundary conditions in the form of Eq~\eqref{eq:boundaryConditionSeparatorFlux} are perhaps interesting future endeavors.

\section{Inclusion motion due to Electromigration}

Electromigration-induced morphological evolution of inclusions (voids, precipitates, and homoepitaxial islands) is considered for a numerical study, which has acquired recent scrutiny for the efficient design of the interconnects and surface nanopatterns. In Chapter~\ref{chapter:phaseFieldModelElectromigration}, a phase-field method is derived from the Cahn-Hilliard equation coupled with the Laplace equation of the electric field. This allows the account of different electrical conductivities of inclusions and the matrix. {Distinct} conductivities alter the local distribution of the electric field in the vicinity of the inclusion. {Longitudinal} component of the electric field enhances up to two times, while the perpendicular component forms a quadrupole pattern around the inclusion. Furthermore, the developed model is employed to investigate diffusional isotropy and anisotropy of inclusions migrating in \{110\}, \{100\} and \{111\} planes of face-centered-cubic crystal, which resembles twofold, fourfold, and sixfold symmetry respectively.

\subsection{Isotropic inclusions}

{Phase}-field study in Chapter~\ref{chapter:EM1} is intended to study isotropic inclusions, which successfully corroborate the findings from the linear stability
analysis. In addition, the numerical results can estimate the shape of the inclusion after diversion from circular shapes, which is {arduous} to obtain from analytical theories. The transition of a circular inclusion to a finger-like slit is elucidated. Following an initial transient regime, the inclusion attains an equilibrium shape with a narrow parallel slit-like body, which contains a circular rear end, and a parabolic tip. It is identified that the subsequent drift of the inclusion is characterized by shape invariance. In addition, {selection} of steady-state shape and velocity are implicitly related. However, the steady-state slit width and velocity are determined to scale with the applied electric field as E$_{\infty}^{-1/2}$ and E$_{\infty}^{3/2}$, respectively. 

{Velocity} of a slit with half-slit width $u$ modifies by a factor of $\xi$ compared to circular inclusion with a radius of $u$. In addition, the slit characteristics are significantly influenced by three non-dimensional parameters, namely, the ratio of conductor line width to inclusion radius, $\Lambda_L= w/R_i$, velocity discrepancy coefficient $\xi$, and competition between the electric field and capillarity, $\eta$. {Effect} of these parameters on slit profiles are analyzed and the results obtained from phase-field simulations are critically compared with {sharp}-interface solution. {These} results reveal a good agreement across all values. Repercussions of the study, in terms of prediction of inclusion migration in {the} flip-chip Sn-Ag-Cu solder bumps and {the} fabrication of channels with desired micro/nanodimensions, are discussed.

The presented results are mostly focused on the case where an initial circular void transforms into a single slit. It is also possible for an initially circular inclusion to break up into a number of independent slits, as shown in Figure~\ref{fig:EMLSA}. The reason for this is due to higher electromigration flux. For instance, consider the constricted concave neck region in Figure~\ref{fig:EMCapillaryComp}. {Capillary} flux from both neighboring convex regions is able to flatten the neck. But, {higher} electromigration flux transports material from the right to the neck and draining it from the left. Eventually, this mass transport causes a thinning of the neck, which would lead to a pinch-off event. The daughter slit subsequently propagates independently from the parent inclusion. The trailing parent inclusion may again undergo a shape bifurcation, ejecting a relay of daughter slits each of which may or may not propagate with the same width and velocity. Consequently, coalescence and coarsening become important events. Preliminary phase-field simulations indicate the possibility of such events, as shown in top insets of Figure~\ref{fig:EMLSA}. However, a detailed exposition requires further investigations and should be reported in future publications.

\subsection{Anisotropic inclusions}

Several inclusion morphologies are observed during its electromigration-induced motion under anisotropic surface diffusion. Morphological maps are constructed from {microstructural} evolution obtained from {phase}-field study. These are categorized as: Firstly, steady-state inclusions are the stable faceted morphologies attained after initial alterations. Secondly, time-periodic oscillations are the repetition of the sequence of morphology with a fixed period. Thirdly, in contrast to time-periodic oscillations where one straight edge remains unaffected, zig-zag oscillations are a result of the facet associated with straight edge changes from upper to lower facet and reverses repeatedly. Finally, inclusions breakup are the result of the rupture or disintegration of the inclusion while propagating under the external electric field.

\paragraph*{2-fold symmetry:}

Chapter~\ref{chapter:EM2} exhibits a study to delineate the effect of conductivity contrast ($\beta$) and misorientation of surface diffusion anisotropy ($\varpi$) on inclusion propagating in a \{110\}-oriented single crystal of face-centered-cubic metals. When high diffusivity sites align perpendicular to the electric field ($\varpi \approx 0^\circ$), then the inclusions are likely to breakup for all conductivity contrast. {Higher} values of $\varpi$ establish a steady-state morphology, while intermediate $\varpi$ exhibits time-periodic oscillation. {Formation} of various migration modes is explained in detail with plausible reasons. For instance, inclusion morphologies for a steady-state are compared with time-periodic oscillations in Figure~\ref{fig:r15} and inclusion breakup in Figure~\ref{fig:apexPlots}. 
 
The conductivity ratio of $\beta$ is found to be influential in determining the migration mode, and more importantly, the shapes while propagation. For instance, the slit forming propensity of the inclusion changes from being along the line to perpendicular to the line as $\beta$ increases for steady-state inclusions. In addition, the results on {steady}-state migration (Figure~\ref{r0BetaVelocityGraph}) are an important extension to the analytical theory of Ho \cite{ho1970motion}, which was developed for inclusion motion under isotropic diffusion. In steady-state morphologies, compared to a circular inclusion due to isotropic diffusion, the triangular and slit-shaped inclusions instigated due to two-fold anisotropic diffusion have about three times lower velocity.

\paragraph*{4-fold and 6-fold symmetries:}

 A phase-field numerical study is presented in Chapter~\ref{chapter:4Fold6FoldEM} to investigate inclusions migrating along \{100\} and \{111\} planes of face-centered-cubic crystal. In particular, the emphasis is laid on understanding the effect of conductivity contrast ($\beta$) between the inclusion and the matrix. {These} results demonstrate that the elevated $\beta$ increases traversal lengths of steady-state inclusions. In addition to that, time-periodic oscillations of inclusions transform their crawling motion to {gliding} kinetics at elevated $\beta$. Furthermore, the breakup of inclusions due to elongation substitutes the retention of species at a higher $\beta$. Finally, inclusions of only lower $\beta$ undergo zigzag motion, while their counterparts of higher values break apart.

\begin{figure}[hbt!]
\begin{center}
\includegraphics[scale=1.4]{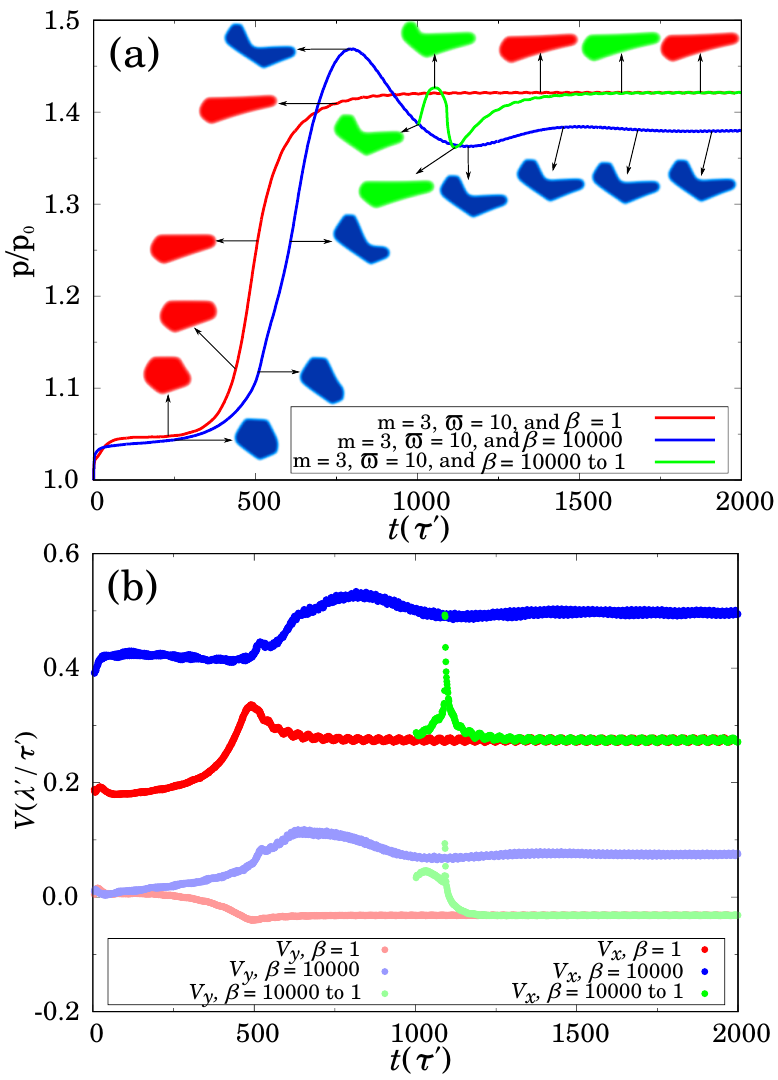}
\caption[Effect due to alteration of the conductivity ratio from $\beta=10000$ to $1$]{The complex shape dynamics of sixfold inclusion at misorientation angle $\varpi = 10^{\circ}$ during the morphological transformation. (a) shows evolution pathway of inclusion on the perimeter curve as a function of time for conductivity ratio, $\beta =1$, $10000$, and the alteration of the conductivity ratio from $\beta=10000$ to $1$ after $t=1000\tau'$. The inset of images show island morphology during propagation. (b) shows evolution history of the velocity components $V_x$ aligned and $V_y$ transverse to the external electric field.}\label{fig:6f_r10}
\end{center}
\end{figure}

{Significance} of $\beta$ can be further extended by adjusting the conductivity contrast during {inclusion} migration. {Electrical} conductivity is a material property, which may vary with external parameters, such as temperature \cite{goli2014thermal, mohsin2017temperature}. Therefore, understanding the effect of conductivity alteration during the inclusion migration has scientific importance. Figure~\ref{fig:6f_r10} depicts morphologies for $\beta=1$ and 10000. When the conductivity ratio changed from 10000 to 1 at time $t=1000\tau'$, the inclusion revamps its longitudinal velocity immediately to match with the newly developed situation. This is in accordance with the inspection during the change in the electric field, as shown in Figure~\ref{fig:potentialChange}. Eventually, the inclusion catches up, in perimeter and shape (green), with the propagation of its counterpart, which is fully developed in the homogeneous system (red). Note that both inclusions (red and green) are identical in shape during its steady-state propagation. However, this is true only for a few cases. For instance, segregated inclusions may not reassemble further. In addition, the inclusion may be in a strange position and reestablishment may lead to further deviation and breakup. These investigations could be performed in future work.

A number of interesting directions can be pursued hereafter.
The nanostructures can be stabilized by reducing the temperature (so that the species diffusivity is inhibited) when the desired morphology has been achieved. {Other} cases would correspond to heteroepitaxial islands i.e. island and the substrate of dissimilar metals. However, {effect} of misfit strain has to be additionally accounted in such cases to faithfully capture the island dynamics similar to the front tracking simulations of Dasgupta et al. \cite{dasgupta2013surface}.

Other driving forces such as thermal gradient can act in conjunction with or counteract the effect of electromigration \cite{chakraborty2018phase,somaiah2016electric}. The Soret effect can be modeled by appending a temperature gradient-dependent species diffusion term in Eq.~\eqref{eq:EMCahnHilliardModified} and supplementing it by a temperature Laplace equation. Thermomigration has been a subject of a number of previous studies \cite{chen2017phase, chen2019phase, zhang2012phase, hu2010phase}. Interestingly boomerang-like topological transition and subsequent splittings as in Figure~\ref{fig:r90} has also been reported during thermomigration \cite{chen2017phase}. A combined understanding of thermomigration and electromigration, however, is still in inchoate stages.

Single crystals have been considered in the present study. Most commercial interconnects are polycrystalline with grain boundary networks. Migrating inclusions can either be pinned to or penetrate into the grain boundary \cite{riege1996influence,ma1993precipitate}. Modeling effort in this direction would require a coupled solution to the Cahn-Hilliard, Allen-Cahn  and Laplace equations as employed in Refs. \cite{mukherjee2016phase, mukherjee2018electromigration}. In addition, isolated inclusions are investigated in the present work. Generally, the metallic conductors are consist of several inclusions simultaneously. The subsequent events of coalescence and splitting between these might lead to shrinkage and/or growth of the inclusions. These issues should be a subject of future work.

{Inclusion} morphologies and dynamics can be tailored in order to form various interesting nanopatterns by monitoring the conductivity contrast. Even though a fixed value of the electric field is employed in the present work, numerous morphologies can be observed. {Variety} of morphologies can be further enhanced by regulating the strength of the electric field or the size of the inclusion. In addition, the incorporation of other fields, such as elasticity flourishes the richness of patterns. {Effect} of these fields could be addressed in upcoming works.

\clearpage

\afterpage{\blankpagewithoutnumberskip}
\clearpage
\newpage
\thispagestyle{empty}
\vspace*{8cm}
\phantomsection\addcontentsline{toc}{chapter}{VI Appendices}
\begin{center}
 \Huge \textbf{Part VI} \\
 \Huge \textbf{Appendices}
\end{center}

\afterpage{\blankpagewithoutnumberskip}
\clearpage
\begin{appendices}

\chapter{Interface profiles}
\label{app:interfaceProfiles}
At equilibrium, the partial derivative of the free energy functional expressed in Eq.~\eqref{eq:EMfreeEnergyFunctional} should vanishes. Consequently, 
\begin{equation}
\frac{\delta F(c,\boldsymbol{\nabla} c)}{\delta c} = \frac{\partial f(c)}{\partial c} - \kappa \boldsymbol{\nabla} \cdot \boldsymbol{\nabla} c =0.
\end{equation}
Assuming that the spatial dependency of the order parameter $c$ is restricted to X-direction, writing one-dimensional form of the above equation, 
\begin{equation}
\frac{\partial f(c)}{\partial c} - \kappa \frac{\textrm{d}^2 c}{\textrm{d} x^2}=0.\label{eq:1DVariationalDerivative}
\end{equation}
Multiplying both sides of Eq.~\eqref{eq:1DVariationalDerivative} by $\textrm{d}c/\textrm{d}x$ and further manipulations would realize, the equation of the form,
\begin{equation}
\frac{\textrm{d}c}{\textrm{d}x} = \sqrt{\frac{2}{\kappa}} \sqrt{f(c)}. \label{eq:1DVariationalDerivativeSimplified}
\end{equation}
This expression is referred as Euler-Lagrange relation. To obtain interface profile of double-well free energy density, consider a simplified form, $f^{\textrm{dw}}(c) = c^2(1-c)^2$. Plugging this into Eq.~\eqref{eq:1DVariationalDerivativeSimplified} and taking integral, the interface profile is expressed as,
\begin{equation}
c = 0.5+0.5\left(\textrm{tanh}\frac{x}{\sqrt{2\kappa}}\right).
\end{equation}
This implies that the well-type free energy assumes hyperbolic-tangent function at the interface. To determine interface profile of double-obstacle, considering the free energy of the form $f^{\textrm{ob}}(c) =X_A c(1-c)$, similar manipulations to the well type, the interface profile is expressed as,
\begin{equation}
c=0.5+0.5 \left(\textrm{cos}\sqrt{\frac{2X_A}{\kappa}}x\right) .
\end{equation}  

\afterpage{\blankpagewithoutnumberskip}
\clearpage

\chapter{Preparation of electrode microstructure with a single particle}
\label{appendix:singleParticleMicrostructure}

\begin{figure}[hbt!]
\begin{center}
\includegraphics[scale=0.60]{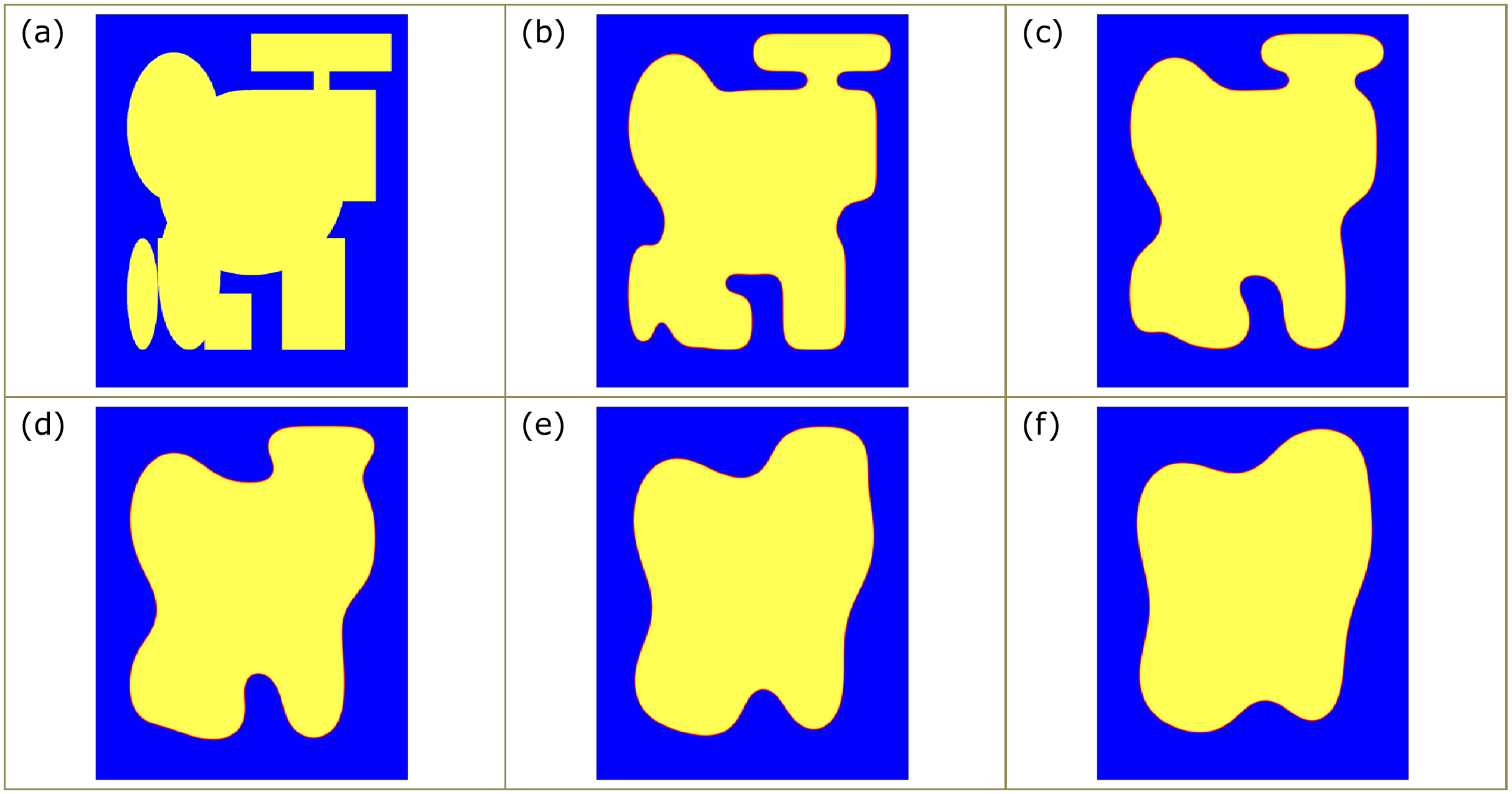}
\end{center}
\caption[Preparation of 2D irregularly shaped particle.]{Morphological evolution of irregularly shaped particle. The time increment follows (a) to (f). The blue region indicates electrolyte ($\phi_1=0$) and yellow color represents electrode particle ($\phi_1=1$). Except (a), other insets from (b) to (f) consist of diffuse interface between the electrode particle and the electrolyte.}\label{fig:AllenCahnParticleEvolution}
\end{figure}

The smooth profile of the domain parameter $\phi_1$ in the simulation domain is obtained by solving the non-conserved Allen-Cahn equation~\cite{allen1979a},
 \begin{equation}
 \tau_p \frac{\partial \phi_1}{\partial t} = - \frac{\partial f^{\textrm{dw}}(\phi_1)}{\partial \phi_1} + \epsilon_p^2  \boldsymbol{\nabla} \cdot \boldsymbol{\nabla} \phi_1 
\label{eq:allencahn}
 \end{equation}
for a few initial time steps, where $\epsilon_p$\nomenclature{$\epsilon_p$}{parameter related to interface thickness in Allen-Cahn equation} is related to the interface thickness, $\tau_p(=L^2/D_1)$ denotes relaxation coefficient, and $f^{\textrm{dw}}(\phi_1)$\nomenclature{$f^{\textrm{dw}}$}{double-well free energy function} is a double-well free energy function expressed as $f^{\textrm{dw}}(\phi_1) = \phi_1^2 (1-\phi_1)^2.$ The irregularly shaped particle is obtained by a strategical distribution of rectangles and ellipses, as shown in Figure~\ref{fig:AllenCahnParticleEvolution}. Furthermore, an originally sharp domain boundary is smoothed by Eq.~(\ref{eq:allencahn}) to yield a diffuse interface with a finite thickness given by $0 <\phi_1<1$. 

Simulations are performed in two independent stages. The Allen-Cahn equation \eqref{eq:allencahn} is solved just initially for a few time steps. The same serves two objectives: Firstly, to obtain the irregularly shaped particle of the desired shape and secondly, to get smooth interface between the cathode particle and the electrolyte. The equation is terminated once the above-mentioned objectives are obtained. Thereafter, the lithiation process starts as described in Chapter~\ref{chapter:NeumannConstantFlux}, where the results related to the lithium insertion are presented. For instance, Figure~\ref{fig:AllenCahnParticleEvolution}(f) can be considered as an input particle profile for the lithiation, while the evolution of order parameter $\phi_1$ is abolished thereafter. To discriminate between the stationary and evolving particles, the evolving parameter $\phi_1$ are referred to stationary parameter $\psi_1$ in the main body of the dissertation (specifically, in Chapters~\ref{chapter:phaseFieldModelLIB}, \ref{chapter:NeumannConstantFlux}, and \ref{chapter:Potentiostatic}).

\chapter{Preparation of electrode microstructure with several particles}
\label{sec:electrodeFormation}
The electrode structure is characterized numerically by employing volume preserved technique on multiphase Allen-Cahn formulation. Each particle is considered as a separate phase freely evolving in the liquid electrolyte phase. A brief procedure is outlined in the following paragraphs.

 The inhomogeneous system consists of different phases whose identity is distinguished by its physical state or orientation. In the present description, a multi-phase Allen Cahn model is considered to study the evolution of multiphase systems whose interfacial energy decreases with preserving their volume. The evolution of a general system containing $N$ phases is governed by Ginzburg-Landau free energy,
\begin{equation}
F=\int_{V_\Omega} f(\boldsymbol{\phi},\boldsymbol{\nabla} \boldsymbol{\phi})\textrm{d}{V_\Omega}=\int_{V_\Omega} \bigg(\epsilon_p a^{\textrm{ac}}(\boldsymbol{\phi},\boldsymbol{\nabla} \boldsymbol{\phi})+\frac{1}{\epsilon_p}w^{\textrm{ac}}(\boldsymbol{\phi})+g^{\textrm{ac}}(\boldsymbol{\phi})\bigg) \textrm{d}{V_\Omega},
\end{equation}
where $f(\boldsymbol{\phi},\boldsymbol{\nabla} \boldsymbol{\phi})$ is the free energy density, ${V_\Omega}$ is the domain under consideration, a vector phase-field parameter \nomenclature{$\boldsymbol{\phi}$}{phase-field parameter set for Allen-Cahn equation}$\boldsymbol{\phi}(\boldsymbol{x},t)=[\phi_1(\boldsymbol{x},t),\ldots,\phi_N(\boldsymbol{x},t)]$, \nomenclature{$\phi_a$}{evolving indicator parameter for phase $a$ in Allen-Cahn equation}where $\phi_a \in [0,1]$, $\forall a \in \{1,2,\ldots,N\}$, $\boldsymbol{x}$ represents spatial coordinates and $\epsilon_p$ denotes the thickness of the diffuse interface. The gradient energy density is expressed as \cite{prajapati2018modeling},
\begin{equation}
a^{\textrm{ac}}(\boldsymbol{\phi},\boldsymbol{\nabla} \boldsymbol{\phi}) = \sum_{\substack{a, b=1 \\ (a < b)}}^{N,N} \gamma_{a b} a{^2_{a b}}(\boldsymbol{\phi}, \boldsymbol{\nabla} \boldsymbol{\phi}) \vert \boldsymbol{q}_{a b}\vert^2
\end{equation}
where $ \gamma_{a b}$ denotes the interfacial energy density between $a$ and $b$ phases\nomenclature{$ \gamma_{a b}$ (or $ \gamma_{a b l}$)}{interfacial energy density between $a$ and $b$ (or $a$, $b$, and $l$) phases}, $\boldsymbol{q}_{a b} = \phi_a \boldsymbol{\nabla} \phi_b - \phi_b \boldsymbol{\nabla} \phi_a$, which defines the gradient vector in the normal direction to $a-b$ interface\nomenclature{$\boldsymbol{q}_{a b}$}{gradient vector in the normal direction to $a-b$ interface}. The function $a{_{a b}}(\boldsymbol{\phi},  \boldsymbol{\nabla} \boldsymbol{\phi})$ represents the interface anisotropy\nomenclature{$a{_{a b}}$}{a-b interface anisotropy function}. For the present study, a faceted anisotropy is considered of the form,
\begin{equation}
a{_{a b}}(\boldsymbol{\phi}, \boldsymbol{\nabla} \boldsymbol{\phi}) = \max_{1\leq i \leq \eta_{a b}} \Bigg\{ \frac{\boldsymbol{q}_{a b}}{\vert \boldsymbol{q}_{a b} \vert} \cdot \boldsymbol{\eta}_{i,a b} \Bigg\}, \label{eq:AllenCahnPositionVectors}
\end{equation}
where $\{\boldsymbol{\eta}_{i,a b} \vert i=1,\ldots, \eta_{a b}\}$\nomenclature{${\boldsymbol{\eta}_{i,a b}}$}{position vectors in the Wulff shape} represents position vectors in the Wulff shape of phase $a$ with respect to phase $b$. The bulk energy density potential $w^{\textrm{ac}}(\boldsymbol{\phi})$\nomenclature{$w^{\textrm{ac}}$}{bulk energy density potential} is assumed to be a multiobstacle type,
\begin{equation}
w^{\textrm{ac}}(\boldsymbol{\phi}) = \frac{16}{\pi^2} \sum_{\substack{a,b=1 \\ (a<b)}}^{N,N} \gamma_{a b} \phi_a \phi_b + \sum_{\substack{a,b,l=1 \\ (a<b<l)}}^{N,N,N} \gamma_{a b l} \phi_a \phi_b \phi_l.
\end{equation}
The higher order interfacial energy term $\gamma_{a b l}$ penalizes the ghost phase occurrences at the interfaces between two phases. The additional bulk free energy density $g^{\textrm{ac}}(\boldsymbol{\phi})$\nomenclature{$g^{\textrm{ac}}$}{additional bulk energy density} is responsible for the volume preservation, can be expressed as,
\begin{equation}
g^{\textrm{ac}}(\boldsymbol{\phi})= \sum_{a=1}^{N} \Upsilon_a h(\phi_a),
\end{equation}
where the function $h(\phi_a)=\phi^3_a (6 \phi_a^2 - 15\phi_a + 10)$ interpolates the free energy density terms $\Upsilon_a$\nomenclature{$\Upsilon_a$}{time-dependent antiforce term for phase $a$} between the bulk phases. The time-dependent antiforce terms $\boldsymbol{\Upsilon}(t)\in \{ \Upsilon_1,\ldots,\Upsilon_N\}$ are calculated at each timestep to maintain the phase volume equal to initially prescribed. For a detailed algorithm, the readers are suggested to Ref. \cite{nestler2008phase}. Finally, the evolution of the phases is governed by a general model for multi-phase Allen-Cahn type equation as,
\begin{equation}
\tau_p \epsilon_p \frac{\partial \phi_a}{\partial t} = -\frac{\delta  f(\boldsymbol{\phi},\boldsymbol{\nabla} \boldsymbol{\phi})}{\delta \phi_a} - \frac{1}{N} \sum_{a=1}^{N} \frac{\delta  f(\boldsymbol{\phi},\boldsymbol{\nabla} \boldsymbol{\phi})}{\delta \phi_a}. \label{eq:AllenCahnEvolution}
\end{equation}  
where $\tau_p$ denotes the relaxation coefficient, $\partial(\bullet)/\partial t$ is partial derivative with time $t$ and $\delta(\bullet)/\delta \phi_a$ denotes functional derivative. The second term in the above equation ensures the constraint $\sum_{a=1}^{N}\phi_a(\boldsymbol{x},t)=1$.

Inherently, the isolate boundary condition implies $\boldsymbol{\nabla} \phi_a \cdot \boldsymbol{n}=0$, where $\boldsymbol{n}$ is the normal vector at the wall of the simulation boundary. In other words, the phases form a $90^\circ$ angle at the boundary. To specify other  contact angles, an extension is made in the form of
\begin{equation}
-\epsilon_p \frac{\partial a(\boldsymbol{\phi}, \boldsymbol{\nabla} \boldsymbol{\phi})}{\partial \boldsymbol{\nabla} \phi_a} \cdot \boldsymbol{n} - \frac{\partial f_b(\boldsymbol{\phi})}{\partial \phi_a} -\Bigg\{\frac{1}{N} \sum_{\phi_a =1}^N\Bigg(-\epsilon_p \frac{\partial a(\boldsymbol{\phi},\boldsymbol{\nabla} \boldsymbol{\phi})}{\partial \boldsymbol{\nabla} \phi_a} \cdot \boldsymbol{n} - \frac{\partial f_b(\boldsymbol{\phi})}{\partial \phi_a}\Bigg) \Bigg\} =0, \label{eq:AllenCahnBoundary}
\end{equation}
where $f_b(\boldsymbol{\phi}) = \sum_{\phi_a =1}^N (\gamma_{a b} h(\phi_a))$ denotes the boundary function and the terms $\gamma_{a b}$ represent the interfacial energy density between $a$ phase and the boundary, the value of these parameters controls the contact angle between the boundary and different phases from $0^\circ$(complete non-wetting) to $180^\circ$ (complete wetting). The term in the parentheses in equation \eqref{eq:AllenCahnBoundary} ensures the sum of all phases at any location in the simulation domain is unity.

\begin{figure}
\centering
\includegraphics[scale=0.5]{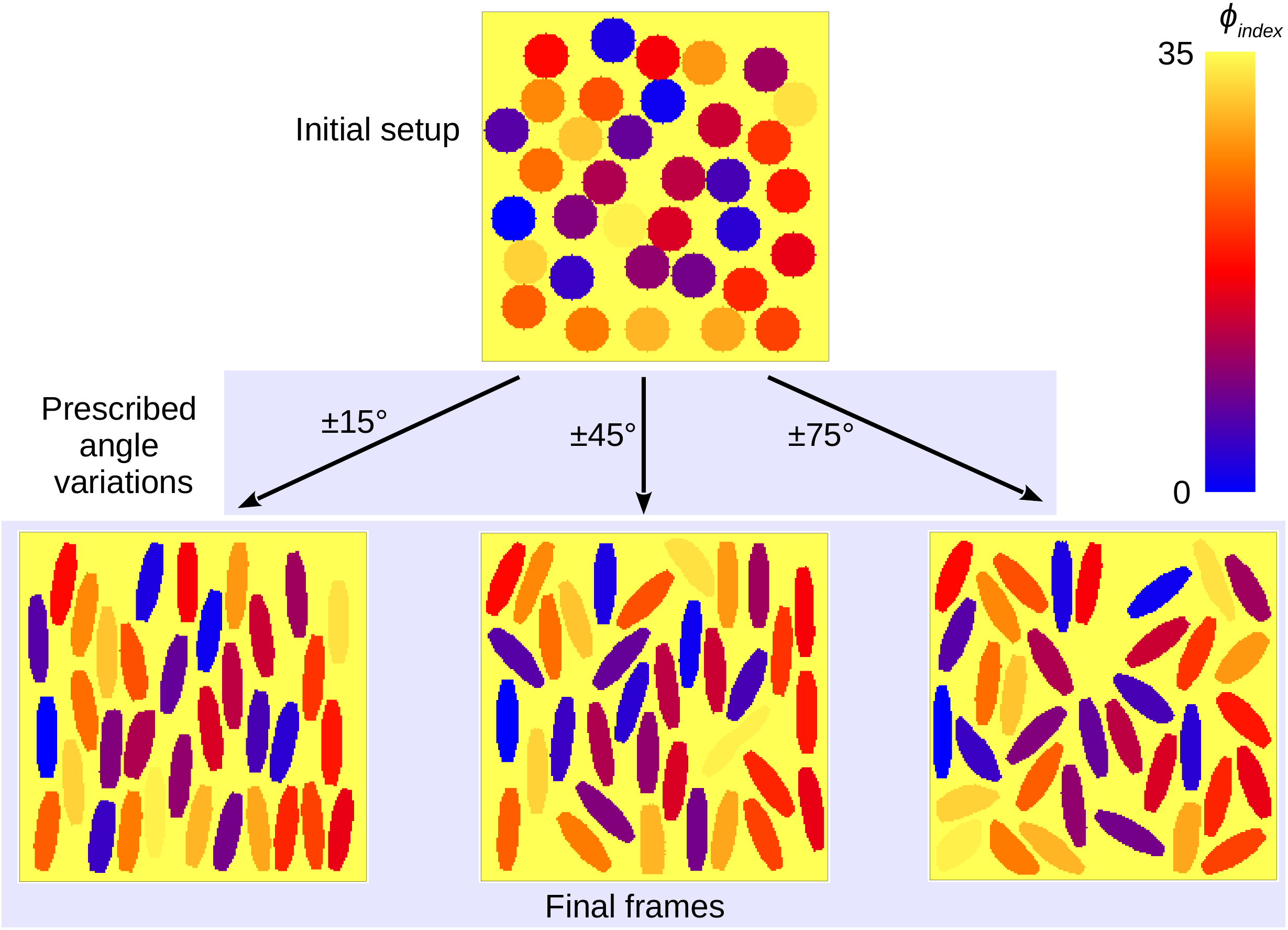}
\caption[Preparation of multiple particles electrode microstructure]{Represetative cases of morphological evolution of initially circular electrode particles to prescribed geometrical shape and orientations. The dominant values of $\phi_{a}$ (i.e. where $\phi_{a}>0.5$) are demontrated in one frame with a parameter $\phi_{\textrm{index}}$, where $\phi_0$ to $\phi_{34}$ are separate electrode particles and $\phi_{35}$ corresponds to the electrolyte.}\label{fig:angleVariations}
\end{figure}

The electrode mictrostructure is obtained from random distribution of $N-1$ circular (or spherical in 3D) particles in the domain of electrolyte. Each particle corresponds to a separate phase $\phi_a$ and the last phase $\phi_N$ denotes the electrolyte. These phases are allowed to evolve under $N$ equations \eqref{eq:AllenCahnEvolution} with volume preserved for non-wetting boundary condition \eqref{eq:AllenCahnBoundary}. The particles evolve to a geometry specified by the position vectors $\boldsymbol{\eta}_{i,a b}$ to the vertices in equation \eqref{eq:AllenCahnPositionVectors}. To obtain the microstructure for the presented work in Chapter~\ref{chapter:Potentiostatic}, four $(\pm0.8,\pm0.1)$ and four $(\pm0.4,\pm0.2)$ vertices for 2D simulations and eight $(\pm0.8,\pm0.1,\pm0.1)$, eight $(\pm0.4,\pm0.1,\pm0.2)$, and eight $(\pm0.4,\pm0.25,\pm0.1)$ vertices for 3D simulations are considered. For each particles, these vertices are rotated at a specified orientation $\theta_1$ in xy plane for 2D ($\theta_1$ in xy, $\theta_2$ in yz, and $\theta_3$ in xz planes for 3D) domain. For instance, Figure~\ref{fig:angleVariations} shows particles are rotated using a random number generator in the range of $\pm15^\circ$, $\pm45^\circ$, and $\pm75^\circ$. The obtained setups are then utilized for the study of insertion of lithium species in the particles completely submerged in the liquid electrolyte, which is described Chapter~\ref{chapter:Potentiostatic}.

\afterpage{\blankpagewithoutnumberskip}
\clearpage

\chapter{Phase separation in Electrode along with homogeneous mixture in Electrolyte}
\label{subsection:ficksLawElectrolyte}
The behavior of two-phase coexistence in the electrode compounds are well known \cite{ohzuku1990electrochemistry, vanderven2000phase, liu1998mechanism, yang1999situ}. However, instead of phase separation, a homogeneous distribution is apparent in the electrolyte. Both behaviors can be achieved by careful consideration of the free energy parameter $\alpha''_{a}$ in Eq.~\eqref{eq:libFreeEnergyDensity} and the gradient energy parameter $\kappa_a$ in Eq.~\eqref{eq:LIBGradientEnergyCoefficeintInterpolation}. The Cahn-Hilliard equation is employed for the study of phase separating behavior in ample literature \cite{zhang2018nonlocal, stein2016effects, dileo2014a, welland2015miscibility, santoki2018phase, huttin2012phase, walk2014comparison}. Therefore, the objective of the present analysis is only to demonstrate that for a specific choice of these parameters, the Cahn-Hilliard equation exactly corresponds to the Fick's dilute solution model \cite{zhang2007numerical}.

 Considering a special case, where any phase $a$ consists the regular solution parameter $\alpha''_{a}=0$ and the gradient energy parameter $\kappa_a=0$. The chemical potential of such system can be expressed as,
\begin{equation}
\mu_a = \alpha'_{a} +  \frac{T}{T_{\textrm{ref}}}(\ln{c} - \ln(1-{c})).
\end{equation}
Here $\mu_a$ is the chemical potential in the phase $a$. The species flux can be written as,
\begin{equation}
J_a = -M_a\nabla \mu_a = -D_a  \frac{T}{T_{\textrm{ref}}} \nabla c. \label{eq:ficksLawEquivalent}
\end{equation}
Here $J_a$ and $M_a$ represent species flux and mobility in the phase $a$ respectively. Eq.~\eqref{eq:ficksLawEquivalent} is equivalent to the equation responsible for the homogeneous mixture. Therefore, parameters $\alpha''_{a}=0$ and $\kappa_a=0$ can be utilized for the regions exhibiting non-phase separating behaviors, such as electrolyte.

\afterpage{\blankpagewithoutnumberskip}
\clearpage
\chapter{Calculation of tortuosity}
\label{sec:calculationOfTortuosity}
The electrode particles are surrounded by the liquid electrolyte, which conducts species flux in the electrode. The species diffusivity of the electrode is much lower than the electrolyte \cite{fuller1994simulation}. Also, the insertion and the extraction of species in the electrode particles takes place at the interface between the electrode and the electrolyte. Therefore, understanding of tortuous pathways in the electrolyte is of scientific interest. In the present study, instead of assuming a Bruggeman relation \cite{bruggeman1935berechnung}, a computational tool is employed to calculate the tortuosity of the electrolyte microstructure considering infinite resistive electrode particles \cite{joos2011reconstruction}. 

\begin{figure}
\centering
\includegraphics[scale=0.12]{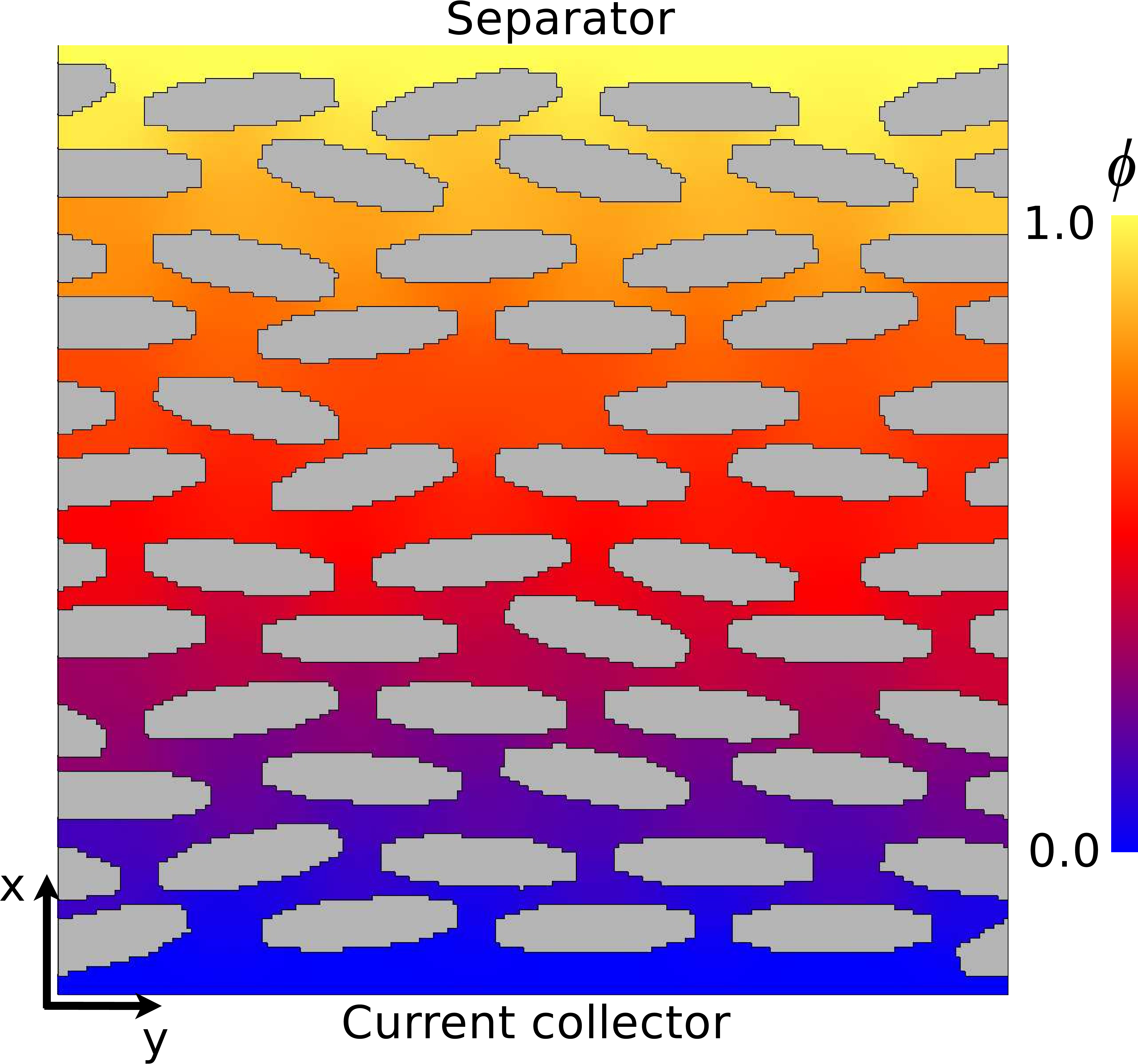}
\caption{Schematic of potential distribution in the electrolyte during the tortuosity calculation.}\label{fig:tortuositySchematic}
\end{figure}

The tortuosity is calculated in the x-direction, where the species flux propagates, from the separator to the current collector. A potential difference $\phi_{\infty}$ is applied between two ends, which is predefined as 1.0V at separator and 0.0V at the current collector for the present investigation as shown in Figure~\ref{fig:tortuositySchematic}. While no-flux boundary conditions applied on the remaining two (four for 3D) boundaries. The potential distribution inside the simulation domain is calculated using the Laplace equation of the form,
\begin{equation}
\boldsymbol{\nabla} \cdot (-\sigma_{\textrm{ety}}\boldsymbol{\nabla} \phi) =0.
\end{equation}
Where $\sigma_{\textrm{ety}}$\nomenclature{$\sigma_{\textrm{ety}}$}{electrical conductivity of the electrolyte} is the electrical conductivity of the electrolyte. Afterwards, the current density $i= \sigma_{\textrm{ety}}\nabla \phi$ can be estimated at the each pixels (voxels in 3D). The total current flow from the cross-section perpendicular to the x-direction can be obtained as a summation,
\begin{equation}
I_{\textrm{m}} = \int_{A_{\perp}} i_{\perp} \,\,\,dA_{\perp}.
\end{equation}
Where $A_{\perp}$ is the area of the cross-section surface under consideration\nomenclature{$I_{\textrm{m}}$ (or $I_{\textrm{i}}$)}{measured (or ideal) current flow}. Note that, as the no-flux conditions prescribed at the remaining boundaries, due to conservation, the measured current flow at any cross-section perpendicular to the x-direction should be nearly equal. If not then the meaningful conclusions can not be drawn from the presented method. For instance, in the case where there is no path for species to flow across when electrode particles completely block the path by forming an entire blockage in the perpendicular direction to the x-axis.

The tortuosity is a measure of conductive pathways deviated from the ideal straight channels of uniform cross-section. The tortuosity can be calculated based on the actual current flow in the microstructure in contemplation to the current flow in the ideal path without any discontinuities,
\begin{equation}
\tau = \frac{I_{\textrm{i}}}{I_{\textrm{m}}},
\end{equation} 
where $I_{\textrm{i}}$ is the ideal current flow, determined from the analytical expression $I_{\textrm{i}}=\sigma_{\textrm{ety}}\phi_{\infty}A_{\perp}/l$ and $l$ is the length of the domain in the direction of the x-axis.

\end{appendices}

\pagestyle{fancy}
\fancyhf{}
\lhead[\thepage]{List of symbols and abbreviations}      
\rhead[\thesection]{\thepage}
\cleardoublepage\phantomsection\addcontentsline{toc}{chapter}{List of Symbols and Abbreviations}

\begin{thenomenclature} 
	\nomgroup{A}
	\item [{$ \Delta x $, $\Delta y $, and $\Delta z$}]\begingroup simulation cell widths in $x$-, $y$- and $z$-directions respectively\nomeqref {3.28}\nompageref{47}
	\item [{$ \gamma_{a b}$ (or $ \gamma_{a b l}$)}]\begingroup interfacial energy density between $a$ and $b$ (or $a$, $b$, and $l$) phases\nomeqref {C.2}\nompageref{181}
	\item [{$\alpha$}]\begingroup index of a lattice site\nomeqref {2.0}\nompageref{22}
	\item [{$\alpha'$}]\begingroup the neighboring interstitial lattice site\nomeqref {2.1}\nompageref{23}
	\item [{$\alpha'_a$ and $\alpha''_a$}]\begingroup regular solution parameters of phase $a$\nomeqref {3.3}\nompageref{39}
	\item [{$\beta$}]\begingroup conductivity ratio, $\sigma_{\textrm{mat}}/\sigma_{\textrm{icl}}$\nomeqref {7.20}\nompageref{106}
	\item [{$\boldsymbol{\phi}$}]\begingroup phase-field parameter set for Allen-Cahn equation\nomeqref {C.1}\nompageref{181}
	\item [{$\boldsymbol{\psi}$}]\begingroup indicator parameter set\nomeqref {3.1}\nompageref{38}
	\item [{$\boldsymbol{J}$}]\begingroup effective diffusional mass fluxes\nomeqref {2.26}\nompageref{32}
	\item [{$\boldsymbol{J}_A$}]\begingroup mass flux of A species\nomeqref {2.28}\nompageref{32}
	\item [{$\boldsymbol{J}_B$}]\begingroup mass flux of B species\nomeqref {2.28}\nompageref{32}
	\item [{$\boldsymbol{J}_i$}]\begingroup flux of the diffusing species $i$\nomeqref {4.5}\nompageref{54}
	\item [{$\boldsymbol{J}_{e}$}]\begingroup flux of the charge carrier\nomeqref {4.11}\nompageref{56}
	\item [{$\boldsymbol{n}$}]\begingroup inward pointing unit normal to the surface\nomeqref {3.6}\nompageref{40}
	\item [{$\boldsymbol{q}_{a b}$}]\begingroup gradient vector in the normal direction to $a-b$ interface\nomeqref {C.2}\nompageref{181}
	\item [{$\boldsymbol{x}_\alpha$}]\begingroup spatial position of the $\alpha$ site\nomeqref {2.15}\nompageref{28}
	\item [{$\boldsymbol{x}_{\alpha B}$}]\begingroup neighbor on the back side of the $\alpha$ site\nomeqref {2.17}\nompageref{28}
	\item [{$\boldsymbol{x}_{\alpha F}$}]\begingroup neighbor on the front side of the $\alpha$ site\nomeqref {2.17}\nompageref{28}
	\item [{$\boldsymbol{x}_{\alpha L}$}]\begingroup neighbor on the left side of the $\alpha$ site\nomeqref {2.17}\nompageref{28}
	\item [{$\boldsymbol{x}_{\alpha O}$}]\begingroup neighbor on the bottom side of the $\alpha$ site\nomeqref {2.17}\nompageref{28}
	\item [{$\boldsymbol{x}_{\alpha R}$}]\begingroup neighbor on the right side of the $\alpha$ site\nomeqref {2.17}\nompageref{28}
	\item [{$\boldsymbol{x}_{\alpha T}$}]\begingroup neighbor on the top side of the $\alpha$ site\nomeqref {2.17}\nompageref{28}
	\item [{$\boldsymbol{x}_{\alpha'}$}]\begingroup spatial position of the $\alpha'$ site\nomeqref {2.16}\nompageref{28}
	\item [{$\chi$}]\begingroup dimensionless number associated with the steady state shapes of an island\nomeqref {7.8}\nompageref{103}
	\item [{$\chi_0$}]\begingroup dimensionless number associated with the stability of circular island/inclusion\nomeqref {7.13}\nompageref{104}
	\item [{$\chi_{0\textrm{ct}}$}]\begingroup critical value of $\chi_0$, above which circular island/inclusion is unstable\nomeqref {7.13}\nompageref{104}
	\item [{$\delta c_\alpha$ (or $\delta c_{\alpha'}$)}]\begingroup fluctuation of $c_\alpha$ (or $c_{\alpha'}$) compared to the average\nomeqref {2.6}\nompageref{24}
	\item [{$\Delta H$}]\begingroup change in internal energy of the system\nomeqref {2.13}\nompageref{25}
	\item [{$\Delta S$}]\begingroup change in entropy of the system\nomeqref {2.13}\nompageref{25}
	\item [{$\Delta {t}$}]\begingroup time step increment\nomeqref {3.31}\nompageref{48}
	\item [{$\delta_s$}]\begingroup interface width\nomeqref {4.15}\nompageref{58}
	\item [{$\epsilon$}]\begingroup interface thickness coefficient\nomeqref {5.2}\nompageref{71}
	\item [{$\epsilon_p$}]\begingroup parameter related to interface thickness in Allen-Cahn equation\nomeqref {B.1}\nompageref{179}
	\item [{$\eta$}]\begingroup slit characteristic parameter\nomeqref {7.48}\nompageref{121}
	\item [{$\gamma$}]\begingroup aspect ratio of ellipse\nomeqref {5.2}\nompageref{71}
	\item [{$\gamma_s$}]\begingroup interfacial energy\nomeqref {4.15}\nompageref{58}
	\item [{$\hat{\bullet}$}]\begingroup normalized quantity of the respective entities\nomeqref {3.25}\nompageref{46}
	\item [{$\hat{\kappa}_s$}]\begingroup non-dimensional curvature of the island surface\nomeqref {7.7}\nompageref{103}
	\item [{$\hat{M}_{\mathrm{max}}$}]\begingroup maximum of dimensionless mobility $\hat{M}$\nomeqref {3.32}\nompageref{49}
	\item [{$\kappa$}]\begingroup gradient energy coefficient\nomeqref {2.23}\nompageref{30}
	\item [{$\kappa_a$}]\begingroup gradient energy coefficient of phase $\alpha$\nomeqref {3.4}\nompageref{39}
	\item [{$\kappa_s$}]\begingroup curvature along the island/inclusion surface\nomeqref {7.4}\nompageref{103}
	\item [{$\lambda'$}]\begingroup reference length scale for normalization\nomeqref {4.23}\nompageref{60}
	\item [{$\Lambda_L$}]\begingroup ratio of line width $w$ to the initial inclusion radius $R_i$\nomeqref {8.0}\nompageref{132}
	\item [{$\mathcal{O}$}]\begingroup order of approximation\nomeqref {3.28}\nompageref{47}
	\item [{$\mathring{a}$}]\begingroup lattice parameter\nomeqref {2.19}\nompageref{28}
	\item [{$\mu$}]\begingroup chemical potential\nomeqref {2.2}\nompageref{23}
	\item [{$\mu^{\mathrm{chem}}$}]\begingroup chemical potential corresponding to the bulk free energy density\nomeqref {3.8}\nompageref{41}
	\item [{$\mu^{\mathrm{grad}}$}]\begingroup chemical potential corresponding to the gradient free energy density\nomeqref {3.9}\nompageref{41}
	\item [{$\mu_A$}]\begingroup chemical potentials of the A species\nomeqref {2.30}\nompageref{32}
	\item [{$\mu_B$}]\begingroup chemical potentials of the B species\nomeqref {2.30}\nompageref{32}
	\item [{$\mu_{a}^{\mathrm{chem}}$}]\begingroup chemical potential corresponding to the bulk free energy density of $a$ phase\nomeqref {3.8}\nompageref{41}
	\item [{$\mu_{a}^{\mathrm{grad}}$}]\begingroup chemical potential corresponding to the gradient free energy density of $a$ phase\nomeqref {3.9}\nompageref{41}
	\item [{$\nu$}]\begingroup denominator coefficient\nomeqref {3.22}\nompageref{45}
	\item [{$\Omega$}]\begingroup atomic volume\nomeqref {4.21}\nompageref{59}
	\item [{$\partial V_\Omega$}]\begingroup simulation domain boundary surfaces\nomeqref {3.23}\nompageref{45}
	\item [{$\phi _\infty $}]\begingroup electrical potential boundary condition\nomeqref {4.0}\nompageref{52}
	\item [{$\phi$}]\begingroup local electrical potential\nomeqref {4.6}\nompageref{55}
	\item [{$\phi_a$}]\begingroup evolving indicator parameter for phase $a$ in Allen-Cahn equation\nomeqref {C.1}\nompageref{181}
	\item [{$\pi$}]\begingroup Archimedes' constant\nomeqref {4.17}\nompageref{58}
	\item [{$\psi$}]\begingroup orthogonal function to the electrical potential\nomeqref {7.23}\nompageref{107}
	\item [{$\psi_a$}]\begingroup stationary indicator parameter for phase $a$\nomeqref {3.1}\nompageref{38}
	\item [{$\rho $}]\begingroup porosity\nomeqref {6.0}\nompageref{90}
	\item [{$\rho_{\textrm{mat}}$}]\begingroup electrical resistivity of the matrix\nomeqref {4.23}\nompageref{60}
	\item [{$\sigma$}]\begingroup electrical conductivity dependent on order parameter $c$\nomeqref {4.14}\nompageref{56}
	\item [{$\sigma_{\textrm{ety}}$}]\begingroup electrical conductivity of the electrolyte\nomeqref {E.1}\nompageref{187}
	\item [{$\sigma_{\textrm{icl}}$}]\begingroup conductivity of an inclusion\nomeqref {4.14}\nompageref{56}
	\item [{$\sigma_{\textrm{mat}}$}]\begingroup conductivity of the matrix\nomeqref {4.14}\nompageref{56}
	\item [{$\tau $}]\begingroup tortuosity\nomeqref {6.0}\nompageref{90}
	\item [{$\tau'$}]\begingroup reference time scale for normalization\nomeqref {4.23}\nompageref{60}
	\item [{$\tau_p$}]\begingroup relaxation coefficient\nomeqref {2.34}\nompageref{34}
	\item [{$\theta$}]\begingroup angle formed by the local tangent at the inclusion surface\nomeqref {4.11}\nompageref{56}
	\item [{$\Upsilon_a$}]\begingroup time-dependent antiforce term for phase $a$\nomeqref {C.5}\nompageref{182}
	\item [{$\varepsilon$}]\begingroup perturbation parameter\nomeqref {7.10}\nompageref{104}
	\item [{$\varepsilon_t$}]\begingroup difference between the A--B bond energy and the average of A-A and B-B bond energies\nomeqref {2.14}\nompageref{25}
	\item [{$\varepsilon_{AA}$}]\begingroup bond energy between A--A species\nomeqref {2.13}\nompageref{25}
	\item [{$\varepsilon_{AB}$}]\begingroup bond energy between A--B species\nomeqref {2.13}\nompageref{25}
	\item [{$\varepsilon_{BB}$}]\begingroup bond energy between B--B species\nomeqref {2.13}\nompageref{25}
	\item [{$\varkappa$}]\begingroup perturbation function\nomeqref {7.10}\nompageref{104}
	\item [{$\varpi$}]\begingroup misorientation angle\nomeqref {4.11}\nompageref{56}
	\item [{$\xi$}]\begingroup discrepancy coefficient\nomeqref {7.45}\nompageref{120}
	\item [{$\zeta$}]\begingroup transformed fraction\nomeqref {6.2}\nompageref{92}
	\item [{$A$}]\begingroup strength of anisotropy\nomeqref {4.11}\nompageref{56}
	\item [{$a$}]\begingroup index of the phase\nomeqref {3.1}\nompageref{38}
	\item [{$A_0$}]\begingroup amplitude of the perturbation at time $t=0$\nomeqref {4.21}\nompageref{60}
	\item [{$a_n$}]\begingroup surface perturbation coefficient corresponds to n$^{\textrm{th}}$-term\nomeqref {7.24}\nompageref{107}
	\item [{$A_t$}]\begingroup amplitude of the perturbation at time $t$\nomeqref {4.22}\nompageref{60}
	\item [{$a{_{a b}}$}]\begingroup a-b interface anisotropy function\nomeqref {C.2}\nompageref{181}
	\item [{$B$}]\begingroup Mullins' constant\nomeqref {4.21}\nompageref{59}
	\item [{$B_{0m}$}]\begingroup the internal system energy parameter\nomeqref {2.1}\nompageref{23}
	\item [{$B_{m}$}]\begingroup two-body interaction field from neighboring lattices\nomeqref {2.1}\nompageref{23}
	\item [{$C$}]\begingroup C-rate\nomeqref {5.1}\nompageref{68}
	\item [{$c$}]\begingroup concentration/order parameter\nomeqref {2.5}\nompageref{23}
	\item [{$c^H$}]\begingroup higher concentration level of a phase-separated state\nomeqref {5.3}\nompageref{74}
	\item [{$c^L$}]\begingroup lower concentration level of a phase-separated state\nomeqref {5.3}\nompageref{74}
	\item [{$c_\alpha$}]\begingroup state of the lattice site $\alpha$\nomeqref {2.0}\nompageref{22}
	\item [{$c_A$}]\begingroup concentration of A species\nomeqref {2.28}\nompageref{32}
	\item [{$c_B$}]\begingroup concentration of B species\nomeqref {2.28}\nompageref{32}
	\item [{$c_i$}]\begingroup initial concentration filling\nomeqref {5.2}\nompageref{71}
	\item [{$c_n$}]\begingroup Dirichlet concentration boundary condition at the particle surface\nomeqref {3.20}\nompageref{44}
	\item [{$c_s$}]\begingroup concentration at the surface of the particle\nomeqref {5.1}\nompageref{68}
	\item [{$c_{\alpha'}$}]\begingroup state of {the} neighboring interstitial lattice state $\alpha'$\nomeqref {2.1}\nompageref{23}
	\item [{$d$}]\begingroup dimensions of the simulation study\nomeqref {3.32}\nompageref{49}
	\item [{$D_a$}]\begingroup diffusion coefficient of phase $a$\nomeqref {3.12}\nompageref{42}
	\item [{$D_s$}]\begingroup surface diffusion coefficient\nomeqref {4.10}\nompageref{55}
	\item [{$e$}]\begingroup electron charge\nomeqref {4.12}\nompageref{56}
	\item [{$E_t$}]\begingroup tangential component of the electric field along the island/inclusion surface\nomeqref {7.1}\nompageref{103}
	\item [{$E_x, E_y$}]\begingroup local electric field components in x and y-directions\nomeqref {7.39}\nompageref{110}
	\item [{$E_{\infty}$}]\begingroup applied electric field\nomeqref {7.2}\nompageref{103}
	\item [{$F$}]\begingroup system free energy\nomeqref {2.5}\nompageref{23}
	\item [{$f$}]\begingroup system free energy density\nomeqref {2.12}\nompageref{25}
	\item [{$f^{\textrm{dw}}$}]\begingroup double-well free energy function\nomeqref {B.1}\nompageref{179}
	\item [{$f^{\textrm{ob}}$}]\begingroup obstacle-type free energy density\nomeqref {4.2}\nompageref{53}
	\item [{$f^{\textrm{S}}$}]\begingroup surface energy density\nomeqref {3.1}\nompageref{38}
	\item [{$f^{\theta}$}]\begingroup anisotropy function\nomeqref {4.10}\nompageref{55}
	\item [{$g$}]\begingroup grand canonical potential function\nomeqref {2.2}\nompageref{23}
	\item [{$g^{\textrm{ac}}$}]\begingroup additional bulk energy density\nomeqref {C.4}\nompageref{182}
	\item [{$g^{\theta \beta}$}]\begingroup function of shape and conductivity contrast\nomeqref {8.1}\nompageref{136}
	\item [{$G_a^{\mathrm{chem}}$}]\begingroup grand-chemical potential of phase $a$\nomeqref {5.3}\nompageref{74}
	\item [{$H$}]\begingroup Hamiltonian of the lattice model\nomeqref {2.0}\nompageref{23}
	\item [{$h$}]\begingroup interpolation function\nomeqref {3.2}\nompageref{39}
	\item [{$I$}]\begingroup indicator function for the obstacle-type free energy density\nomeqref {4.3}\nompageref{54}
	\item [{$i$}]\begingroup complex number\nomeqref {7.21}\nompageref{106}
	\item [{$i,j,k \in \boldsymbol{x}$}]\begingroup spatial location of a simulation grid point in the x-, y-, and z-directions respectively\nomeqref {3.23}\nompageref{45}
	\item [{$I_{\textrm{m}}$ (or $I_{\textrm{i}}$)}]\begingroup measured (or ideal) current flow\nomeqref {E.2}\nompageref{187}
	\item [{$J_n$}]\begingroup Neumann flux boundary condition at the particle surface\nomeqref {3.16}\nompageref{43}
	\item [{$J_s$}]\begingroup surface atomic flux\nomeqref {7.1}\nompageref{102}
	\item [{$j_{\textrm{Cot}}$}]\begingroup Cottrell flux\nomeqref {6.1}\nompageref{91}
	\item [{$J_{ps}$}]\begingroup Neumann flux boundary condition at the separator\nomeqref {3.24}\nompageref{46}
	\item [{$k$}]\begingroup frequency of the sinusoidal perturbation\nomeqref {4.21}\nompageref{60}
	\item [{$k_B$}]\begingroup Boltzmann constant\nomeqref {2.2}\nompageref{23}
	\item [{$k_{\textrm{Cot}}$}]\begingroup Cottrell flux-time proportionality constant\nomeqref {6.1}\nompageref{91}
	\item [{$L$}]\begingroup reference length-scale for normalization\nomeqref {3.25}\nompageref{46}
	\item [{$M$}]\begingroup effective diffusional mobility\nomeqref {2.33}\nompageref{33}
	\item [{$m$}]\begingroup grain symmetry parameter\nomeqref {4.11}\nompageref{56}
	\item [{$M_m$}]\begingroup co-ordination number\nomeqref {2.1}\nompageref{23}
	\item [{$M_{AA}$}]\begingroup mobility of the A species, due to the interaction with the A species\nomeqref {2.30}\nompageref{32}
	\item [{$M_{AB}$}]\begingroup mobility of the A species, due to the interaction with the B species\nomeqref {2.30}\nompageref{32}
	\item [{$M_{BA}$}]\begingroup mobility of the B species, due to the interaction with the B species\nomeqref {2.30}\nompageref{32}
	\item [{$M_{BB}$}]\begingroup mobility of the B species, due to the interaction with the B species\nomeqref {2.30}\nompageref{32}
	\item [{$M_{ee}$}]\begingroup mobility of electrons due to interaction with the electrons\nomeqref {4.12}\nompageref{56}
	\item [{$M_{ei}$}]\begingroup mobility of electrons due to interaction with species $i$\nomeqref {4.12}\nompageref{56}
	\item [{$M_{ie}$}]\begingroup mobility of species $i$ due to the interaction with the electrons\nomeqref {4.6}\nompageref{55}
	\item [{$M_{ii}$}]\begingroup mobility of species $i$ due to the interaction with the species $i$\nomeqref {4.6}\nompageref{55}
	\item [{$N$}]\begingroup total number of phases\nomeqref {3.1}\nompageref{38}
	\item [{$n$}]\begingroup discrete time\nomeqref {3.25}\nompageref{47}
	\item [{$N_B$}]\begingroup the total number of specified species, occupied in the lattice sites\nomeqref {2.0}\nompageref{22}
	\item [{$N_c$}]\begingroup nucleation rate coefficient\nomeqref {6.2}\nompageref{92}
	\item [{$N_m$}]\begingroup total number of lattice sites\nomeqref {2.0}\nompageref{22}
	\item [{$N_x$, $N_y$, and $N_z$}]\begingroup number of grid points in x-, y- and z-directions\nomeqref {3.23}\nompageref{45}
	\item [{$N_{\textrm{ex}}$}]\begingroup characteristic extension\nomeqref {5.1}\nompageref{68}
	\item [{$p$}]\begingroup transitory perimeter of the inclusion\nomeqref {7.39}\nompageref{116}
	\item [{$p_0$}]\begingroup initial perimeter of the inclusion\nomeqref {7.39}\nompageref{116}
	\item [{$P_{AB}$}]\begingroup total number of A--B bonds\nomeqref {2.14}\nompageref{25}
	\item [{$R$}]\begingroup particle whose area is equivalent to the area of the circular particle of radius $R$\nomeqref {6.0}\nompageref{90}
	\item [{$r, \vartheta$}]\begingroup polar coordinates of a point\nomeqref {7.13}\nompageref{105}
	\item [{$R_0$}]\begingroup radius of the reference particle\nomeqref {5.1}\nompageref{68}
	\item [{$R_i$}]\begingroup radius of circular island/inclusion\nomeqref {7.11}\nompageref{104}
	\item [{$S$}]\begingroup displacement function\nomeqref {7.29}\nompageref{108}
	\item [{$s$}]\begingroup arc length along the island/inclusion surface\nomeqref {7.1}\nompageref{103}
	\item [{$S_\Omega$}]\begingroup surface of the domain\nomeqref {3.0}\nompageref{38}
	\item [{$T$}]\begingroup absolute temperature\nomeqref {2.2}\nompageref{23}
	\item [{$t$}]\begingroup time\nomeqref {2.27}\nompageref{32}
	\item [{$t_\mathrm {PE}$}]\begingroup time attained at the end of phase separation\nomeqref {5.3}\nompageref{76}
	\item [{$t_\mathrm {PS}$}]\begingroup time at the onset of phase separation\nomeqref {5.3}\nompageref{76}
	\item [{$T_{\textrm{ref}}$}]\begingroup reference temperature\nomeqref {2.12}\nompageref{25}
	\item [{$u$}]\begingroup half-slit width\nomeqref {7.39}\nompageref{112}
	\item [{$V$}]\begingroup velocity along the external electric field\nomeqref {7.3}\nompageref{103}
	\item [{$V_0$}]\begingroup steady-state velocity of an island/inclusion\nomeqref {7.11}\nompageref{104}
	\item [{$V_\Omega$}]\begingroup volume of the system or simulation domain\nomeqref {2.20}\nompageref{29}
	\item [{$V_n$}]\begingroup velocity along the surface normal\nomeqref {7.2}\nompageref{103}
	\item [{$w$}]\begingroup line width of the conductor\nomeqref {7.39}\nompageref{112}
	\item [{$w^{\textrm{ac}}$}]\begingroup bulk energy density potential\nomeqref {C.3}\nompageref{182}
	\item [{$W_c$}]\begingroup spatially dependent weight for Dirichlet concentration boundary condition\nomeqref {3.22}\nompageref{45}
	\item [{$W_J$}]\begingroup spatially dependent weight for Neumann flux boundary condition\nomeqref {3.22}\nompageref{45}
	\item [{$X_1$}]\begingroup first free energy density parameter\nomeqref {2.12}\nompageref{25}
	\item [{$X_2$}]\begingroup second free energy density parameter\nomeqref {2.12}\nompageref{25}
	\item [{$X_A$}]\begingroup barrier height of the obstacle-type free energy density\nomeqref {4.3}\nompageref{54}
	\item [{$Z_i$}]\begingroup valence of the diffusing metal species $i$\nomeqref {4.6}\nompageref{55}
	\item [{$Z_s$}]\begingroup effective valence\nomeqref {7.1}\nompageref{103}
	\item [{$Z_w$}]\begingroup momentum exchange effect between the electrons and the diffusing species\nomeqref {4.6}\nompageref{55}
	\item [{$Z_{gc}$}]\begingroup grand canonical partition function\nomeqref {2.1}\nompageref{23}
	\item [{${\boldsymbol{\eta}_{i,a b}}$}]\begingroup position vectors in the Wulff shape\nomeqref {C.3}\nompageref{182}
	\item [{${c}_{ps}$}]\begingroup Dirichlet concentration boundary condition at the separator\nomeqref {3.23}\nompageref{45}
	\item [{${f}_{a}^{\mathrm{chem}}$}]\begingroup regular solution free energy density of phase $a$\nomeqref {3.2}\nompageref{39}
	\item [{2D}]\begingroup two-dimensional\nomeqref {5.2}\nompageref{71}
	\item [{3D}]\begingroup three-dimensional\nomeqref {5.2}\nompageref{71}
	\item [{EM}]\begingroup electromigration\nomeqref {1.2}\nompageref{12}
	\item [{GB}]\begingroup grain boundaries\nomeqref {1.2}\nompageref{13}
	\item [{IMC}]\begingroup intermetallic compound\nomeqref {7.39}\nompageref{115}
	\item [{JMA}]\begingroup Johnson-Mehl-Avrami\nomeqref {6.0}\nompageref{90}
	\item [{LFP}]\begingroup lithium iron phosphate, LiFePO$_4$\nomeqref {3.11}\nompageref{41}
	\item [{LMO}]\begingroup lithium manganese oxide, LiMn$_2$O$_4$\nomeqref {5.2}\nompageref{70}
	\item [{LTO}]\begingroup lithium titanate, Li$_4$Ti$_5$O$_{12}$\nomeqref {6.0}\nompageref{85}
	\item [{MPI}]\begingroup message passing interface\nomeqref {5.2}\nompageref{69}
	\item [{PACE3D}]\begingroup Parallel Algorithms for Crystal Evolution in 3D\nomeqref {4.23}\nompageref{61}
	\item [{PFM}]\begingroup phase-field model\nomeqref {6.0}\nompageref{90}
	\item [{PITT}]\begingroup Potentiostatic Intermittent Titration Technique\nomeqref {6.1}\nompageref{91}
	\item [{REV}]\begingroup representative elementary volume\nomeqref {3.1}\nompageref{38}
	\item [{SOC}]\begingroup state of charge\nomeqref {5.2}\nompageref{71}
	\item [{SOC$_\mathrm {PE}$}]\begingroup state of charge attained before the end of phase separation\nomeqref {5.3}\nompageref{76}
	\item [{SOC$_\mathrm{PS}$}]\begingroup state of charge at the onset of phase separation\nomeqref {5.2}\nompageref{73}
	
\end{thenomenclature}

\clearpage
\afterpage{\blankpagewithoutnumberskip}
\clearpage

\makeatletter
     \renewcommand*\l@figure{\@dottedtocline{0}{0em}{3.2em}}
\makeatother
\pagestyle{fancy}
\fancyhf{}
\lhead[\thepage]{List of figures}      
\rhead[\thesection]{\thepage}
\cleardoublepage\phantomsection\addcontentsline{toc}{chapter}{List of Figures}
\listoffigures
\def\pagedeclaration#1{, \dotfill\hyperlink{page.#1}{#1}}
\clearpage
\afterpage{\blankpagewithoutnumberskip}

\pagestyle{fancy}
\fancyhf{}
\lhead[\thepage]{List of tables}      
\rhead[\thesection]{\thepage}
\cleardoublepage\phantomsection\addcontentsline{toc}{chapter}{List of Tables}
\listoftables
\clearpage
\afterpage{\blankpagewithoutnumberskip}
\clearpage

\pagestyle{fancy}
\fancyhf{}
\lhead[\thepage]{Bibliography}      
\rhead[\thesection]{\thepage}
\bibliographystyle{Bst/unsrt}
\cleardoublepage\phantomsection

\clearpage


\end{document}